\documentclass[structabstract]{aa}  
\usepackage{graphicx}
\usepackage[varg]{txfonts}
\usepackage{lscape}
\usepackage{txfonts}
\usepackage{caption}
\usepackage{color}
\usepackage{amsmath}
\usepackage{grffile}
\usepackage{afterpage}
\usepackage{natbib}

\newcommand{\hii}{H\,{\scriptsize II}}
\newcommand{\uchii}{UC-H\,{\scriptsize II}}

\newcommand{\kms}{km s$^{-1}$}

\newcommand{\at}{{ATLASGAL}}

\newcommand{\mum}{$\mu$m}
\newcommand{\msol}{M$_{\odot}$}
\newcommand{\lsol}{L$_{\odot}$}

\newcommand{\nwiseflux}{75}
\newcommand{\nwisemipsoverlap}{21}
\newcommand{\nwisemipsoverlapl}{19}

\newcommand{\nmipsnotdet}{27}
\newcommand{\nirq}{80}

\begin{document}

   \title{The ATLASGAL survey: The sample of young massive cluster progenitors}


   \author{T. Csengeri
          \inst{1}
         \and
           S. Bontemps
          \inst{2}
         \and
           F. Wyrowski
          \inst{1}
          \and
           S. T. Megeath
          \inst{3}
          F. Motte
          \inst{4,5}
          \and
           A. Sanna
         \inst{1}
         \and
          M. Wienen
         \inst{1}
         \and
         K. M. Menten
       \inst{1}
          }

   \institute{Max Planck Institute for Radioastronomy,
              Auf dem H\"ugel 69, 53121 Bonn, Germany
              \email{ctimea@mpifr-bonn.mpg.de}
        \and
           OASU/LAB-UMR5804, CNRS, Universit\'e Bordeaux 1, 33270 Floirac, France    
        \and
          Ritter Astrophysical Research Center, Department of Physics and Astronomy, University of Toledo, 2801 W. Bancroft St., Toledo, OH 43606, USA      
        \and
         Institut de Plan\' etologie et d'Astrophysique de Grenoble, Univ. Grenoble Alpes -- CNRS-INSU, BP 53, 38041 Grenoble Cedex 9, France
        \and
       Laboratoire AIM Paris Saclay, CEA-INSU/CNRS-Universit\'e Paris Diderot, IRFU/SAp CEA-Saclay, 91191 Gif-sur-Yvette, France
             }

   \date{Received ; accepted }

 
  \abstract
   {The progenitors of high-mass stars and clusters are still challenging to recognise. 
     Only unbiased surveys,
     sensitive to compact regions of high dust column density, can unambiguously reveal such a small population 
    of particularly massive and cold clumps.}
     { Here we use the 
   \at\ survey to identify a sample of candidate progenitors of massive clusters in the inner Galaxy. }{We characterise a flux limited sample of compact sources selected from the \at\ survey. Sensitive mid-infrared data at $21-24$~\mum\ from the WISE and MIPSGAL surveys were explored to search for embedded objects, and complementary spectroscopic data were used to investigate their stability and their star formation activity. 
}{We identify an unbiased sample of  infrared-quiet massive clumps in the Galaxy that potentially represent the earliest stages of massive cluster formation. An important fraction of this sample consists of 
  sources that have not been studied in detail before. We first find that clumps hosting more evolved embedded objects and infrared-quiet clumps exhibit similar physical properties in terms of mass and size, suggesting that the sources are not only capable of  forming high-mass stars, but likely also follow a single evolutionary track leading to the formation of massive clusters. The majority of the clumps are likely not in virial-equilibrium, suggesting collapse on the clump scale.   
}{We identify the precursors of the most massive clusters in the Galaxy within our completeness limit, and argue that these objects are undergoing large-scale collapse. This is in line with the  low number of infrared-quiet massive clumps and earlier findings that     star formation, in particular for high-mass objects is a fast, dynamic process.     We propose a scenario in which massive clumps start to fragment and collapse    before their final mass is accumulated indicating that strong self-gravity and global collapse    is needed to build up rich clusters and the most massive stars.}{}


   \maketitle
%

\section{Introduction}

The formation of massive stars and  clusters remains an enigma in modern astrophysics.
To explain the origin of high-mass stars, two theoretical scenarios
are commonly invoked: the turbulent-core model \citep{MT03}
and a more dynamic view with cluster formation  \citep{Bonnell2001,BonnellBate2006}.
The observational challenge to these theories is  a systematic, Galaxy-wide
search for the earliest phases of high-mass stars and massive
protoclusters where the initial conditions can  best be studied. 

With {\it earliest phases} we refer to the starless/pre-stellar, and the subsequent phase
where the submillimetre emission is still dominated by the massive core.  This implies that no luminous
massive young stellar object (MYSO) has yet emerged, and no
ionising emission from embedded {\uchii} regions.
This phase is difficult to observe, however, mainly because of
the {short duration of less than $7.5\times10^4$\,years} at the scale of $\sim$0.3\,pc (\citealp{Csengeri2014}),
and also because of  the high complexity of active sites of high-mass star formation. 
Only a few candidates of massive starless cores have been 
claimed so far in the literature with masses up to 60\,\msol\ in a size of $\simeq$0.1\,pc \citep{Tan2013,Cyganowski2014},
although towards the former object outflow activity has recently been discovered \citep{Tan2016}, 
and the absence of any molecular emission towards the latter object casts doubts on its Galactic origin.
Confirmed examples of massive pre-stellar cores, 
as well as exploration of the even higher mass regime, are of extreme importance in order to constrain star formation models.

Early searches for the birthplaces of high-mass stars
 focused on IRAS selected  infrared-bright sources, where low- and
high-luminosity sources were defined \citep{Molinari1996}.
High-luminosity sources were associated with {\uchii} regions, i.e.
the final stage of high-mass star formation, while the lower luminosity sources
were associated with MYSOs.  {\uchii}
regions have typical bolometric luminosities of $> 10^4$ L$_{\odot}$, while (M)YSOs
show a broad distribution peaking between 
$3\times10^3- 2\times10^4$~L$_{\odot}$ (e.g.\,\citealp{Molinari2000,Mottram2011b}). 
In  fact, {\uchii} regions and lower luminosity sources fill a distinct space
in a mass-luminosity diagram, strongly suggesting an evolutionary
trend of more luminous objects with similar masses being more evolved
than their lower luminosity counterparts (see\,\citealp{Molinari1996,Sridharan2002, Elia2013}).

Only with the advance of sensitive and large format detector arrays have (sub)millimetre surveys of entire molecular clouds 
and larger areas of the Galaxy revealed lower luminosity objects 
associated with very massive gas reservoirs (e.g.\,\citealp{Motte2003,Mookerjea2004,M07,Lopez2011,Russeil2010,Tackenberg2012,Traficante2015,Svoboda2016}). Systematic searches for
the massive but colder and lower luminosity sources have begun. In an unbiased
survey, \citet{M07} has identified
a sample of massive dense cores (MDCs) with masses between $40-200$\,\msol, sizes of $\sim$0.1\,pc  and low luminosities ($<10^3$~L$_{\odot}$) in the
Cygnus-X region. Their high concentration of mass and  low bolometric luminosity suggest that they host
high-mass protostars, potential precursors to MYSOs. 
High angular-resolution observations of \citet{Bontemps2010} indeed
identified a handful of individual
high-mass protostars within MDCs with envelope masses in the range of $8-50$\,{\msol}.
Studies of the outflow activity further suggest that these protostars are the high-mass equivalents
of Class~0 protostars \citep{Duarte2013}.
Altogether, this is strong evidence that low luminosity MDCs
host objects in the earliest stages of high-mass protostellar evolution. 
An unbiased, systematic study
to identify such objects on Galactic scales is the necessary next step
for studying the initial conditions for high-mass star and cluster formation.

In this paper we present the first systematic and unbiased study of the brightest submillimetre 
sources in the inner Galaxy selected from the 870 $\mu$m APEX Telescope LArge Survey of the GALaxy (ATLASGAL, \citealp{Schuller2009, Csengeri2014}); we estimate their physical 
properties and search for the most massive clumps still in an early evolutionary stage among them. 
This allows us to identify and study the properties of the youngest precursors of massive clusters in the Galaxy
which potentially host high-mass protostars. 
Various samples have been studied from the \at\ survey,
mostly based on associations with {\hii} regions and sources from the Red MSX Survey (RMS, \citealp{Lumsden2013}), and their flux densities at 8\,$\mu$m (GLIMPSE, \citealp{Benjamin2003}), and 24\,$\mu$m (MIPSGAL, \citealp{Carey2009}). 
Focusing on the $I^{st}$ Galactic quadrant, \citet{Csengeri2016a} present the largest sample to date of 430 clumps in different evolutionary stages over a wide mass range. 
The sample studied in \citet{Giannetti2014} represents a subsample of this source selection limited to the hundred brightest sources in different evolutionary stages in both the $I^{st}$ and the $IV^{th}$ Galactic quadrant. This strategic selection of ATLASGAL clumps over various evolutionary stages has been targeted by extensive spectroscopic follow-up observations; see details in \citet{Csengeri2016a}.
Other samples using selective star formation tracers of more evolved objects, such as clumps with {\hii} regions, RMS sources, and associations with masers from the Methanol MultiBeam survey, have been discussed in a series of papers by \citet{Urquhart2013a,Urquhart2013b,Urquhart2014}. 

The paper is organised as follows. 
In Sect.\,\ref{sec:obs} we describe the flux limited sample of the brightest submillimetre
sources from the \at\ survey and investigate
their statistical properties in Sect.\,\ref{sec:statistics}. 
Using complementary 21-24~\mum\ data, we identify
a sample of massive clumps which are potentially the youngest massive objects 
known so far. In Sect.\,\ref{sec:form} we study their physical properties
and discuss their potential to form high-mass stars.
%

\section{Brightest submillimetre clumps from the \at\ survey}\label{sec:obs}

The \at\ survey \citep{Schuller2009} imaged the Galactic plane between Galactic longitude 
$\ell=\pm60^\circ$
and Galactic latitude $b=\pm1.5^\circ$ at 870~$\mu$m with the 
LABOCA camera \citep{siringo2009} on the APEX Telescope \citep{gusten2006}. 
In an extension of the survey, the $-80^\circ < \ell < -60^\circ$ and $-2^\circ < b < +1^\circ$
region was also  mapped.
Thus, all together the \at\ survey imaged
$\sim$420 sq. deg. of the Galactic plane at a 19.2{\arcsec} spatial resolution.
The catalogue  of compact sources is 97\% complete down to $\sim260$~\msol\ at the distance of the Galactic  centre, i.e., 8 kpc, and 
1~\msol\ at 500 pc for cold clumps \citep{Csengeri2014}.

We study here a flux limited sample of the \at\ compact sources 
within the highest sensitivity part of the survey, corresponding
to a Galactic longitude range of $|\ell|\leq60^\circ$. 
The initial sample was selected using a 5\,Jy beam-averaged flux density threshold,
which translates to $\sim$650\,\msol\ (see \citealp{Csengeri2014}) 
at an average distance of \at\ sources of $d\simeq$\,4.5\,kpc \citep{Wienen2015}.
We estimate the corresponding column density by 
\begin{equation}\label{eq:nh2}
N({\rm H_2})=\frac{F_\nu\,R}{B_\nu(T_d)\,\Omega\,\kappa_{\nu}\,\mu_{\rm H_2}\,m_{\rm H}}
\rm{[cm^{-2}]
},\end{equation}
where {$F_\nu$ is the beam-averaged peak flux density}; $\Omega$ is the solid angle
of the beam calculated by $\Omega = 1.13\times \Theta^2$, where 
$\Theta$ is 19\rlap{.}{''}2; 
 $\mu_{\rm H_2}$ is 
 the mean molecular weight per 
hydrogen molecule and is equal to 2.8 \citep{Kauffmann2008};
 $m_{\rm H}$ is the mass of a hydrogen atom; and $R$ corresponds to the gas-to-dust ratio of 100. 
Our flux density threshold thus corresponds to a column density (${N_{\rm H_2}}$) 
of $4.9-40.8\times10^{23}$~cm$^{-2}$ using  
a centre frequency of 345~GHz at our resolution of 19\rlap{.}{''}2,  
and dust temperatures of $10-40$~K.
For the highest column density sources this translates to an extremely high visual extinction of 
$A_{\rm V}>{521\,mag}$ \footnote{For the conversion we use 
$A_{\rm V}=N_{\rm H_2}/0.94\times10^{21}$~cm$^{-2}$~\citep{Bohlin1978}.}
(see details in Sect.\,\ref{sec:mass}).

From the \at\ compact source catalogue 
we find a total of  219 sources above the 
5~Jy beam-averaged flux density threshold \citep{Csengeri2014}. 
In the subsequent steps we first provide distance estimates 
(Sect.\,\ref{sec:distance}), 
and then determine their physical properties 
(Sect.\,\ref{sec:temperatures} and \ref{sec:mass}). 
Using mid-infrared data between 21--24~\mum\ from 
the WISE point source catalogue~\citep{Wright2010} and photometry based on
the MIPSGAL survey \citep{Carey2009} and complemented with classification
from the MYSO and \uchii\ regions from the RMS \citep{Urquhart2007} and CORNISH surveys \citep{Hoare2012,Purcell2013}, 
we then estimate the evolutionary stage based on the luminosity of the embedded
protostars associated with the dust peaks (Sects.\,\ref{sec:ev} and \ref{sec:ev2}).
This leads to the identification of infrared-quiet massive clumps in Sect.\,\ref{sec:definition}.
In Sect.\,\ref{sec:sio87} we present molecular line data to characterise their 
associations with shocks from outflow activity to probe ongoing star formation.

\subsection{Distances and association with massive molecular complexes}\label{sec:distance}

The large statistics of the \at\ survey revealed that
most of the compact sources, and correspondingly most of the 
star formation activity is associated with giant molecular
complexes (GMCs)~(e.g.\,\citealp{Csengeri2014}).
As a first step, therefore, we associated the sample with known GMCs 
(see Table\,\ref{tab:complexes}). 
In fact, our selection corresponds to the brightest submillimetre sources 
in the inner Galactic plane, and it covers 
well-characterised massive star-forming 
complexes such as Sgr B2, W51 Main, W49A, W33, and W43. 
Other examples are found at the edge of expanding bubbles like
G348.1825+0.4829, which is the brightest compact source in the vicinity of 
RCW 120 \citep{Zavagno2010}.
We find a few examples of relatively isolated sources, which are not associated
with any known GMCs, but smaller molecular clouds,  such as 
G0.5464-0.8521, G333.0164+0.7654, and G43.7950-0.1270, towards which
our distance estimates may be less reliable than for the majority of the sample. 

We collected distance estimates from the literature using primarily 
maser parallax measurements (see \citealp{Reid2014} for an overview) 
or spectrophotometric distance estimates (e.g.\,\citealp{Russeil2003,Moises2011})
for the known GMCs. For the remaining unassociated sources
we used the high-density tracer NH$_3$ to determine kinematic distances from
\citet{Wienen2012,Wienen2015}. In short, the rest velocity of the source is compared to the Galactic rotation curve using the parameters derived by 
\citet{Reid2009} for the I${st}$ and \citet{BB1993} for the IV${th}$ Galactic quadrants, respectively. 
We attempt to solve the distance ambiguity by making use of 
H{\scriptsize I} emission and 21\,cm continuum absorption against bright
{\hii} regions (see \citealp{Wienen2015} for details). We then visually inspected
near-IR extinction maps \citep{Schneider2010} and data from the Galactic
Ring Survey (GRS, \citealp{Jackson2006}) to
look for spatial and kinematic associations with nearby molecular clouds 
also considering the larger scale
Galactic structure such as spiral arms, and associations with 
extinction features as well.
We also inspected sensitive mid-infrared data from
GLIMPSE and MIPSGAL
to look for possible associations with infrared-dark dust lanes appearing
in high contrast absorption at mid-infrared wavelengths.

\begin{figure*}[!htpb]
\centering
\includegraphics[width=0.8\linewidth, angle=0]{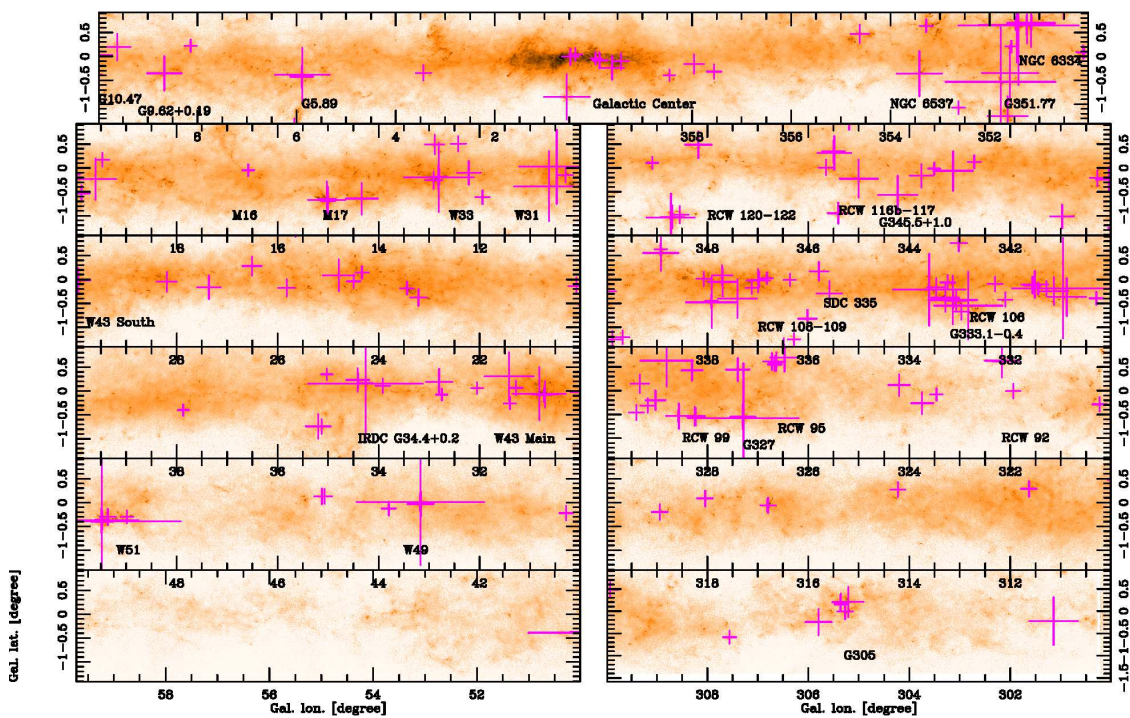}
\caption{Overview of the Galactic distribution of the sample and the associated molecular clouds. 
Figure adopted from \citet{Csengeri2016b}
showing the ATLASGAL-Planck combined view of the inner galaxy. Magenta crosses show the location of the sources.
The size of the symbols scales linearly with the peak flux density.} 
\label{fig:overview}
\end{figure*}

In total we associate 78 (36\%) sources with
available maser parallax measurements.
In the I${st}$  Galactic quadrant, such 
measurements are available for 59 out of 85 sources (70\%), providing a robust 
estimate of their physical parameters. Sources in the IV${th}$ Galactic quadrant
have less reliable distance measurements owing to the lack of currently available
maser parallax distances, except for G339.8841-1.2568 \citep{KRISHNAN2015}. 
We were able to provide distance estimates to all but one source in the sample 
(see Table~\ref{tab:table-dust}). For simplicity, in the following analysis we exclude the 9 sources associated with the Galactic centre, leaving a total of 210 sources in our sample. An overview of their Galactic distribution is shown in Fig.\,\ref{fig:overview}.

\begin{table}
\caption{Association of the brightest {ATLASGAL} sources with GMCs.}\label{tab:complexes}
\begin{tabular}{lllrrrrrrrrr}
\hline\hline
Associated GMC or & $d$ & number  \\
molecular cloud &  [kpc] & of sources\\
\hline
W51 & 5.41 \tablefootmark{b} & 8   \\
W49 & 11.11  \tablefootmark{c} & 3  \\
IRDC G34.4+0.2 & 1.56\tablefootmark{p} & 3  \\
W43  & 5.49\tablefootmark{d} & 4 &  \\
W31  &  4.95\tablefootmark{f} & 3  \\
W33  & 2.4\tablefootmark{e} & 3  \\
M17          & 1.98 \tablefootmark{m} & 4   \\
G5.89  & 1.28\tablefootmark{r} & 2   \\
Galactic centre & 8.5\tablefootmark{j}  & 9 \\
NGC~6334 &  1.35 \tablefootmark{i} & 9  \\ 
RCW~122  & 1.8\tablefootmark{i,j} & 4 &  \\ 
RCW~108-109& 1.8\tablefootmark{a,l} &  3  &  \\ 
G333.1-0.4  & 4.2\tablefootmark{a,g} & 6  \\
RCW106  & 4.2\tablefootmark{a} & 5  \\ 
G327  & 2.5\tablefootmark{a} & 3  \\ 
RCW~95 &  2.5 \tablefootmark{a,g,\textdagger} &5 & \\ 
RCW~92 & 3.2\tablefootmark{a} & 2  \\ 
G305 & 3.5\tablefootmark{q} & 2 \\
G10.47 & 8.47\tablefootmark{f} & 1 \\
RCW~120  & 1.3\tablefootmark{a} & 1  \\
RCW~131/NGC 6537 &  1.7\tablefootmark{h} & 1   \\
G345.5+1.0 & 1.8\tablefootmark{a,s} & 1 \\
RCW~116b& 1.8\tablefootmark{a,g} & 1   \\
RCW 117 &   1.8\tablefootmark{a,k} & 1   \\
RCW~99 & 2.5\tablefootmark{a} & 1 \\
G9.62+0.19 & 5.15\tablefootmark{n} & 1 \\
G351.77 & 1.0\tablefootmark{o} & 1 \\
SDC335 & 3.8\tablefootmark{a,$\ddagger$} & 1   \\
\hline
\end{tabular}
 \tablefoot{ 
\tablefoottext{a}{This work.}
\tablefoottext{b}{\citet{Sato}}
\tablefoottext{c}{\citet{Zhang2013}}
\tablefoottext{d}{\citet{Zhang2014}}
\tablefoottext{e}{\citet{Immer}}
\tablefoottext{f}{\citet{Sanna2014}}
\tablefoottext{g}{\citet{Moises2011}}
\tablefoottext{\textdagger}{We revised the spectrophotometric distance of 1.81\,kpc by \citet{Moises2011}.}
\tablefoottext{h}{\citet{Wu2014}}
\tablefoottext{i}{\citet{Wu2012}}
\tablefoottext{j}{\citet{Reid2014}}
\tablefoottext{k}{\citet{Dutra2003}}
\tablefoottext{l}{\citet{Straw1987}}
\tablefoottext{m}{\citet{Xu2011}}
\tablefoottext{n}{\citet{Sanna2009}}
\tablefoottext{o}{\citet{Leurini2011}}
\tablefoottext{p}{\citet{Kurayama2012}}
\tablefoottext{q}{\citet{Russeil2003}}
\tablefoottext{r}{\citet{Motogi}}
\tablefoottext{s}{\citet{Lopez2011}}
\tablefoottext{$\ddagger$}{\citet{Peretto2013} adopts 3.25\,kpc.}
}
\end{table}

\subsection{Temperature estimates}\label{sec:temperatures}

Our sample includes clumps in various evolutionary stages
such as active star-forming sites with clusters of 
known (UC-){\hii} regions (e.g.\,W51 Main, W49 N, G5.89)
 and typical hot molecular cores (HMCs) (e.g.\,G10.47+0.3, G34.26+0.15) associated with infrared-bright MYSOs.
Over such a large star formation activity, the estimated physical parameters from the 870\,$\mu$m dust emission 
can be highly sensitive to the used dust temperature.
As a simple first-order approximation, a single dust temperature is commonly used for which a
$T_{\rm d}$ of 18~K is a reasonable choice (see e.g.\,\citealp{Csengeri2014}) 
based on the {\sl Herschel} measured 
dust temperatures dominated by the interstellar radiation-field
\citep{Bernard2010}. 
However,
the brightest submillimetre sources are more likely to host 
embedded heating source(s), which could imply higher dust temperatures. 
This is supported by statistical studies; for instance, 
\citet{Battersby2011} show that tracers of star formation are more likely associated with 
higher dust temperatures, while 
 \citet{Konig2016} show an increase in dust temperature with evolutionary stage by fitting the spectral energy distribution
of 
ATLASGAL clumps over a range of evolutionary stages.  
Hence, 
a temperature correction is necessary
for clumps hosting {\uchii} regions 
and infrared-bright protostars, and we discuss below our choice of a higher dust temperature. 
 
At the high densities probed by the \at\ survey it is reasonable to assume
that the gas and dust are thermalised. 
Therefore, to investigate which dust temperature is appropriate for the warmer clumps, we can also rely on complementary molecular line data available in the literature. The 
NH$_3$ measurements by 
\citet{Wienen2012} give a gas kinetic temperature  
between $12-35$~K for a large sample of \at\ sources within a beam of $\sim$40\arcsec.
Similarly, \citet{Dunham2011} derive a mean gas kinetic temperature of 17~K
for a sample of 199 sources from the Bolocam Galactic Plane 
Survey (BGPS) with a beam of 31\arcsec. This corresponds well to the average dust temperature
also used in various studies of ATLASGAL selected sources (e.g.\,\citealt{Csengeri2016a}). 

A total of  127 sources (corresponding to 58\% of the sample) 
have kinetic temperature estimates from \citet{Wienen2012} and 
\citet[submitted]{Wienen2016}, and the bulk of these temperature measurements 
range between $17$ and $35$\,K. 
Since these measurements do not cover the full sample, 
they only allow us to estimate  a reasonable 
temperature correction.
Furthermore, as these studies mostly rely on the (1,1) to (2,2) inversion transitions of
NH$_3$, they are more sensitive to the cold gas rather than the hot gas component. 
Therefore, for the more evolved objects these temperatures should be considered as lower limits on the dust temperature, and thus an upper limit on the mass.
 When discussing the statistical properties of the sample (Sect.\,\ref{sec:statistics}),
 as a second-order approximation, we account for the increasing temperature towards sources with embedded massive stars using a higher $T_{\rm d}$ of 30~K. 
This is consistent with the  typical values used so far of 15-20\,K  for cold sources and up to 30-40\,K 
for warm sources (e.g.\,\citealp{M07,Russeil2010}).

\subsection{Estimating physical parameters: mass, size, density, and surface density}\label{sec:mass}

We use the same formula as described in \citet{Schuller2009} and \citet{Csengeri2014} 
to estimate the gas mass from the optically thin 870~\mum\ dust emission 
\begin{multline}
     M_{\rm gas} = \frac{S_\nu\,R\,d^2}{B_\nu(T_d)\,\kappa_{\nu}} \simeq \\
     \frac{M_{\rm gas}}{[M_{\odot}]} = C\times \frac{S_{\nu}}{[\rm Jy]}\times~\Big(\frac{d}{\rm [kpc]}\Big)^2, 
\end{multline}
where {$S_\nu$ is the integrated flux} and $d$ corresponds to the distance of the sources. 
The numerical constant, $C_{\rm 18\,K}=6.33$ for $\nu=345$\,GHz, $T_{\rm dust}=18$\,K, and
$\kappa_\nu = 0.0185$\,cm$^{2}$\,g$^{-1}$ from \citet{OH1994} accounting for a gas-to-dust ratio (R) of 100. We note that recent studies suggest a higher gas-to-dust ratio of 150 \citep{Draine2011}, which would directly increase our mass estimates by 50\%. 
 \citet{Peretto2013} adopt a 50\% lower $\kappa_\nu$, which would also lead to an increase in the estimated mass.
Following the analysis in 
Sect.\,\ref{sec:temperatures}, we introduce a temperature corrected mass estimate ($M_{\rm corr}$) for the more evolved and thus likely warmer objects adopting $T_{\rm dust}=30$\,K, corresponding to $C_{\rm 30K}=3.08$ from Sect.\,\ref{sec:statistics}. 

An additional factor of uncertainty in the mass determination
originates from optically thick free-free emission which may contribute to the
870~\mum\ flux density towards the clumps hosting {\uchii} regions. 
For a uniform density 
{\hii} region this contribution can be negligible because the thermal dust
emission dominates the spectral energy distribution \citep{Hunter2000}. 
However, for a non-uniform density source with high temperature and density, a more 
significant contribution from free-free emission is possible~(e.g.\,\citealp{Beuther2004,Keto2002,Keto2003}). This
may lead to an overestimate of the mass. 
\citet{Schuller2009}, however, estimate
this contribution to be considerably lower than the thermal emission from dust
at 345~GHz, not more than 20\% for the most extreme cases.
In addition, the already formed ionising sources are 
still compact and therefore are expected
to have a minor influence on the mass estimates.
To conclude, for sources
associated with known embedded {\uchii} regions we conservatively 
consider 
the derived masses as upper limits (Table~\ref{tab:table-dust}).

In \citet{Csengeri2014} the ATLASGAL sources have been fitted with a 2D Gaussian, from 
which we estimate the source size following the method of \citet{Bontemps2010}. We
define the source radius ($R_{90}$) which contains 90\% of the total mass, assuming a 2D Gaussian profile, by $R_{90} = 1.95\times FWHM_{\rm deconvolved}/\sqrt{8\,\rm ln (2)}$, where $FWHM_{\rm deconvolved}$ corresponds to the deconvolved source size\footnote{We use 19\rlap{.}{''}2 for the resolution of the \at\ maps.}. From the mass and size, we use the
following equations to calculate the mean volume density and surface
density:

\begin{equation}
   \frac{\bar{n}}{[{\rm cm^{-3}}]} = \frac{M_{\rm gas}}{\frac{4}{3}\,R_{90}^3\,\pi}
   \end{equation}

\begin{equation}
  \frac{\Sigma}{[\rm g~cm^{-2}]} = \frac{M_{\rm gas}}{R_{90}^2\,\pi} 
  .\end{equation}
The estimated physical parameters are listed in Table\,\ref{tab:table-dust}.

\subsection{Mid-infrared tracers of embedded protostars}\label{sec:ev}

Here we investigate the properties of the embedded YSOs  
closely associated with the dust peaks in order to reveal their star formation activity
and, in particular, the youngest clumps within
the sample. 
To classify an embedded source as
a high-mass young stellar object, 
we rely on the IRAS-based  colour properties for embedded high-mass stars from \citet{Wood1989}
following the method of
\citet{M07} and \citet{Russeil2010}. 
Based on stellar models we 
adopt a luminosity threshold corresponding to 
objects with $L_{\rm bol} >10^4\, \rm L_{\odot}$, which potentially
host a $>15$~\msol\ zero age main sequence (ZAMS)  star
with a spectral type earlier than B0.5 (see also \citealp{Mottram2011b}).
In practice, using the expression from \citet{M07}, we translate this luminosity to a flux density threshold of
$\sim$289\,Jy at 21\,\mum\ flux for an object at 1\,kpc. This limit is a factor of ten higher 
than  was used for Cygnus-X (\citealp{M07}) and NGC6334 \citep{Russeil2010}, 
and we evaluate the implication of the higher luminosity limit in Sect.\,\ref{sec:ev2}.
We extrapolate between the $22$\,\mum\  and $24$\,\mum\ data 
assuming blackbody radiation
for the embedded protostar following \citet{Russeil2010}.

Below this luminosity threshold we find high-mass protostars or their potential precursors,
which are still deeply embedded in their mass reservoir, while
their energetics is dominated
by accretion processes. 
Although theoretical models are strongly dependent on the 
accretion rates, they predict Class~0-like high-mass protostars
up to $L_{\rm bol}<10^4\, \rm L_{\odot}$ with an envelope mass
of $>50$~\msol. Our threshold
is also consistent with evolutionary tracks
of high-mass stars assuming constant or varying accretion
with rates of $10^{-4}-10^{-3}$\,\msol\,yr$^{-1}$\,\citep{Duarte2013,Kuiper2013}.
In fact, the potentially highest mass protostar in Cygnus-X, N63,
has an envelope mass of $\sim44$\,\msol\ and $L_{\rm bol}\sim339\,\rm L_{\odot}$ estimated from
SED models using Herschel data \citep{Duarte2013}.

To extract the mid-infrared flux density of the sample, we first use the 
most sensitive and highest angular-resolution data from
the MIPSGAL survey \citep{Carey2009}. 
The MIPS detector saturates at 2\,Jy for point sources, and
confusion due to nearby point sources or bright extended emission 
may hinder the photometry at the position of the dust peaks. 
Considering only the point source saturation limit,
the MIPSGAL survey reveals B3 type (or earlier) stars only beyond 3.8\,kpc. 
Less luminous but closer objects are likely to appear saturated in the MIPS images. Therefore, in order to cover 
a continuous sensitivity range it is necessary to complement the MIPSGAL data with data from the WISE satellite. The PSF photometry of the WISE point source catalogue is sensitive up to a $\sim$330\,Jy
flux density threshold \citep{Cutri2012}.

The MIPSGAL survey at 24~\mum\ has a 
3$\sigma$ point source sensitivity of 2 mJy at a 6\arcsec\
angular resolution, while the WISE survey provides a 5$\sigma$ point source sensitivity
of 6~mJy at a 12\arcsec\ resolution at 22~{\mum} in unconfused regions of the Galaxy. 
Owing to extended emission from nebulosity and polycyclic aromatic hydrocarbons (PAHs),
these sensitivity levels may be
higher towards confused regions. 
The factor of 4 difference in beam areas may also result in different flux densities determined
for the same source from the MIPSGAL and WISE data. In Appendix\,\ref{app:mips_photometry}
we discuss the cross-calibration between the MIPS and WISE photometry.

Although ultimately the sensitivity threshold is only a factor of few better  for the MIPSGAL
survey, its  angular-resolution (6\arcsec), which is more than three times higher, is a key benefit in 
our study. 
This angular resolution corresponds to a physical scale of $\sim$0.14\,pc
at a 4.5\,kpc distance, which is the size scale of massive dense cores embedded in 
massive clumps \citep{M07,Zhang2009,Csengeri2011b}. This angular resolution allows us
to resolve and distinguish between infrared-bright and -quiet massive dense cores for objects
located at distances closer than $\sim$4.5\,kpc.

All sources were therefore first visually inspected using the publicly available calibrated data tiles
from the MIPSGAL survey, and where no point source was found 
within 6\arcsec\  of the \at\ peak position, we extracted 
an upper limit on the flux density (for more details see Appendix\,\ref{sec:mipsu}).
Aperture photometry was performed on
point sources where no saturated pixels are present (Appendix\,\ref{sec:mips}). Where the
MIPSGAL data was not useable, we used the flux density from the 
WISE point source catalogue \citet{Wright2010} (Appendix\,\ref{sec:crosscal}) \footnote{The WISE 22~$\mu$m flux density is calculated from the W4 band
magnitudes by using the formula F$_{\rm 22\mu\,m}=8.2839\times10^{mag_{W4}/-2.5}$.}.

\begin{figure}
\centering
  \includegraphics[width=9.0cm]{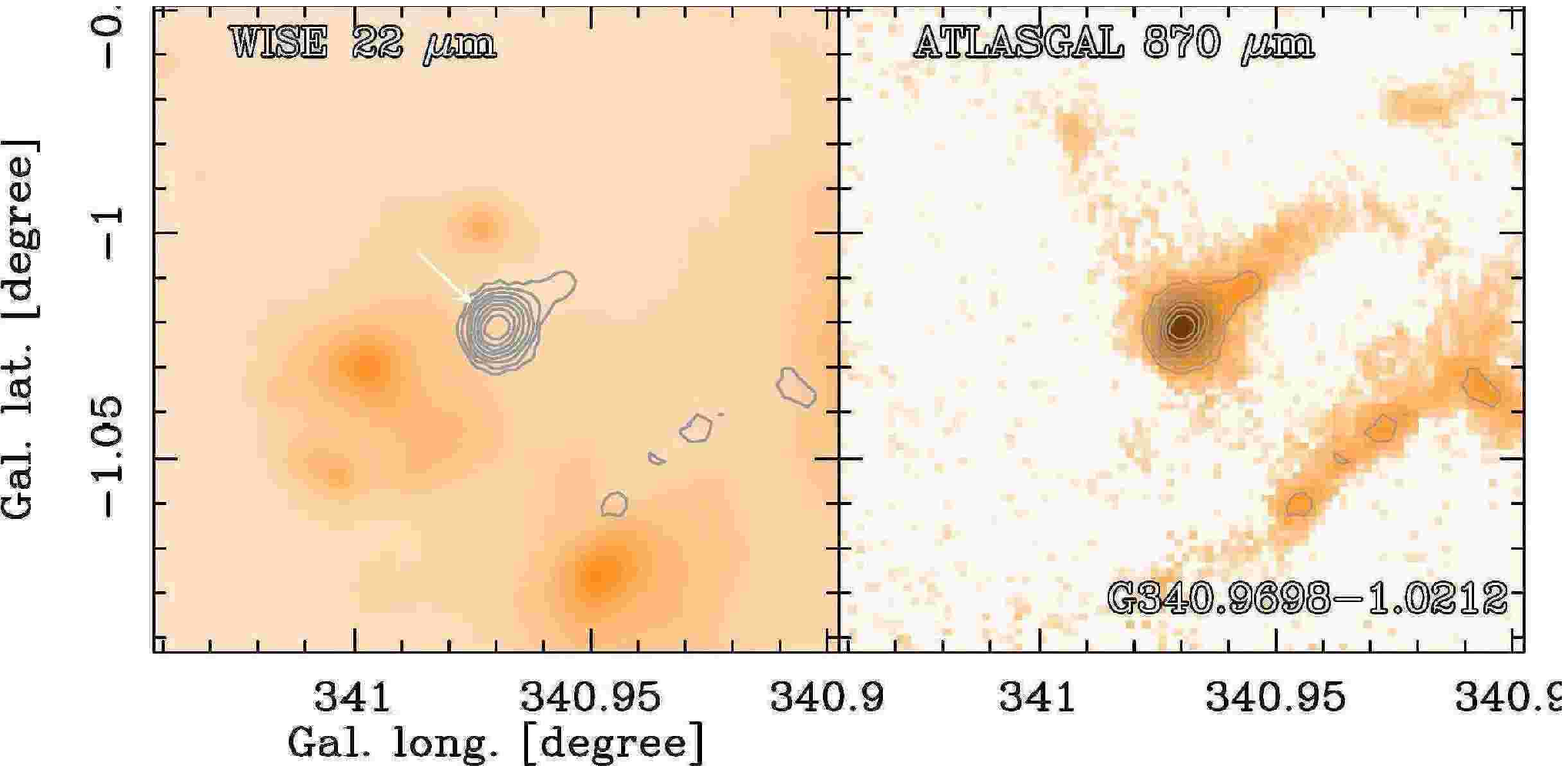}
  \includegraphics[width=9.0cm]{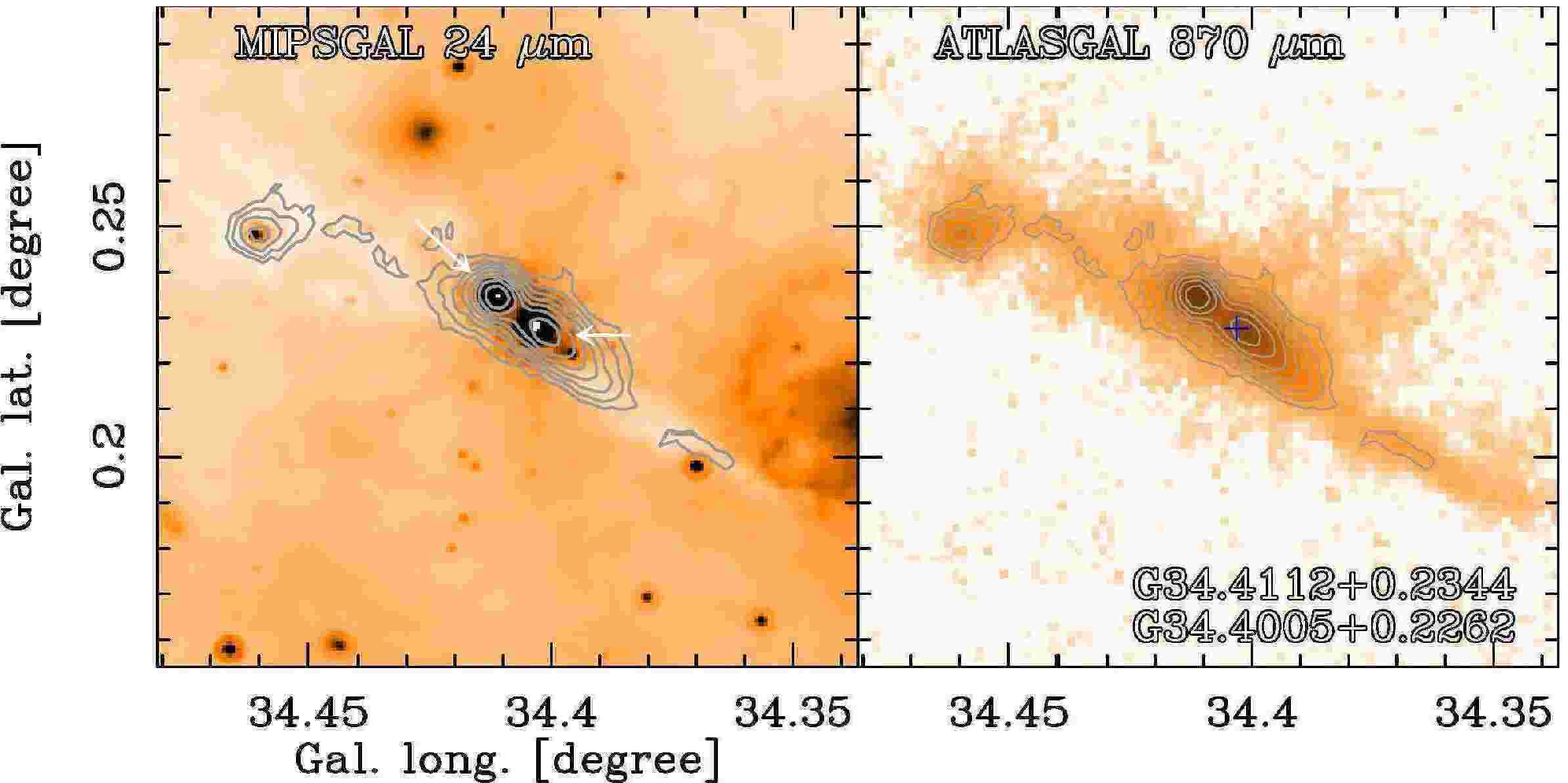}
 \caption{{\bf Top:} { Example of an infrared-quiet massive clump with a single dust peak. The left panel shows the $22-24$\,$\mu$m image from WISE/MIPSGAL with contours from ATLASGAL. The white arrows mark the position of infrared-quiet ATLASGAL clumps (Sect.\,\ref{sec:definition}). The right panel shows the 870\,$\mu$m dust emission from ATLASGAL with the same contours as in the left panel.} 
               {\bf Bottom:} Example of a massive clump towards IRDC 34.43+00.24 \citep{Garay2004} with a positional shift between the dust and the 24\,\mum\ point source in the MIPSGAL image, which also coincides with a {\uchii} region from CORNISH marked by a blue cross. }
 \label{fig:example}
 \end{figure}

In Fig.\,\ref{fig:example} we show two examples of 
massive clumps with no (or weak) embedded mid-infrared sources, the upper panel showing
the WISE\,22\,$\mu$m image of G340.9698-1.0212 
and the lower panel showing the MIPSGAL\,24\,$\mu$m image towards 
G34.43+00.24  \citep{Garay2004,Rathborne2005}; the rest of the sample is shown in Appendix\,\ref{app:irq}, Fig.\,\ref{app:fig1}. 
This particular example shows the importance of angular resolution because this region hosts an already formed {\uchii} region in the vicinity of G34.4+0.2-MM1 and MM2, 
while the brightest dust peak itself is clearly shifted from the position of the {\uchii} region \citep{Rathborne2005}.
The other embedded dust peaks in this clump are active sites of high-mass star formation in their earliest phase \citep{Rathborne2008}.

\begin{figure}
\centering
  \includegraphics[width=7.0cm,angle=90,clip]{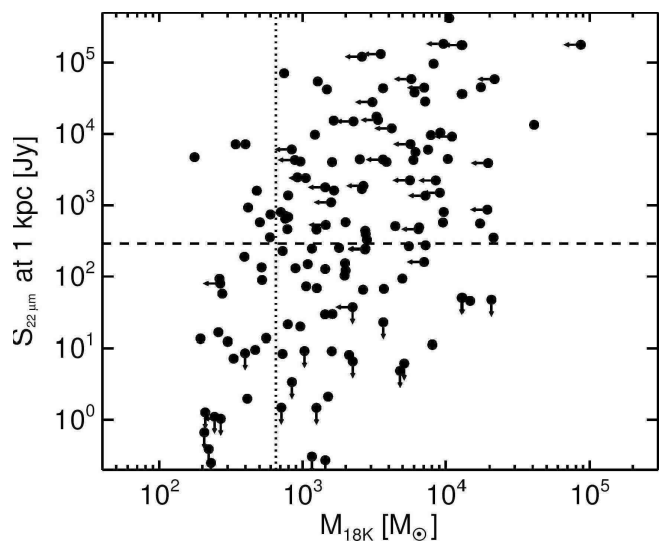}
 \caption{Distance scaled 22\,$\mu$m  flux density based on {\sl Spitzer/}MIPS photometry and the WISE point source catalogue versus mass estimated from the \at\ flux density. 
 The mid-infrared flux densities have been scaled to a common wavelength. Symbols with  arrows pointing left 
 are sources with embedded {\uchii} regions and upper limits for their mass estimate. 
 Symbols with arrows pointing down correspond to upper limits for the 22\,$\mu$m photometry.
 The dashed line shows the 289 Jy flux density threshold discussed in Sect.\,\ref{sec:definition}, and the dotted line
 corresponds to 650\,\msol.}
 \label{fig:mass_lum}
 \end{figure}

In Fig.\,\ref{fig:mass_lum}, we show the resulting photometry measurements 
versus clump mass using $T_{\rm d}=18$\,K (see Sect.\,
\ref{sec:mass}). The photometry is
scaled to a common wavelength of 22\,$\mu$m and to a distance of 1\,kpc. Using a 
 homogeneous dust temperature facilitates the comparison with the larger sample
of sources discussed in Sect.\,\ref{sec:ev2}.
While the brightest mid-infrared sources correspond to active high-mass star-forming sites in the inner Galaxy,
we reveal a population of massive clumps which are clearly fainter at mid-infrared wavelengths (see also  Sect.\,\ref{sec:definition}).

\subsection{Tracers of embedded high-mass stars}\label{sec:ev2}

 According to stellar evolution models, stars of types B3--B0.5  may also have luminosities between $L_{\rm bol}=10^3-10^4\, \rm L_{\odot}$, hence they fall below our threshold introduced in Sect.\,\ref{sec:ev}. 
We assess their potential contamination of the sample of the youngest massive clumps 
by using sensitive radio-surveys such as CORNISH \citep{Hoare2012,Purcell2013} and \citet{Urquhart2007} to search for ionising emission from stars in our sample of types later than B0.5. 
Since there is no homogeneous high angular-resolution radio dataset available covering the entire sample towards the 
$IV^{th}$ Galactic quadrant, we 
estimate a contamination level by performing the same analysis as above for a larger sample of CORNISH sources.

\begin{figure}[!htpb]
\centering
\includegraphics[width=6.5cm, angle=90,clip]{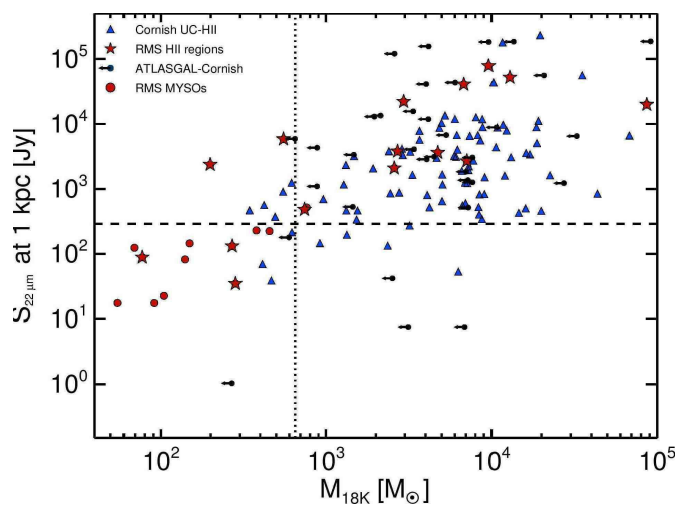}
\caption{Mass versus 22~\mum\ flux density of the sources
covered by the CORNISH survey normalised to a distance of 1\,kpc. 
Black filled circles show our sample
with a left arrow indicating upper limits for the masses. }
\label{fig:flux_mass_cornish}
\end{figure}

 In Fig.\,\ref{fig:flux_mass_cornish} 
 we show the same plot of
 distance scaled photometry at 22\,$\mu$m versus mass as in Fig.\,\ref{fig:mass_lum} 
(see Sect.\,\ref{sec:ev}), but for a larger sample of \at\ sources hosting embedded \uchii\ regions from \citet{UCornish2013}, as well as MYSOs from the RMS survey \citep{Lumsden2013}.
This sample has been studied in multiwavelength follow-up observations
(for a summary, see \citealp{Lumsden2013}), and the sources have been classified according whether they
host (M)YSOs, {\hii} regions, or a mixture of both. 
We extracted the 22~\mum\ 
fluxes for these samples from the WISE catalogue considering matches between the ATLASGAL dust peaks and mid-infrared point sources with a small angular separation ($<6$\arcsec). We considered only good quality flag photometry and
a high signal-to-noise ratio of $>10$ in the 22\,$\mu$m band. More importantly,
for the RMS and CORNISH sources 
we only show those with the most reliable distance estimates based
on maser parallax measurements \citep{Reid2014}.

We find that massive clumps ($M_{\rm clump}>650$\,\msol) hosting 
embedded {\uchii} regions occupy a well-defined region, and mostly lie above our threshold of $S_{22\mu\,m} > 289$\,Jy at 1\,kpc.
Out of the 78 sources of the shown ATLASGAL-CORNISH matches, 5 lie below this threshold, corresponding to 6.4\%
of the entire sample. 
Such weak, compact radio continuum emission has been found towards
low bolometric luminosity ($L_{\rm bol}\le5\times10^4L_{\odot}$) 
MYSOs (see e.g.\,\citealp{Guzman2010,Lumsden2013,Avison2015}). They
are likely associated with
shock-ionisation from an early outflow activity (e.g.\,\citealp{HF2007}).

Our flux limited selection
of ATLASGAL sources with embedded {\uchii} regions are also shown as a comparison,
and find that 3 sources out of the 31  fall below our mid-infrared threshold, which is 9.7\%.
We inspected these sources visually, and find that one source lies in a confused region
where either the mid-infrared photometry may be affected by extended emission, 
or the radio continuum emission may be associated with the extended {\hii} region rather than the dust peak.
The rest could correspond to the above discussed genuine 
low bolometric luminosity radio sources associated with MYSOs.
To summarise, the lowest luminosity massive clumps 
likely lack embedded {\uchii} regions and bright MYSOs, 
and their contamination below our luminosity threshold is
estimated to be less than ten per cent.

\subsection{Classification of the sample}\label{sec:definition}

Using the 
information described in the previous sections,
 we classify 
our sample primarily based on their mid-infrared flux density and  indicate the presence of
{\uchii} regions if radio continuum data is available.
We do not rely on radio continuum data, however,  because sensitive unbiased surveys are not yet available in the $IV{th}$ quadrant, 
and thus the sample of \uchii\ regions associated with our clumps is incomplete.
We introduced the threshold of 289\,Jy at 22\,$\mu$m at 1\,kpc as discussed in Sect.\,\ref{sec:ev},
and as empirically justified in Sect.\,\ref{sec:ev2}.
Sources above and below this threshold are termed  `infrared bright' and  `infrared quiet', respectively. 
This classification is 
consistent, but simplified compared to previous works (e.g.\,\citealp{Csengeri2014, Giannetti2014, Csengeri2016a}). In addition, it quantitatively selects the young clumps  
based on the luminosity of the embedded protostar(s). Contamination of more evolved objects is estimated to be less than 10\% (Sect.\,\ref{sec:ev2}). Nevertheless, in the following we indicate
  the subsample of massive clumps hosting genuine {\uchii} regions for the infrared-bright clumps. 

From the sample of 210 sources
we were able to classify 209 sources: 139 (66\%) are infrared-bright including
51 (24\%) unambiguously associated with {\uchii} regions, and \nirq\ (34\%) are infrared  quiet. 
Considering  associations with infrared sources only from the WISE catalogue,
we find 56 sources (26\%) with no counterpart; the remaining 74\% are
associated with a mid-infrared source, which is consistent with 
earlier findings from \citet{Csengeri2014}. Some objects are located in
regions with complex extended emission at mid-infrared wavelengths, which may affect the photometry, 
hence their nature being quiet or bright may remain somewhat uncertain.
As shown before, this definition of mid-infrared-quiet clumps picks up
massive clumps hosting high-mass protostars or their precursors; 
where no ancillary radio data is available, 
the contamination from $8-15$ \msol\ main sequence stars 
is minor. We will  investigate the nature of these sources further in Sect.\,\ref{sec:form}.
We show the images of infrared-quiet sources in  Appendix\,\ref{app:irq}.

\subsection{SiO (8--7) emission towards infrared-quiet massive clumps}\label{sec:sio87}

To further investigate the properties of the infrared-quiet clumps (see also Sect.\,\ref{sec:definition2}),
we used spectroscopic observations of a typical shock tracer, SiO, as an indirect probe of  ongoing accretion processes.
Of the \nirq\ infrared-quiet clumps, we selected 46 objects to follow up
with the FLASH$^+$ receiver
\citep{Klein_2014} at the APEX telescope \citep{gusten2006}. 
The source selection was driven by the distance distribution of the sample 
to favour close-by objects ($d\leq$4.5\,kpc), and focused on the newly discovered objects (see Sect.\,\ref{sec:stat}). 

Using the APEX/FLASH$^+$ receiver we covered 8\,GHz bandwidth simultaneously centred at 
347.15\,GHz in USB set-up, from which we exploit here the SiO (8--7) transition at 347.33058\,GHz. 
The observations were carried out  
in 2013 and 2014 August, and we obtained $90\arcsec \times 90\arcsec$ OTF maps. The 
reference position was offset by 10\arcmin\ in right ascension and declination. Pointing and 
focus were regularly checked during the observations.  

The data have been reduced using the GILDAS software following standard procedures.  
The spectra at the centre, i.e.\,the dust peak position, have been extracted from the maps. 
Baselines of order zero have been subtracted and the data have been scaled to the $T_{\rm mb}$ temperature  
using 73\% efficiency\footnote{http://www.apex-telescope.org/telescope/efficiency/}. 
The average noise level is 48\,mK in 1.3\,\kms\ resolution.

A total of  27 out of the 46 objects (59\%) show emission of this line above 4$\sigma$ 
as measured on the line area with Gaussian fitting. We show an example in Fig.\,\ref{fig:sio},
while all the detections are shown in Appendix\,\ref{app:sio}, Fig.\,\ref{app:fig2}. The observed detection rate is similar to that of
\citet{Klaassen2012} despite the lower sensitivity, and is somewhat lower than reported by \citet{Leurini2014},
who had, on the other hand, a factor of $\sim$\,2 higher sensitivity for the same transition. These studies
focused on more evolved, mostly infrared-bright clumps, many of them hosting {\uchii} regions. The fact that
the high detection rates are broadly similar towards the less luminous thus likely younger clumps is consistent
with the findings of \citet{Csengeri2016a} using lower-J SiO lines.

%
\begin{figure}[!htpb]
\centering
\includegraphics[width=7.5cm,angle=0]{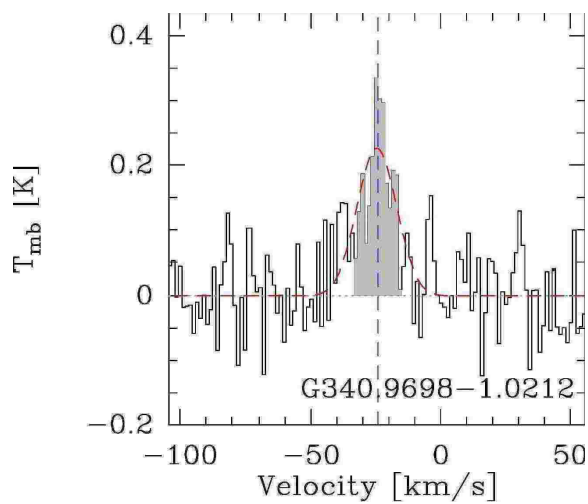}
\caption{ SiO (8--7) line observed towards the infrared-quiet massive clump G340.9698-1.0212. The dashed line shows the rest velocity of the source from the C$^{17}$O (3--2) observations at 337.06110\,GHz from the same dataset. }\label{fig:sio}
\end{figure}

\section{Statistical properties of the brightest submillimetre sources}\label{sec:statistics}

In the following we investigate the distribution and the statistics of the estimated physical
properties of the entire sample. As defined in Sect.\,\ref{sec:definition} we distinguish between 
infrared-quiet massive clumps and those classified as infrared-bright, and also show the subsample 
hosting MYSOs and/or {\uchii} regions. 
Their average physical properties are summarised in Table\,\ref{tab:prop}. 

\subsection{Mass$-$distance distribution and completeness}\label{sec:completeness}

\begin{figure}[!htpb]
\centering
\includegraphics[width=6.5cm, angle=90]{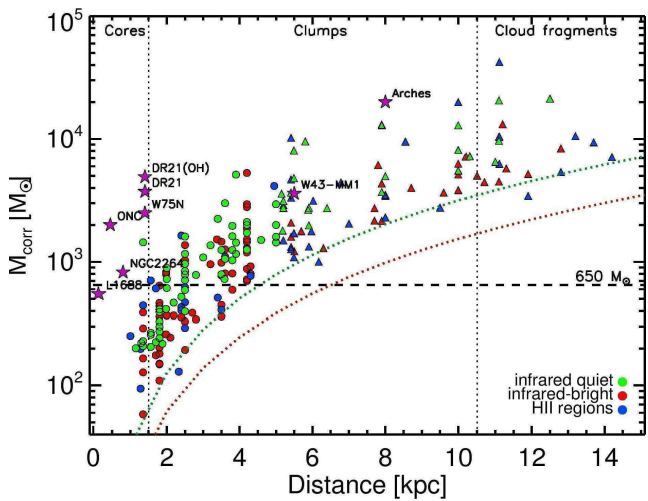}
\caption{Temperature corrected mass versus distance distribution. 
Dotted lines show the mass sensitivity corresponding to the flux density threshold of 5\,Jy using
$T_{\rm d}$=30\,K  for infrared-bright (red), and $T_{\rm d}$=18\,K for infrared-quiet clumps (green).
Coloured symbols are shown in the figure legend; filled circles correspond to sources where our selection
is complete, while filled triangles show sources where the sample is incomplete above 650\,\msol.
The dashed horizontal line shows the 650\,\msol\ limit for massive clumps potentially capable of forming 
high-mass stars from \citet{Csengeri2014}. Known (proto)clusters 
from the literature are labelled with magenta stars, as are the estimated initial clump mass at half-radius 
for the galactic (super) star clusters like the Orion Nebula Cluster (ONC, e.g.\,\citealp{DaRio2012}) and Arches \citep{Habibi2013}, 
following \citet{Tan2014}. The approximate distance limit for detecting sources corresponding to cores,
clumps, and cloud fragments is shown as a dotted line \citep{BT2007}.} 
\label{fig:dist-mass2}
\end{figure}

In Fig.\,\ref{fig:dist-mass2}
we show the distance distribution
of the sample. As discussed in Sect.\,\ref{sec:distance}, a substantial fraction of the sources
are part of active star-forming complexes. 
Their typical distance is around $3-4$~kpc, close to the 
average distance of 4.5\,kpc for \at\ clumps \citep{Wienen2012}.
Overall 60\% of the sample is located within 4.5\,kpc; a total of  86 sources are located at greater distances, of which 29 are infrared quiet. The more distant sources are therefore dominated by infrared-bright clumps.
We also show a threshold of 650\,\msol, introduced in \citet{Csengeri2014}, which is an extrapolation of the massive dense core definition by \citet{M07} to the $\sim0.3$\,pc scale of ATLASGAL clumps. Empirically this selection has proved to select cores hosting high-mass protostars by \citet{Bontemps2010}.  

The mass estimation corresponding to the threshold of 5\,Jy used for our flux limit is indicated in Fig.\,\ref{fig:dist-mass2} as green and red dotted lines for infrared-bright ($T_{\rm d}$=30\,K) and infrared-quiet clumps ($T_{\rm d}$=18\,K), respectively.  
It shows that our selection is complete for infrared-quiet clumps above 650\,\msol\ up to a distance of 4.5\,kpc, while for infrared-bright clumps above 310\,\msol. 
Since we expect the more evolved clumps to have
lost a part of their mass in the central region through cluster
formation and dissipation, we can argue
that this sample is roughly complete for all clumps having 650\,\msol\ 
at birth at all their stages of evolution
from infrared quiet to infrared bright and possibly also \uchii\
regions. A total of 28 infrared-quiet and 55 infrared-bright clumps are found above 650\,\msol\  and 
310\,\msol\ at a distance closer than 4.5\,kpc. 

The area covered by the sector of our Galaxy from $\ell = -60^{\circ}$ to 60$^{\circ}$ and for a distance lower than 4.5\,kpc is found to represent $\sim 10\,\%$ of the total area of the galactic disk within the solar circle after removing the inner 3\,kpc. In a rough approximation, we can consider this 10$\,\%$
as the fraction of star formation in the Galactic disk which is occurring in our distance limited sample.  

\subsection{Temperature distribution}\label{sec:disc-temperature}

\begin{figure}[!htpb]
\centering
\includegraphics[width=6.5cm, angle=90]{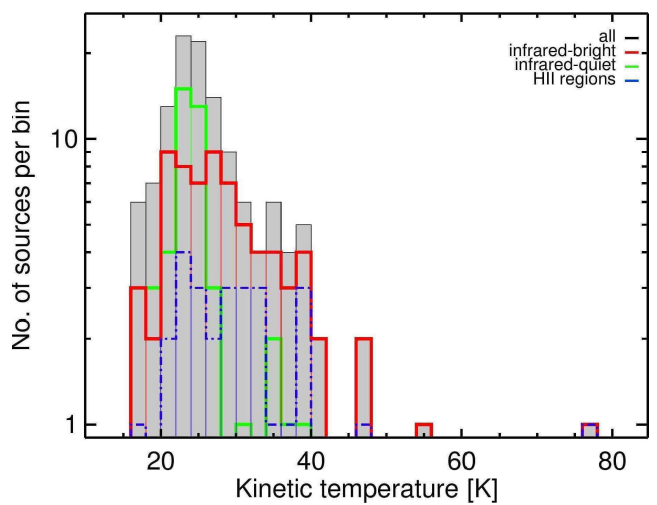}
\caption{Distribution of kinetic temperature from NH$_3$ (1,1) and (2,2) measurements from \citet{Wienen2012} and \citet[submitted]{Wienen2016}. The infrared-bright and infrared-quiet sources are shown in different colours, as explained
in the figure legend. We also show massive clumps hosting {\uchii} regions as a subsample of the infrared-bright sources.}\label{fig:histo_temp}
\end{figure}

From the NH$_3$ data discussed in Sect.\,\ref{sec:temperatures}
we determine a median kinetic temperature of 25.5\,K for 60\% of the sample.
Figure\,\ref{fig:histo_temp} shows the temperature distribution of these sources.
In this sample 28 have a firm association with {\uchii}-regions, and
the majority of the sources are classified as infrared bright, in total 71.
 These types of sources have a median kinetic temperature of 29.6 and 27.2\,K, 
with a standard deviation of 8-11\,K. Towards infrared-quiet sources we find a median of 23.8\,K
with a standard deviation of 4.7\,K. This suggests a trend of lower temperatures towards younger
sources; however, given their large standard deviation, this difference is not statistically
significant. 
Following our argumentation in Sect.\,\ref{sec:ev}, and considering the similar temperatures,
here we do not make a distinction between massive clumps with {\uchii} regions and those
hosting infrared-bright sources.
A more accurate temperature assignment to the group of infrared-quiet massive clumps 
still remains uncertain since the sources with available NH$_3$ measurements
are dominated by the more evolved sources, and the targeted infrared-quiet clumps
may not be representative of the entire subsample.

\subsection{{\bf Distribution of physical properties}}\label{sec:mass-stat}
\begin{figure}[!htpb] 
\centering
\includegraphics[width=6.5cm, angle=90]{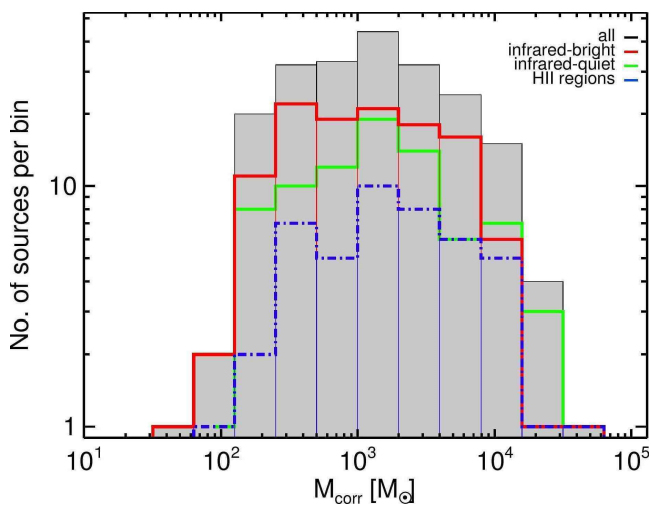}
\caption{Temperature corrected mass distribution (using $T_{\rm dust}=30$\,K for the infrared-bright clumps). Colours are the same as in Fig.\,\ref{fig:histo_temp}. The clumps hosting genuine {\uchii} regions are a subsample of the infrared-bright clumps.}\label{fig:histo_mass}
\end{figure}

In Fig.\,\ref{fig:histo_mass} we show the mass distribution of the sample, where we adopted
$T_{\rm dust} = 30$\,K for the infrared-bright sources and kept $T_{\rm dust} = 18$\,K for the infrared-quiet clumps (see Sect.\,\ref{sec:mass}). We give an overview of the statistical properties of the physical parameters of all sources and the two source types in Table.\,\ref{tab:prop}. 
The global statistical properties of the infrared-bright and -quiet samples are similar, and
we find a remarkably flat mass distribution for all source types, 
the majority of the sources lying between $650$\,\msol\ and $8000$\,\msol.
 Lower mass objects correspond to
nearby massive dense cores with masses between $\sim$$86-120$\,\msol. 
Towards sources at greater distances we find masses of $\sim$\,$10^5$\,\msol\ corresponding to cloud fragments rather than clumps.  The mean clump mass is $\sim$$4.5\times10^3$\,\msol, which is about a factor of two higher compared to the star formation selected larger sample of ATLASGAL clumps studied by 
\citet{Urquhart2014}.

\begin{figure}[!htpb]
\centering
\includegraphics[width=6.5cm, angle=90]{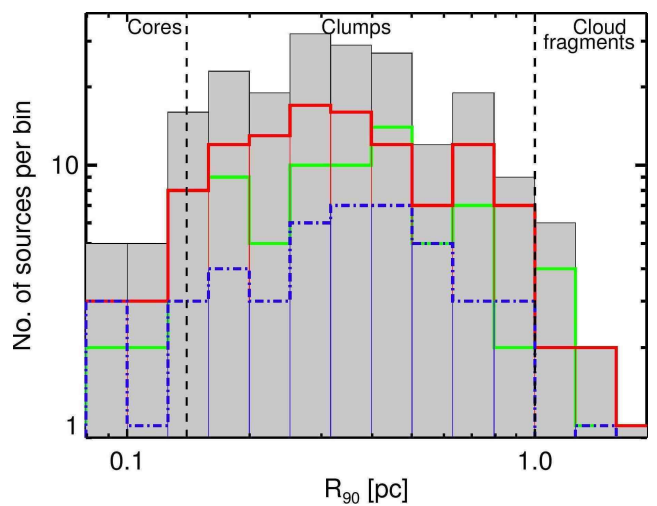}
\caption{Distribution of the deconvolved physical size ($R_{\rm 90}$) of the sample. Colours are the same as in Fig.\,\ref{fig:histo_temp}. The dashed lines indicate an approximate distinction between cores, clumps, and cloud fragments \citep{BT2007}.}\label{fig:histo_size}
\end{figure}

The size distribution of the sample is shown in Fig.\,\ref{fig:histo_size}, and 
 as above, there is  no significant difference between the different source types.
Adopting the nomenclature of \citet{Williams2000} (see also \citealp{BT2007}), we find a small fraction of 
sources that correspond to cores or entire cloud fragments, while the majority of the sources correspond
to clumps. 
They have typical sizes of 0.3--0.4\,pc, which is comparable to the 
mean size of $\sim$\,0.36 pc found by \citet{Russeil2010}
in NGC~6334/NGC~6357.  Compared to the population of 
massive dense cores in Cygnus-X, the sizes are $\sim$$3$ times larger, while the masses are on average $\sim5$ times larger, suggesting their potential to form high-mass stars.

\begin{figure}[!htpb]
\centering
\includegraphics[width=6.5cm, angle=90]{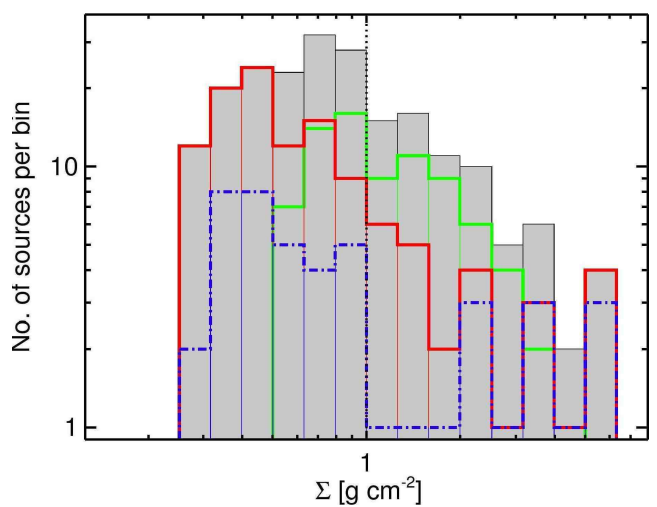}
\caption{Mass surface density ($\Sigma$) distribution of the sample using $T_{\rm dust}=30$\,K for the infrared-bright clumps.
Colours and symbols are the same as in Fig.\,\ref{fig:histo_temp}.}\label{fig:histo_surfdens}
\end{figure}

In Fig.\,\ref{fig:histo_surfdens} we show the mass surface density distribution of the sample. 
{Lower surface densities are found towards the infrared-bright
sources,  a considerable fraction of which host {\uchii} regions. 
This is likely due to an observational bias, driven both by
a  larger fraction of such sources at a larger distance, 
which then correspond to larger structures 
with lower average surface density, and the lower mass selection threshold
for the infrared-bright sources. }
The mean surface density is above the theoretical
value of $\Sigma_{\rm cl}\geq1$\,g\,cm$^{-2}$ by \citet{Krumholz2008}, suggesting that a substantial fraction of our sources have the potential to form high-mass stars. 
Correspondingly, the median volume-averaged density of $3.13\times10^5$ cm$^{-3}$ is higher
than that found in other samples in the literature (e.g.\,\citealt{Russeil2010}), and suggests that we 
probe here high density, star-forming gas. 

\begin{table*}
\centering
\caption{Statistics of the physical properties of all sources, and the infrared-bright and infrared-quiet samples.
}\label{tab:prop}
\begin{tabular}{lrrrrrrrrrrr}
\hline\hline
& \multicolumn{5}{c}{All sources}\\
        & min & max  & mean  & median & stdev \\
\hline   
Distance [kpc] & 1.00  &  15.6 &   4.9 &    3.9 & 3.30 \\
$R_{\rm 90}$ [pc]          & 0.07  & 1.63   & 0.40  & 0.32  & 0.28 \\
$M_{\rm 18\,K}$ [\msol]& 120 &    $0.9\times10^5$   &   $4.5\times10^3$   &   $1.8\times10^3$ & $7.9\times10^3$  \\
$M_{\rm corr}$ [\msol]& 58 &    $4.2\times10^5$   &   $2.8\times10^3$   &   $1.3\times10^3$ & $4.6\times10^3$ \\
$\Sigma_{\rm corr}$ [g~cm$^{-2}$] &    0.27   & 5.9 & 1.11   &    0.75 & 1.02 \\
$\bar{n}_{\rm corr}$ [$\times10^5$ cm$^{-3}$] & $0.14$  & $44.15$  &  $3.17$ &  $1.56$ & 5.25 \\
\hline   
\hline
& \multicolumn{5}{c}{Infrared-bright massive clumps} \\
\hline
Distance [kpc] & 1.0  &  14.20 &   5.14 &    4.20 & 3.43 \\
$R_{\rm 90}$          & 0.07  & 1.41   & 0.42  & 0.35 & 0.27 \\
Mass$_{\rm 18\,K}$ [\msol]& 120 &    $8.7\times10^4$   &   $5.2\times10^3$   &   $2.4\times10^3$  &   $9.3\times10^3$\\
Mass$_{\rm corr}$ [\msol]& 58 &    $4.2\times10^4$   &   $2.7\times10^3$   &   $1.2\times10^3$ &   $4.7\times10^3$ \\
$\Sigma_{\rm 18\,K}$ [g~cm$^{-2}$] &    0.27   & 6.00 & 0.99   &    0.56 & 1.14\\
$\bar{n}_{\rm 18\,K}$ [$\times10^5$ cm$^{-3}$] & $0.14$ & $44.15$  &  $2.78$ &  $1.03$  & 5.89 \\
\hline   
\hline
& \multicolumn{5}{c}{Infrared-quiet massive clumps} \\
\hline
Distance [kpc] & 1.15  &  15.60 &   4.47 &    3.70 & 3.05 \\
$R_{\rm 90}$         & 0.08  & 1.63   & 0.36  & 0.28 & 0.28 \\
Mass$_{\rm 18\,K}$ [\msol]& 200 &    $2.1\times10^4$   &   $3.1\times10^3$   &   $1.4\times10^3$ &   $4.5\times10^3$ \\
$\Sigma_{\rm 18\,K}$ [g~cm$^{-2}$] &    0.51   & 4.05 & 1.31   &    1.03 & 0.74\\
$\bar{n}_{\rm 18\,K}$ [$\times10^5$ cm$^{-3}$] & $0.20$ & $18.46$  &  $3.83$ &  $2.39$  & 3.91 \\
\hline
\end{tabular}
\end{table*} 



\section{ Nature of the brightest submillimetre sources in the inner Galaxy}\label{sec:form}

The brightest submillimetre sources in the inner Galaxy are generally associated with active sites
of star formation.  Here we first investigate the star formation potential and evolutionary aspect of the entire 
sample (Sect.\,\ref{sec:definition2}), and argue that the majority of the infrared-quiet massive clumps are
potential precursors
of massive stars and/or clusters. We then show that our sample of infrared-quiet clumps 
contains sources previously not recognised in other surveys using infrared emission, 
or absorption (Sect.\,\ref{sec:stat}). Finally, we search for signposts of
embedded protostars in the infrared-quiet subsample to assess the stage of ongoing star formation
(Sect.\,\ref{sec:tracers}).
We then discuss the stability of the entire sample (Sect.\,\ref{sec:stability}), 
and provide predictions for their small-scale fragmentation (Sect.\,\ref{sec:phys-selected}).

\subsection{Precursors of massive clusters}\label{sec:definition2}

\begin{figure}[!htpb]
\centering
\includegraphics[width=6.5cm, angle=90]{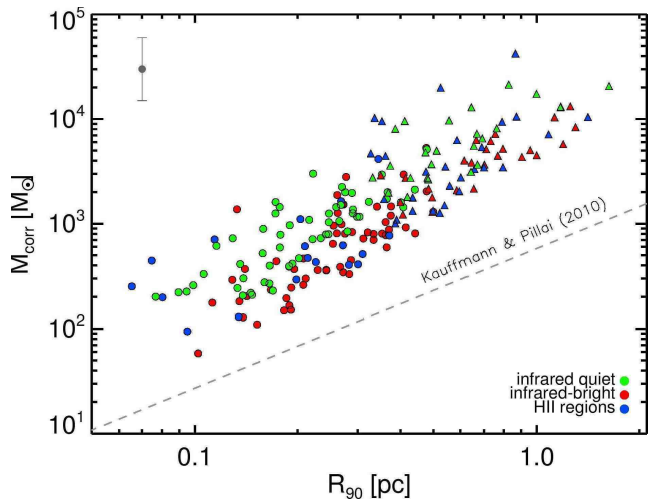}
\caption{Temperature corrected mass estimates (using $T_{\rm dust}=30$\,K for the infrared-bright clumps) versus deconvolved physical size ($R_{\rm 90}$) from Sect.\,\ref{sec:mass}. Dashed lines show the relation of \citet{KP2010} scaled to the \citet{OH1994} dust opacity following \citet{Dunham2011}. In the upper left corner we show the factor of two uncertainty in the mass estimate.
Colours are the same as in Fig.\,\ref{fig:histo_temp}.}
\label{fig:mass_size}
\end{figure}

Comparing the physical properties in terms of mass, size, and density, in Table\,\ref{tab:prop} 
we show that our 
sample exhibits rather homogeneous properties, and no statistically significant differences are found between the two source types. 
This includes  
well-studied Galactic mini-starburst regions, for instance the W43 ridge (e.g.\,\citealp{Motte2003,Bally2010,Quang2011}), W51 Main (e.g.\,\citealp{Mufson1979,Zapata2008,Ginsburg2015}), and W49N (e.g.\,\citealp{Peng2010,Galvn-Madrid2013}). 
While indications of ongoing star formation is prevalent in the sample,  are all clumps prone to form mini-starbursts and rich clusters?
We compare our results with the empirical relation of \citet{Kauffmann2010} in Fig.\,\ref{fig:mass_size}, which 
suggests that all of our sources have the potential to form high-mass stars. 
While the comparative statistical analysis of infrared-quiet and -bright sources reveals similar 
physical characteristics in terms of mass and size, they seem to differ only in their embedded (proto-)stellar population. 
Low bolometric luminosity sources with similar physical characteristics to their more evolved counterparts have 
been frequently interpreted as being in an earlier evolutionary stage (e.g.\,\citealp{M07,Molinari2008}).
This suggests that infrared-quiet clumps are the precursors of the infrared-bright clumps and {\uchii} regions,
 and the sample can be arranged along a similar evolutionary path.

The large mass reservoir of infrared-bright sources suggests that they likely continue producing 
high-mass stars. Unless specific conditions are required for the formation of rich 
clusters, we shall have selected here their true precursors.

\subsection{Massive cluster progenitors in isolation?}
\label{sec:stat}

Because the clustered nature of high-mass star formation was recognised early on, 
several studies searched 
for younger objects in the vicinity of bright, luminous sources (e.g.\,\citealp{Garay2004, Thompson2005}). 
To look for neighbouring, more evolved objects with {\uchii} regions in the close vicinity of the infrared-quiet sample,
we used the CS (2--1) survey of IRAS sources \citep{Bronfman1996}, which is complete towards 
IRAS selected {\uchii} regions in the Galactic plane. 
In total we found 92 associations (42\%) within 0.8\arcmin\ offset, which 
corresponds to 0.26\,pc for the nearest source  
and 2.9\,pc for the most distant source in our sample. However, in the infrared-quiet sample only 29 (36\%)  
are found in the vicinity of an IRAS source, and their offsets correspond to a physical separation between $0.3-16.3$\,pc.
Compared with the more sensitive and higher angular resolution 
Red MSX survey (RMS, \citealp{Lumsden2013}),
we find six infrared-quiet clumps in the vicinity of a bright MYSO with $L_{\rm bol}>2\times10^4$\,\lsol.

Surveys sensitive to  cold dust are better suited to directly probe the youngest clumps.
To evaluate the fraction of previously identified infrared-quiet clumps, we first looked
for associations with cloud structures 
identified as infrared-dark clouds (IRDCs) by
correlating the peak dust positions with the 
catalogues of IRDCs revealed by MSX \citep{Jackson2008,Simon2006} and 
Spitzer \citep{Peretto2009}. 
In total we find 20 associations with an  MSX Dark Cloud within 120\arcsec,   15 of which are 
infrared-quiet sources. For the association with Spitzer Dark Clouds we
used the Herschel-confirmed clouds from \citet{Peretto2016} with a smaller positional offset
of 50\arcsec\ corresponding to their average equivalent size of
$\sim$25\arcsec+1$\sigma$, where $\sigma$ is their standard deviation.
In total, 65 of our clumps have associations,  31 of which are infrared-quiet clumps.
This means that $\sim$61\% of the infrared-quiet clumps have not been recognised before in
these surveys.

Thus, we find that the infrared-quiet sample is found in relative isolation compared
to the bright clumps which already undergo active star formation. 
Less than half of the ATLASGAL selected massive clumps have IRDC counterparts,
and comparison with the star formation selected ATLASGAL clumps of \citet{Urquhart2014}
shows that 25 of the infrared-quiet sources are newly identified. In summary, 
we find that $\sim$23\% of our infrared-quiet sample has not been recognised by these previous studies.

\subsection{Signposts of already formed embedded high-mass protostars}
\label{sec:tracers}

To tackle the stage of ongoing star formation, here we compare our sample with tracers of high-mass star formation.  We found a total of 23 clumps coinciding with extended green objects (EGOs); (\citealt{Cyganowski2008} and \citealt{Chen2013}) most of them  are  infrared-quiet clumps, 2 sources are infrared-bright, and 3 are {\uchii} regions. Since EGOs are believed to trace powerful outflows mostly from (presumably massive) Class 0 YSOs, it confirms that the infrared-quiet clumps host the youngest protostellar phases. However, only 25\% of the infrared-quiet clumps are associated with EGOs. This may indicate that the EGO phase is shorter than the infrared-quiet clump phase, and therefore does not represent well the entire early evolutionary phase of massive clumps. 
The few EGOs associated with infrared-bright clumps may indicate that young massive protostars can still form at the time of the infrared-bright phase, and even perhaps at the {\uchii} phase if star formation is a continuous process in clumps.

On the other hand, the lack of EGOs does not necessarily imply a lack of
outflow activity, since high extinction, even at 4.5\,$\mu$m, may prevent their detection. In our SiO (8--7) spectroscopic study, $\sim$\,60\% of infrared-quiet clumps show emission, which is a much larger fraction than that of EGOs. Using the lower energy SiO (2--1) line, \citet{Csengeri2016a} find that an even higher fraction of the infrared-quiet massive clumps  exhibit emission in this shocked gas tracer. This is consistent with a picture that infall and accretion processes may already be occurring 
in at least  some of the infrared-quiet clumps, indirectly pointing at an underlying population of deeply embedded
protostars (see the example of NGC 6334 I(N) by \citealp{Megeath1999}).

Another tracer of the nature of embedded protostars is
Class\,II methanol maser emission. These masers are excited by radiative pumping 
requiring the intense mid-infrared emission from the warm, dense material
of deeply embedded hot-cores or {\uchii} regions \citep{Menten1991,Sobolev1997}. 
We therefore cross-correlated our sample
with the Methanol MultiBeam survey \citep{Caswell2010_MMBI,Green2010_MMBII,Caswell2011_MMBIII,Breen2015_MMBV},
and found 138 (66\%) associations within 20\arcsec, of which 57 are infrared-quiet massive clumps corresponding to $\sim$\,71\%
of the subsample. Methanol masers thus seem to trace a larger fraction of the infrared-quiet phase, although less selectively since they are also found towards the more evolved stages of clump evolution. 
Altogether we find that a considerable fraction of the infrared-quiet sample show signs of embedded (high-mass) protostars
and ongoing star formation process.


\subsection{Stability estimates and timescales}\label{sec:stability}

Using high-density tracers we estimate here the virial mass 
and compare it to the mass estimates based on the dust emission. 
As a first step we used the NH$_3$ (1,1) 
line width measurements of \citet{Wienen2012} and \citet[submitted]{Wienen2016} where available,
and then complemented these data with measurements of the CS (2-1) line
from \citet{Bronfman1996}, and finally of the H$^{13}$CO$^+$ (1--0) line 
from recent surveys using the Mopra (Wyrowski et al., in prep.) and the IRAM~30m
telescopes \citep{Csengeri2016a}.  We have line width
estimates for a total of  173 sources, 80\% of the whole sample.  
While this is a smaller sample than studied in \citet{Wienen2012} and \citet{Urquhart2014},
it is more focused on the brightest ATLASGAL clumps.

We estimate the virial mass assuming a density profile of $n(r)\sim r^{-2}$, and
using $M_{\rm Vir}=3\times \frac{R_{\rm 90}\times\sigma_{\rm tot}^2}{G}$~\citep{BertoldiMcKee1992}, where 
$R_{\rm 90}$ is the radius of a sphere, $\sigma_{\rm tot}$ is the velocity 
dispersion of the gas, and $G$ is the gravitational constant, 
\begin{equation}
\frac{M_{\rm Vir}}{[\rm M_{\odot}]}= 697 \, \frac{R_{90}}{[\rm pc]} \, \frac{\sigma_{\rm tot}^2}{[\rm km~s^{-1}]}.
\end{equation}

\begin{figure}[!htpb]
\centering
\includegraphics [width=6.5cm, angle=90]{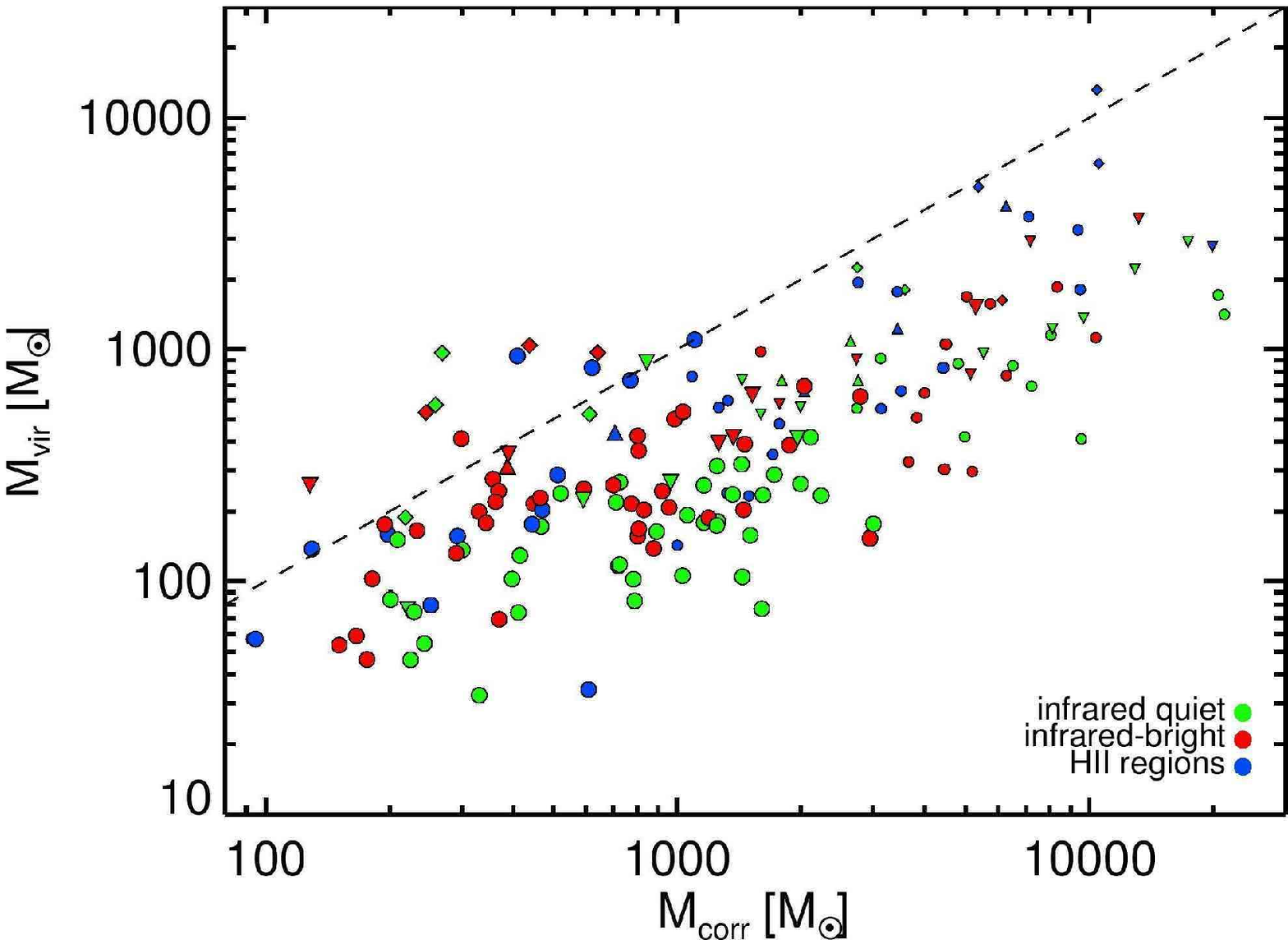}
\caption{Temperature corrected clump mass estimates compared to the virial mass.
The dashed line indicates
where the virial mass equals  the dust mass.
Colours are the same as in Fig.\,\ref{fig:histo_temp}, filled circles correspond to 
line width estimates from \citet{Wienen2012,Wienen2015}, filled upward triangle 
to that of \citet{Csengeri2016a}, filled downward triangle to that of Wyrowski et al. (in prep.),
filled diamonds to that of \citet{Bronfman1996}. Sources with $d>$5 kpc are
plotted in smaller symbols.}
\label{fig:mass-mvir}
\end{figure}

 As shown in Fig.\,\ref{fig:mass-mvir}, we find surprisingly small $M_{\rm Vir}/M$ ratios towards the majority of the sources, 
on average 0.47 with a median of 0.29, strongly suggesting that they are gravitationally unstable. 
While the used
molecular tracers selectively probe the high-density gas, 
uncertainties due to the different tracers and beam sizes could affect our
virial mass estimates. However, despite the various datasets used, we see no systematics
in the estimated virial masses. 
As Fig.\,\ref{fig:mass-mvir} shows, 
155 sources (89\%) have a smaller virial mass than their estimated gas mass. 
We find the same trend for the more distant
sources of the sample, which are beyond our completeness limit. 
 Uncertainties in the mass estimates due to the dust opacity and gas-to-dust ratio would 
lower the $M_{\rm Vir}/M$ ratio by a factor of two. 
Even considering the case of a uniform $n(r)\sim r^{0}$ density profile,
we would still find a large fraction (67\%) of the clumps having $M_{\rm Vir}/M<1$, which  could thus still be considered as
gravitationally unstable.
Magnetic fields could, however, slow down the rate of collapse. For a large number
of clouds \citet{Kauffmann2013} provides estimates of the virial parameter defined by $M_{\rm Vir}/M$,
and find similarly low values. They suggest that magnetic fields as strong as 1\,mG would be required 
to slow down the collapse, otherwise they undergo rapid collapse.

The size-scale of clumps, 0.2--1\,pc, is large compared to the scale of MDCs ($<0.15$\,pc) and cores ($<0.05$\,pc)
for which collapse signatures are frequently observed. Our results therefore
suggest that these massive cluster progenitors are already collapsing on a {global scale}, i.e.\,a clump scale.
This is particularly intriguing because it suggests that
 gravity could be the dominant force in the mass assembly process over size-scales 
of entire clumps, which are larger than previously considered (e.g.\,\citealp{McKee2007}). 

We also calculate the free-fall time of these clumps by $\tau_{\rm ff}=\sqrt{3\pi/32\,G\,n}$,
where $G$ is the gravitational constant, and $n$ is the volume density. We find on average $5.64\times10^4$\,yr,
which is somewhat shorter than the lifetime estimate of $7.5\times10^4$\,yr
by \citet{Csengeri2014}. Such relative collapse 
timescales (i.e.\,a few times the free-fall timescale) are typically observed towards low-mass 
cores (see \citealp{Ward-Thompso2007}, and references therein), 
and have recently been suggested for the formation of  
high-mass cores by \citet{Duarte2013}.

Collapse timescales of a few times $10^4$\,yr suggests that all massive clumps, precursors of 
massive clusters, are transient entities and should disappear on short timescales, typically less than $10^5$\,yr  (see also \citealp{Zinnecker2007,M07}). Embedded MYSOs and {\uchii}  evolve, however, on longer timescales, 
of the order of $10^5-10^6$\,yr according to stellar evolution models with accretion rates of $M_{\rm acc}\sim10^{-3}$\,\msol\,yr$^{-1}$ (e.g.\,\citealp{Churchwell2002,Hoare2007,Kuiper2013}). 
This is at least up to an order of magnitude 
longer  than estimated for the collapse of their parent clump. 
This means that the clump material should have already  collapsed  leaving a considerably less dense and less massive clump 
 towards the more evolved objects. This apparent contradiction to the observations can only be explained if 
there is mass replenishment onto the clump on large scales fuelling a continuous star formation, 
or if we selectively pick  here the youngest of the galactic {\uchii} regions still surrounded by their natal clump. 
Case studies of massive cluster progenitors
show evidence both for global collapse and for flows of dense gas (e.g.\,  
\citealp{Schneider2010,Csengeri2011a,Csengeri2011b,Peretto2013,Tackenberg2014}). 
This points to a scenario  
where massive clumps undergo star formation, while
gas replenishment arrives on larger scales, 
and maintains the observed similar mass range for these clumps.

\subsection{Initial stages of massive proto-clusters: Prediction for fragmentation properties}\label{sec:phys-selected}
We overview the physical properties of our infrared-quiet sample 
in comparison with typical IRDCs \citep{Rathborne2006}, Herschel clumps embedded in IRDCs \citep{Traficante2015},
and clumps in the Cygnus-X high-mass star-forming region \citep{M07}
in  Table\,\ref{tab:comp}.
While IRDCs are typically more extended structures, they have lower masses
and exhibit smaller volume densities. Herschel selected clumps span a much broader range of physical conditions and
 cover similarly high-mass clumps as our sample. They are found to be, however, more extended and are therefore on average lower density structures.
The ATLASGAL counterparts of IRDCs selected by \citet{Kainulainen2013}
correspond to sources below the flux density threshold used for this study. 
The   \at\  sources used here are the brightest and therefore the most massive among the compact millimetre sources in the inner Galaxy. Their typical size of $\sim 0.4$\,pc corresponds to protoclusters. 
We can therefore discuss our sample in terms of the precursors of the richest clusters in the Galaxy.

\begin{figure}[!htpb]
\centering
\includegraphics[width=6.5cm,angle=90]{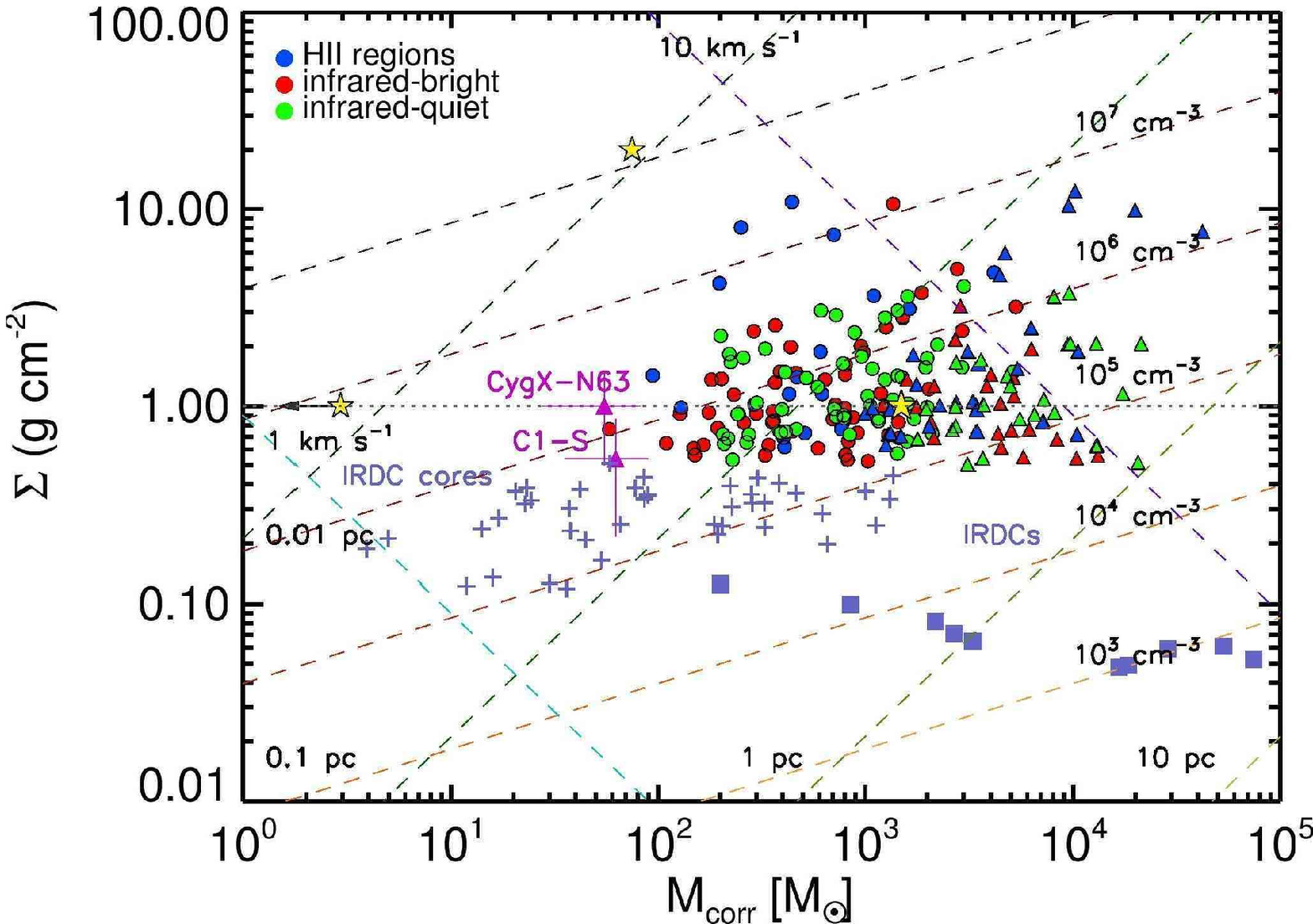}
\caption{
Surface-density, $\Sigma$, versus temperature corrected mass estimates (using $T_{\rm dust}=30$\,K for the infrared-bright clumps). The coloured dashed lines in different shades show constant radius (green), $n_{\rm H}$ number density (red), and escape speed (blue) (see\,\citealp{Tan2014}). Colours and symbols are the same as in Fig.\,\ref{fig:histo_mass}.
We mark two massive cores with $M_{\rm MDC}=60$\,\msol\ (C1-S; \citealt{Tan2013}) and 55\,\msol\ (CygX-N63; \citealp{Bontemps2010}). For comparison IRDC clumps \citep{Kainulainen2013} and cores are shown \citep{BT2012}.
The dotted line marks the
1\,g\,cm$^{-3}$ value of \citet{Krumholz2008}.}
\label{fig:mass-surface}
\end{figure}

In Fig.\,\ref{fig:mass-surface} we show the temperature corrected mass versus the surface 
density distribution of the entire sample following \citet{Tan2013}, with different colours corresponding to the
infrared-bright and infrared-quiet subsamples.
It is clear that the sample discussed here lies, on average, close to the theoretical limit
for high-mass star formation around $\Sigma\simeq$1\,g\,cm$^{-2}$ \citep{Krumholz2008} supporting the scenario that it is  capable of forming high-mass stars and clusters.  Taking the two extreme profiles of constant density $n(r)\sim r^0$ and power law with $n(r)\sim r^{-2}$,
we mark with stars the predicted fragmentation on scales smaller by a factor of ten, approaching the size-scales of a few thousand AU corresponding to individual collapsing objects (e.g.\,\citealp{Bontemps2010}). 
The uniform density profile, where the surface density stays constant, predicts fragments below  solar mass,
which would correspond to clumps forming a rich cluster of low-mass stars. The case of the steeper density profile 
predicts more massive structures on smaller scales, reaching a mass range of several tens of solar masses.
Envelope masses of this order are expected to form high-mass stars. 

Some of the well-studied sources of our selection
have been subject to high angular resolution follow-ups. 
The extreme clumps of the W43 complex, MM1 to MM3~\citep{Motte2003,Sridharan2014} 
 hosts one of the most extreme sites of active high-mass star formation in 
terms of mass and youth\,\citep{Louvet2014}. 
Massive envelopes have been identified towards other
clumps as well, for example G35.03+0.35 (G35.0250+0.3501) \citep{Beltran2014}, G35.20-0.74 (G35.1976$-$0.7427) \citep{SM2013},
and SDC335-MM1(G335.5857$-$0.2906) \citep{Peretto2013}.
High angular resolution observations with ALMA towards a large fraction of  the infrared-quiet sources of the presented sample 
indeed reveal a significant population of massive dense cores \citep[submitted]{Csengeri2016c}.

\begin{table*}
\caption{Comparison of the properties of the infrared-quiet sources with the literature.
}\label{tab:comp}
\centering
\begin{tabular}{lccccrrrrrr}
\hline\hline
&{ATLASGAL infrared-quiet}\tablefootmark{\textdagger} & Herschel-IRDC\tablefootmark{a} & IRDC & CygX  \\
&clumps   & clumps & clumps & clumps\\
\hline
Distance [kpc] & $1.15 -15.60 (4.47)$ & $0.12-16.45 (4.18)$ & $1.8-7.1 (3.7) $\tablefootmark{b} & 1.4\tablefootmark{d} \\
Size [pc]          & $0.08-1.63  (0.36)$ &  $0.02-2.6 (0.6) $ & $1-3$ \tablefootmark{b}& 0.56\tablefootmark{e}\\
$M$ [\msol]{$^\ddagger$}& $200-21000$   ($      1400$) & $0.11-16000$   ($      700$)   &$120-16000 (950)$\tablefootmark{b} & 680\tablefootmark{e}\\
$\bar{n}_{\rm corr}$ [$\times10^5$ cm$^{-3}$] & $0.20-18.46$  ($2.39$)  & $0-0.58$ (0.01) & $0.001-0.1$\tablefootmark{b} & 0.12\tablefootmark{e}\\
\hline
\end{tabular}
\tablefoot{
\tablefoottext{\textdagger}{Range of minimum to maximum values,  the median is given between parentheses.} 
\tablefoottext{$^\ddagger$}{For the adopted dust temperature see their original work.} 
\tablefoottext{a}{Adopted from the values listed in Table\,2 from \citet{Traficante2015}.}
\tablefoottext{b}{Adopted from the values listed in Table\,3 from \citet{Rathborne2006}.}
\tablefoottext{c}{Adopted from the values listed in Table\,1 from \citet{Rathborne2006}.}
\tablefoottext{d}{Distance from \citet{Rygl2012}.}
\tablefoottext{e}{Scaled to 1.4\,kpc from \citet{M07}.}}
\end{table*}

\section{Summary and conclusions}\label{sec:sum}

We investigate the properties of a flux limited sample of massive clumps from the \at\ survey. 
Based on mid-infrared photometry, we divide the sample into infrared-quiet and infrared-bright clumps. 
Infrared-bright clumps peak on embedded objects with $L_{\rm bol}>10^4$\,\lsol\ 
corresponding to MYSOs or high-mass stars with {\uchii} regions, 
while infrared-quiet clumps lack such bright embedded objects.
 From the initial sample of 210 sources above a 5\,Jy beam-averaged peak flux density, 
we identify \nirq\ infrared-quiet massive clumps, of which 50 are located within $d<$ 4.5\,kpc. 
About one-fourth of them have not been recognised by other surveys. 

We estimate their physical properties based on the dust 
emission, and find that statistically both samples exhibit similar properties in terms of mass and size.
Only their embedded (proto)stellar content is different, the infrared-quiet sample lacks luminous embedded objects.
This suggests  that infrared-quiet clumps are the precursors of the bright clumps, and thus they 
likely follow the same evolutionary path.
A substantial fraction of infrared-quiet clumps exhibit signatures of ongoing star formation activity such as
emission from shocked gas associated with material ejection. In particular,
selective star formation tracers, such as EGO activity and Class II methanol maser emission,
suggest that at least some infrared-quiet massive  
clumps host high-mass protostars. This makes them excellent candidates to be the potential 
precursors of the most massive stars forming in our Galaxy.

The overall low ratio of virial mass to mass  of the sample suggests a collapse on the clump scale, which is
consistent with a fast, dynamic star formation process. In this scenario  massive clumps start to fragment and collapse before their final mass is accumulated indicating that strong self-gravity and global collapse is needed to build up rich clusters and the most massive stars.

High angular-resolution observations of infrared-quiet massive clumps are the next step to reveal their protostellar content 
and uncover the initial conditions for high-mass star formation.

\setlength{\tabcolsep}{4pt}
\fontsize{8pt}{8pt}
\selectfont
\begin{longtab}
\begin{landscape}
\begin{longtable}{lrrrrccccccccccccccc}
\caption{Flux limited sample of ATLASGAL sources with distance estimate and the corresponding dust properties.}\label{tab:table-dust}\\
\hline\hline
\small
ATLASGAL\tablefootmark{a} & Gal. long. & Gal. lat. & $F_\nu$ & $S_\nu$ & $\Theta_{\rm A}\times\Theta_{\rm B}$ & PA & $d$\tablefootmark{b} & $d_{\rm min}$ & $d_{\rm max}$ & Ref. & $M_{\rm corr}$ & $R_{\rm 90}$ & $\Sigma$ & $\bar{n}$  & IRq & GMC/SIRDC & EGO\\
   source name        & [$^\circ$]  & [$^\circ$] & [Jy/beam] & [Jy] &  [\arcsec$\times$\arcsec] & [$^\circ$] & [kpc] &  [kpc] & [kpc] & & [\msol] & [pc] & [g\,cm$^{-2}$] & [$\times10^4$\,cm$^{-3}$] & & & \\
\hline
\endfirsthead
\caption{continued.}\\
\hline\hline
ATLASGAL\tablefootmark{a} & Gal. long. & Gal. lat. & $F_\nu$ & $S_\nu$ & $\Theta_{\rm A}\times\Theta_{\rm B}$ & PA & $d$\tablefootmark{b} & $d_{\rm min}$ & $d_{\rm max}$ & Ref. & $M_{\rm corr}$ & $R_{\rm 90}$ & $\Sigma$ & $\bar{n}$  & IRq & GMC/SIRDC & EGO \\
   source name        & [$^\circ$]  & [$^\circ$] & [Jy/beam] & [Jy] &  [\arcsec$\times$\arcsec] & [$^\circ$] & [kpc] &  [kpc] & [kpc] & & [\msol] & [pc] & [g\,cm$^{-2}$] & [$\times10^4$\,cm$^{-3}$] & & & \\
\hline\endhead
\hline
\endfoot
G300.9688+1.1456   &  300.9688   &      1.1456   &    6.5   &   16.2   &  32.1  $\times$    28.4   &   125   &     4.3   &     1.6   &     7.0   &  (1)   &     920   &     0.4   &     0.4   &     5.3   &  n  &                    --  &                    --  \\
G301.1365-0.2256$^\star$   &  301.1365   &     -0.2256   &   21.4   &   34.5   &  26.3  $\times$    22.6   &   102   &     4.2   &     3.5   &    10.4   &  (1)   &    1877   &     0.3   &     1.8   &    43.6   &  n  &         301.147-0.228  &                    --  \\
G305.2083+0.2063$^\star$   &  305.2083   &      0.2063   &   13.5   &   26.1   &  29.7  $\times$    24.0   &    15   &     3.5   &     1.5   &     6.6   &  (17)   &     985   &     0.3   &     0.9   &    21.0   &  n  &  G305/ SIRDC 305.209+0.205  &                    --  \\
G305.2703-0.0090   &  305.2703   &     -0.0090   &    6.7   &   18.8   &  37.3  $\times$    27.9   &   126   &     3.5   &     1.3   &     4.9   &  (17)   &     710   &     0.4   &     0.3   &     5.6   &  n  &  G305/ SIRDC 305.270-0.018  &                    --  \\
G305.3570+0.2033   &  305.3570   &      0.2033   &    7.6   &   20.5   &  35.0  $\times$    28.4   &    66   &     3.5   &     1.5   &     6.6   &  (17)   &     774   &     0.4   &     0.4   &     6.7   &  n  &                  G305  &                    --  \\
G305.3617+0.1514   &  305.3617   &      0.1514   &    5.4   &    9.6   &  29.1  $\times$    22.3   &    80   &     3.5   &     1.5   &     6.6   &  (17)   &     361   &     0.2   &     0.4   &    10.5   &  n  &                  G305  &                    --  \\
G305.3666+0.2130   &  305.3666   &      0.2130   &    5.7   &   10.2   &  26.5  $\times$    24.8   &    80   &     3.5   &     1.3   &     6.6   &  (17)   &     792   &     0.2   &     0.9   &    22.0   &  y  &  G305/ SIRDC 305.364+0.215  &                    --  \\
\hline
\end{longtable}
 \tablefoot{The full table is available only in electronic form at the CDS via anonymous ftp to cdsarc.u-strasbg.fr (130.79.125.5) or via http://cdsweb.u-strasbg.fr/cgi-bin/qcat?J/A\&A/. Column 1 gives the corresponding ATLASGAL source name from \citet{Csengeri2014} including the Galactic coordinates, while column 2 listst the assigned GMC complex, where appropriate. Column 3-4 give the peak flux density, and the integrated flux density assuming a Gaussian flux distribution from \citet{Csengeri2014}, respectively. Column 56 gives the beam averaged peak H$_2$ column density using Eq.\,\ref{eq:nh2} in Sect.\,\ref{sec:mass}. Columns 6-8 list the major and minor axis of the fitted Gaussian FWHM to the source, and the position angle. Column 9 gives the used distance estimate. Column 10-11 give the minimum and the maximum distances. Column 12 gives the reference for the distance. 
\tablefoottext{\textdagger}{We revised the spectrophotometric distance of 1.81\,kpc by \citep{Moises2011}.
{$\star$}{Sources also included in the ATLASGAL top100 selection \citet{Giannetti2014}.}}}
\tablebib{
(1)~{This work.}
(2)~\citet{Sato};
(3)~\citet{Zhang2013};
(4)~\citet{Zhang2014};
(5)~\citet{Immer};
(6)~\citet{Sanna2014};
(7)~\citet{Moises2011};
(8)~\citet{Wu2014};
(9)~\citet{Wu2012};
(10)~\citet{Reid2014};
(11)~\citet{Dutra2003};
(12)~\citet{Straw1987};
(13)~\citet{Xu2011};
(14)~\citet{Sanna2009};
(15)~\citet{Leurini2011};
(16)~\citet{Kurayama2012};
(17)~\citet{Russeil2003};
(18)~\citet{Reid2009};
(19)~\citet{Brunthaler2009}
(20)~\citet{Urquhart2013};
(21)~\citet{KRISHNAN2015};
(22)~\citet{Motogi}}
\end{landscape}
\end{longtab}

\begin{acknowledgement}
We thank the referee for the careful reading of the manuscript.
This work was partially funded by the ERC Advanced Investigator Grant GLOSTAR (247078).
T.Cs. acknowledges support from the \emph{Deut\-sche For\-schungs\-ge\-mein\-schaft, DFG\/}  via the SPP (priority programme) 1573 `Physics of the ISM'.  T.Cs. thanks R. Gutermuth for making available the PhotVis tool.
This paper is based on data acquired with the Atacama Pathfinder EXperiment (APEX). APEX
is a collaboration between the Max Planck Institute for Radioastronomy, the European
Southern Observatory, and the Onsala Space Observatory.
This research made use of data products from the Midcourse Space Experiment. Processing of the data was funded by the Ballistic Missile Defense Organization with additional support from NASA Office of Space Science. This research has also made use of the NASA/ IPAC Infrared Science Archive, which is operated by the Jet Propulsion Laboratory, California Institute of Technology, under contract with the National Aeronautics and Space Administration. 
This publication makes use of data products from the Wide-field Infrared Survey Explorer, which is a joint project of the University of California, Los Angeles, and the Jet Propulsion Laboratory/California Institute of Technology, funded by the National Aeronautics and Space Administration.
This research made use of Montage, funded by the National Aeronautics and Space Administration's Earth Science Technology Office, Computation Technologies Project, under Cooperative Agreement Number NCC5-626 between NASA and the California Institute of Technology. Montage is maintained by the NASA/IPAC Infrared Science Archive.
\end{acknowledgement}

\bibliographystyle{aa}
\bibliography{ag-extr}

\begin{thebibliography}{144}
\expandafter\ifx\csname natexlab\endcsname\relax\def\natexlab#1{#1}\fi

\bibitem[{{Avison} {et~al.}(2015){Avison}, {Peretto}, {Fuller},
  {Duarte-Cabral}, {Traficante}, \& {Pineda}}]{Avison2015}
{Avison}, A., {Peretto}, N., {Fuller}, G.~A., {et~al.} 2015, \aap, 577, A30

\bibitem[{{Bally} {et~al.}(2010){Bally}, {Anderson}, {Battersby}, {Calzoletti},
  {Digiorgio}, {Faustini}, {Ginsburg}, {Li}, {Nguyen Luong}, {Molinari},
  {Motte}, {Pestalozzi}, {Plume}, {Rodon}, {Schilke}, {Schlingman},
  {Schneider-Bontemps}, {Shirley}, {Stringfellow}, {Testi}, {Traficante},
  {Veneziani}, \& {Zavagno}}]{Bally2010}
{Bally}, J., {Anderson}, L.~D., {Battersby}, C., {et~al.} 2010, \aap, 518, L90

\bibitem[{{Battersby} {et~al.}(2011){Battersby}, {Bally}, {Ginsburg},
  {Bernard}, {Brunt}, {Fuller}, {Martin}, {Molinari}, {Mottram}, {Peretto},
  {Testi}, \& {Thompson}}]{Battersby2011}
{Battersby}, C., {Bally}, J., {Ginsburg}, A., {et~al.} 2011, \aap, 535, A128

\bibitem[{{Beltr{\'a}n} {et~al.}(2014){Beltr{\'a}n}, {S{\'a}nchez-Monge},
  {Cesaroni}, {Kumar}, {Galli}, {Walmsley}, {Etoka}, {Furuya}, {Moscadelli},
  {Stanke}, {van der Tak}, {Vig}, {Wang}, {Zinnecker}, {Elia}, \&
  {Schisano}}]{Beltran2014}
{Beltr{\'a}n}, M.~T., {S{\'a}nchez-Monge}, {\'A}., {Cesaroni}, R., {et~al.}
  2014, \aap, 571, A52

\bibitem[{{Benjamin} {et~al.}(2003){Benjamin}, {Churchwell}, {Babler}, {Bania},
  {Clemens}, {Cohen}, {Dickey}, {Indebetouw}, {Jackson}, {Kobulnicky},
  {Lazarian}, {Marston}, {Mathis}, {Meade}, {Seager}, {Stolovy}, {Watson},
  {Whitney}, {Wolff}, \& {Wolfire}}]{Benjamin2003}
{Benjamin}, R.~A., {Churchwell}, E., {Babler}, B.~L., {et~al.} 2003, \pasp,
  115, 953

\bibitem[{{Bergin} \& {Tafalla}(2007)}]{BT2007}
{Bergin}, E.~A. \& {Tafalla}, M. 2007, \araa, 45, 339

\bibitem[{{Bernard} {et~al.}(2010){Bernard}, {Paradis}, {Marshall}, {Montier},
  {Lagache}, {Paladini}, {Veneziani}, {Brunt}, {Mottram}, {Martin},
  {Ristorcelli}, {Noriega-Crespo}, {Compi{\`e}gne}, {Flagey}, {Anderson},
  {Popescu}, {Tuffs}, {Reach}, {White}, {Benedetti}, {Calzoletti}, {Digiorgio},
  {Faustini}, {Juvela}, {Joblin}, {Joncas}, {Mivilles-Deschenes}, {Olmi},
  {Traficante}, {Piacentini}, {Zavagno}, \& {Molinari}}]{Bernard2010}
{Bernard}, J.-P., {Paradis}, D., {Marshall}, D.~J., {et~al.} 2010, \aap, 518,
  L88

\bibitem[{{Bertoldi} \& {McKee}(1992)}]{BertoldiMcKee1992}
{Bertoldi}, F. \& {McKee}, C.~F. 1992, \apj, 395, 140

\bibitem[{{Beuther} {et~al.}(2004){Beuther}, {Zhang}, {Greenhill}, {Reid},
  {Wilner}, {Keto}, {Marrone}, {Ho}, {Moran}, {Rao}, {Shinnaga}, \&
  {Liu}}]{Beuther2004}
{Beuther}, H., {Zhang}, Q., {Greenhill}, L.~J., {et~al.} 2004, \apjl, 616, L31

\bibitem[{{Bohlin} {et~al.}(1978){Bohlin}, {Savage}, \& {Drake}}]{Bohlin1978}
{Bohlin}, R.~C., {Savage}, B.~D., \& {Drake}, J.~F. 1978, \apj, 224, 132

\bibitem[{{Bonnell} \& {Bate}(2006)}]{BonnellBate2006}
{Bonnell}, I.~A. \& {Bate}, M.~R. 2006, \mnras, 370, 488

\bibitem[{{Bonnell} {et~al.}(2001){Bonnell}, {Bate}, {Clarke}, \&
  {Pringle}}]{Bonnell2001}
{Bonnell}, I.~A., {Bate}, M.~R., {Clarke}, C.~J., \& {Pringle}, J.~E. 2001,
  \mnras, 323, 785

\bibitem[{{Bontemps} {et~al.}(2010){Bontemps}, {Motte}, {Csengeri}, \&
  {Schneider}}]{Bontemps2010}
{Bontemps}, S., {Motte}, F., {Csengeri}, T., \& {Schneider}, N. 2010, \aap,
  524, A18

\bibitem[{{Brand} \& {Blitz}(1993)}]{BB1993}
{Brand}, J. \& {Blitz}, L. 1993, \aap, 275, 67

\bibitem[{{Breen} {et~al.}(2015){Breen}, {Fuller}, {Caswell}, {Green},
  {Avison}, {Ellingsen}, {Gray}, {Pestalozzi}, {Quinn}, {Richards}, {Thompson},
  \& {Voronkov}}]{Breen2015_MMBV}
{Breen}, S.~L., {Fuller}, G.~A., {Caswell}, J.~L., {et~al.} 2015, \mnras, 450,
  4109

\bibitem[{{Bronfman} {et~al.}(1996){Bronfman}, {Nyman}, \&
  {May}}]{Bronfman1996}
{Bronfman}, L., {Nyman}, L.-A., \& {May}, J. 1996, \aaps, 115, 81

\bibitem[{{Butler} \& {Tan}(2012)}]{BT2012}
{Butler}, M.~J. \& {Tan}, J.~C. 2012, \apj, 754, 5

\bibitem[{{Carey} {et~al.}(2009){Carey}, {Noriega-Crespo}, {Mizuno}, {Shenoy},
  {Paladini}, {Kraemer}, {Price}, {Flagey}, {Ryan}, {Ingalls}, {Kuchar},
  {Pinheiro Gon{\c c}alves}, {Indebetouw}, {Billot}, {Marleau}, {Padgett},
  {Rebull}, {Bressert}, {Ali}, {Molinari}, {Martin}, {Berriman}, {Boulanger},
  {Latter}, {Miville-Deschenes}, {Shipman}, \& {Testi}}]{Carey2009}
{Carey}, S.~J., {Noriega-Crespo}, A., {Mizuno}, D.~R., {et~al.} 2009, \pasp,
  121, 76

\bibitem[{{Caswell} {et~al.}(2010){Caswell}, {Fuller}, {Green}, {Avison},
  {Breen}, {Brooks}, {Burton}, {Chrysostomou}, {Cox}, {Diamond}, {Ellingsen},
  {Gray}, {Hoare}, {Masheder}, {McClure-Griffiths}, {Pestalozzi}, {Phillips},
  {Quinn}, {Thompson}, {Voronkov}, {Walsh}, {Ward-Thompson}, {Wong-McSweeney},
  {Yates}, \& {Cohen}}]{Caswell2010_MMBI}
{Caswell}, J.~L., {Fuller}, G.~A., {Green}, J.~A., {et~al.} 2010, \mnras, 404,
  1029

\bibitem[{{Caswell} {et~al.}(2011){Caswell}, {Fuller}, {Green}, {Avison},
  {Breen}, {Ellingsen}, {Gray}, {Pestalozzi}, {Quinn}, {Thompson}, \&
  {Voronkov}}]{Caswell2011_MMBIII}
{Caswell}, J.~L., {Fuller}, G.~A., {Green}, J.~A., {et~al.} 2011, \mnras, 417,
  1964

\bibitem[{{Chen} {et~al.}(2013){Chen}, {Gan}, {Ellingsen}, {He}, {Shen}, \&
  {Titmarsh}}]{Chen2013}
{Chen}, X., {Gan}, C.-G., {Ellingsen}, S.~P., {et~al.} 2013, \apjs, 206, 9

\bibitem[{{Churchwell}(2002)}]{Churchwell2002}
{Churchwell}, E. 2002, \araa, 40, 27

\bibitem[{{Csengeri} {et~al.}(2011{\natexlab{a}}){Csengeri}, {Bontemps},
  {Schneider}, {Motte}, \& {Dib}}]{Csengeri2011a}
{Csengeri}, T., {Bontemps}, S., {Schneider}, N., {Motte}, F., \& {Dib}, S.
  2011{\natexlab{a}}, \aap, 527, A135

\bibitem[{{Csengeri} {et~al.}(2011{\natexlab{b}}){Csengeri}, {Bontemps},
  {Schneider}, {Motte}, {Gueth}, \& {Hora}}]{Csengeri2011b}
{Csengeri}, T., {Bontemps}, S., {Schneider}, N., {et~al.} 2011{\natexlab{b}},
  \apjl, 740, L5

\bibitem[{{Csengeri} {et~al.}(2016{\natexlab{a}}){Csengeri}, {Bontemps}, \& {et
  al.}}]{Csengeri2016c}
{Csengeri}, T., {Bontemps}, S., S., \& {et al.} 2016{\natexlab{a}}, \aap

\bibitem[{{Csengeri} {et~al.}(2016{\natexlab{b}}){Csengeri}, {Leurini},
  {Wyrowski}, {Urquhart}, {Menten}, {Walmsley}, {Bontemps}, {Wienen},
  {Beuther}, {Motte}, {Nguyen-Luong}, {Schilke}, {Schuller}, {Zavagno}, \&
  {Sanna}}]{Csengeri2016a}
{Csengeri}, T., {Leurini}, S., {Wyrowski}, F., {et~al.} 2016{\natexlab{b}},
  \aap, 586, A149

\bibitem[{{Csengeri} {et~al.}(2014){Csengeri}, {Urquhart}, {Schuller}, {Motte},
  {Bontemps}, {Wyrowski}, {Menten}, {Bronfman}, {Beuther}, {Henning}, {Testi},
  {Zavagno}, \& {Walmsley}}]{Csengeri2014}
{Csengeri}, T., {Urquhart}, J.~S., {Schuller}, F., {et~al.} 2014, \aap, 565,
  A75

\bibitem[{{Csengeri} {et~al.}(2016{\natexlab{c}}){Csengeri}, {Weiss},
  {Wyrowski}, {Menten}, {Urquhart}, {Leurini}, {Schuller}, {Beuther},
  {Bontemps}, {Bronfman}, {Henning}, \& {Schneider}}]{Csengeri2016b}
{Csengeri}, T., {Weiss}, A., {Wyrowski}, F., {et~al.} 2016{\natexlab{c}}, \aap,
  585, A104

\bibitem[{{Cutri} {et~al.}(2012){Cutri}, {Wright}, {Conrow}, {Bauer},
  {Benford}, {Brandenburg}, {Dailey}, {Eisenhardt}, {Evans}, {Fajardo-Acosta},
  {Fowler}, {Gelino}, {Grillmair}, {Harbut}, {Hoffman}, {Jarrett},
  {Kirkpatrick}, {Leisawitz}, {Liu}, {Mainzer}, {Marsh}, {Masci}, {McCallon},
  {Padgett}, {Ressler}, {Royer}, {Skrutskie}, {Stanford}, {Wyatt}, {Tholen},
  {Tsai}, {Wachter}, {Wheelock}, {Yan}, {Alles}, {Beck}, {Grav}, {Masiero},
  {McCollum}, {McGehee}, {Papin}, \& {Wittman}}]{Cutri2012}
{Cutri}, R.~M., {Wright}, E.~L., {Conrow}, T., {et~al.} 2012, {Explanatory
  Supplement to the WISE All-Sky Data Release Products}, Tech. rep.

\bibitem[{{Cyganowski} {et~al.}(2014){Cyganowski}, {Brogan}, {Hunter},
  {Graninger}, {{\"O}berg}, {Vasyunin}, {Zhang}, {Friesen}, \&
  {Schnee}}]{Cyganowski2014}
{Cyganowski}, C.~J., {Brogan}, C.~L., {Hunter}, T.~R., {et~al.} 2014, \apjl,
  796, L2

\bibitem[{{Cyganowski} {et~al.}(2008){Cyganowski}, {Whitney}, {Holden},
  {Braden}, {Brogan}, {Churchwell}, {Indebetouw}, {Watson}, {Babler},
  {Benjamin}, {Gomez}, {Meade}, {Povich}, {Robitaille}, \&
  {Watson}}]{Cyganowski2008}
{Cyganowski}, C.~J., {Whitney}, B.~A., {Holden}, E., {et~al.} 2008, \aj, 136,
  2391

\bibitem[{{Da Rio} {et~al.}(2012){Da Rio}, {Robberto}, {Hillenbrand},
  {Henning}, \& {Stassun}}]{DaRio2012}
{Da Rio}, N., {Robberto}, M., {Hillenbrand}, L.~A., {Henning}, T., \&
  {Stassun}, K.~G. 2012, \apj, 748, 14

\bibitem[{{Draine}(2011)}]{Draine2011}
{Draine}, B.~T. 2011, {Physics of the Interstellar and Intergalactic Medium}

\bibitem[{{Duarte-Cabral} {et~al.}(2013){Duarte-Cabral}, {Bontemps}, {Motte},
  {Hennemann}, {Schneider}, \& {Andr{\'e}}}]{Duarte2013}
{Duarte-Cabral}, A., {Bontemps}, S., {Motte}, F., {et~al.} 2013, \aap, 558,
  A125

\bibitem[{{Dunham} {et~al.}(2011){Dunham}, {Rosolowsky}, {Evans}, {Cyganowski},
  \& {Urquhart}}]{Dunham2011}
{Dunham}, M.~K., {Rosolowsky}, E., {Evans}, II, N.~J., {Cyganowski}, C., \&
  {Urquhart}, J.~S. 2011, \apj, 741, 110

\bibitem[{{Dutra} {et~al.}(2003){Dutra}, {Bica}, {Soares}, \&
  {Barbuy}}]{Dutra2003}
{Dutra}, C.~M., {Bica}, E., {Soares}, J., \& {Barbuy}, B. 2003, \aap, 400, 533

\bibitem[{{Elia} {et~al.}(2013){Elia}, {Molinari}, {Fukui}, {Schisano}, {Olmi},
  {Veneziani}, {Hayakawa}, {Pestalozzi}, {Schneider}, {Benedettini}, {di
  Giorgio}, {Ikhenaode}, {Mizuno}, {Onishi}, {Pezzuto}, {Piazzo}, {Polychroni},
  {Rygl}, {Yamamoto}, \& {Maruccia}}]{Elia2013}
{Elia}, D., {Molinari}, S., {Fukui}, Y., {et~al.} 2013, \apj, 772, 45

\bibitem[{{Engelbracht} {et~al.}(2007){Engelbracht}, {Blaylock}, {Su}, {Rho},
  {Rieke}, {Muzerolle}, {Padgett}, {Hines}, {Gordon}, {Fadda},
  {Noriega-Crespo}, {Kelly}, {Latter}, {Hinz}, {Misselt}, {Morrison},
  {Stansberry}, {Shupe}, {Stolovy}, {Wheaton}, {Young}, {Neugebauer},
  {Wachter}, {P{\'e}rez-Gonz{\'a}lez}, {Frayer}, \&
  {Marleau}}]{Engelbracht2007}
{Engelbracht}, C.~W., {Blaylock}, M., {Su}, K.~Y.~L., {et~al.} 2007, \pasp,
  119, 994

\bibitem[{{Galv{\'a}n-Madrid} {et~al.}(2013){Galv{\'a}n-Madrid}, {Liu},
  {Zhang}, {Pineda}, {Peng}, {Zhang}, {Keto}, {Ho}, {Rodr{\'{\i}}guez},
  {Zapata}, {Peters}, \& {De Pree}}]{Galvn-Madrid2013}
{Galv{\'a}n-Madrid}, R., {Liu}, H.~B., {Zhang}, Z.-Y., {et~al.} 2013, \apj,
  779, 121

\bibitem[{{Garay} {et~al.}(2004){Garay}, {Fa{\'u}ndez}, {Mardones}, {Bronfman},
  {Chini}, \& {Nyman}}]{Garay2004}
{Garay}, G., {Fa{\'u}ndez}, S., {Mardones}, D., {et~al.} 2004, \apj, 610, 313

\bibitem[{{Giannetti} {et~al.}(2014){Giannetti}, {Wyrowski}, {Brand},
  {Csengeri}, {Fontani}, {Walmsley}, {Nguyen Luong}, {Beuther}, {Schuller},
  {G{\"u}sten}, \& {Menten}}]{Giannetti2014}
{Giannetti}, A., {Wyrowski}, F., {Brand}, J., {et~al.} 2014, \aap, 570, A65

\bibitem[{{Ginsburg} {et~al.}(2015){Ginsburg}, {Bally}, {Battersby},
  {Youngblood}, {Darling}, {Rosolowsky}, {Arce}, \& {Lebr{\'o}n
  Santos}}]{Ginsburg2015}
{Ginsburg}, A., {Bally}, J., {Battersby}, C., {et~al.} 2015, \aap, 573, A106

\bibitem[{{Green} {et~al.}(2010){Green}, {Caswell}, {Fuller}, {Avison},
  {Breen}, {Ellingsen}, {Gray}, {Pestalozzi}, {Quinn}, {Thompson}, \&
  {Voronkov}}]{Green2010_MMBII}
{Green}, J.~A., {Caswell}, J.~L., {Fuller}, G.~A., {et~al.} 2010, \mnras, 409,
  913

\bibitem[{{G{\"u}sten} {et~al.}(2006){G{\"u}sten}, {Nyman}, {Schilke},
  {Menten}, {Cesarsky}, \& {Booth}}]{gusten2006}
{G{\"u}sten}, R., {Nyman}, L.~{\AA}., {Schilke}, P., {et~al.} 2006, \aap, 454,
  L13

\bibitem[{{Gutermuth} {et~al.}(2008){Gutermuth}, {Myers}, {Megeath}, {Allen},
  {Pipher}, {Muzerolle}, {Porras}, {Winston}, \& {Fazio}}]{Gutermuth2008}
{Gutermuth}, R.~A., {Myers}, P.~C., {Megeath}, S.~T., {et~al.} 2008, \apj, 674,
  336

\bibitem[{{Guzm{\'a}n} {et~al.}(2010){Guzm{\'a}n}, {Garay}, \&
  {Brooks}}]{Guzman2010}
{Guzm{\'a}n}, A.~E., {Garay}, G., \& {Brooks}, K.~J. 2010, \apj, 725, 734

\bibitem[{{Habibi} {et~al.}(2013){Habibi}, {Stolte}, {Brandner}, {Hu{\ss}mann},
  \& {Motohara}}]{Habibi2013}
{Habibi}, M., {Stolte}, A., {Brandner}, W., {Hu{\ss}mann}, B., \& {Motohara},
  K. 2013, \aap, 556, A26

\bibitem[{{Hoare} \& {Franco}(2007)}]{HF2007}
{Hoare}, M.~G. \& {Franco}, J. 2007, Astrophysics and Space Science
  Proceedings, 1, 61

\bibitem[{{Hoare} {et~al.}(2007){Hoare}, {Kurtz}, {Lizano}, {Keto}, \&
  {Hofner}}]{Hoare2007}
{Hoare}, M.~G., {Kurtz}, S.~E., {Lizano}, S., {Keto}, E., \& {Hofner}, P. 2007,
  Protostars and Planets V, 181

\bibitem[{{Hoare} {et~al.}(2012){Hoare}, {Purcell}, {Churchwell}, {Diamond},
  {Cotton}, {Chandler}, {Smethurst}, {Kurtz}, {Mundy}, {Dougherty}, {Fender},
  {Fuller}, {Jackson}, {Garrington}, {Gledhill}, {Goldsmith}, {Lumsden},
  {Mart{\'{\i}}}, {Moore}, {Muxlow}, {Oudmaijer}, {Pandian}, {Paredes},
  {Shepherd}, {Spencer}, {Thompson}, {Umana}, {Urquhart}, \&
  {Zijlstra}}]{Hoare2012}
{Hoare}, M.~G., {Purcell}, C.~R., {Churchwell}, E.~B., {et~al.} 2012, \pasp,
  124, 939

\bibitem[{{Hunter} {et~al.}(2000){Hunter}, {Churchwell}, {Watson}, {Cox},
  {Benford}, \& {Roelfsema}}]{Hunter2000}
{Hunter}, T.~R., {Churchwell}, E., {Watson}, C., {et~al.} 2000, \aj, 119, 2711

\bibitem[{{Immer} {et~al.}(2013){Immer}, {Reid}, {Menten}, {Brunthaler}, \&
  {Dame}}]{Immer}
{Immer}, K., {Reid}, M.~J., {Menten}, K.~M., {Brunthaler}, A., \& {Dame}, T.~M.
  2013, \aap, 553, A117

\bibitem[{{Jackson} {et~al.}(2008){Jackson}, {Finn}, {Rathborne}, {Chambers},
  \& {Simon}}]{Jackson2008}
{Jackson}, J.~M., {Finn}, S.~C., {Rathborne}, J.~M., {Chambers}, E.~T., \&
  {Simon}, R. 2008, \apj, 680, 349

\bibitem[{{Jackson} {et~al.}(2006){Jackson}, {Rathborne}, {Shah}, {Simon},
  {Bania}, {Clemens}, {Chambers}, {Johnson}, {Dormody}, {Lavoie}, \&
  {Heyer}}]{Jackson2006}
{Jackson}, J.~M., {Rathborne}, J.~M., {Shah}, R.~Y., {et~al.} 2006, \apjs, 163,
  145

\bibitem[{{Kainulainen} \& {Tan}(2013)}]{Kainulainen2013}
{Kainulainen}, J. \& {Tan}, J.~C. 2013, \aap, 549, A53

\bibitem[{{Kauffmann} {et~al.}(2008){Kauffmann}, {Bertoldi}, {Bourke}, {Evans},
  \& {Lee}}]{Kauffmann2008}
{Kauffmann}, J., {Bertoldi}, F., {Bourke}, T.~L., {Evans}, II, N.~J., \& {Lee},
  C.~W. 2008, \aap, 487, 993

\bibitem[{{Kauffmann} \& {Pillai}(2010)}]{Kauffmann2010}
{Kauffmann}, J. \& {Pillai}, T. 2010, \apjl, 723, L7

\bibitem[{{Kauffmann} {et~al.}(2013){Kauffmann}, {Pillai}, \&
  {Goldsmith}}]{Kauffmann2013}
{Kauffmann}, J., {Pillai}, T., \& {Goldsmith}, P.~F. 2013, \apj, 779, 185

\bibitem[{{Kauffmann} {et~al.}(2010){Kauffmann}, {Pillai}, {Shetty}, {Myers},
  \& {Goodman}}]{KP2010}
{Kauffmann}, J., {Pillai}, T., {Shetty}, R., {Myers}, P.~C., \& {Goodman},
  A.~A. 2010, \apj, 712, 1137

\bibitem[{{Keto}(2002)}]{Keto2002}
{Keto}, E. 2002, \apj, 580, 980

\bibitem[{{Keto}(2003)}]{Keto2003}
{Keto}, E. 2003, \apj, 599, 1196

\bibitem[{{Klaassen} {et~al.}(2012){Klaassen}, {Testi}, \&
  {Beuther}}]{Klaassen2012}
{Klaassen}, P.~D., {Testi}, L., \& {Beuther}, H. 2012, \aap, 538, A140

\bibitem[{Klein {et~al.}(2014)Klein, Ciechanowicz, Leinz, Heyminck, Gusten,
  Kasemann, Wunsch, Maier, \& Sekimoto}]{Klein_2014}
Klein, T., Ciechanowicz, M., Leinz, C., {et~al.} 2014, Terahertz Science and
  Technology, IEEE Transactions on, 4, 588

\bibitem[{{K{\"o}nig} {et~al.}(2016){K{\"o}nig}, {Urquhart}, {Csengeri},
  {Leurini}, {Wyrowski}, {Giannetti}, {Wienen}, {Pillai}, {Kauffmann},
  {Menten}, \& {Schuller}}]{Konig2016}
{K{\"o}nig}, C., {Urquhart}, J.~S., {Csengeri}, T., {et~al.} 2016, ArXiv
  e-prints

\bibitem[{{Krishnan} {et~al.}(2015){Krishnan}, {Ellingsen}, {Reid},
  {Brunthaler}, {Sanna}, {McCallum}, {Reynolds}, {Bignall}, {Phillips},
  {Dodson}, {Rioja}, {Caswell}, {Chen}, {Dawson}, {Fujisawa}, {Goedhart},
  {Green}, {Hachisuka}, {Honma}, {Menten}, {Shen}, {Voronkov}, {Walsh}, {Xu},
  {Zhang}, \& {Zheng}}]{KRISHNAN2015}
{Krishnan}, V., {Ellingsen}, S.~P., {Reid}, M.~J., {et~al.} 2015, \apj, 805,
  129

\bibitem[{{Krumholz} \& {McKee}(2008)}]{Krumholz2008}
{Krumholz}, M.~R. \& {McKee}, C.~F. 2008, \nat, 451, 1082

\bibitem[{{Kuiper} \& {Yorke}(2013)}]{Kuiper2013}
{Kuiper}, R. \& {Yorke}, H.~W. 2013, \apj, 772, 61

\bibitem[{{Kurayama} {et~al.}(2011){Kurayama}, {Nakagawa}, {Sawada-Satoh},
  {Sato}, {Honma}, {Sunada}, {Hirota}, \& {Imai}}]{Kurayama2012}
{Kurayama}, T., {Nakagawa}, A., {Sawada-Satoh}, S., {et~al.} 2011, \pasj, 63,
  513

\bibitem[{{Leurini} {et~al.}(2014){Leurini}, {Codella}, {L{\'o}pez-Sepulcre},
  {Gusdorf}, {Csengeri}, \& {Anderl}}]{Leurini2014}
{Leurini}, S., {Codella}, C., {L{\'o}pez-Sepulcre}, A., {et~al.} 2014, \aap,
  570, A49

\bibitem[{{Leurini} {et~al.}(2011){Leurini}, {Codella}, {Zapata},
  {Beltr{\'a}n}, {Schilke}, \& {Cesaroni}}]{Leurini2011}
{Leurini}, S., {Codella}, C., {Zapata}, L., {et~al.} 2011, \aap, 530, A12

\bibitem[{{L{\'o}pez} {et~al.}(2011){L{\'o}pez}, {Bronfman}, {Nyman}, {May}, \&
  {Garay}}]{Lopez2011}
{L{\'o}pez}, C., {Bronfman}, L., {Nyman}, L.-{\AA}., {May}, J., \& {Garay}, G.
  2011, \aap, 534, A131

\bibitem[{{Louvet} {et~al.}(2014){Louvet}, {Motte}, {Hennebelle}, {Maury},
  {Bonnell}, {Bontemps}, {Gusdorf}, {Hill}, {Gueth}, {Peretto},
  {Duarte-Cabral}, {Stephan}, {Schilke}, {Csengeri}, {Nguyen Luong}, \&
  {Lis}}]{Louvet2014}
{Louvet}, F., {Motte}, F., {Hennebelle}, P., {et~al.} 2014, \aap, 570, A15

\bibitem[{{Lumsden} {et~al.}(2013){Lumsden}, {Hoare}, {Urquhart}, {Oudmaijer},
  {Davies}, {Mottram}, {Cooper}, \& {Moore}}]{Lumsden2013}
{Lumsden}, S.~L., {Hoare}, M.~G., {Urquhart}, J.~S., {et~al.} 2013, \apjs, 208,
  11

\bibitem[{{McKee} \& {Ostriker}(2007)}]{McKee2007}
{McKee}, C.~F. \& {Ostriker}, E.~C. 2007, \araa, 45, 565

\bibitem[{{McKee} \& {Tan}(2003)}]{MT03}
{McKee}, C.~F. \& {Tan}, J.~C. 2003, \apj, 585, 850

\bibitem[{{Megeath} {et~al.}(2012){Megeath}, {Gutermuth}, {Muzerolle},
  {Kryukova}, {Flaherty}, {Hora}, {Allen}, {Hartmann}, {Myers}, {Pipher},
  {Stauffer}, {Young}, \& {Fazio}}]{Megeath2012}
{Megeath}, S.~T., {Gutermuth}, R., {Muzerolle}, J., {et~al.} 2012, \aj, 144,
  192

\bibitem[{{Megeath} \& {Tieftrunk}(1999)}]{Megeath1999}
{Megeath}, S.~T. \& {Tieftrunk}, A.~R. 1999, \apjl, 526, L113

\bibitem[{{Menten}(1991)}]{Menten1991}
{Menten}, K.~M. 1991, \apjl, 380, L75

\bibitem[{{Mois{\'e}s} {et~al.}(2011){Mois{\'e}s}, {Damineli}, {Figuer{\^e}do},
  {Blum}, {Conti}, \& {Barbosa}}]{Moises2011}
{Mois{\'e}s}, A.~P., {Damineli}, A., {Figuer{\^e}do}, E., {et~al.} 2011,
  \mnras, 411, 705

\bibitem[{{Molinari} {et~al.}(1996){Molinari}, {Brand}, {Cesaroni}, \&
  {Palla}}]{Molinari1996}
{Molinari}, S., {Brand}, J., {Cesaroni}, R., \& {Palla}, F. 1996, \aap, 308,
  573

\bibitem[{{Molinari} {et~al.}(2000){Molinari}, {Brand}, {Cesaroni}, \&
  {Palla}}]{Molinari2000}
{Molinari}, S., {Brand}, J., {Cesaroni}, R., \& {Palla}, F. 2000, \aap, 355,
  617

\bibitem[{{Molinari} {et~al.}(2008){Molinari}, {Pezzuto}, {Cesaroni}, {Brand},
  {Faustini}, \& {Testi}}]{Molinari2008}
{Molinari}, S., {Pezzuto}, S., {Cesaroni}, R., {et~al.} 2008, \aap, 481, 345

\bibitem[{{Mookerjea} {et~al.}(2004){Mookerjea}, {Kramer}, {Nielbock}, \&
  {Nyman}}]{Mookerjea2004}
{Mookerjea}, B., {Kramer}, C., {Nielbock}, M., \& {Nyman}, L.-{\AA}. 2004,
  \aap, 426, 119

\bibitem[{{Motogi} {et~al.}(2011){Motogi}, {Sorai}, {Habe}, {Honma},
  {Kobayashi}, \& {Sato}}]{Motogi}
{Motogi}, K., {Sorai}, K., {Habe}, A., {et~al.} 2011, \pasj, 63, 31

\bibitem[{{Motte} {et~al.}(2007){Motte}, {Bontemps}, {Schilke}, {Schneider},
  {Menten}, \& {Brogui{\`e}re}}]{M07}
{Motte}, F., {Bontemps}, S., {Schilke}, P., {et~al.} 2007, \aap, 476, 1243

\bibitem[{{Motte} {et~al.}(2003){Motte}, {Schilke}, \& {Lis}}]{Motte2003}
{Motte}, F., {Schilke}, P., \& {Lis}, D.~C. 2003, \apj, 582, 277

\bibitem[{{Mottram} {et~al.}(2011){Mottram}, {Hoare}, {Davies}, {Lumsden},
  {Oudmaijer}, {Urquhart}, {Moore}, {Cooper}, \& {Stead}}]{Mottram2011b}
{Mottram}, J.~C., {Hoare}, M.~G., {Davies}, B., {et~al.} 2011, \apjl, 730, L33

\bibitem[{{Mufson} \& {Liszt}(1979)}]{Mufson1979}
{Mufson}, S.~L. \& {Liszt}, H.~S. 1979, \apj, 232, 451

\bibitem[{{Nguyen Luong} {et~al.}(2011){Nguyen Luong}, {Motte}, {Schuller},
  {Schneider}, {Bontemps}, {Schilke}, {Menten}, {Heitsch}, {Wyrowski},
  {Carlhoff}, {Bronfman}, \& {Henning}}]{Quang2011}
{Nguyen Luong}, Q., {Motte}, F., {Schuller}, F., {et~al.} 2011, \aap, 529, A41

\bibitem[{{Ossenkopf} \& {Henning}(1994)}]{OH1994}
{Ossenkopf}, V. \& {Henning}, T. 1994, \aap, 291, 943

\bibitem[{{Peng} {et~al.}(2010){Peng}, {Wyrowski}, {van der Tak}, {Menten}, \&
  {Walmsley}}]{Peng2010}
{Peng}, T.-C., {Wyrowski}, F., {van der Tak}, F.~F.~S., {Menten}, K.~M., \&
  {Walmsley}, C.~M. 2010, \aap, 520, A84

\bibitem[{{Peretto} \& {Fuller}(2009)}]{Peretto2009}
{Peretto}, N. \& {Fuller}, G.~A. 2009, \aap, 505, 405

\bibitem[{{Peretto} {et~al.}(2013){Peretto}, {Fuller}, {Duarte-Cabral},
  {Avison}, {Hennebelle}, {Pineda}, {Andr{\'e}}, {Bontemps}, {Motte},
  {Schneider}, \& {Molinari}}]{Peretto2013}
{Peretto}, N., {Fuller}, G.~A., {Duarte-Cabral}, A., {et~al.} 2013, \aap, 555,
  A112

\bibitem[{{Peretto} {et~al.}(2016){Peretto}, {Lenfestey}, {Fuller},
  {Traficante}, {Molinari}, {Thompson}, \& {Ward-Thompson}}]{Peretto2016}
{Peretto}, N., {Lenfestey}, C., {Fuller}, G.~A., {et~al.} 2016, \aap, 590, A72

\bibitem[{{Purcell} {et~al.}(2013){Purcell}, {Hoare}, {Cotton}, {Lumsden},
  {Urquhart}, {Chandler}, {Churchwell}, {Diamond}, {Dougherty}, {Fender},
  {Fuller}, {Garrington}, {Gledhill}, {Goldsmith}, {Hindson}, {Jackson},
  {Kurtz}, {Mart{\'{\i}}}, {Moore}, {Mundy}, {Muxlow}, {Oudmaijer}, {Pandian},
  {Paredes}, {Shepherd}, {Smethurst}, {Spencer}, {Thompson}, {Umana}, \&
  {Zijlstra}}]{Purcell2013}
{Purcell}, C.~R., {Hoare}, M.~G., {Cotton}, W.~D., {et~al.} 2013, \apjs, 205, 1

\bibitem[{{Rathborne} {et~al.}(2005){Rathborne}, {Jackson}, {Chambers},
  {Simon}, {Shipman}, \& {Frieswijk}}]{Rathborne2005}
{Rathborne}, J.~M., {Jackson}, J.~M., {Chambers}, E.~T., {et~al.} 2005, \apjl,
  630, L181

\bibitem[{{Rathborne} {et~al.}(2006){Rathborne}, {Jackson}, \&
  {Simon}}]{Rathborne2006}
{Rathborne}, J.~M., {Jackson}, J.~M., \& {Simon}, R. 2006, \apj, 641, 389

\bibitem[{{Rathborne} {et~al.}(2008){Rathborne}, {Jackson}, {Zhang}, \&
  {Simon}}]{Rathborne2008}
{Rathborne}, J.~M., {Jackson}, J.~M., {Zhang}, Q., \& {Simon}, R. 2008, \apj,
  689, 1141

\bibitem[{{Reid} {et~al.}(2014){Reid}, {Menten}, {Brunthaler}, {Zheng}, {Dame},
  {Xu}, {Wu}, {Zhang}, {Sanna}, {Sato}, {Hachisuka}, {Choi}, {Immer},
  {Moscadelli}, {Rygl}, \& {Bartkiewicz}}]{Reid2014}
{Reid}, M.~J., {Menten}, K.~M., {Brunthaler}, A., {et~al.} 2014, \apj, 783, 130

\bibitem[{{Reid} {et~al.}(2009){Reid}, {Menten}, {Zheng}, {Brunthaler},
  {Moscadelli}, {Xu}, {Zhang}, {Sato}, {Honma}, {Hirota}, {Hachisuka}, {Choi},
  {Moellenbrock}, \& {Bartkiewicz}}]{Reid2009}
{Reid}, M.~J., {Menten}, K.~M., {Zheng}, X.~W., {et~al.} 2009, \apj, 700, 137

\bibitem[{{Russeil}(2003)}]{Russeil2003}
{Russeil}, D. 2003, \aap, 397, 133

\bibitem[{{Russeil} {et~al.}(2010){Russeil}, {Zavagno}, {Motte}, {Schneider},
  {Bontemps}, \& {Walsh}}]{Russeil2010}
{Russeil}, D., {Zavagno}, A., {Motte}, F., {et~al.} 2010, \aap, 515, A55

\bibitem[{{Rygl} {et~al.}(2012){Rygl}, {Brunthaler}, {Sanna}, {Menten}, {Reid},
  {van Langevelde}, {Honma}, {Torstensson}, \& {Fujisawa}}]{Rygl2012}
{Rygl}, K.~L.~J., {Brunthaler}, A., {Sanna}, A., {et~al.} 2012, \aap, 539, A79

\bibitem[{{S{\'a}nchez-Monge} {et~al.}(2013){S{\'a}nchez-Monge}, {Cesaroni},
  {Beltr{\'a}n}, {Kumar}, {Stanke}, {Zinnecker}, {Etoka}, {Galli}, {Hummel},
  {Moscadelli}, {Preibisch}, {Ratzka}, {van der Tak}, {Vig}, {Walmsley}, \&
  {Wang}}]{SM2013}
{S{\'a}nchez-Monge}, {\'A}., {Cesaroni}, R., {Beltr{\'a}n}, M.~T., {et~al.}
  2013, \aap, 552, L10

\bibitem[{{Sanna} {et~al.}(2014){Sanna}, {Reid}, {Menten}, {Dame}, {Zhang},
  {Sato}, {Brunthaler}, {Moscadelli}, \& {Immer}}]{Sanna2014}
{Sanna}, A., {Reid}, M.~J., {Menten}, K.~M., {et~al.} 2014, \apj, 781, 108

\bibitem[{{Sanna} {et~al.}(2009){Sanna}, {Reid}, {Moscadelli}, {Dame},
  {Menten}, {Brunthaler}, {Zheng}, \& {Xu}}]{Sanna2009}
{Sanna}, A., {Reid}, M.~J., {Moscadelli}, L., {et~al.} 2009, \apj, 706, 464

\bibitem[{{Sato} {et~al.}(2010){Sato}, {Reid}, {Brunthaler}, \&
  {Menten}}]{Sato}
{Sato}, M., {Reid}, M.~J., {Brunthaler}, A., \& {Menten}, K.~M. 2010, \apj,
  720, 1055

\bibitem[{{Schneider} {et~al.}(2010){Schneider}, {Csengeri}, {Bontemps},
  {Motte}, {Simon}, {Hennebelle}, {Federrath}, \& {Klessen}}]{Schneider2010}
{Schneider}, N., {Csengeri}, T., {Bontemps}, S., {et~al.} 2010, \aap, 520, A49

\bibitem[{{Schuller} {et~al.}(2009){Schuller}, {Menten}, {Contreras},
  {Wyrowski}, {Schilke}, {Bronfman}, {Henning}, {Walmsley}, {Beuther},
  {Bontemps}, {Cesaroni}, {Deharveng}, {Garay}, {Herpin}, {Lefloch}, {Linz},
  {Mardones}, {Minier}, {Molinari}, {Motte}, {Nyman}, {Reveret}, {Risacher},
  {Russeil}, {Schneider}, {Testi}, {Troost}, {Vasyunina}, {Wienen}, {Zavagno},
  {Kovacs}, {Kreysa}, {Siringo}, \& {Wei{\ss}}}]{Schuller2009}
{Schuller}, F., {Menten}, K.~M., {Contreras}, Y., {et~al.} 2009, \aap, 504, 415

\bibitem[{{Simon} {et~al.}(2006){Simon}, {Rathborne}, {Shah}, {Jackson}, \&
  {Chambers}}]{Simon2006}
{Simon}, R., {Rathborne}, J.~M., {Shah}, R.~Y., {Jackson}, J.~M., \&
  {Chambers}, E.~T. 2006, \apj, 653, 1325

\bibitem[{{Siringo} {et~al.}(2009){Siringo}, {Kreysa}, {Kov{\'a}cs},
  {Schuller}, {Wei{\ss}}, {Esch}, {Gem{\"u}nd}, {Jethava}, {Lundershausen},
  {Colin}, {G{\"u}sten}, {Menten}, {Beelen}, {Bertoldi}, {Beeman}, \&
  {Haller}}]{siringo2009}
{Siringo}, G., {Kreysa}, E., {Kov{\'a}cs}, A., {et~al.} 2009, \aap, 497, 945

\bibitem[{{Sobolev} {et~al.}(1997){Sobolev}, {Cragg}, \&
  {Godfrey}}]{Sobolev1997}
{Sobolev}, A.~M., {Cragg}, D.~M., \& {Godfrey}, P.~D. 1997, \mnras, 288, L39

\bibitem[{{Sridharan} {et~al.}(2002){Sridharan}, {Beuther}, {Schilke},
  {Menten}, \& {Wyrowski}}]{Sridharan2002}
{Sridharan}, T.~K., {Beuther}, H., {Schilke}, P., {Menten}, K.~M., \&
  {Wyrowski}, F. 2002, \apj, 566, 931

\bibitem[{{Sridharan} {et~al.}(2014){Sridharan}, {Rao}, {Qiu}, {Cortes}, {Li},
  {Pillai}, {Patel}, \& {Zhang}}]{Sridharan2014}
{Sridharan}, T.~K., {Rao}, R., {Qiu}, K., {et~al.} 2014, \apjl, 783, L31

\bibitem[{{Straw} {et~al.}(1987){Straw}, {Hyland}, {Jones}, {Harvey},
  {Wilking}, \& {Joy}}]{Straw1987}
{Straw}, S., {Hyland}, A.~R., {Jones}, T.~J., {et~al.} 1987, \apj, 314, 283

\bibitem[{{Svoboda} {et~al.}(2016){Svoboda}, {Shirley}, {Battersby},
  {Rosolowsky}, {Ginsburg}, {Ellsworth-Bowers}, {Pestalozzi}, {Dunham},
  {Evans}, {Bally}, \& {Glenn}}]{Svoboda2016}
{Svoboda}, B.~E., {Shirley}, Y.~L., {Battersby}, C., {et~al.} 2016, \apj, 822,
  59

\bibitem[{{Tackenberg} {et~al.}(2014){Tackenberg}, {Beuther}, {Henning},
  {Linz}, {Sakai}, {Ragan}, {Krause}, {Nielbock}, {Hennemann}, {Pitann}, \&
  {Schmiedeke}}]{Tackenberg2014}
{Tackenberg}, J., {Beuther}, H., {Henning}, T., {et~al.} 2014, \aap, 565, A101

\bibitem[{{Tackenberg} {et~al.}(2012){Tackenberg}, {Beuther}, {Henning},
  {Schuller}, {Wienen}, {Motte}, {Wyrowski}, {Bontemps}, {Bronfman}, {Menten},
  {Testi}, \& {Lefloch}}]{Tackenberg2012}
{Tackenberg}, J., {Beuther}, H., {Henning}, T., {et~al.} 2012, \aap, 540, A113

\bibitem[{{Tan} {et~al.}(2014){Tan}, {Beltr{\'a}n}, {Caselli}, {Fontani},
  {Fuente}, {Krumholz}, {McKee}, \& {Stolte}}]{Tan2014}
{Tan}, J.~C., {Beltr{\'a}n}, M.~T., {Caselli}, P., {et~al.} 2014, Protostars
  and Planets VI, 149

\bibitem[{{Tan} {et~al.}(2013){Tan}, {Kong}, {Butler}, {Caselli}, \&
  {Fontani}}]{Tan2013}
{Tan}, J.~C., {Kong}, S., {Butler}, M.~J., {Caselli}, P., \& {Fontani}, F.
  2013, \apj, 779, 96

\bibitem[{{Tan} {et~al.}(2016){Tan}, {Kong}, {Zhang}, {Fontani}, {Caselli}, \&
  {Butler}}]{Tan2016}
{Tan}, J.~C., {Kong}, S., {Zhang}, Y., {et~al.} 2016, \apjl, 821, L3

\bibitem[{{Thompson} {et~al.}(2005){Thompson}, {Gibb}, {Hatchell}, {Wyrowski},
  \& {Pestalozzi}}]{Thompson2005}
{Thompson}, M.~A., {Gibb}, A.~G., {Hatchell}, J.~H., {Wyrowski}, F., \&
  {Pestalozzi}, M.~R. 2005, in Protostars and Planets V Posters, 8109

\bibitem[{{Traficante} {et~al.}(2015){Traficante}, {Fuller}, {Peretto},
  {Pineda}, \& {Molinari}}]{Traficante2015}
{Traficante}, A., {Fuller}, G.~A., {Peretto}, N., {Pineda}, J.~E., \&
  {Molinari}, S. 2015, \mnras, 451, 3089

\bibitem[{{Urquhart} {et~al.}(2007){Urquhart}, {Busfield}, {Hoare}, {Lumsden},
  {Clarke}, {Moore}, {Mottram}, \& {Oudmaijer}}]{Urquhart2007}
{Urquhart}, J.~S., {Busfield}, A.~L., {Hoare}, M.~G., {et~al.} 2007, \aap, 461,
  11

\bibitem[{{Urquhart} {et~al.}(2014){Urquhart}, {Moore}, {Csengeri}, {Wyrowski},
  {Schuller}, {Hoare}, {Lumsden}, {Mottram}, {Thompson}, {Menten}, {Walmsley},
  {Bronfman}, {Pfalzner}, {K{\"o}nig}, \& {Wienen}}]{Urquhart2014}
{Urquhart}, J.~S., {Moore}, T.~J.~T., {Csengeri}, T., {et~al.} 2014, \mnras,
  443, 1555

\bibitem[{{Urquhart} {et~al.}(2013{\natexlab{a}}){Urquhart}, {Moore},
  {Schuller}, {Wyrowski}, {Menten}, {Thompson}, {Csengeri}, {Walmsley},
  {Bronfman}, \& {K{\"o}nig}}]{Urquhart2013a}
{Urquhart}, J.~S., {Moore}, T.~J.~T., {Schuller}, F., {et~al.}
  2013{\natexlab{a}}, \mnras, 431, 1752

\bibitem[{{Urquhart} {et~al.}(2013{\natexlab{c}}){Urquhart}, {Thompson},
  {Moore}, {Purcell}, {Hoare}, {Schuller}, {Wyrowski}, {Csengeri}, {Menten},
  {Lumsden}, {Kurtz}, {Walmsley}, {Bronfman}, {Morgan}, {Eden}, \&
  {Russeil}}]{UCornish2013}
{Urquhart}, J.~S., {Thompson}, M.~A., {Moore}, T.~J.~T., {et~al.}
  2013{\natexlab{c}}, ArXiv e-prints

\bibitem[{{Urquhart} {et~al.}(2013{\natexlab{b}}){Urquhart}, {Thompson},
  {Moore}, {Purcell}, {Hoare}, {Schuller}, {Wyrowski}, {Csengeri}, {Menten},
  {Lumsden}, {Kurtz}, {Walmsley}, {Bronfman}, {Morgan}, {Eden}, \&
  {Russeil}}]{Urquhart2013b}
{Urquhart}, J.~S., {Thompson}, M.~A., {Moore}, T.~J.~T., {et~al.}
  2013{\natexlab{b}}, \mnras, 435, 400

\bibitem[{{Ward-Thompson} {et~al.}(2007){Ward-Thompson}, {Andr{\'e}},
  {Crutcher}, {Johnstone}, {Onishi}, \& {Wilson}}]{Ward-Thompso2007}
{Ward-Thompson}, D., {Andr{\'e}}, P., {Crutcher}, R., {et~al.} 2007, Protostars
  and Planets V, 33

\bibitem[{{Wienen} {et~al.}(2016){Wienen}, {Wyrowski}, \&
  {Menten}}]{Wienen2016}
{Wienen}, M., {Wyrowski}, F., \& {Menten}. 2016, \aap

\bibitem[{{Wienen} {et~al.}(2015){Wienen}, {Wyrowski}, {Menten}, {Urquhart},
  {Csengeri}, {Walmsley}, {Bontemps}, {Russeil}, {Bronfman}, {Koribalski}, \&
  {Schuller}}]{Wienen2015}
{Wienen}, M., {Wyrowski}, F., {Menten}, K.~M., {et~al.} 2015, \aap, 579, A91

\bibitem[{{Wienen} {et~al.}(2012){Wienen}, {Wyrowski}, {Schuller}, {Menten},
  {Walmsley}, {Bronfman}, \& {Motte}}]{Wienen2012}
{Wienen}, M., {Wyrowski}, F., {Schuller}, F., {et~al.} 2012, \aap, 544, A146

\bibitem[{{Williams} {et~al.}(2000){Williams}, {Blitz}, \&
  {McKee}}]{Williams2000}
{Williams}, J.~P., {Blitz}, L., \& {McKee}, C.~F. 2000, Protostars and Planets
  IV, 97

\bibitem[{{Wood} \& {Churchwell}(1989)}]{Wood1989}
{Wood}, D.~O.~S. \& {Churchwell}, E. 1989, \apj, 340, 265

\bibitem[{{Wright} {et~al.}(2010){Wright}, {Eisenhardt}, {Mainzer}, {Ressler},
  {Cutri}, {Jarrett}, {Kirkpatrick}, {Padgett}, {McMillan}, {Skrutskie},
  {Stanford}, {Cohen}, {Walker}, {Mather}, {Leisawitz}, {Gautier}, {McLean},
  {Benford}, {Lonsdale}, {Blain}, {Mendez}, {Irace}, {Duval}, {Liu}, {Royer},
  {Heinrichsen}, {Howard}, {Shannon}, {Kendall}, {Walsh}, {Larsen}, {Cardon},
  {Schick}, {Schwalm}, {Abid}, {Fabinsky}, {Naes}, \& {Tsai}}]{Wright2010}
{Wright}, E.~L., {Eisenhardt}, P.~R.~M., {Mainzer}, A.~K., {et~al.} 2010, \aj,
  140, 1868

\bibitem[{{Wu} {et~al.}(2014){Wu}, {Sato}, {Reid}, {Moscadelli}, {Zhang}, {Xu},
  {Brunthaler}, {Menten}, {Dame}, \& {Zheng}}]{Wu2014}
{Wu}, Y.~W., {Sato}, M., {Reid}, M.~J., {et~al.} 2014, \aap, 566, A17

\bibitem[{{Wu} {et~al.}(2012){Wu}, {Xu}, {Menten}, {Zheng}, \& {Reid}}]{Wu2012}
{Wu}, Y.~W., {Xu}, Y., {Menten}, K.~M., {Zheng}, X.~W., \& {Reid}, M.~J. 2012,
  in IAU Symposium, Vol. 287, IAU Symposium, ed. R.~S. {Booth}, W.~H.~T.
  {Vlemmings}, \& E.~M.~L. {Humphreys}, 425--426

\bibitem[{{Xu} {et~al.}(2011){Xu}, {Moscadelli}, {Reid}, {Menten}, {Zhang},
  {Zheng}, \& {Brunthaler}}]{Xu2011}
{Xu}, Y., {Moscadelli}, L., {Reid}, M.~J., {et~al.} 2011, \apj, 733, 25

\bibitem[{{Zapata} {et~al.}(2008){Zapata}, {Palau}, {Ho}, {Schilke}, {Garrod},
  {Rodr{\'{\i}}guez}, \& {Menten}}]{Zapata2008}
{Zapata}, L.~A., {Palau}, A., {Ho}, P.~T.~P., {et~al.} 2008, \aap, 479, L25

\bibitem[{{Zavagno} {et~al.}(2010){Zavagno}, {Russeil}, {Motte}, {Anderson},
  {Deharveng}, {Rod{\'o}n}, {Bontemps}, {Abergel}, {Baluteau}, {Sauvage},
  {Andr{\'e}}, {Hill}, \& {White}}]{Zavagno2010}
{Zavagno}, A., {Russeil}, D., {Motte}, F., {et~al.} 2010, \aap, 518, L81

\bibitem[{{Zhang} {et~al.}(2014){Zhang}, {Moscadelli}, {Sato}, {Reid},
  {Menten}, {Zheng}, {Brunthaler}, {Dame}, {Xu}, \& {Immer}}]{Zhang2014}
{Zhang}, B., {Moscadelli}, L., {Sato}, M., {et~al.} 2014, \apj, 781, 89

\bibitem[{{Zhang} {et~al.}(2013){Zhang}, {Reid}, {Menten}, {Zheng},
  {Brunthaler}, {Dame}, \& {Xu}}]{Zhang2013}
{Zhang}, B., {Reid}, M.~J., {Menten}, K.~M., {et~al.} 2013, \apj, 775, 79

\bibitem[{{Zhang} {et~al.}(2009){Zhang}, {Wang}, {Pillai}, \&
  {Rathborne}}]{Zhang2009}
{Zhang}, Q., {Wang}, Y., {Pillai}, T., \& {Rathborne}, J. 2009, \apj, 696, 268

\bibitem[{{Zinnecker} \& {Yorke}(2007)}]{Zinnecker2007}
{Zinnecker}, H. \& {Yorke}, H.~W. 2007, \araa, 45, 481

\end{thebibliography}


\appendix
\section{Mid-infrared photometry}\label{app:mips_photometry}
\subsection{MIPSGAL upper limits}\label{sec:mipsu}

Using the publicly available calibrated MIPSGAL tiles,  we searched for embedded compact
objects spatially coinciding with the \at\ dust peaks. From the 219 brightest submm sources
we find that \nmipsnotdet\ do not show any clear point source within 5{\arcsec} from the dust peaks.
After a visual inspection, we extracted an upper limit for these sources by taking the median of the pixel values
within a radius of 3{\arcsec}, i.e. half of the FWHM of the PSF at 24~\mum. We then multiply this value
 by the beam area assuming a Gaussian beam with a FWHM of 6\arcsec\ to obtain an upper
limit on the source flux. We estimate the error by calculating the
REDMSQ\footnote{$REDMSQ=\sqrt{MEDIAN((S[i,j]-median(S[i,j])^2)}$.} 
value in the same area, in the footprint of the PSF,  as was defined in \citet{Megeath2012}. 
We extract upper
limits between $0.041 - 1.368$~Jy/beam with an average of 0.4~Jy/beam.

\subsection{MIPSGAL aperture photometry at 24~\mum}\label{sec:mips}

We performed aperture photometry towards the dust peaks 
that are clearly associated with a $24\,\mu$m 
point source
and without saturated pixels. 
In greater  detail, we extracted the flux density of point sources from the MIPSGAL data 
using the PhotVis v1.10 tool \citep{Gutermuth2008} positionally coinciding with the \at\ source.
We adopted a zero point magnitude of 12.448, which was calculated assuming
a pixel size of 1\rlap{.}"25, an aperture correction of 2.05, which is used for
apertures of 7\arcsec\ with a sky annulus between 20\arcsec\ to 32{\arcsec} 
\citep{Engelbracht2007}\footnote{We used the following formula to calculate the zero point magnitude, 
mag$_0=2.5\times LOG_{10}\Big(\frac{F_{\nu0}}{10^6\times\Omega\times app_{corr}}\Big)$, where $\Omega$
corresponds to the pixel size in steradian and F$_{\nu0}$ is 7.17 Jy.}. We chose these radii because the background emission 
varies substantially and a larger sky annulus
helps to avoid contamination by nearby bright sources. However, our background values still show
variations and so may still be contaminated
by extended emission. 
Therefore, we caution that our values are subject to uncertainties. 

We then converted  the measured magnitudes to flux density 
using F$_{\nu}=\rm c_{corr} \times F_{\nu0}\times10^{(-mag/2.5)}$, where F$_{\nu0}$ corresponds to
the zero magnitude flux level, which is 7.17 Jy and $mag$ is the magnitude given by PhtoVis.  
The targeted sample consists of deeply embedded objects, where $\beta \sim$1-2 corresponds to
rising SEDs towards longer wavelengths. 
Therefore, we applied a colour correction (c$_{corr}$) of
0.981 for objects with $\beta \sim$1 \citep{Engelbracht2007}. 

\subsection{Comparing MIPSGAL and WISE flux densities}\label{sec:crosscal}

We measure 24~\mum\ flux densities (and upper limits) between 0.041 and 6.88 Jy for these sources, 
with a mean of 1.84 and a median of 0.81 Jy. 
For {\nwiseflux} sources in the sample we also have photometry from the WISE point source catalogue \citep{Wright2010}. 
Since there is an overlap in the wavelength ranges of the WISE band at 22~\mum\ and
the MIPS detector at 24~{\mum}, we attempted to cross-calibrate the photometry of the two measurements.
We  found that numerous sources appearing in the WISE catalogue and not saturated with
MIPS are flagged as likely spurious detections. This is likely due to source confusion and extended 
nebulosity surrounding the majority of our weak sources, which results in bad quality flags for
these sources in the WISE point source catalogue.
Therefore, we only have \nwisemipsoverlap\ sources
with reliable flux measurement from WISE, of which only \nwisemipsoverlapl\  have
smaller than 5\arcsec\ angular offset from the \at\ peak position.

\begin{figure}[!htpb]
\centering
\includegraphics[width=6cm,angle=90]{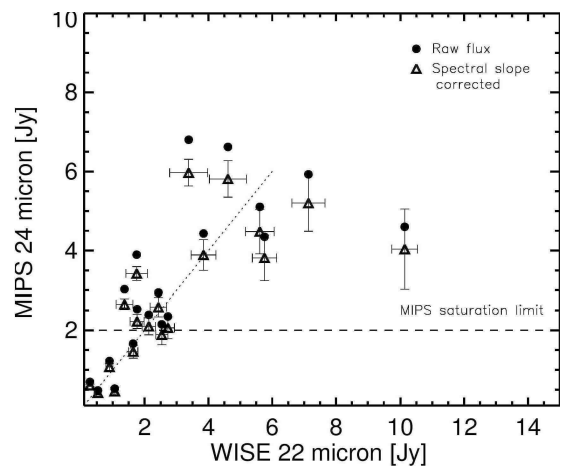}
\caption{Cross-calibration between flux densities from the WISE point source catalogue and the MIPSGAL survey.}\label{fig:mips-wise}
\end{figure}

Our attempt to cross-calibrate the two instruments is shown in Fig.\,\ref{fig:mips-wise},
where we also correct for a spectral index of 1.5 that has been found between MSX and MIPSGAL
according to the MIPSGAL data delivery document\footnote{http://irsa.ipac.caltech.edu/data/SPITZER/MIPSGAL/images/mipsgal\_delivery\_guide\_v3\_29aug08.pdf}.
With some scatter, we find similar flux density measurements below $\sim$4~Jy with the two instruments.
Since the MIPS detector starts to saturate at 2~Jy, the flattening of the MIPSGAL fluxes between $\sim2-4$~Jy is likely
due to the saturation or non-linearity of the detector. In addition, since WISE has a beam that is two times larger  (the MIPS PSF is 
6\arcsec\ at 24~{\mum}, while WISE has a 12{\arcsec} angular resolution), it may include
emission from the surrounding diffuse emission, not only the protostar. Since saturation and non-linearity of the detector may 
significantly influence the determined values, in our analysis for sources above 2~Jy we rely on the WISE
point source catalogue values.

%

\section{Infrared-quiet massive clumps}\label{app:irq}

\begin{figure*}
\centering
  \includegraphics[width=0.45\linewidth,clip]{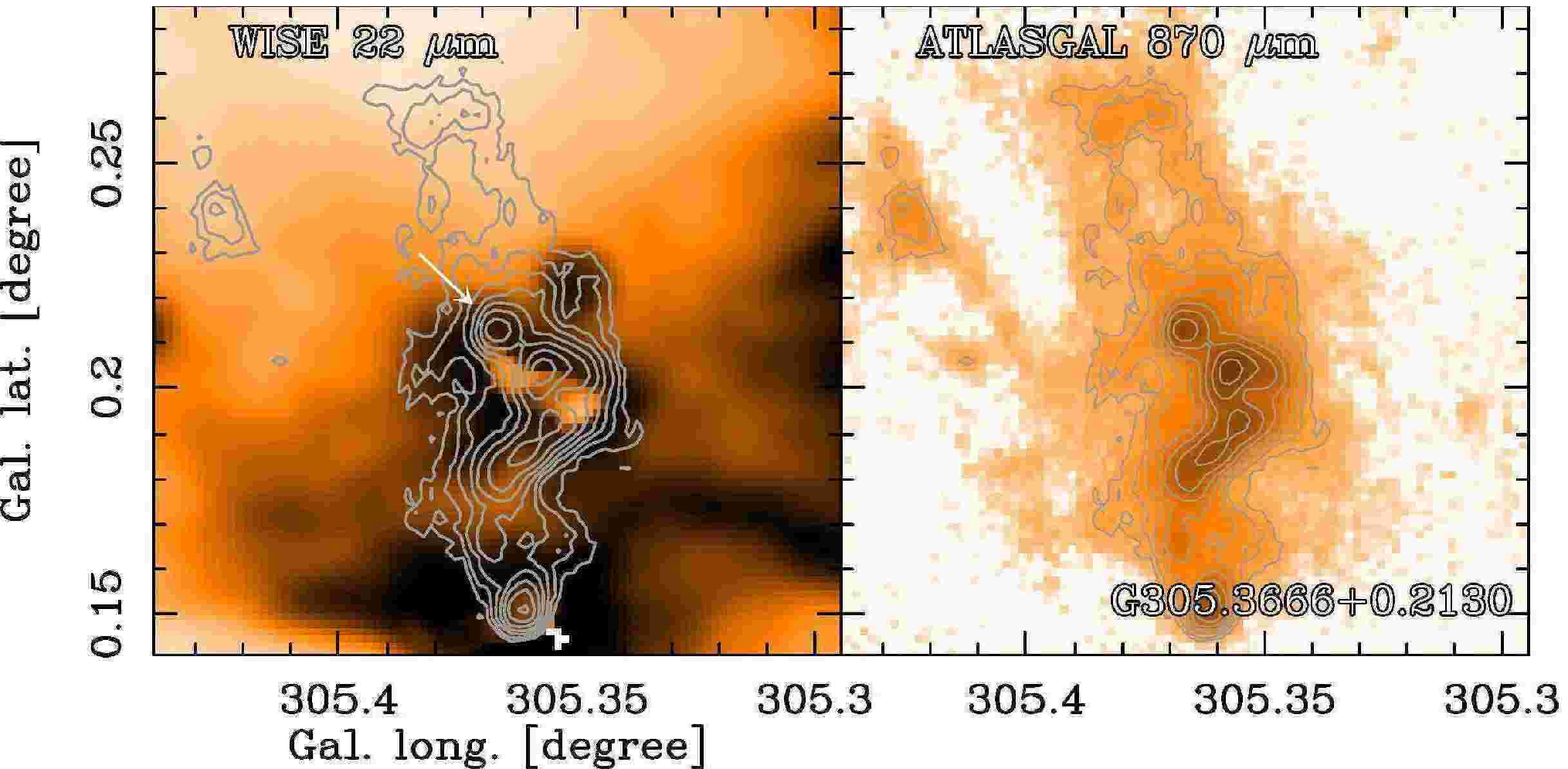} 
  \includegraphics[width=0.45\linewidth,clip]{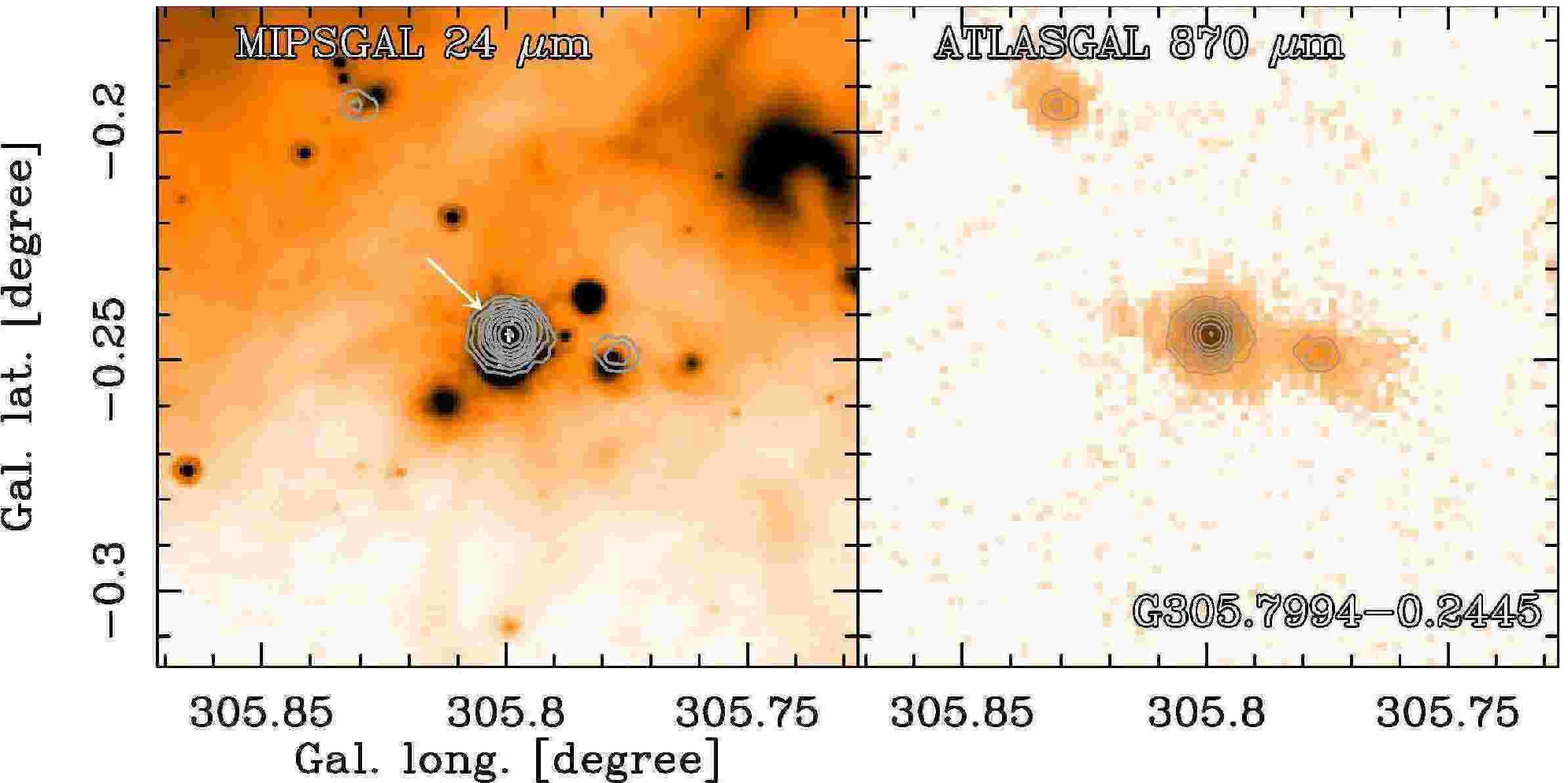} 
  \includegraphics[width=0.45\linewidth,clip]{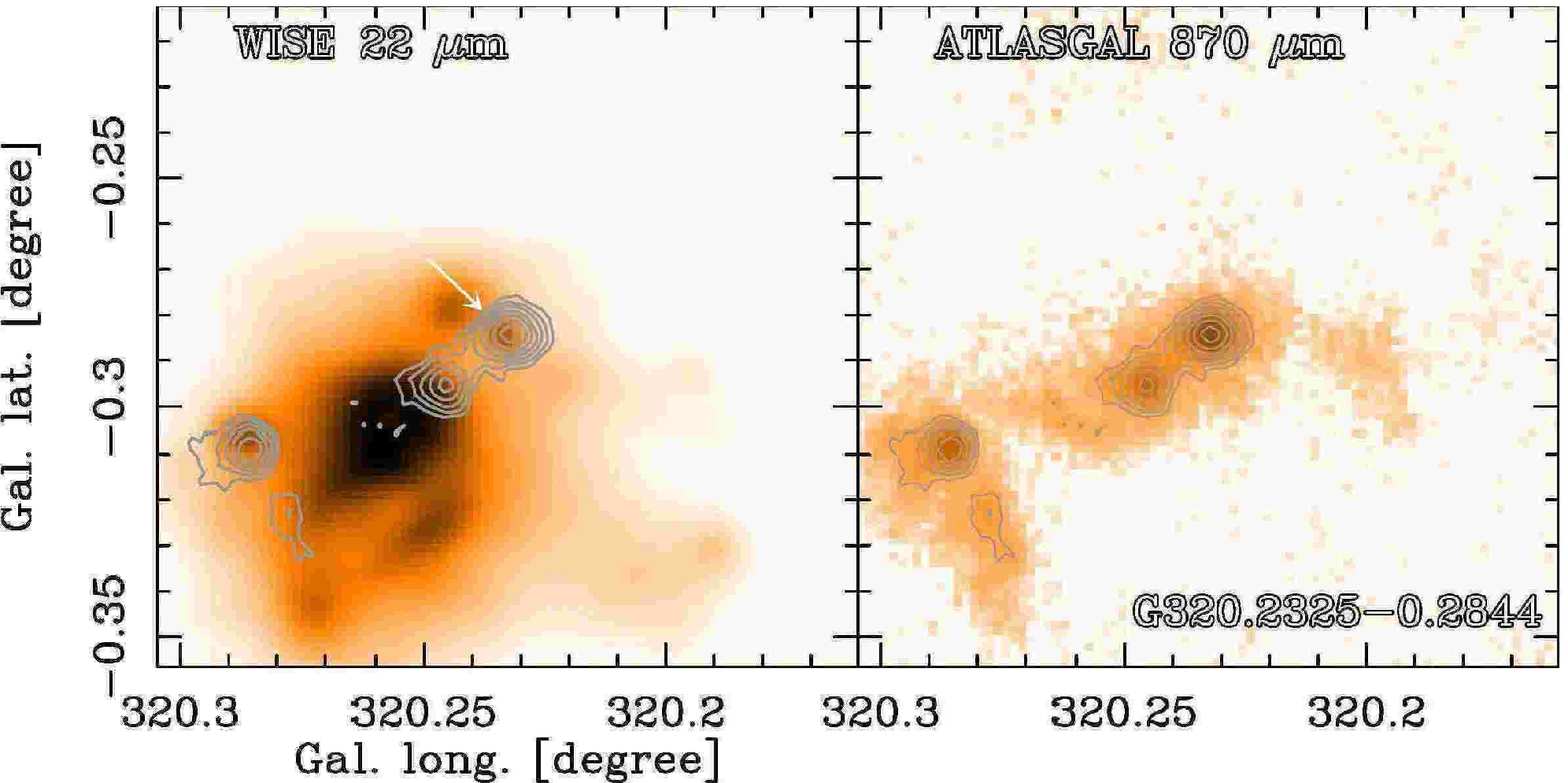} 
  \includegraphics[width=0.45\linewidth,clip]{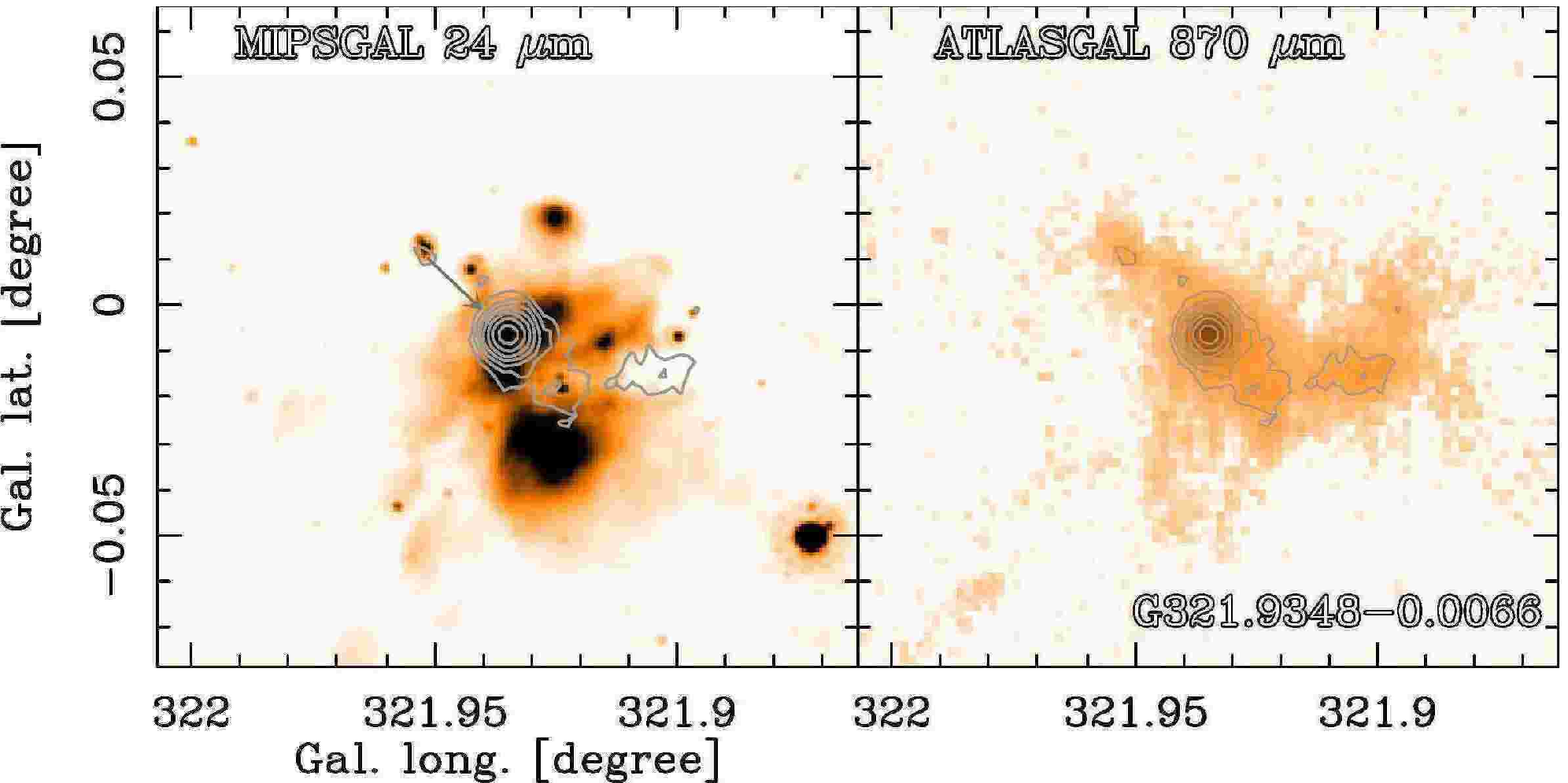} 
  \includegraphics[width=0.45\linewidth,clip]{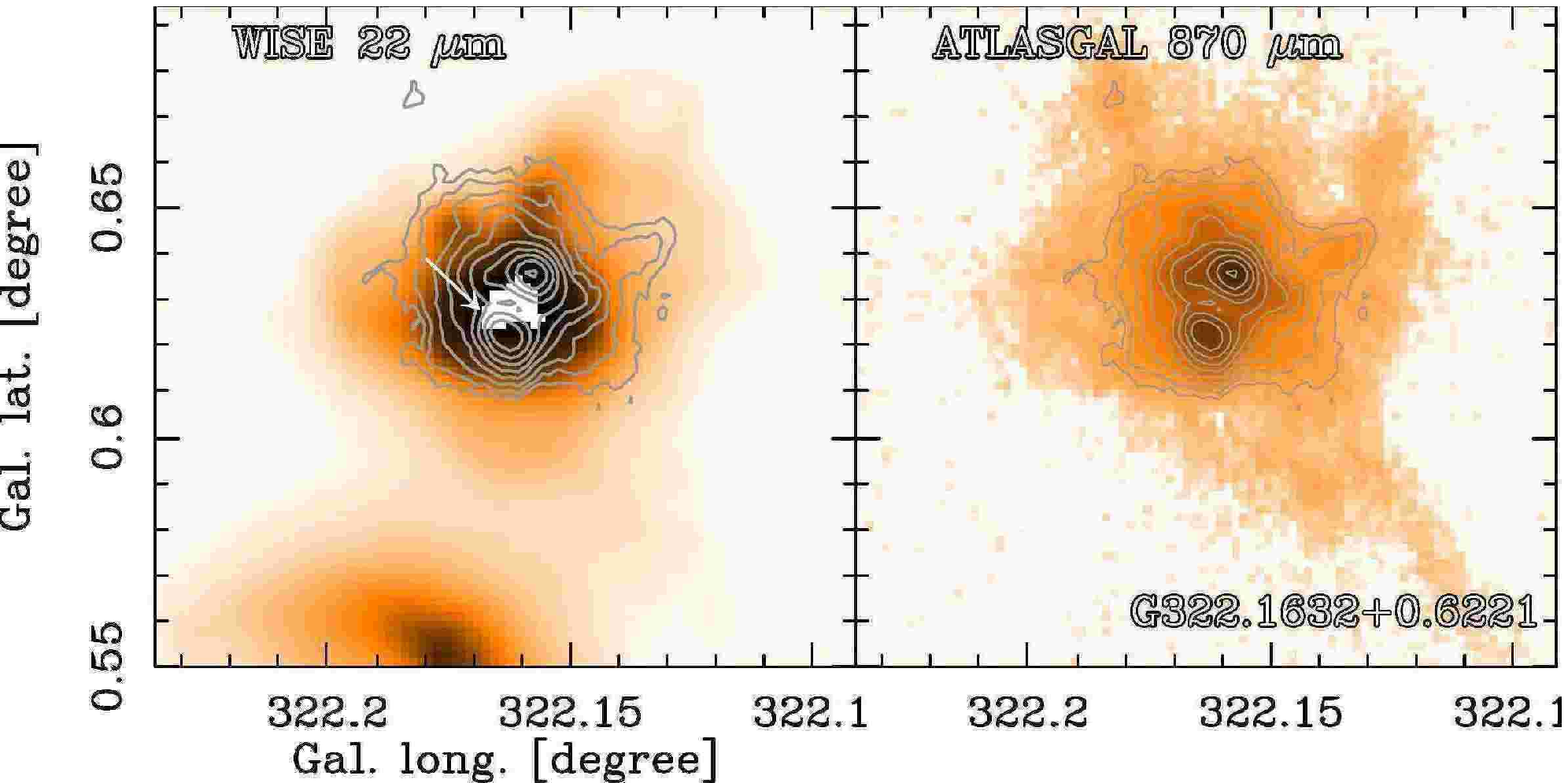} 
  \includegraphics[width=0.45\linewidth,clip]{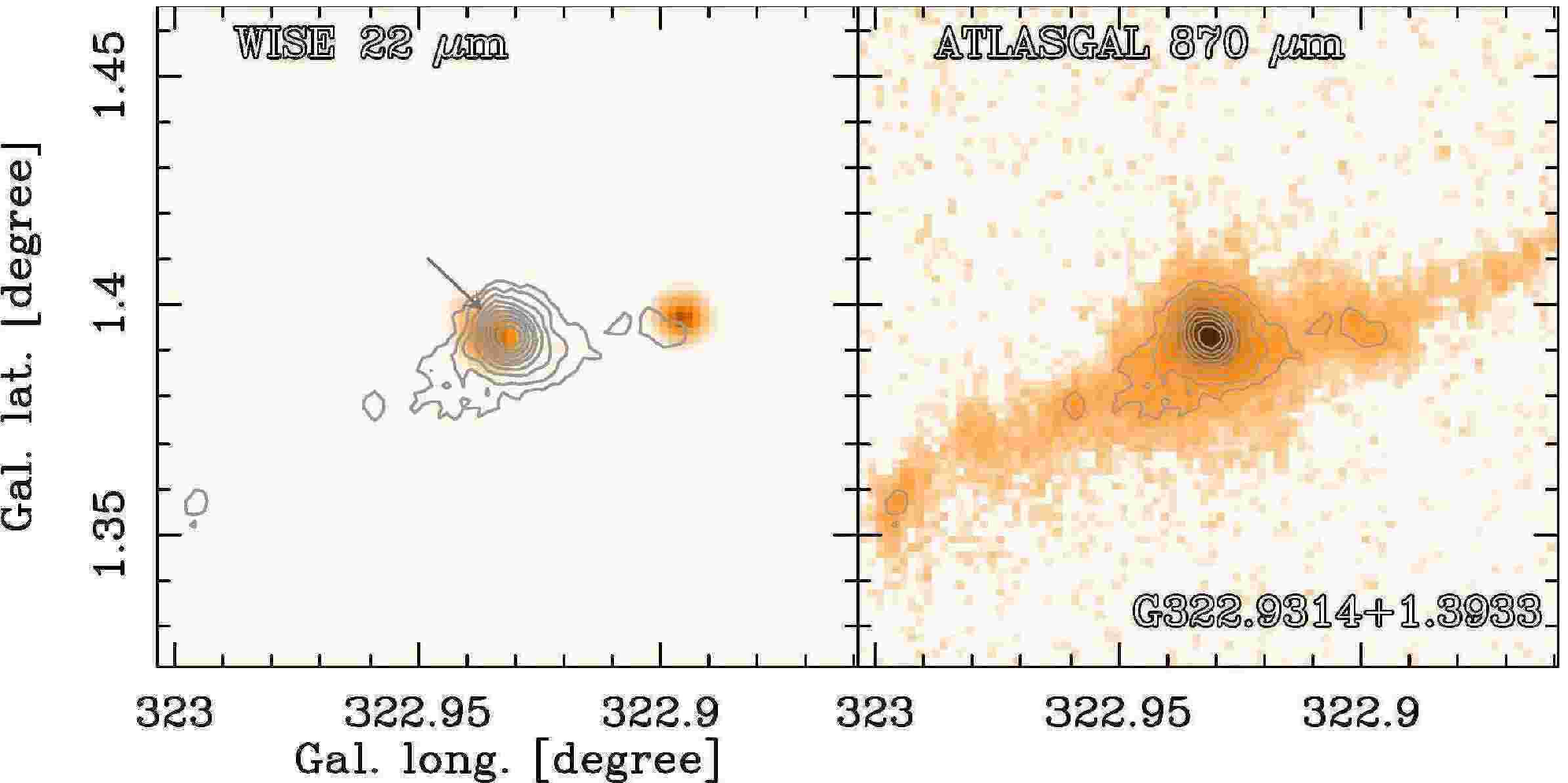} 
  \includegraphics[width=0.45\linewidth,clip]{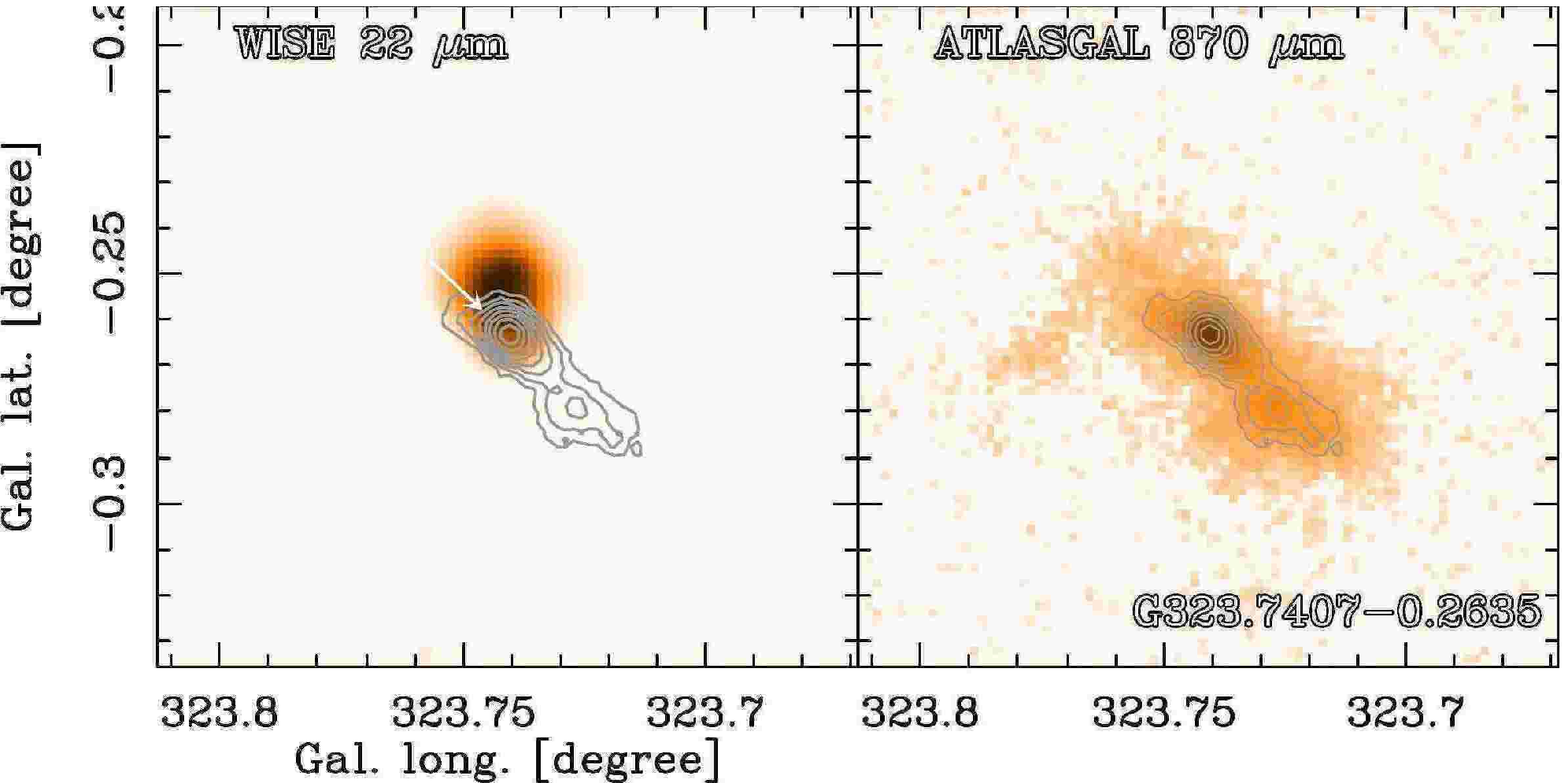} 
  \includegraphics[width=0.45\linewidth,clip]{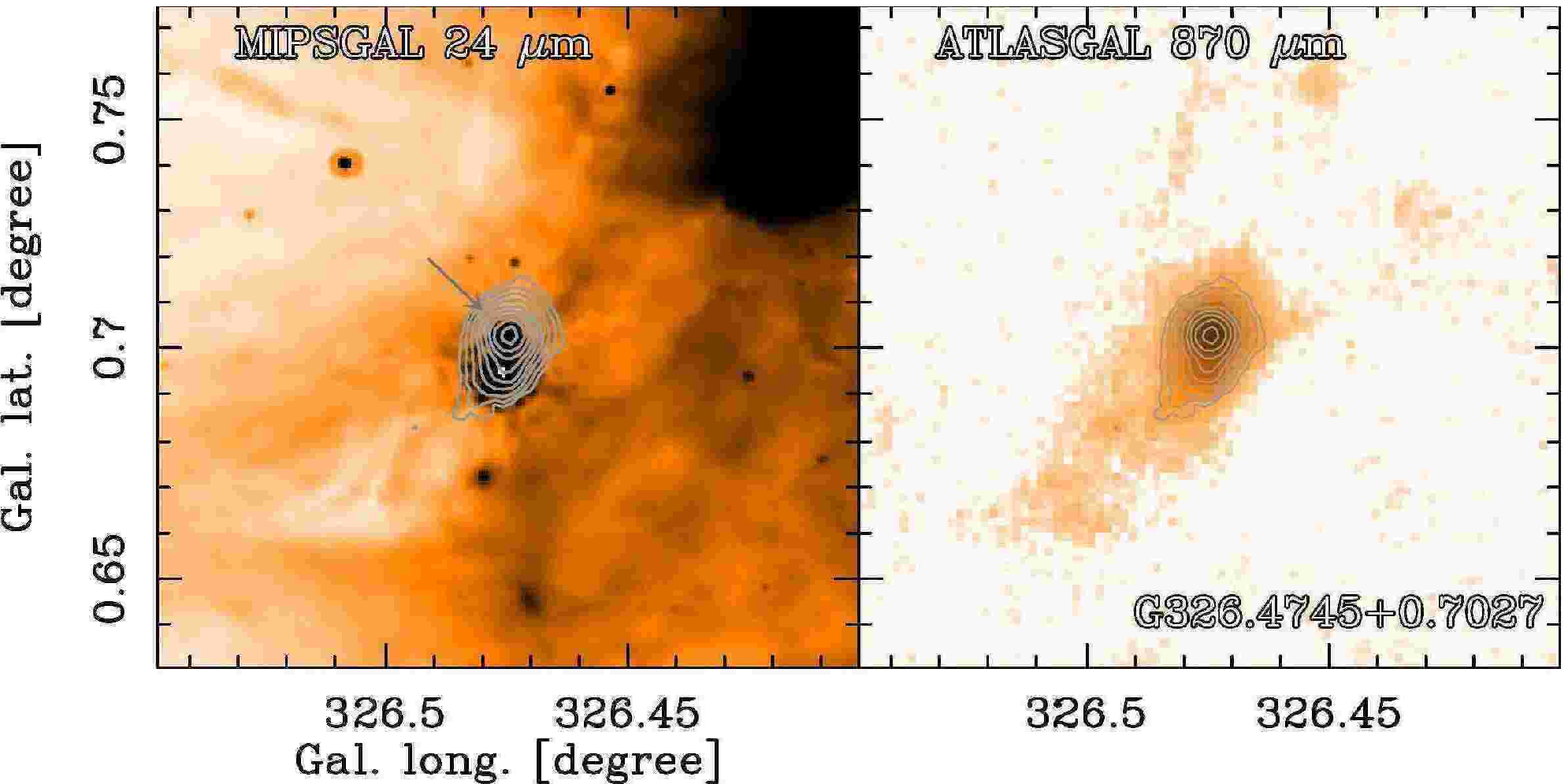} 
  \includegraphics[width=0.45\linewidth,clip]{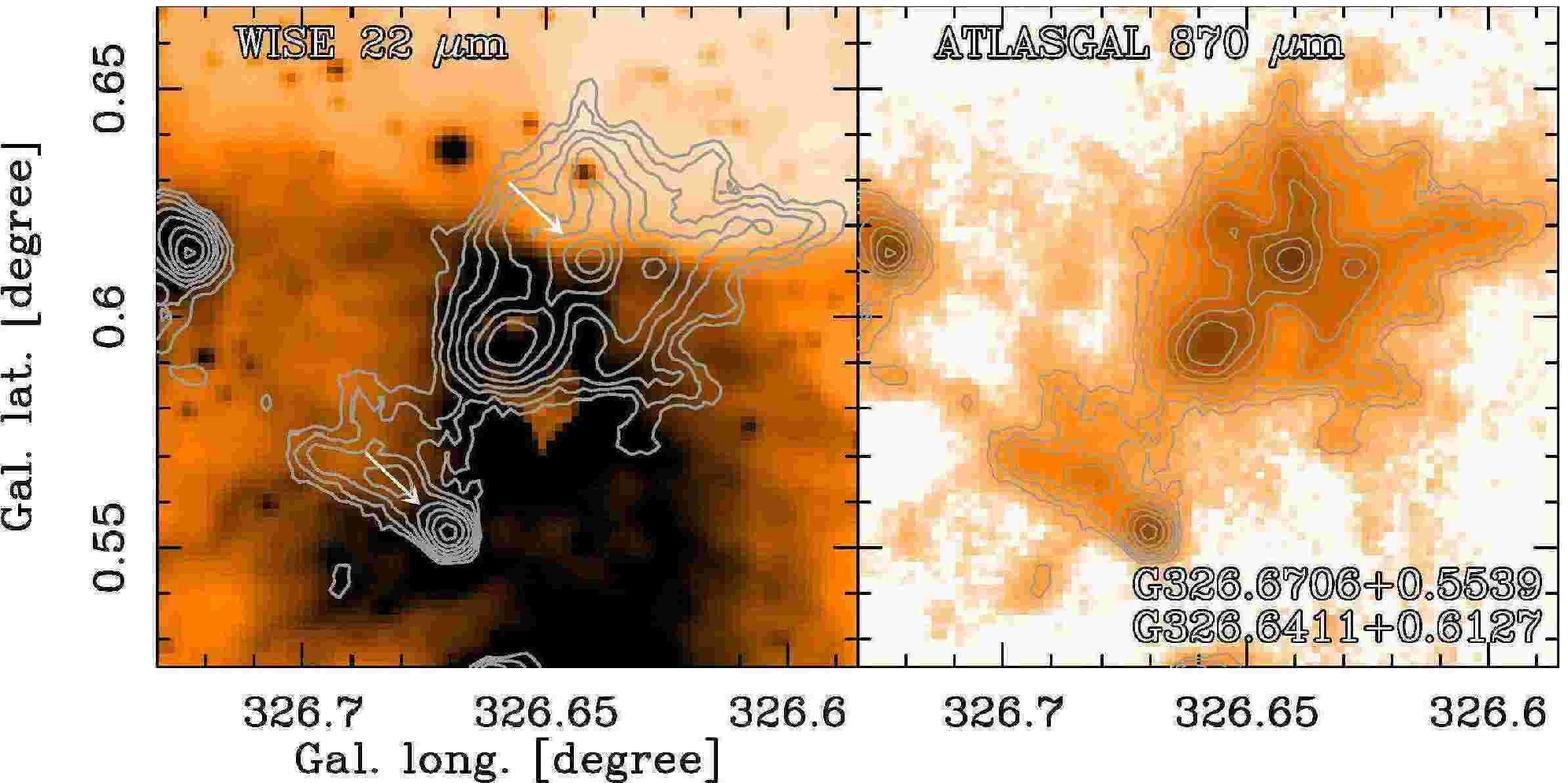} 
  \includegraphics[width=0.45\linewidth,clip]{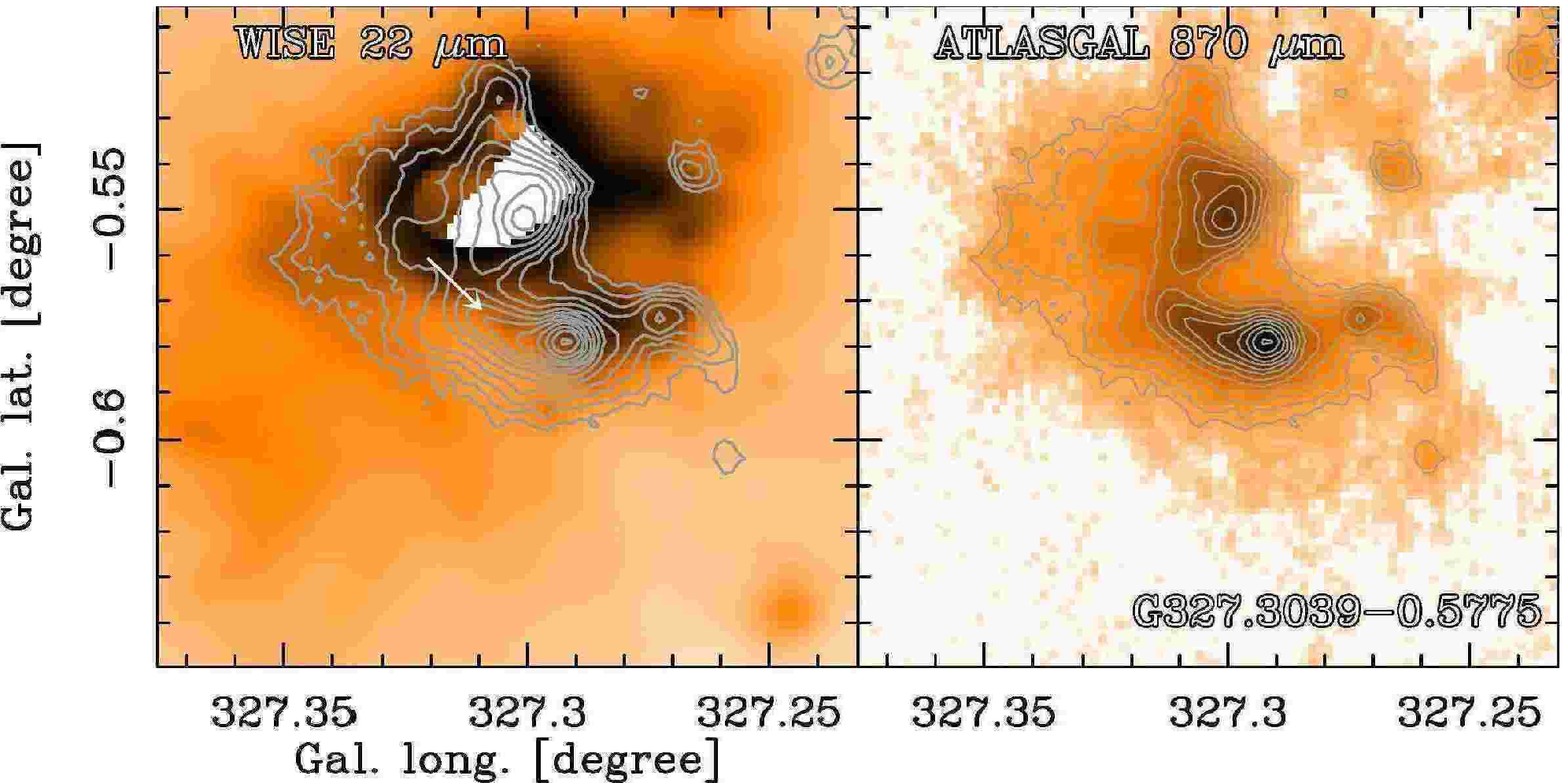} 
   \includegraphics[width=0.45\linewidth,clip]{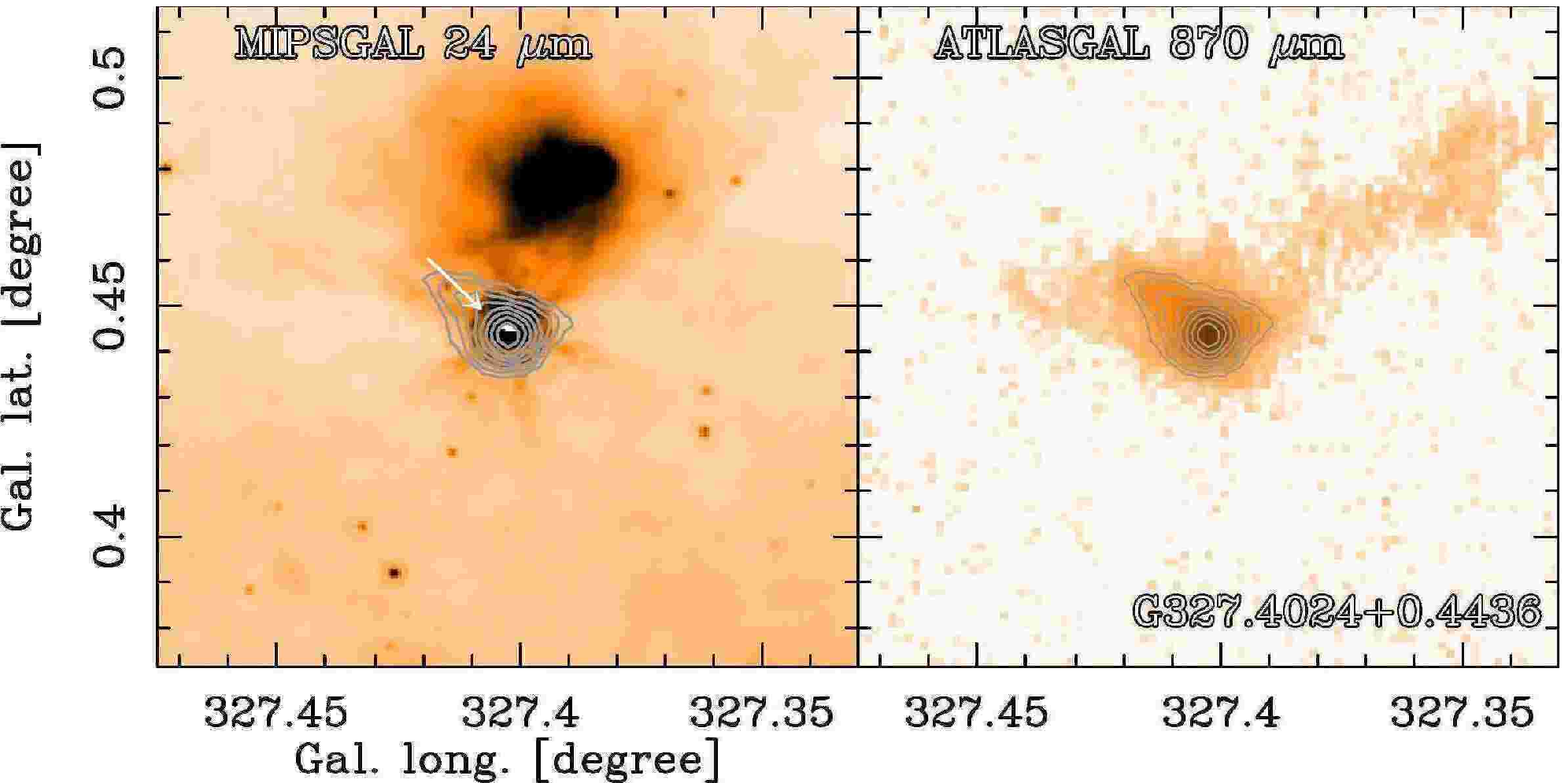} 
  \includegraphics[width=0.45\linewidth,clip]{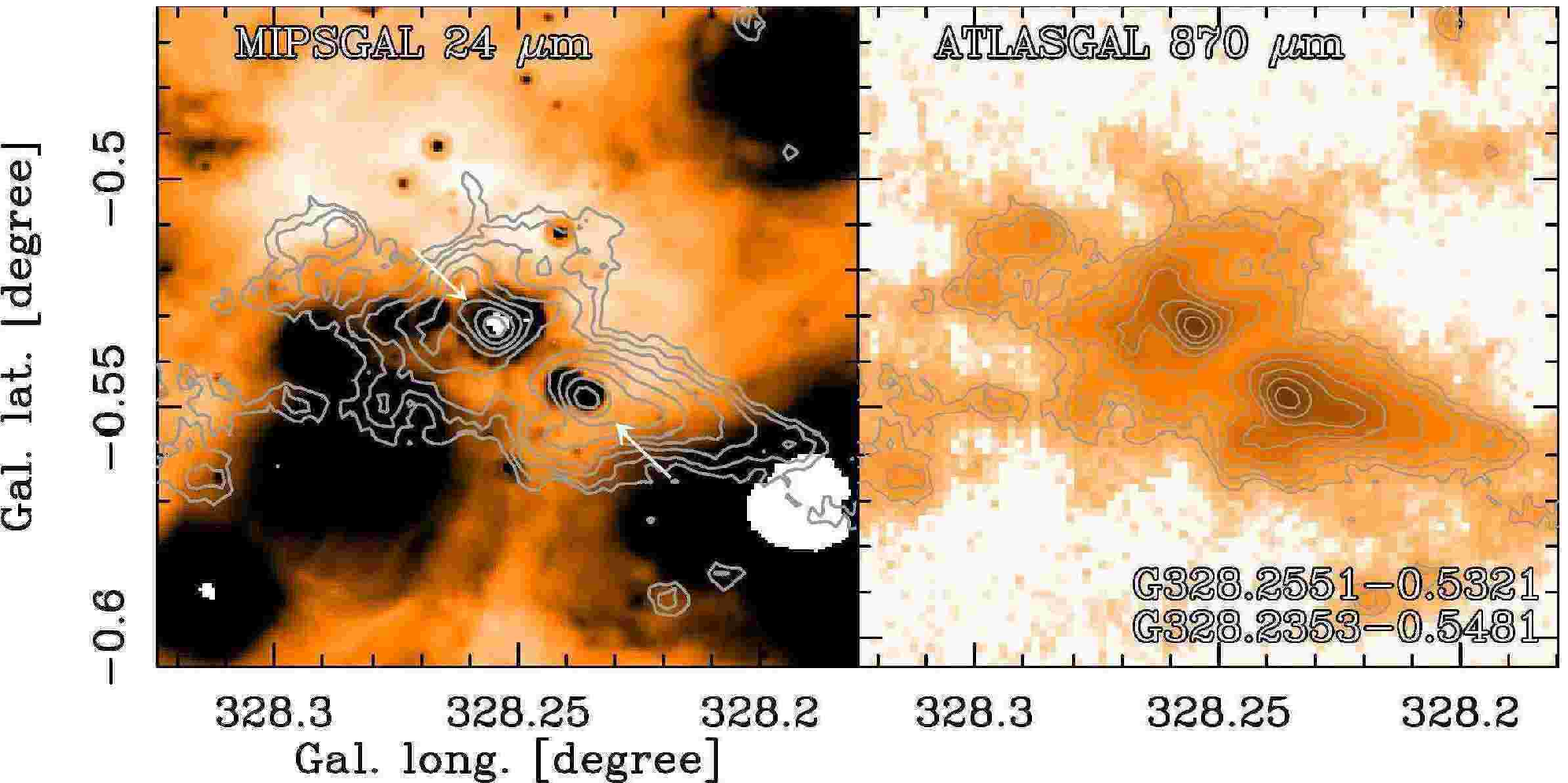} 
   \caption{Left:  Mid-infrared emission from MIPSGAL at 24\,$\mu$m or WISE at 22\,$\mu$m in colour scale with the 870\,$\mu$m contours from ATLASGAL. The arrows mark the position of infrared-quiet clumps. Right:  870\,$\mu$m emission with contours from ATLASGAL. The blue crosses mark the positions of {\uchii} regions from CORNISH. The figure label shows the corresponding ATLASGAL source name.}\label{app:fig1}
 \end{figure*}
\begin{figure*}
\centering
\ContinuedFloat
  \includegraphics[width=0.45\linewidth,clip]{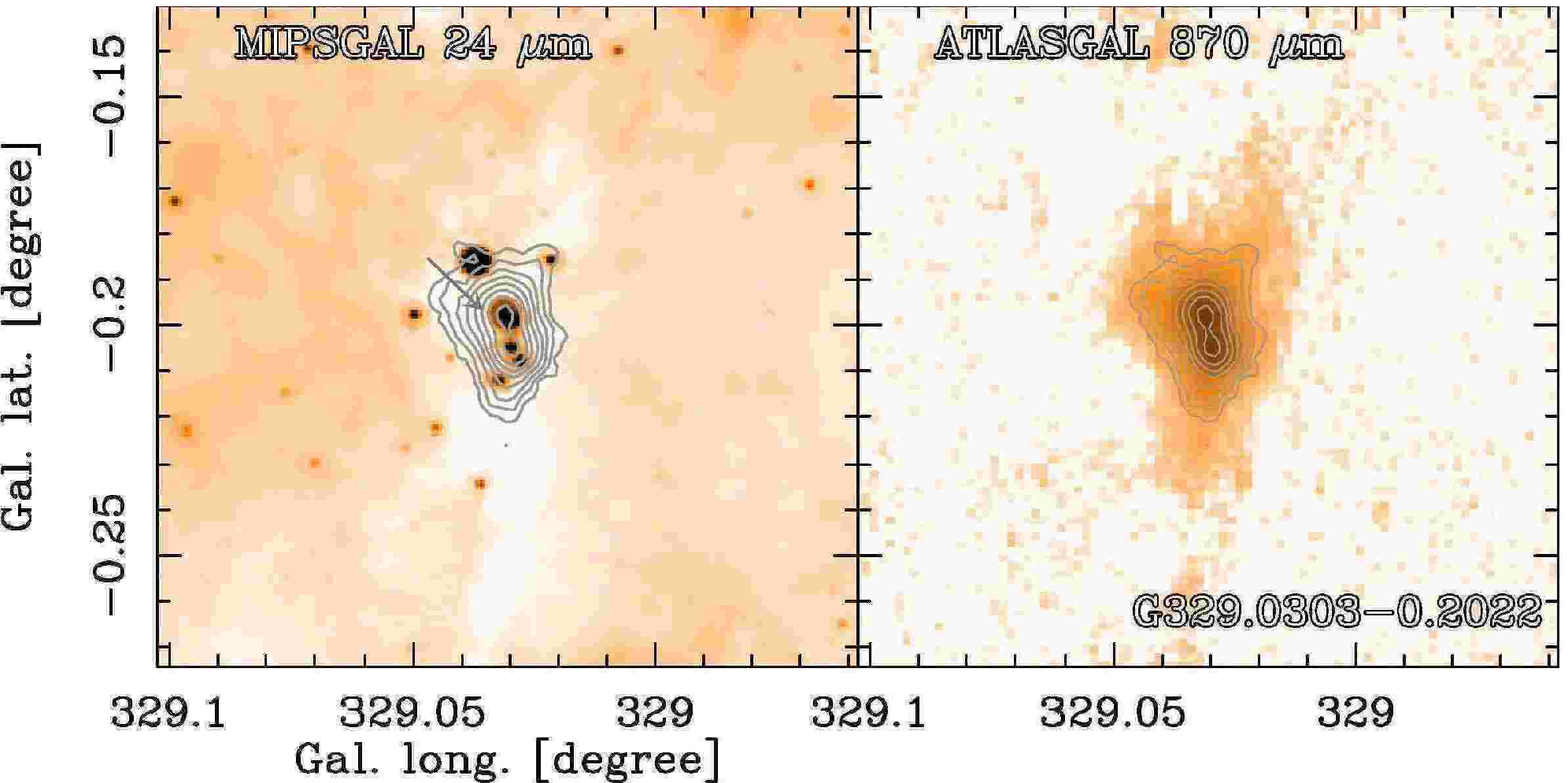} 
  \includegraphics[width=0.45\linewidth,clip]{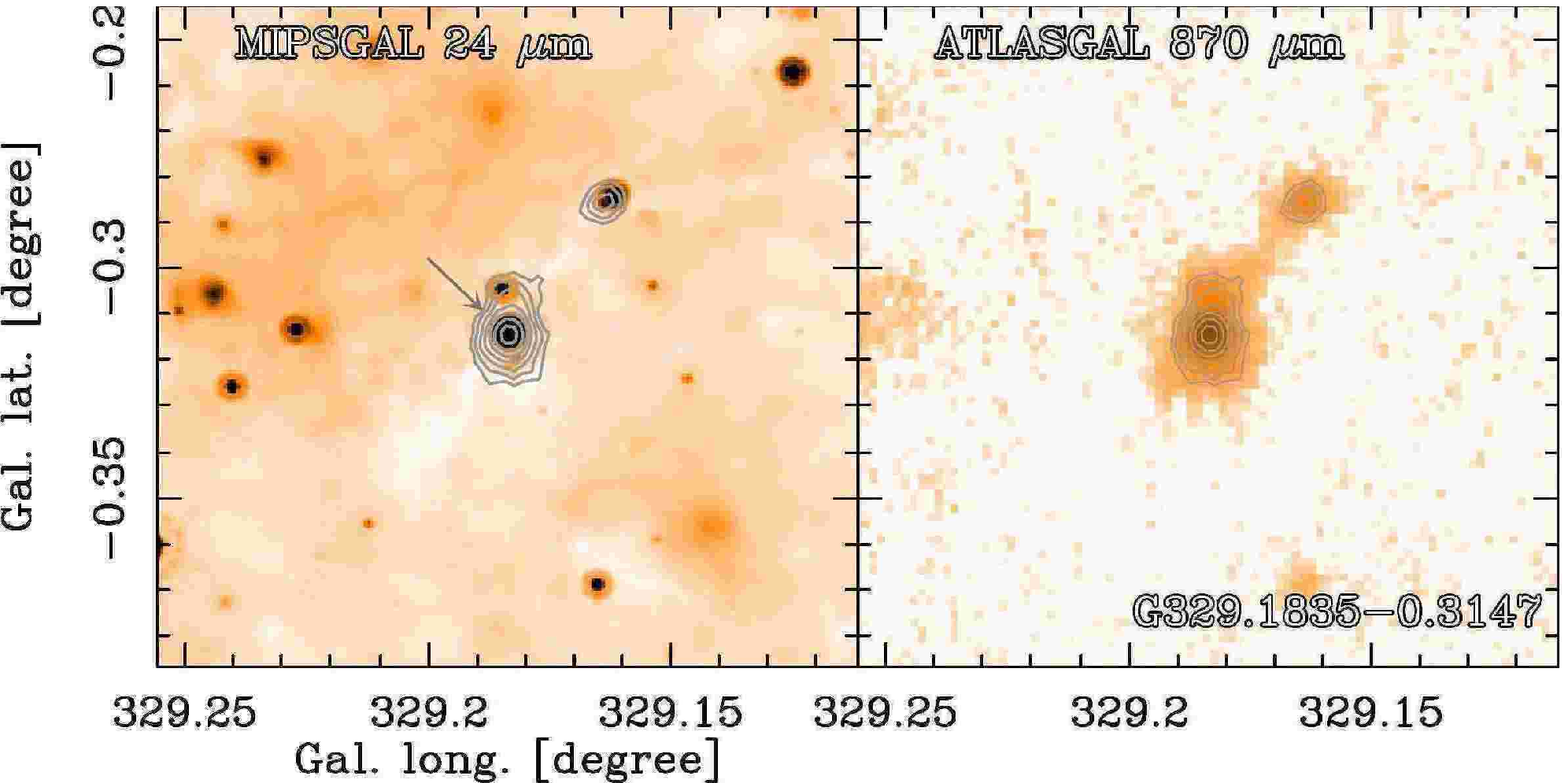} 
   \includegraphics[width=0.45\linewidth,clip]{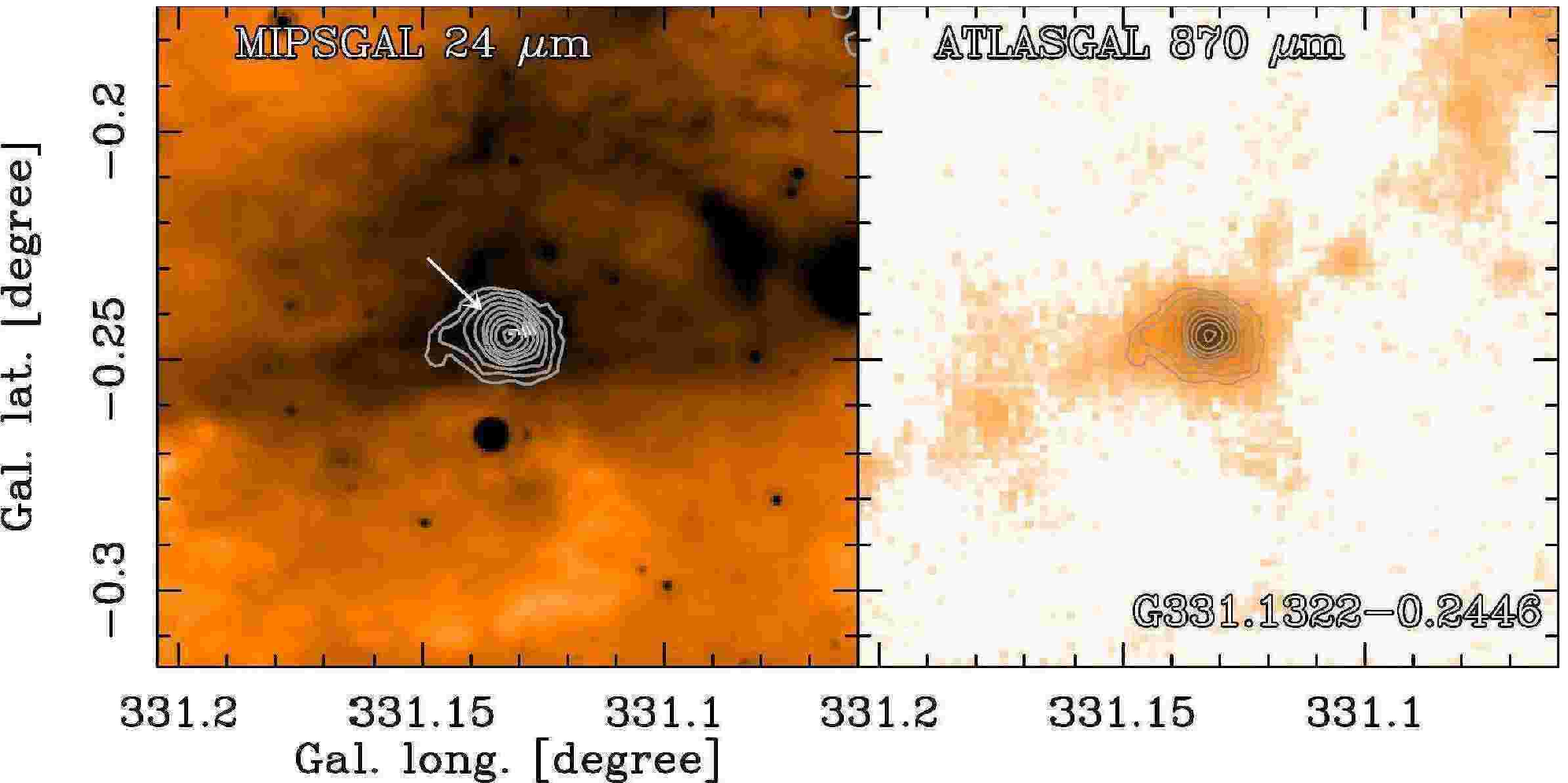} 
   \includegraphics[width=0.45\linewidth,clip]{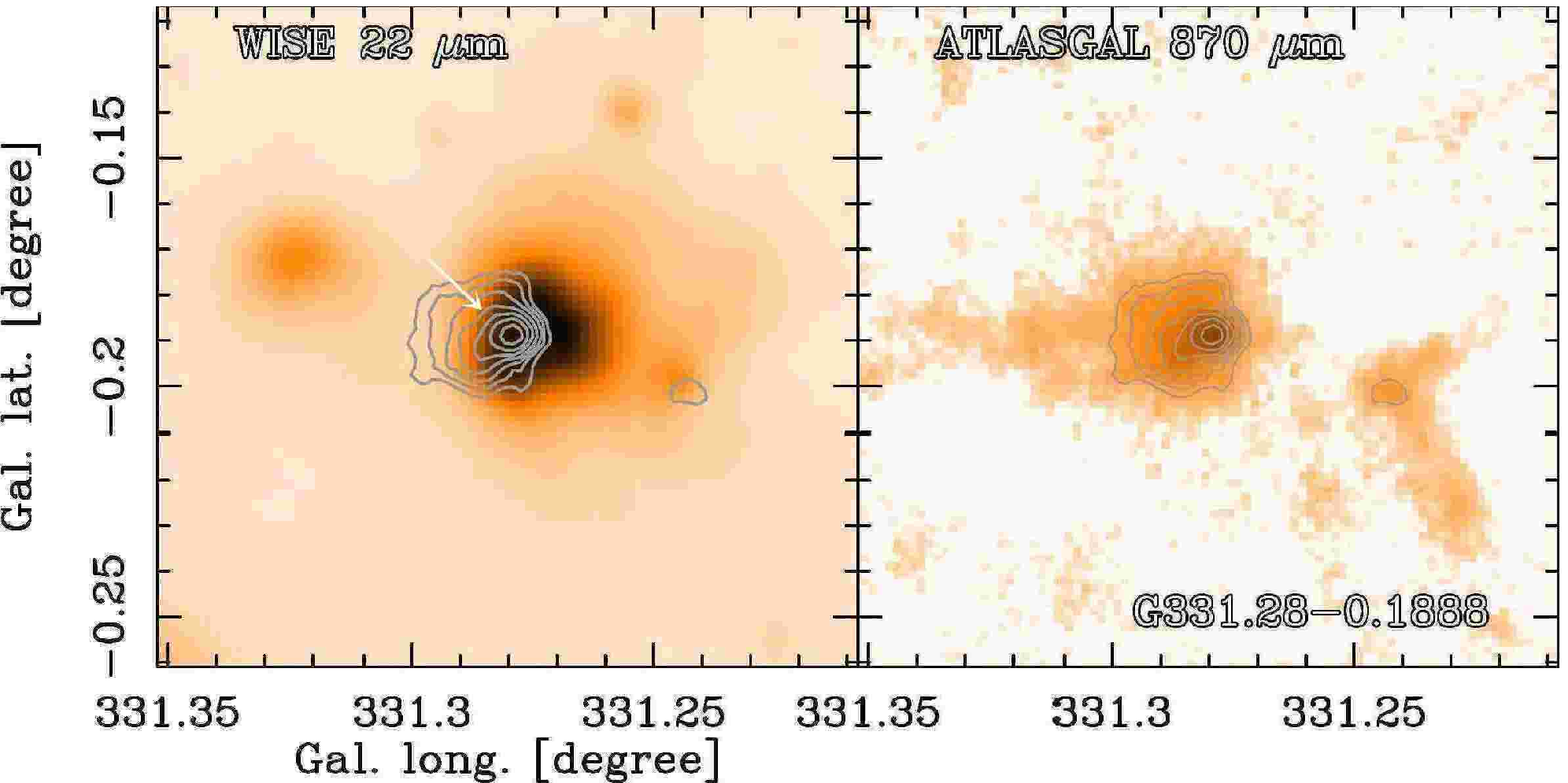} 
     \includegraphics[width=0.45\linewidth,clip]{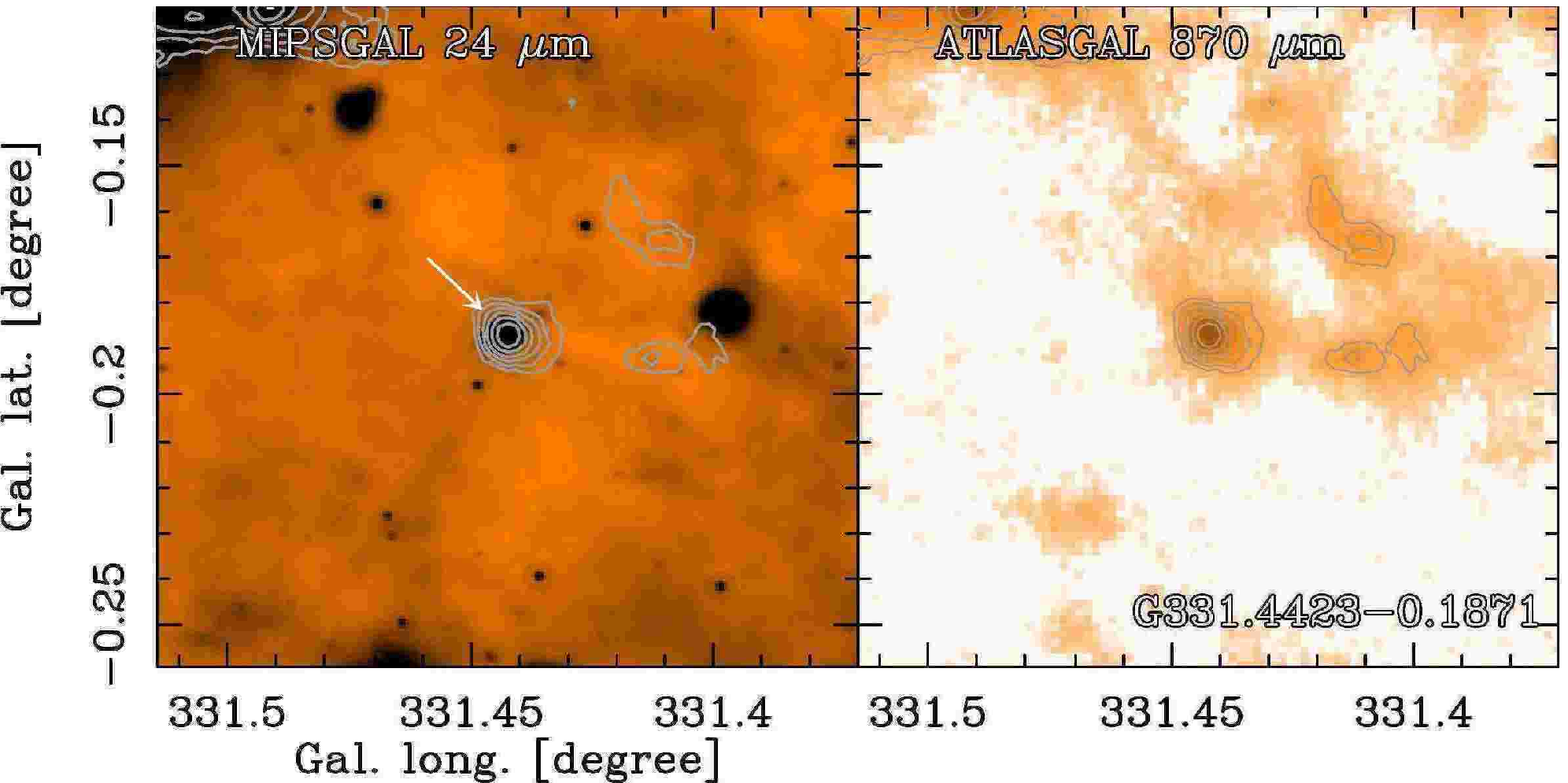} 
  \includegraphics[width=0.45\linewidth,clip]{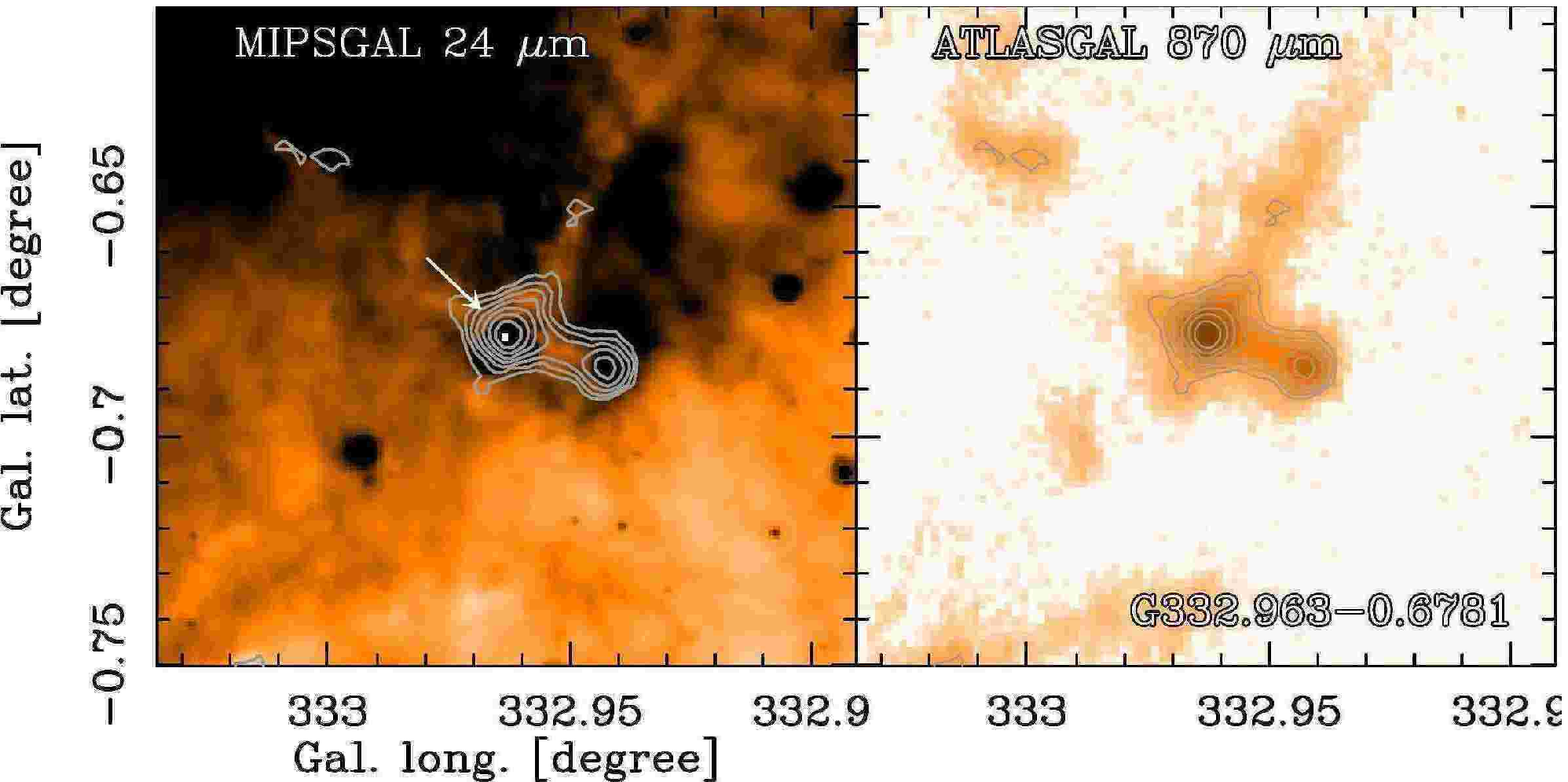} 
  \includegraphics[width=0.45\linewidth,clip]{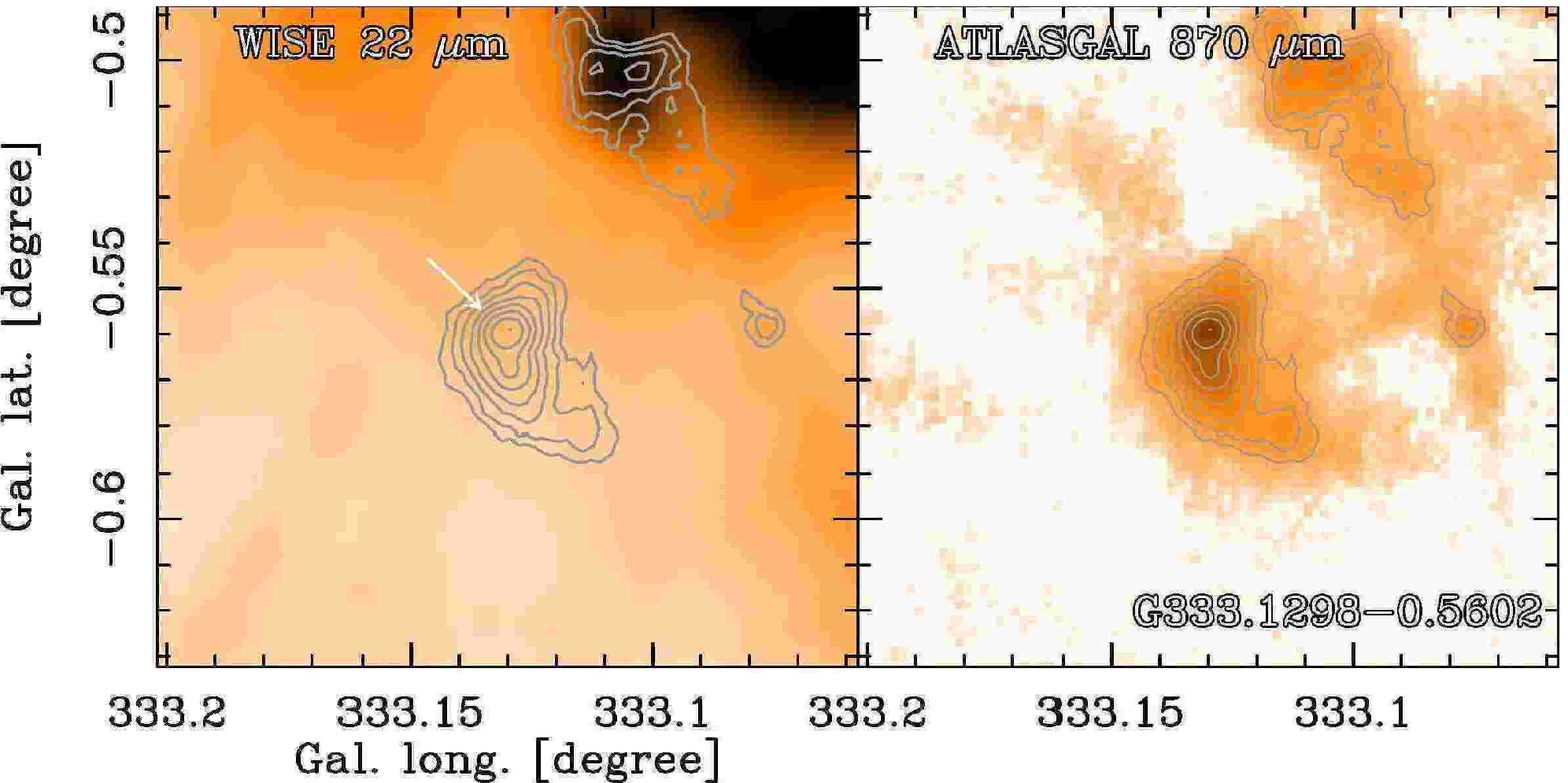} 
    \includegraphics[width=0.45\linewidth,clip]{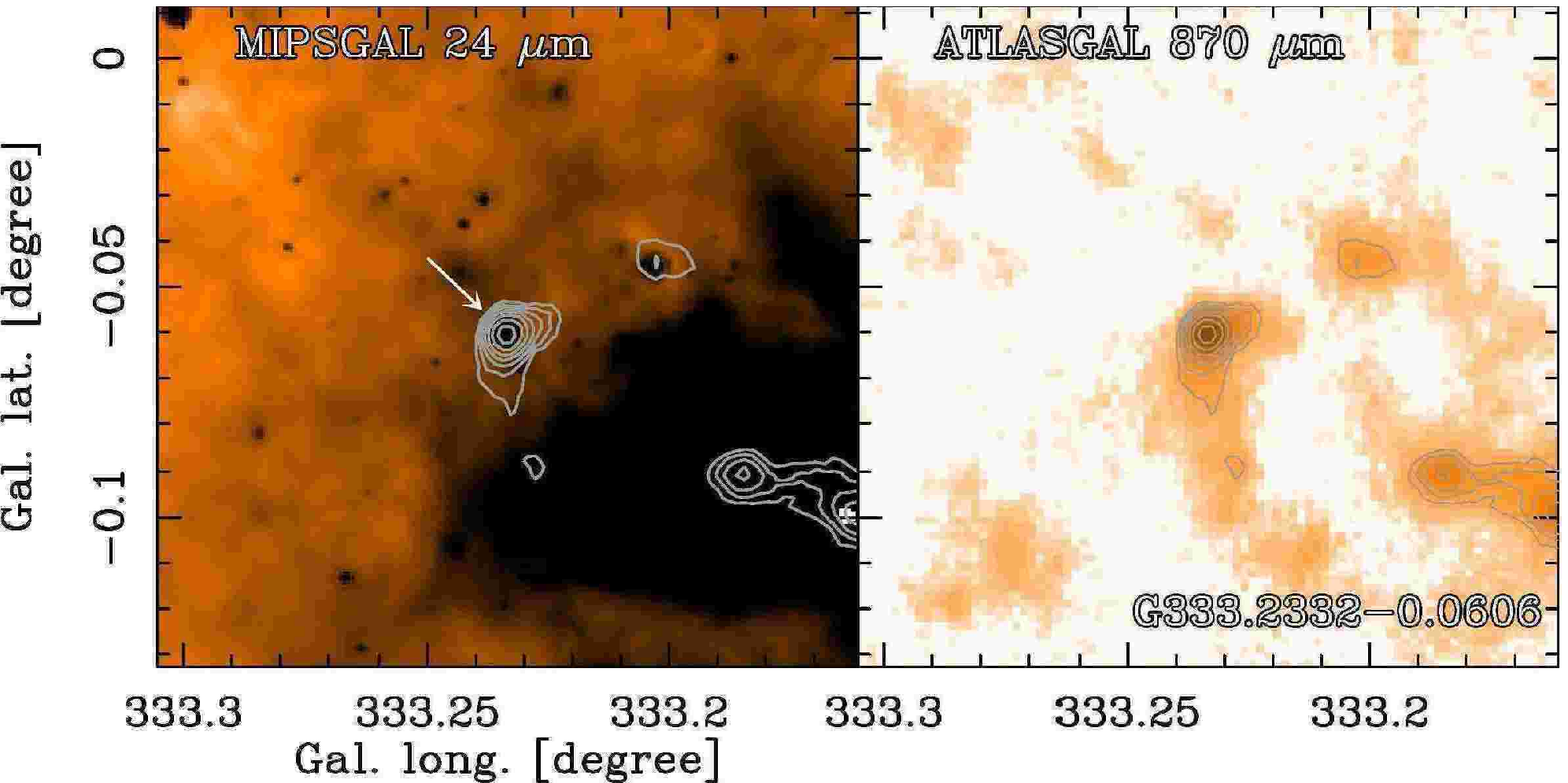} 
   \includegraphics[width=0.45\linewidth,clip]{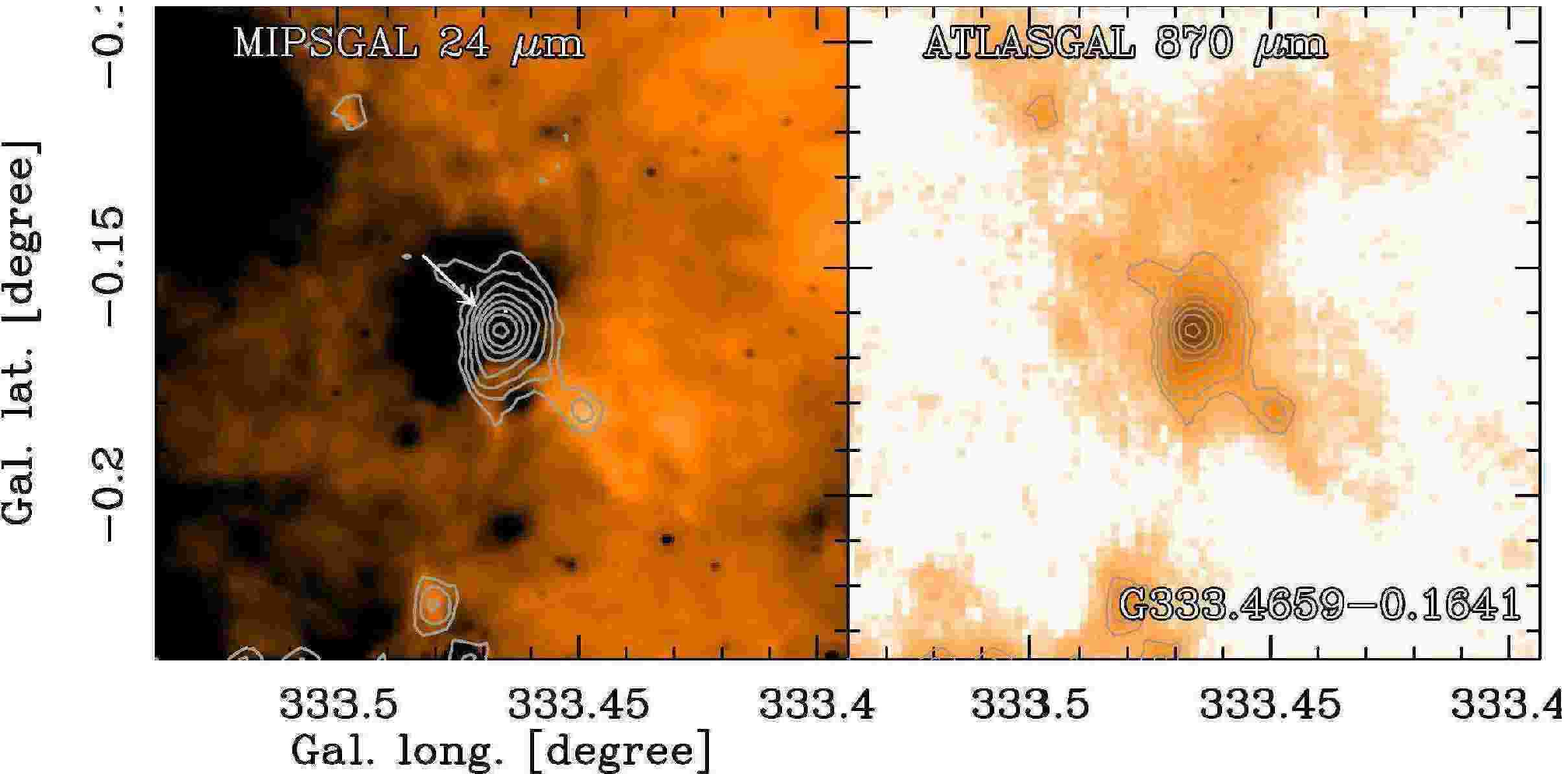} 
  \includegraphics[width=0.45\linewidth,clip]{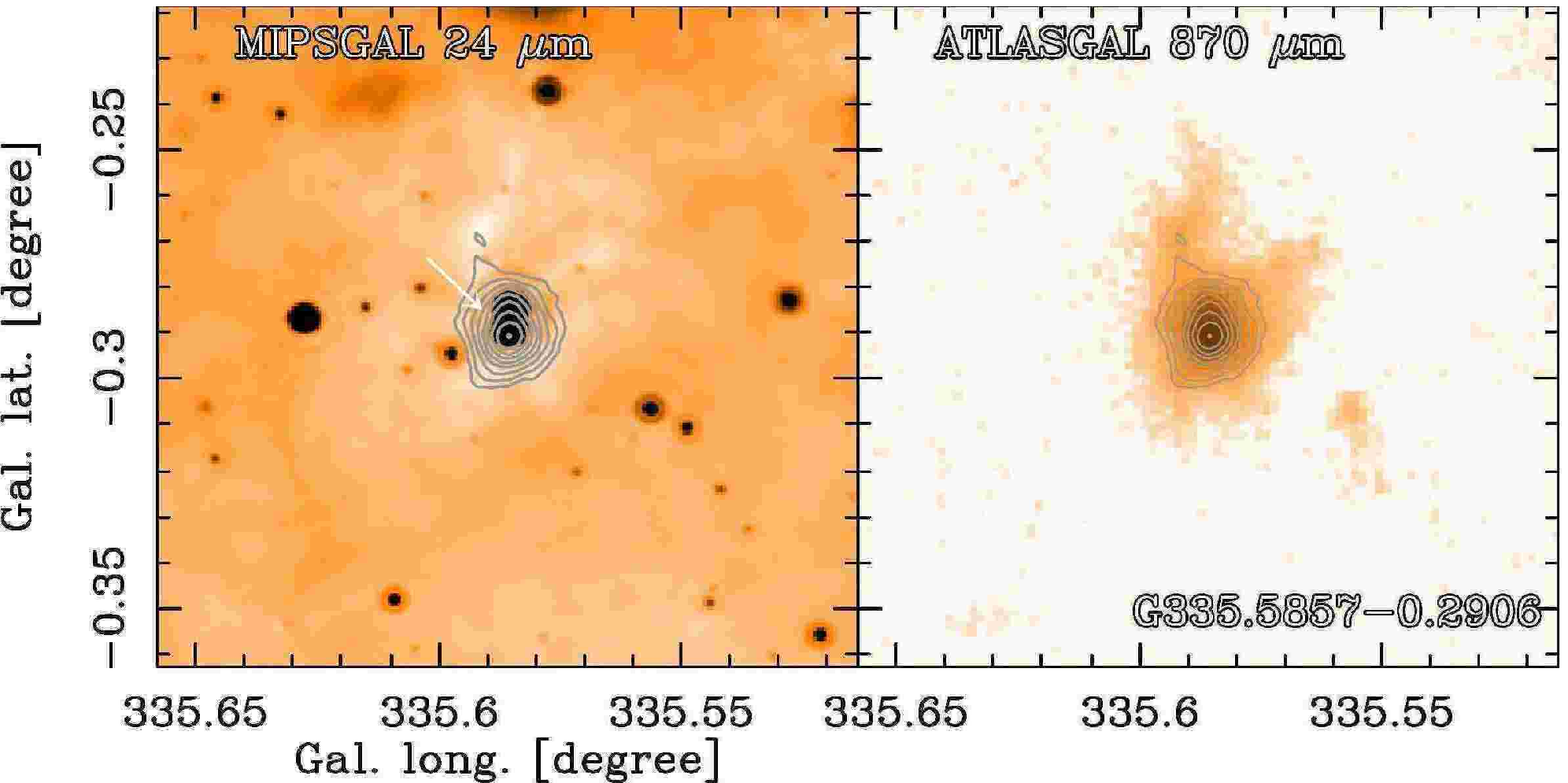} 
  \includegraphics[width=0.45\linewidth,clip]{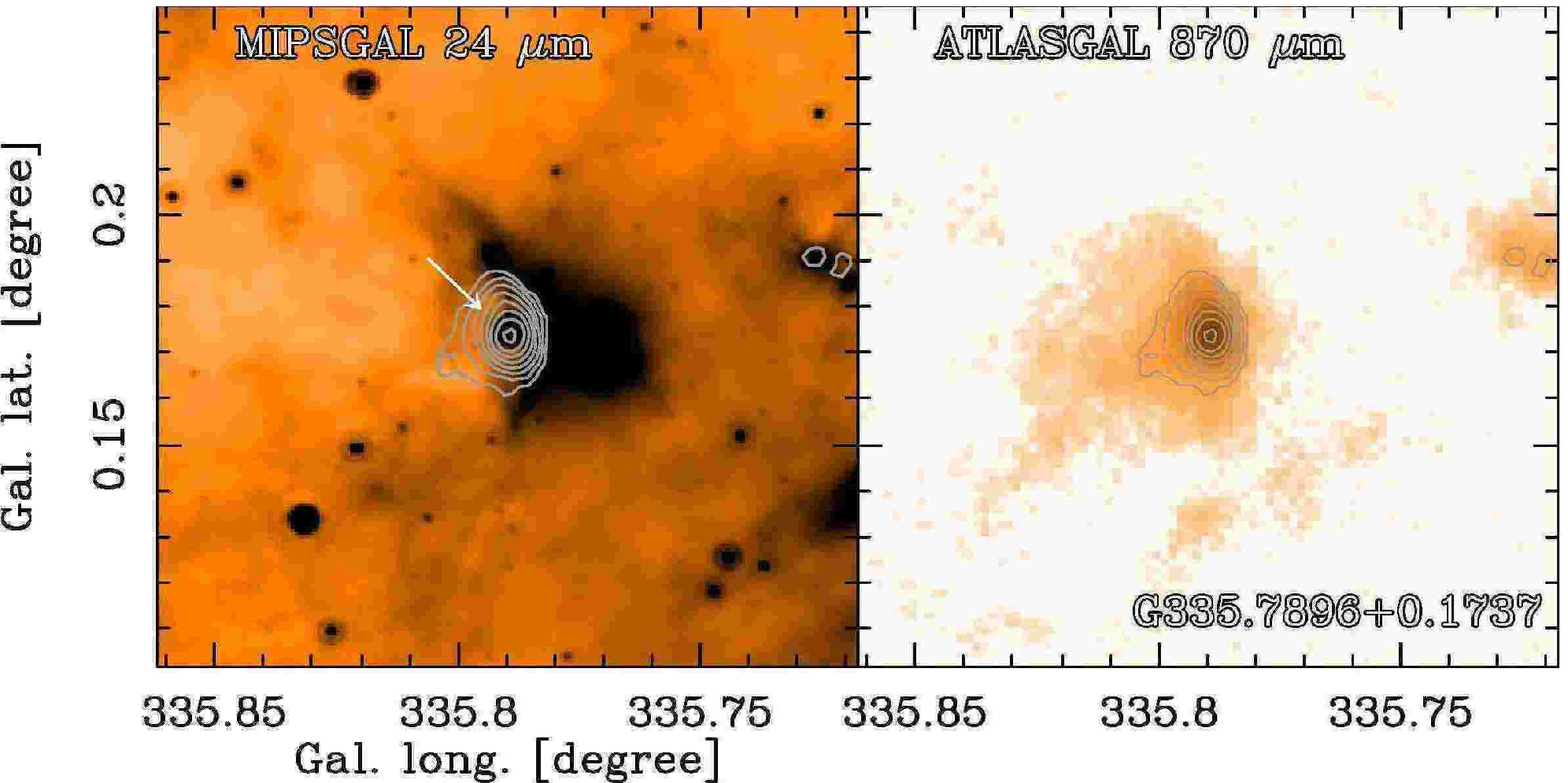} 
  \includegraphics[width=0.45\linewidth,clip]{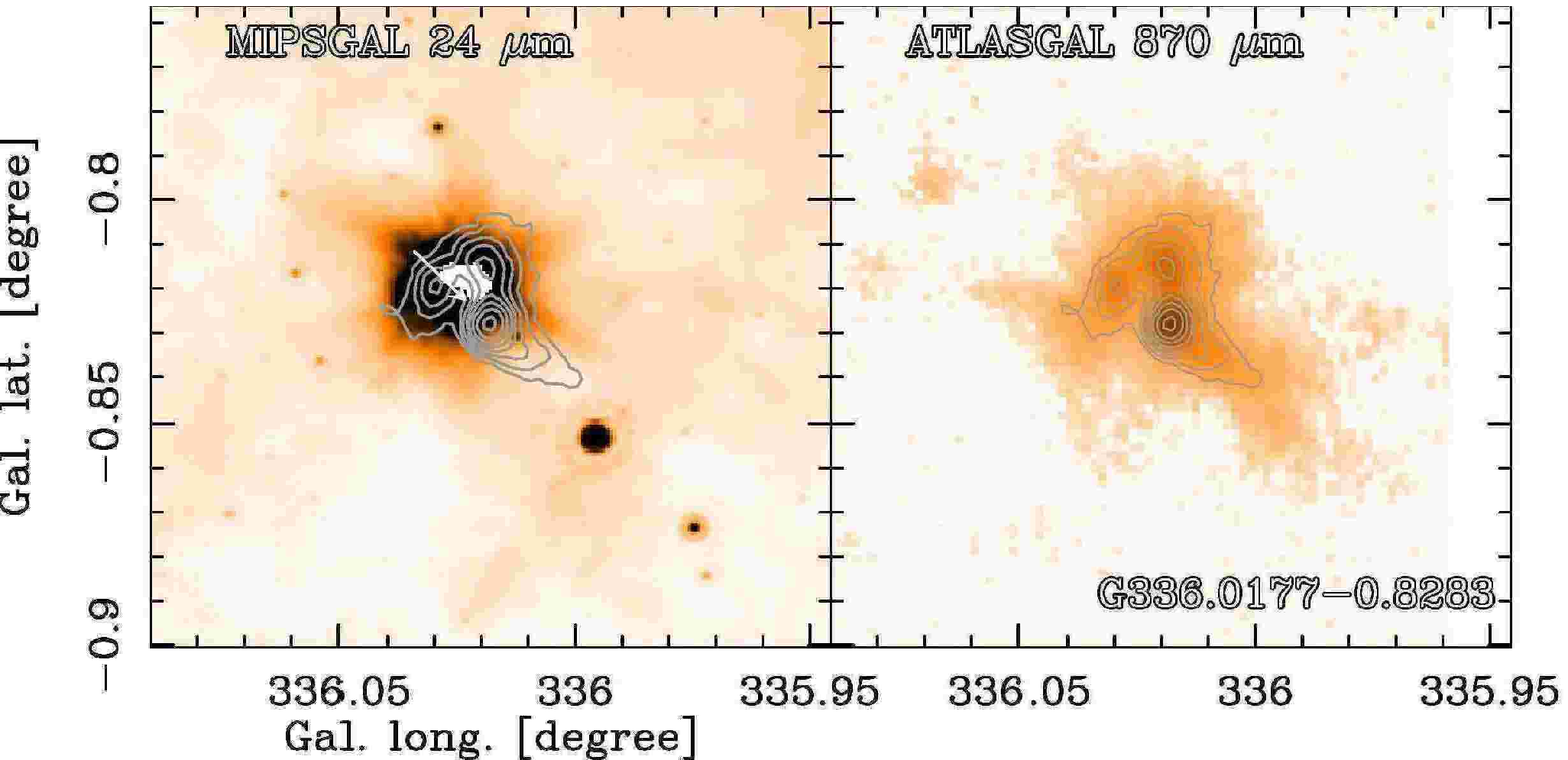} 
 \caption{Continued.}
 \end{figure*}
\begin{figure*}
\centering
\ContinuedFloat
  \includegraphics[width=0.45\linewidth,clip]{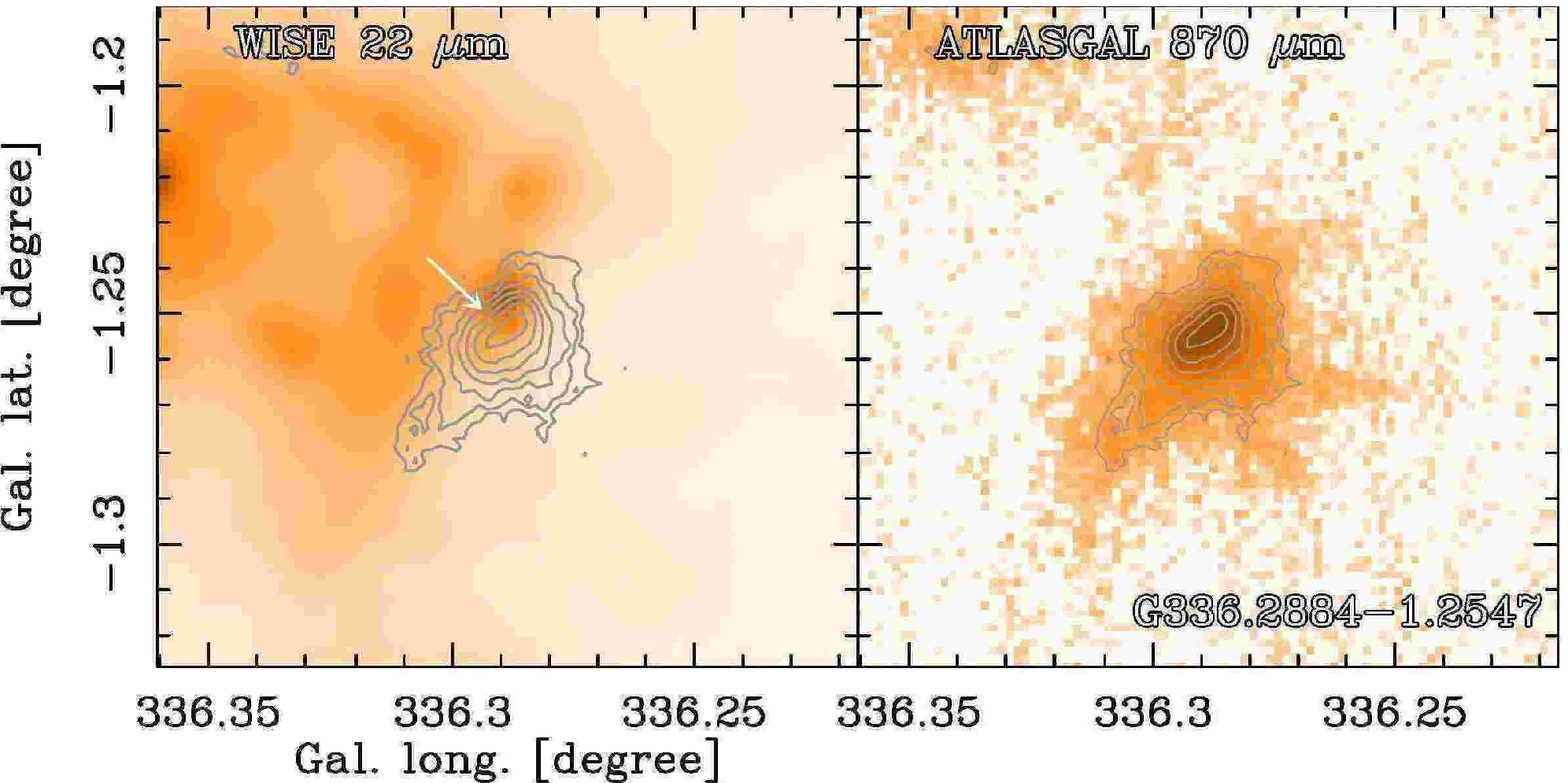} 
  \includegraphics[width=0.45\linewidth,clip]{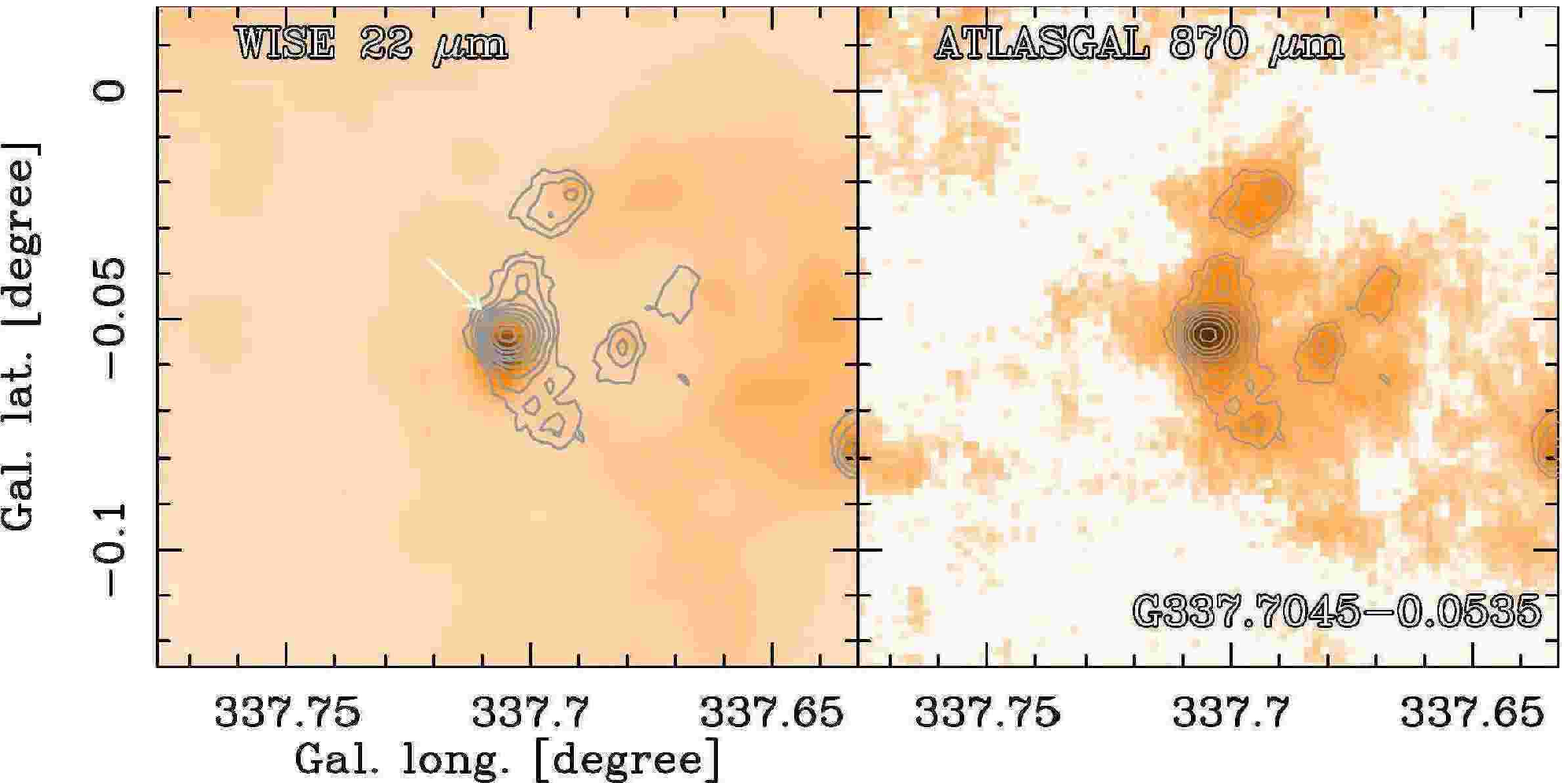} 
  \includegraphics[width=0.45\linewidth,clip]{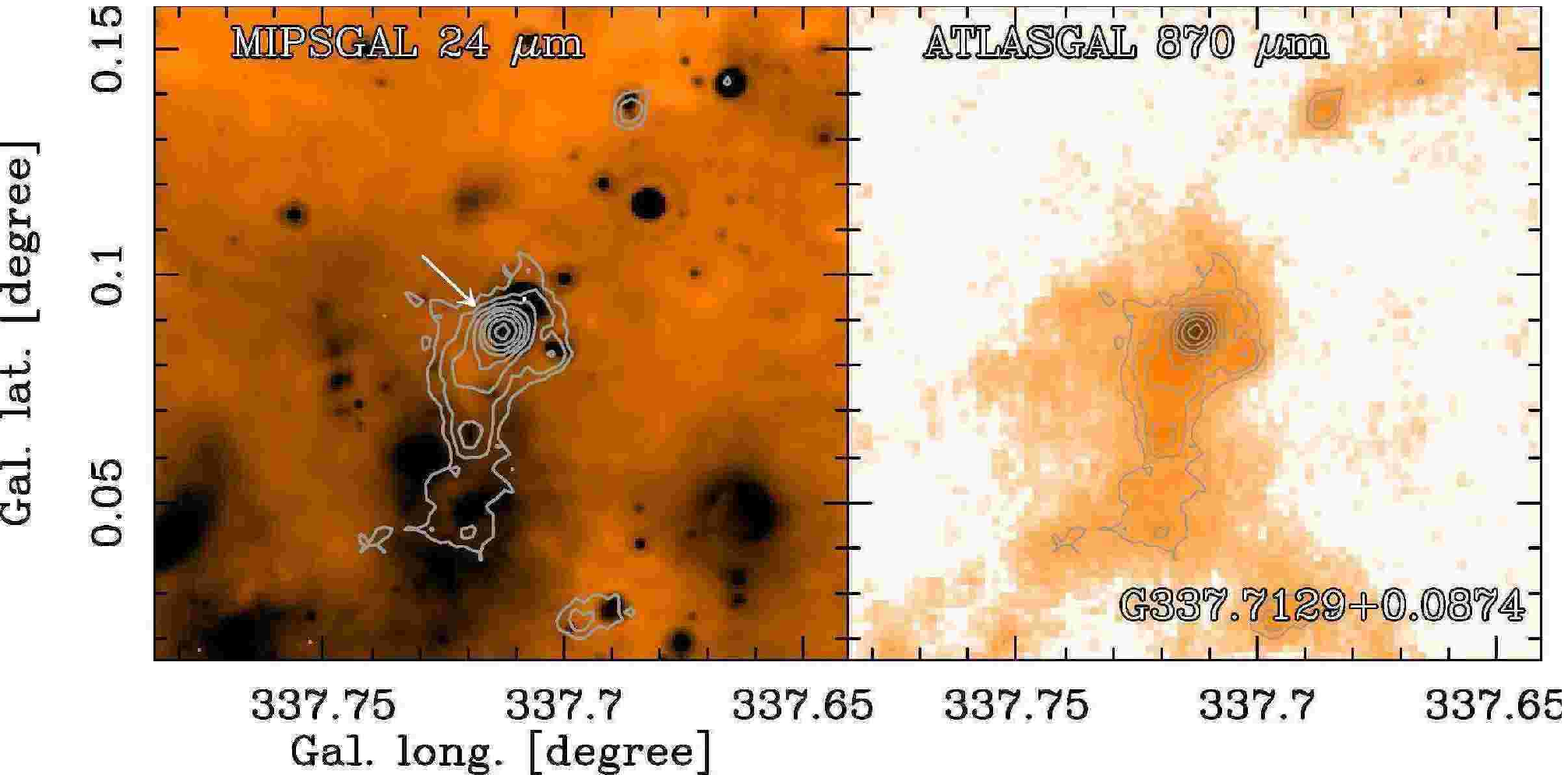} 
  \includegraphics[width=0.45\linewidth,clip]{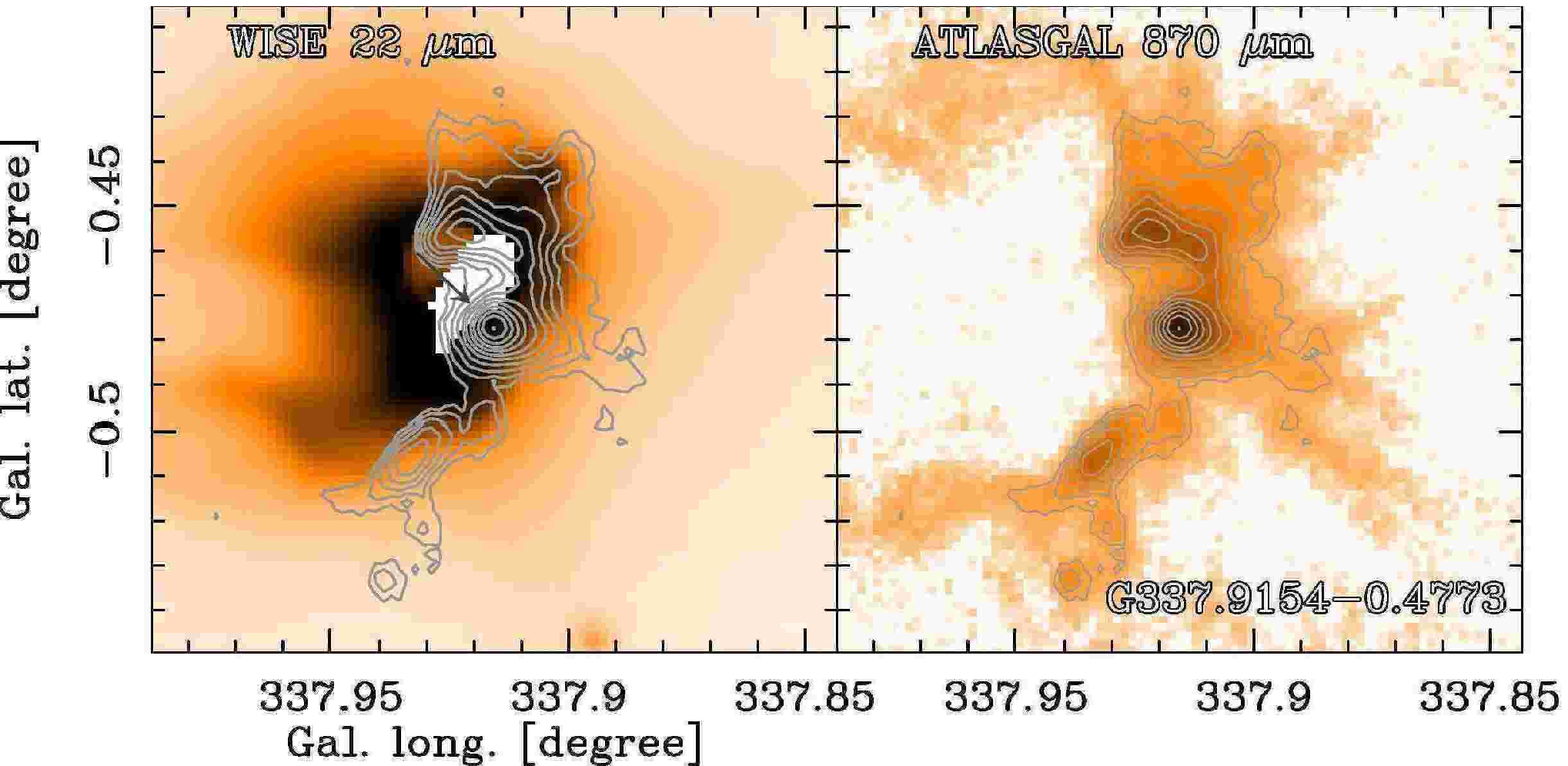} 
  \includegraphics[width=0.45\linewidth,clip]{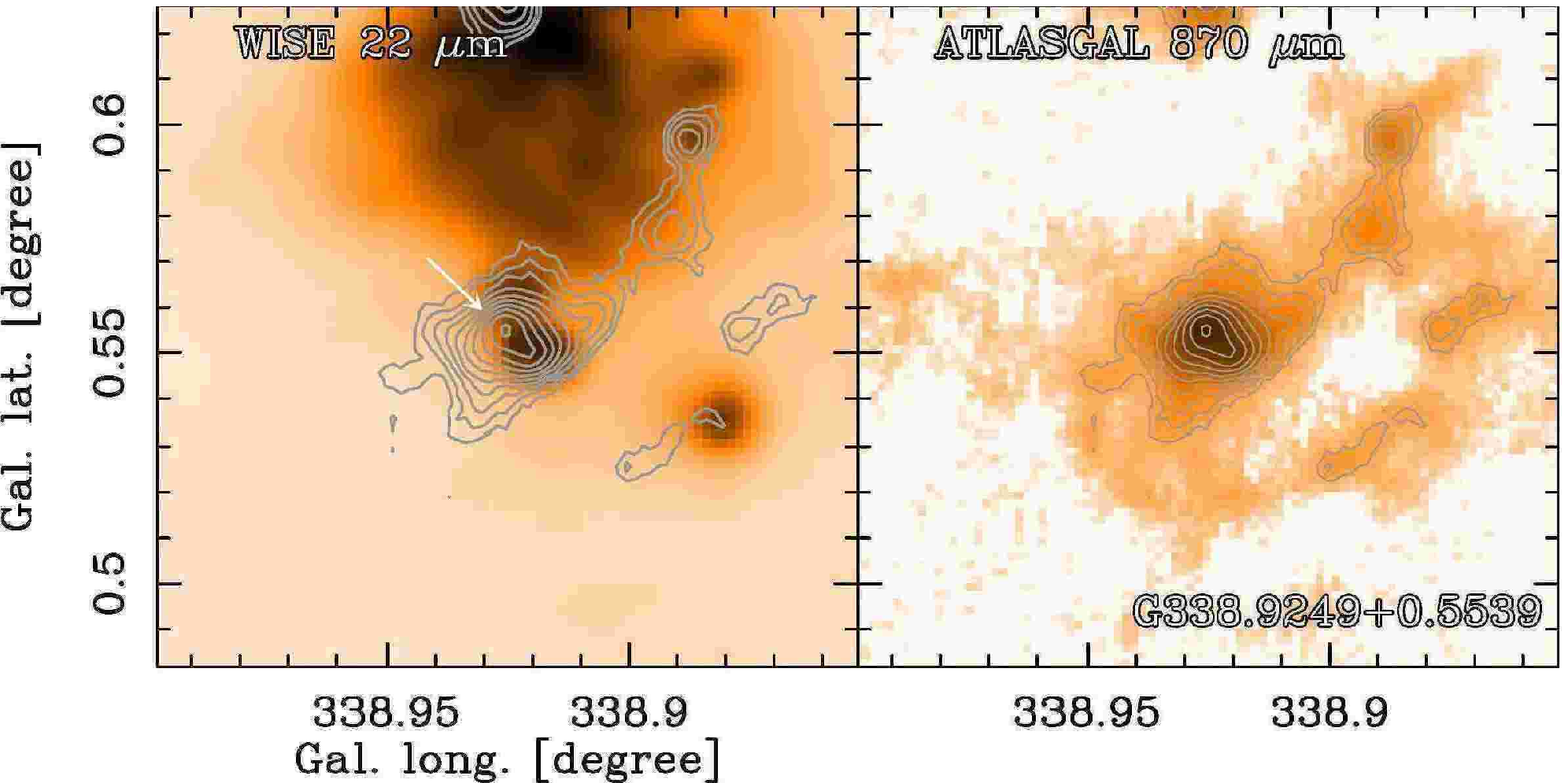} 
 \includegraphics[width=0.45\linewidth,clip]{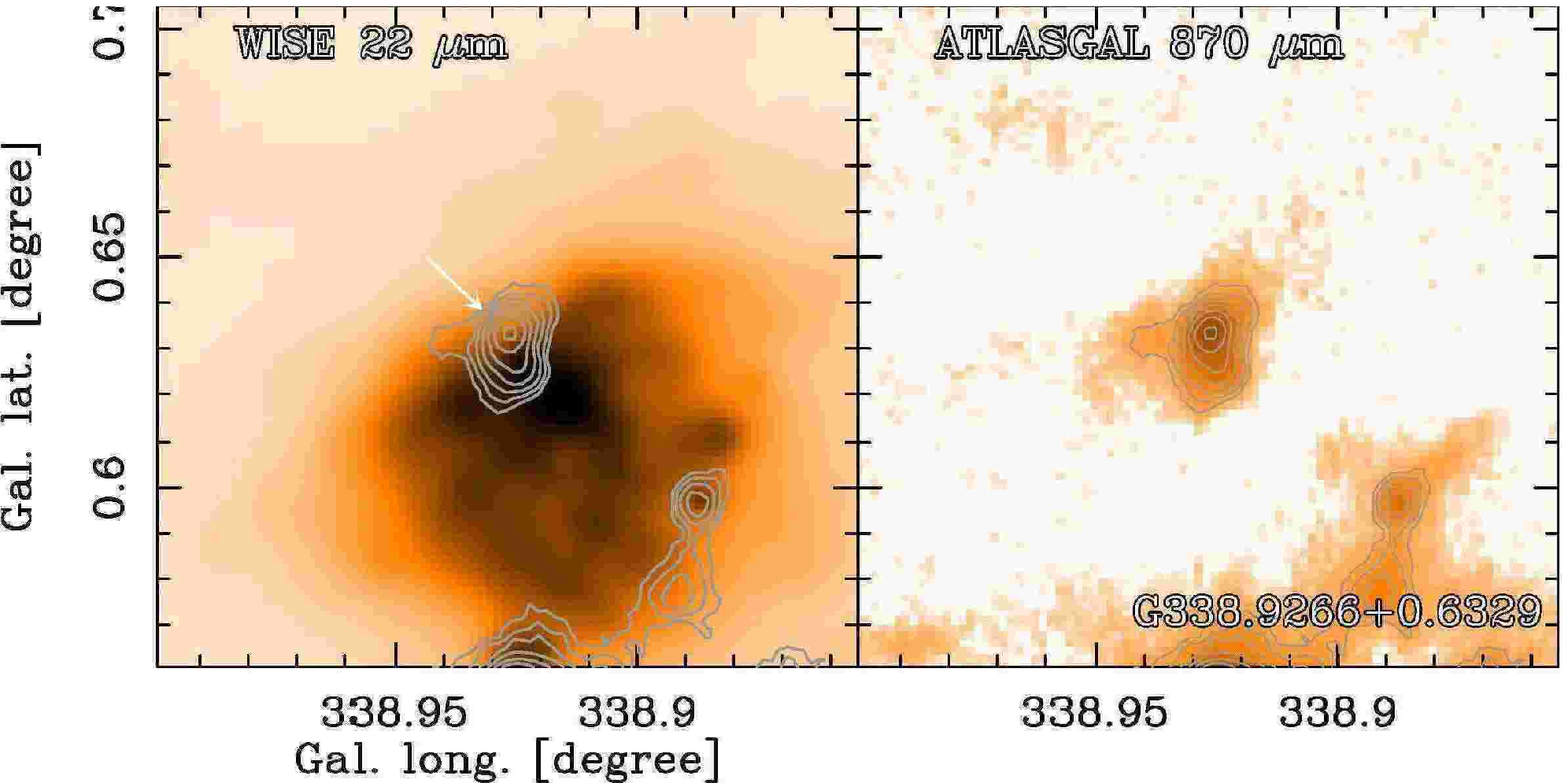} 
  \includegraphics[width=0.45\linewidth,clip]{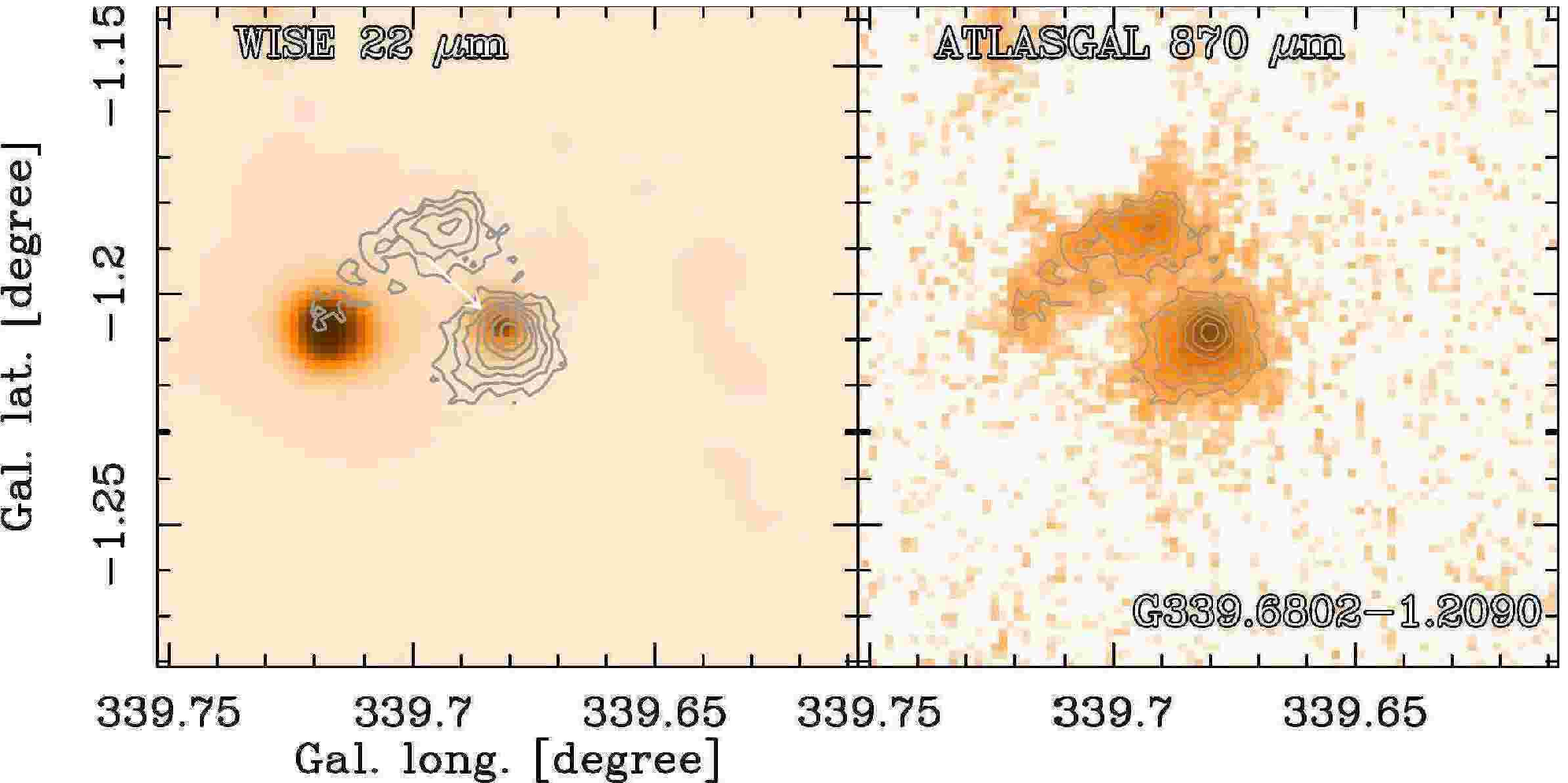} 
  \includegraphics[width=0.45\linewidth,clip]{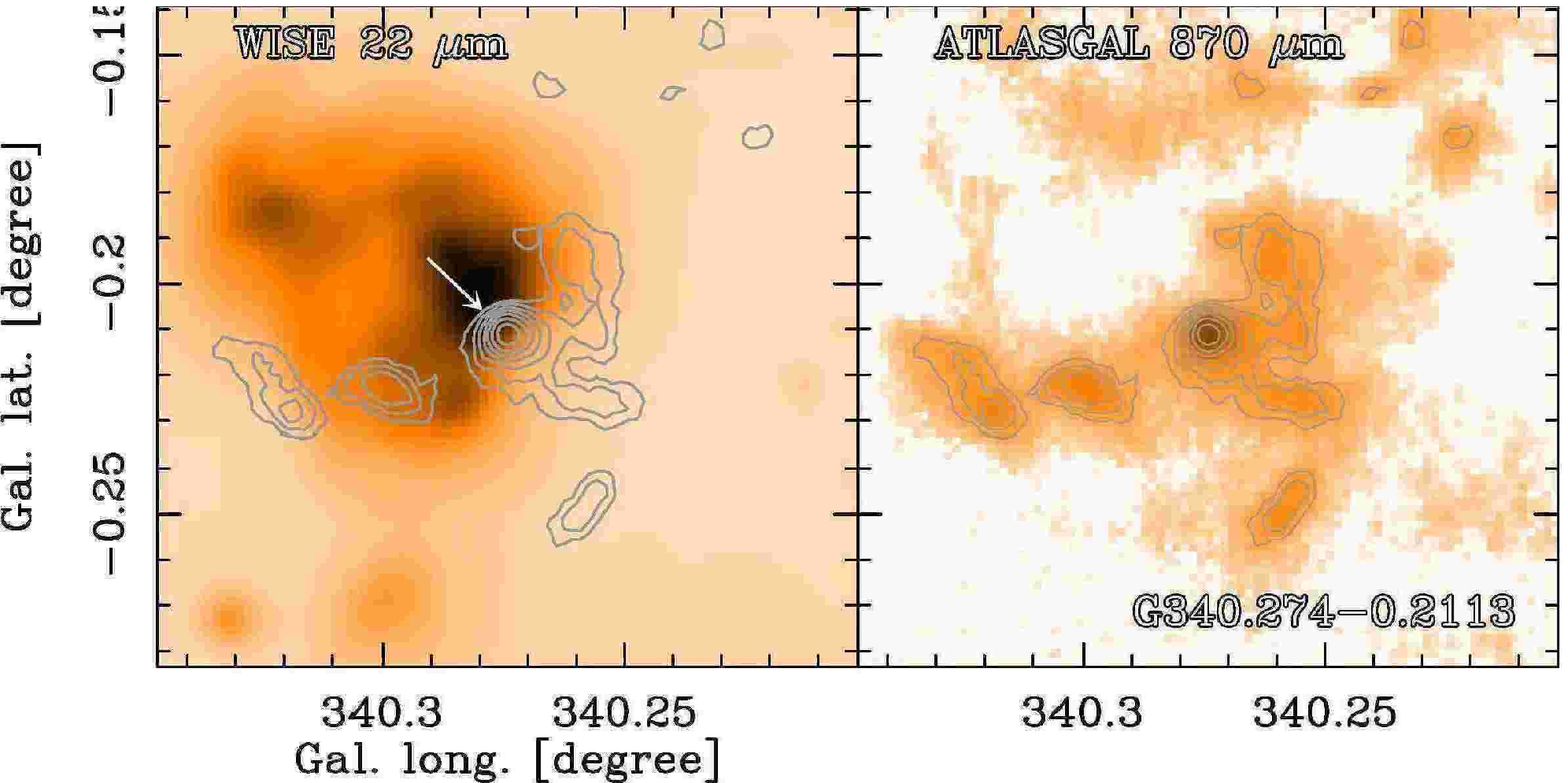} 
   \includegraphics[width=0.45\linewidth,clip]{FIG_ASTRO_PH/OUT_PAPER_WISE/70.eps} 
 \includegraphics[width=0.45\linewidth,clip]{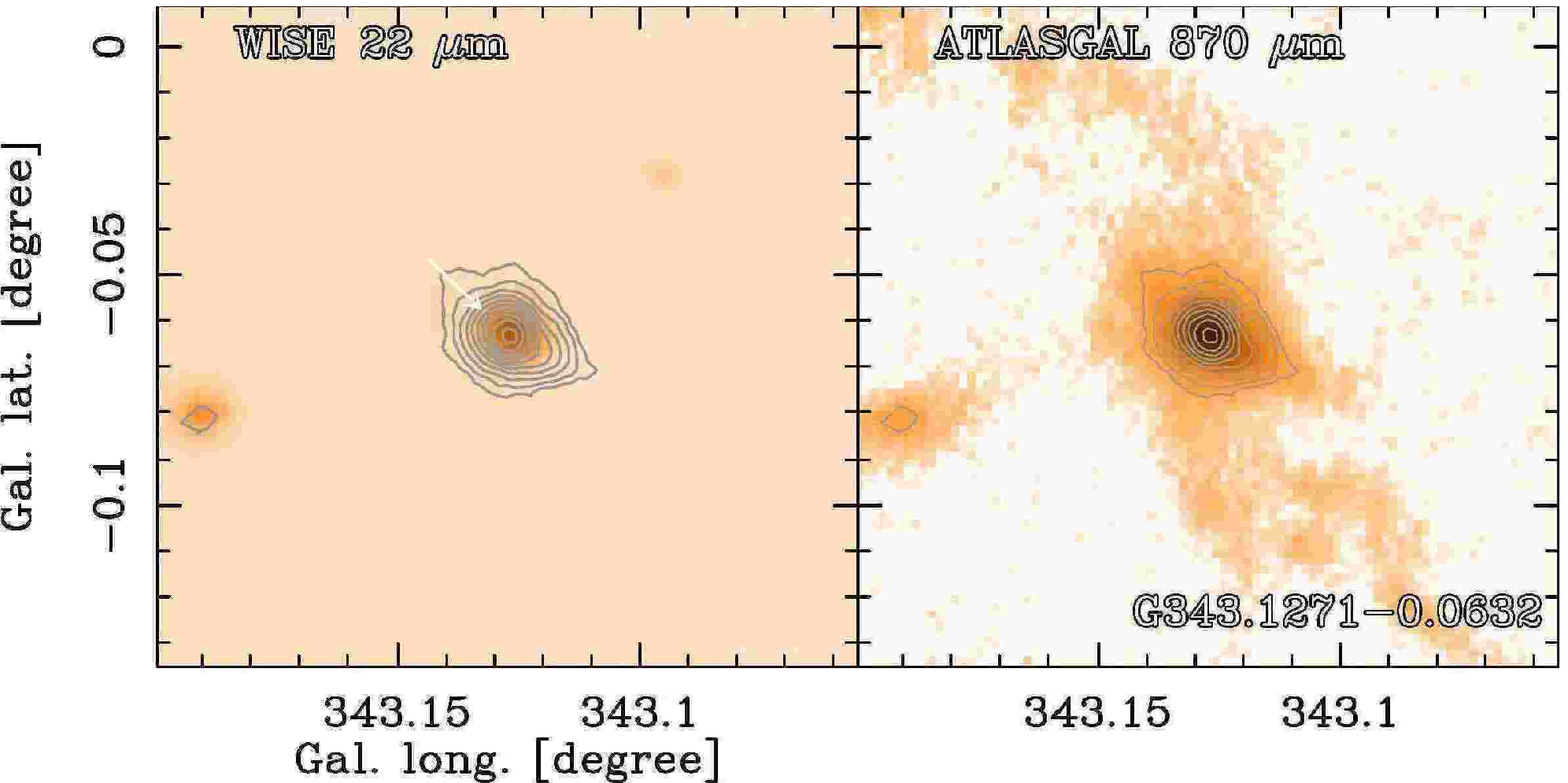} 
   \includegraphics[width=0.45\linewidth,clip]{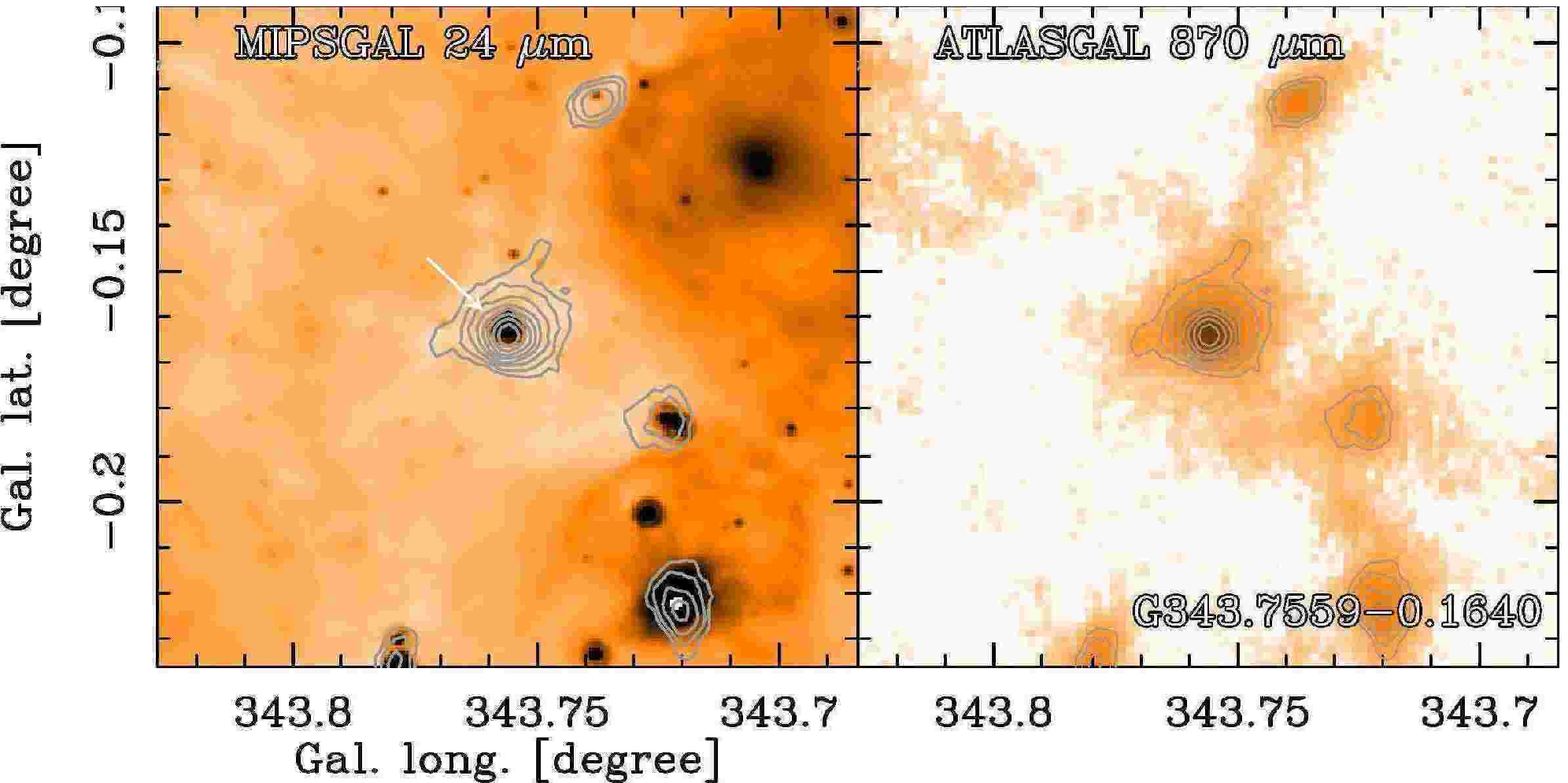} 
 \includegraphics[width=0.45\linewidth,clip]{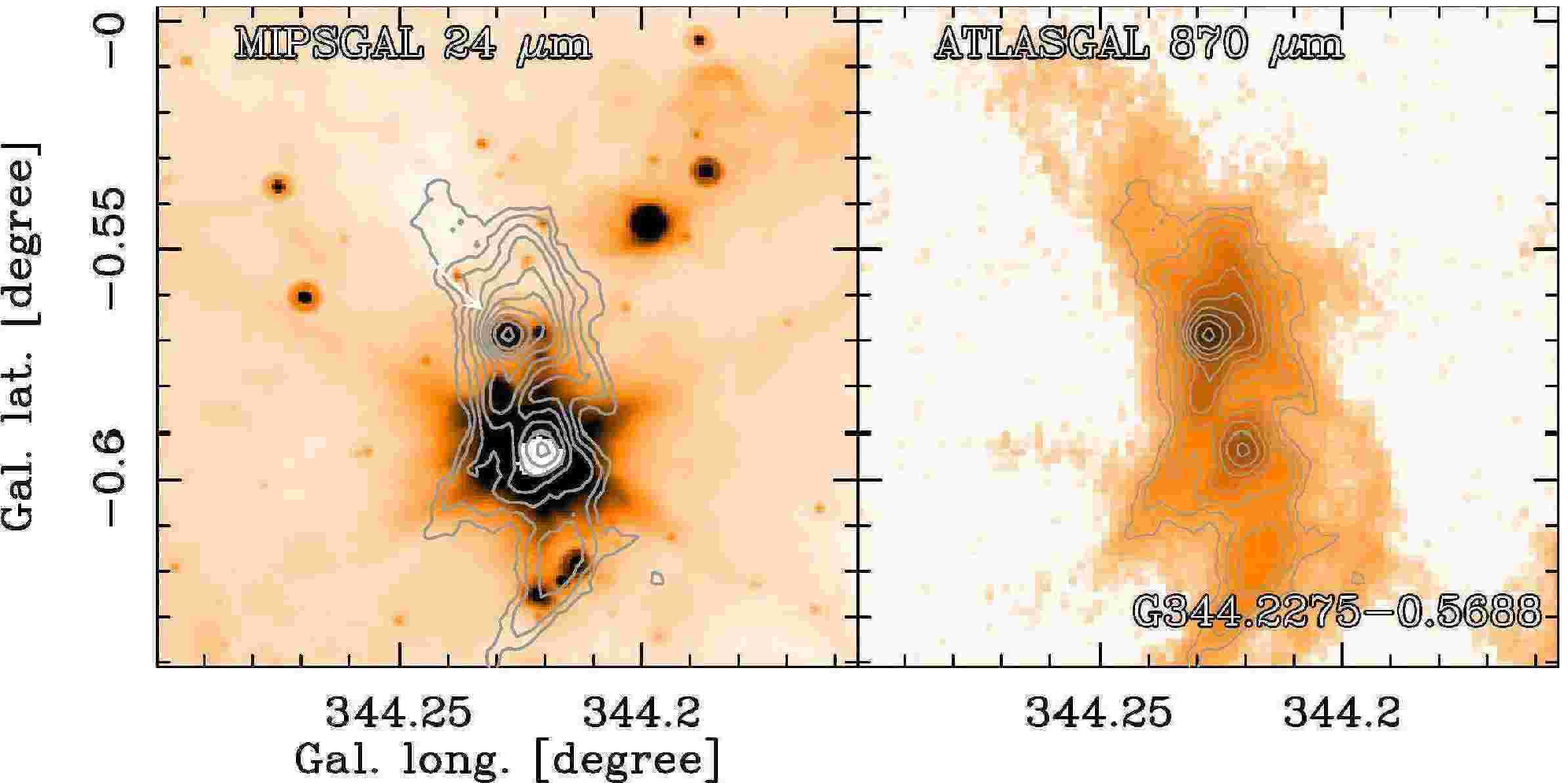} 
  \caption{Continued.}
 \end{figure*}
\begin{figure*}
\centering
\ContinuedFloat
  \includegraphics[width=0.45\linewidth,clip]{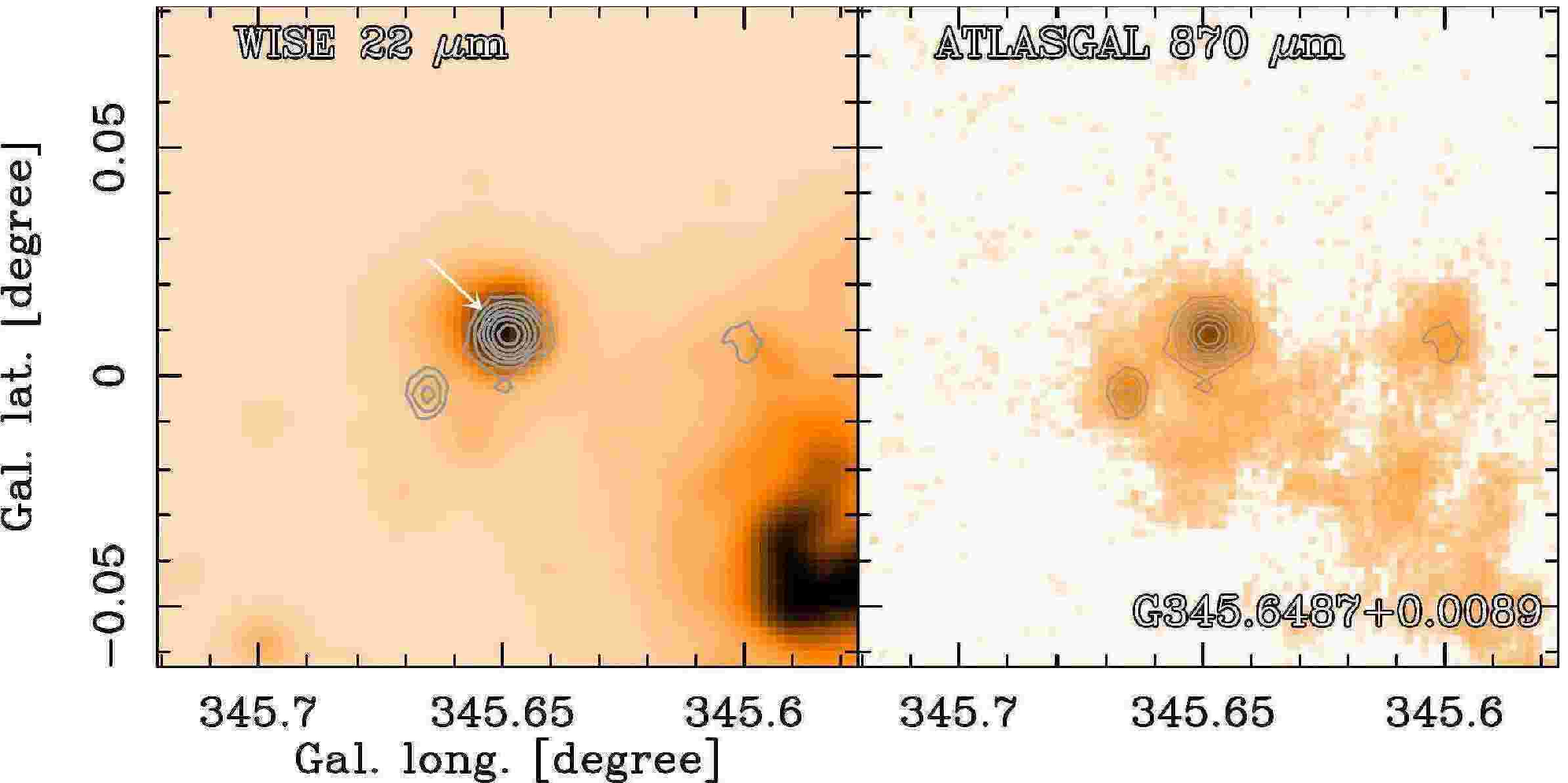} 
 \includegraphics[width=0.45\linewidth,clip]{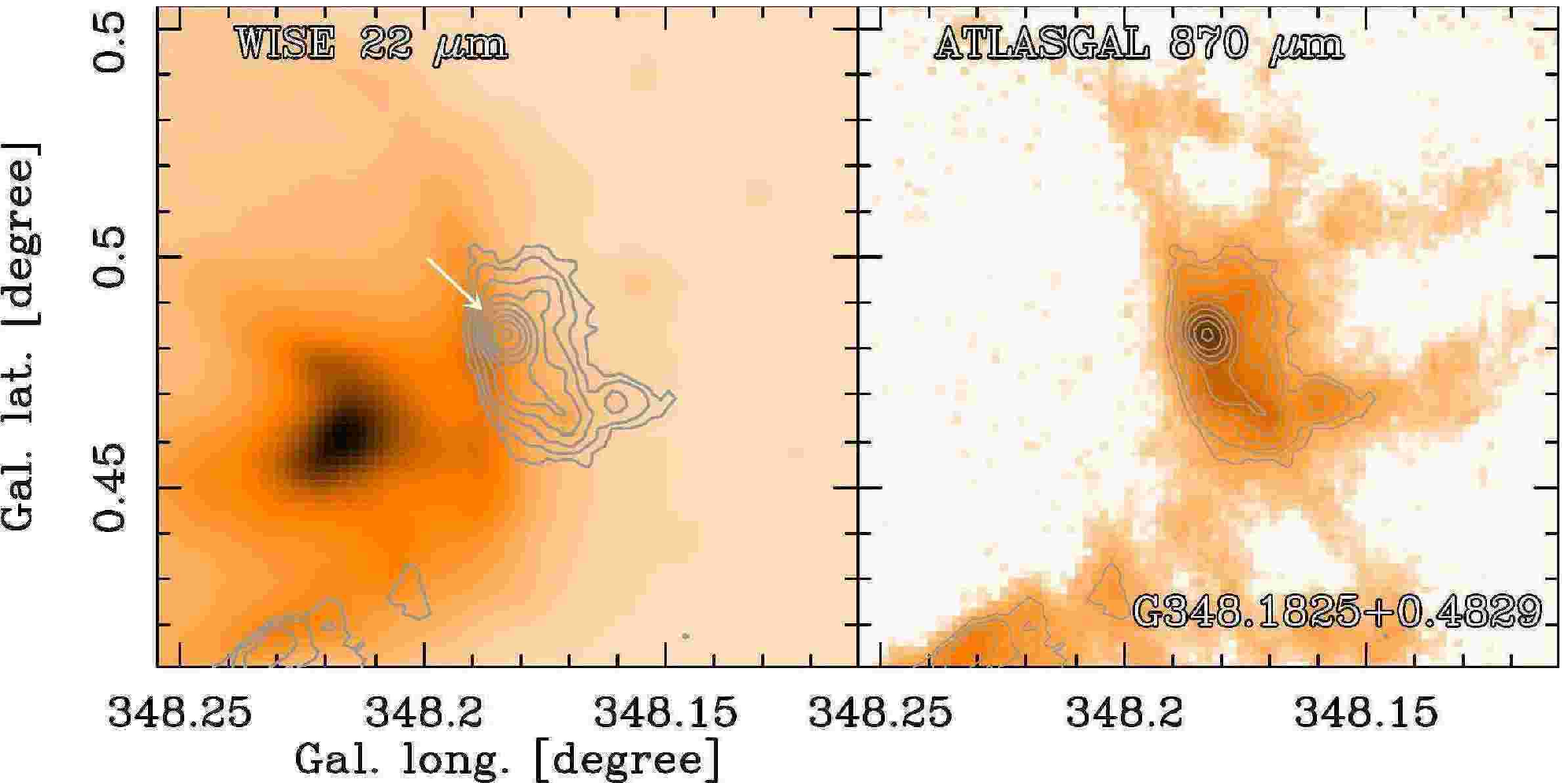} 
 \includegraphics[width=0.45\linewidth,clip]{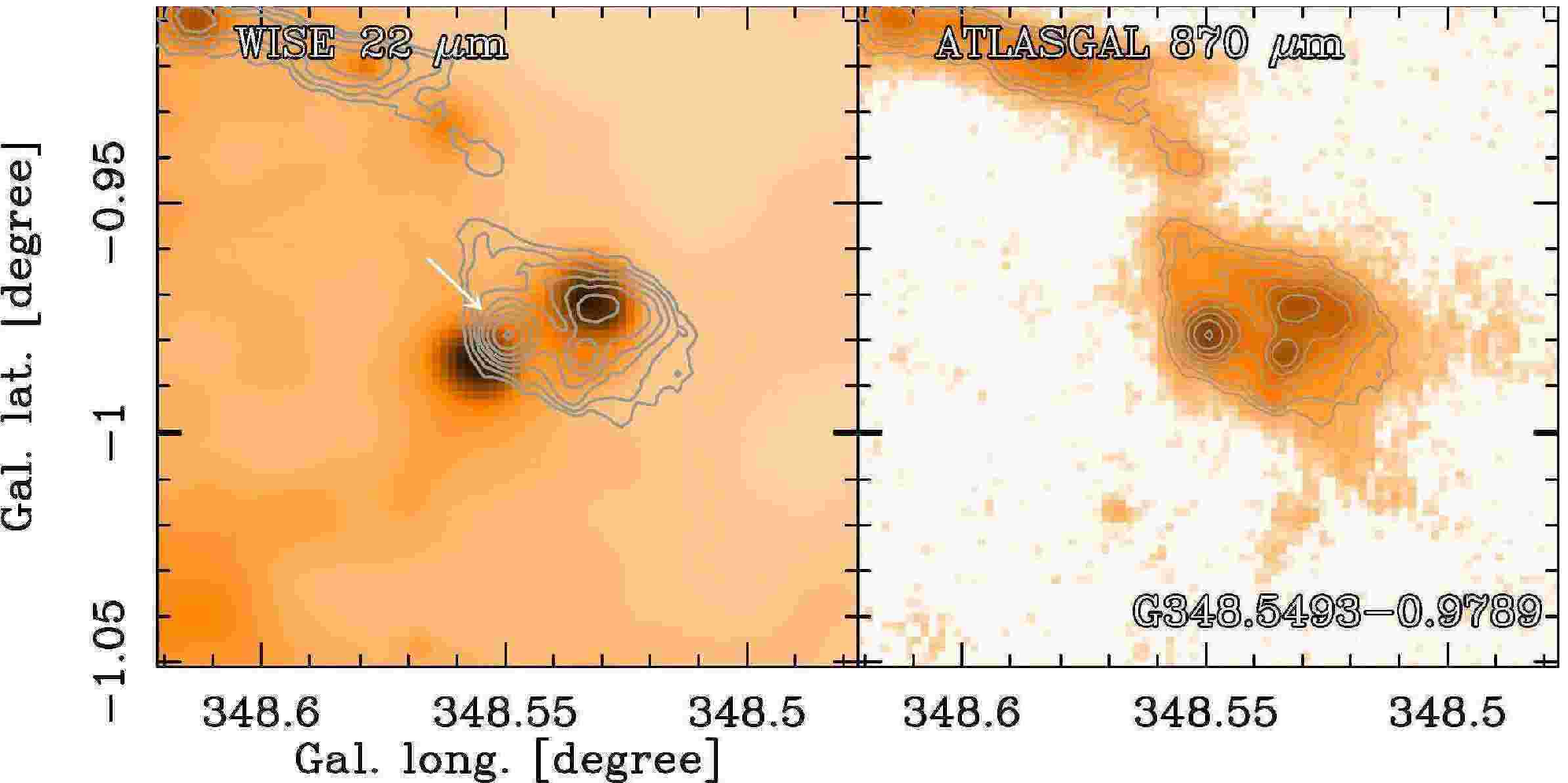} 
   \includegraphics[width=0.45\linewidth,clip]{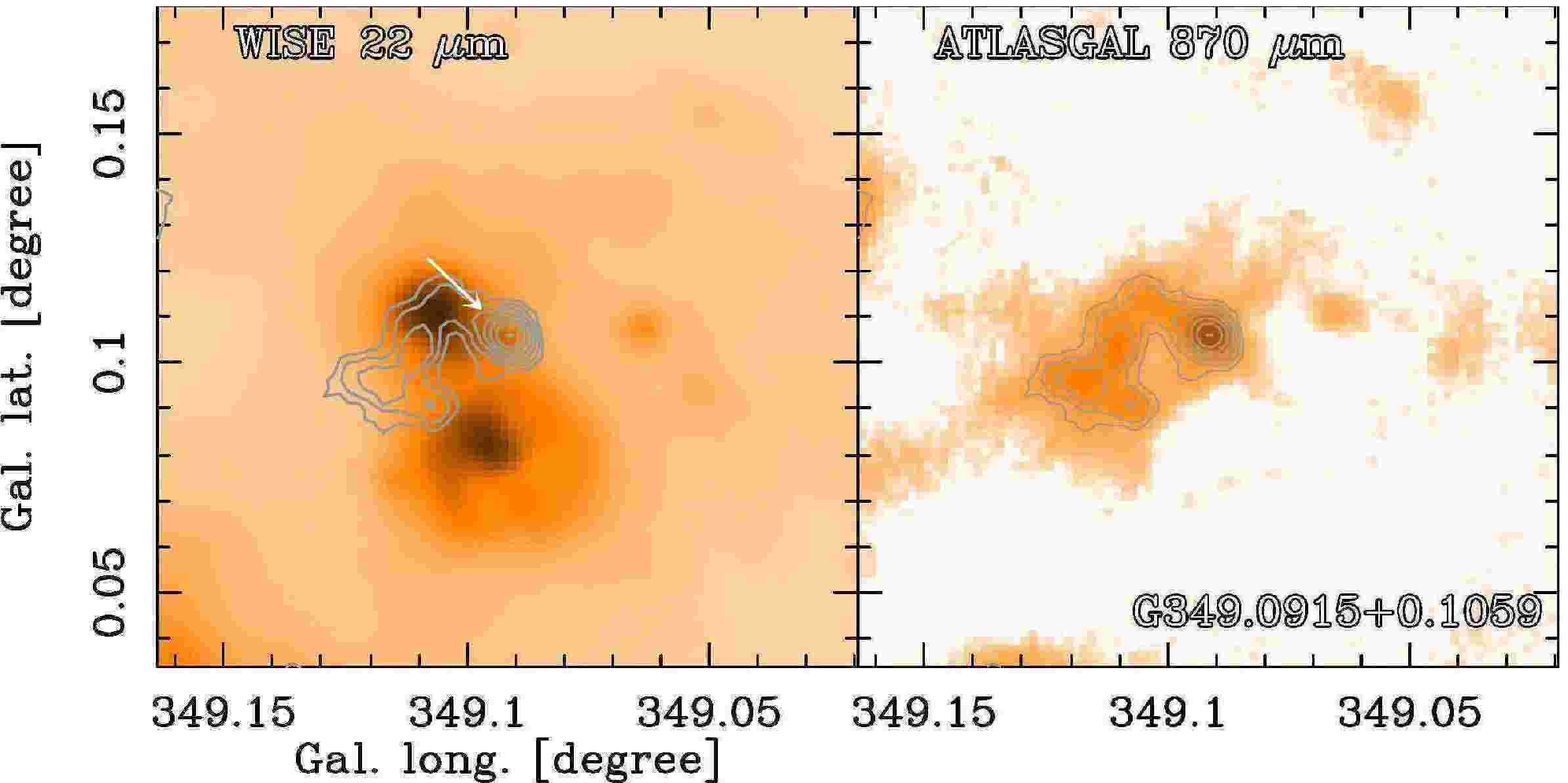} 
 \includegraphics[width=0.45\linewidth,clip]{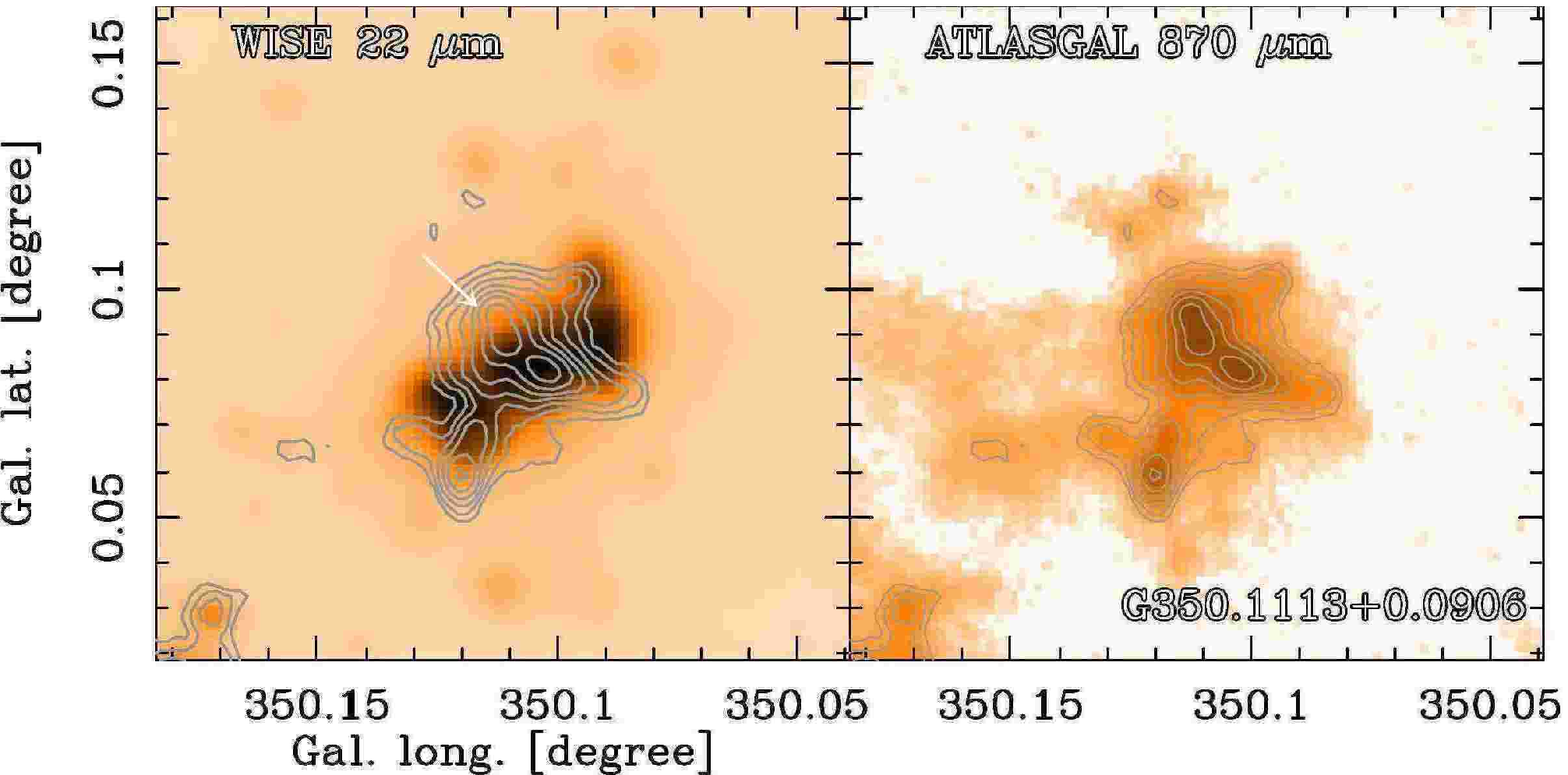} 
 \includegraphics[width=0.45\linewidth,clip]{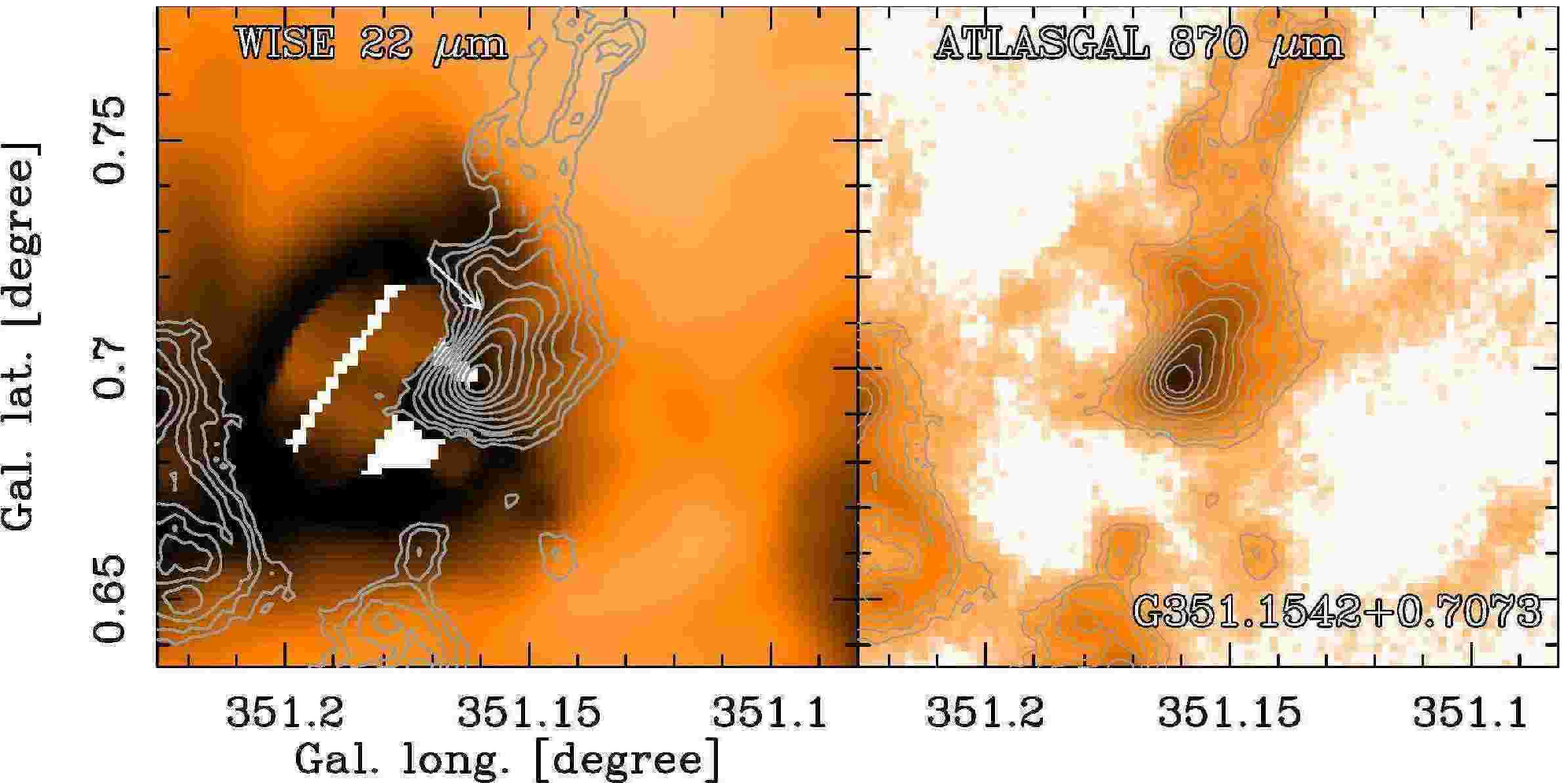} 
   \includegraphics[width=0.45\linewidth,clip]{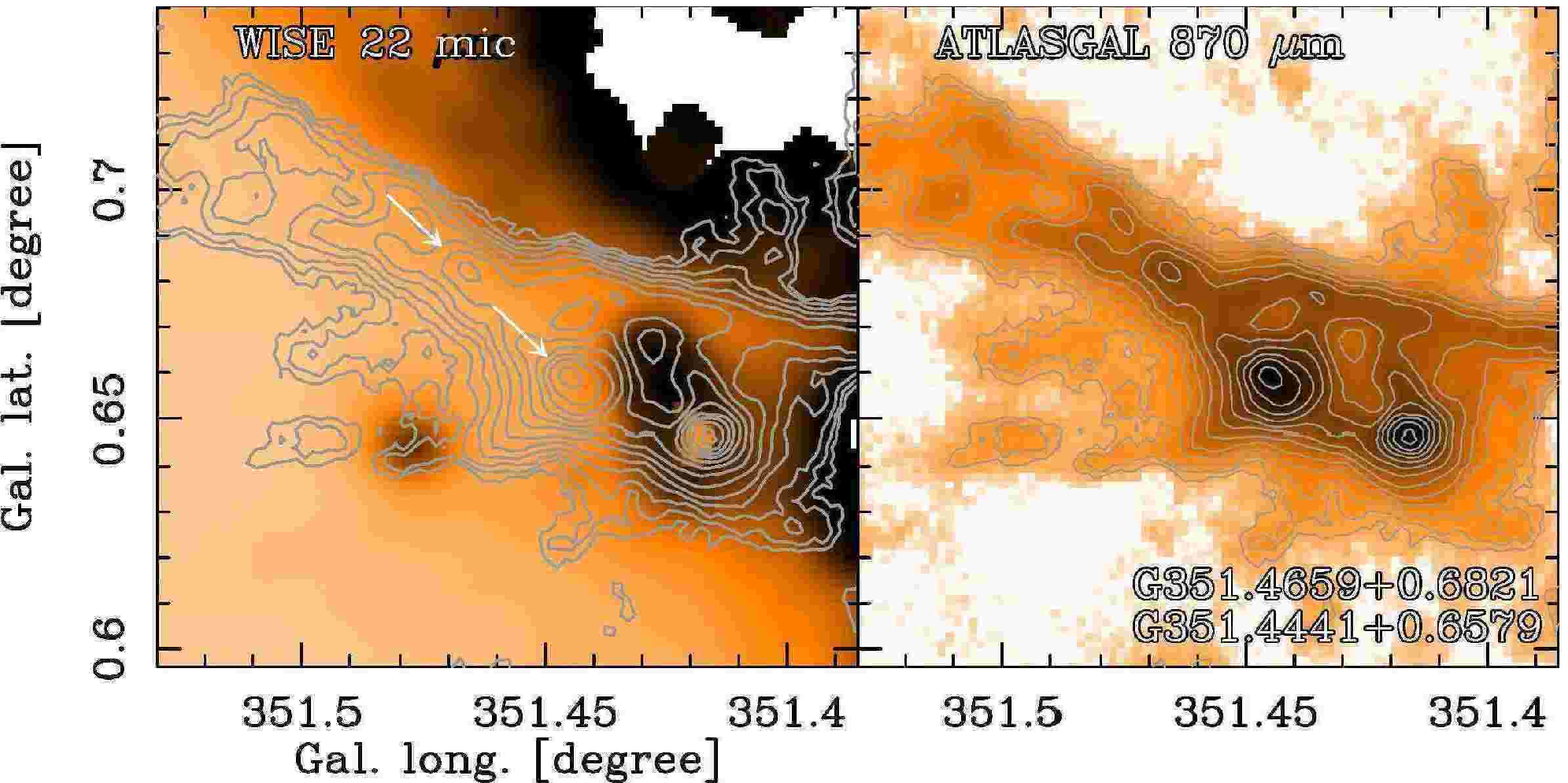} 
   \includegraphics[width=0.45\linewidth,clip]{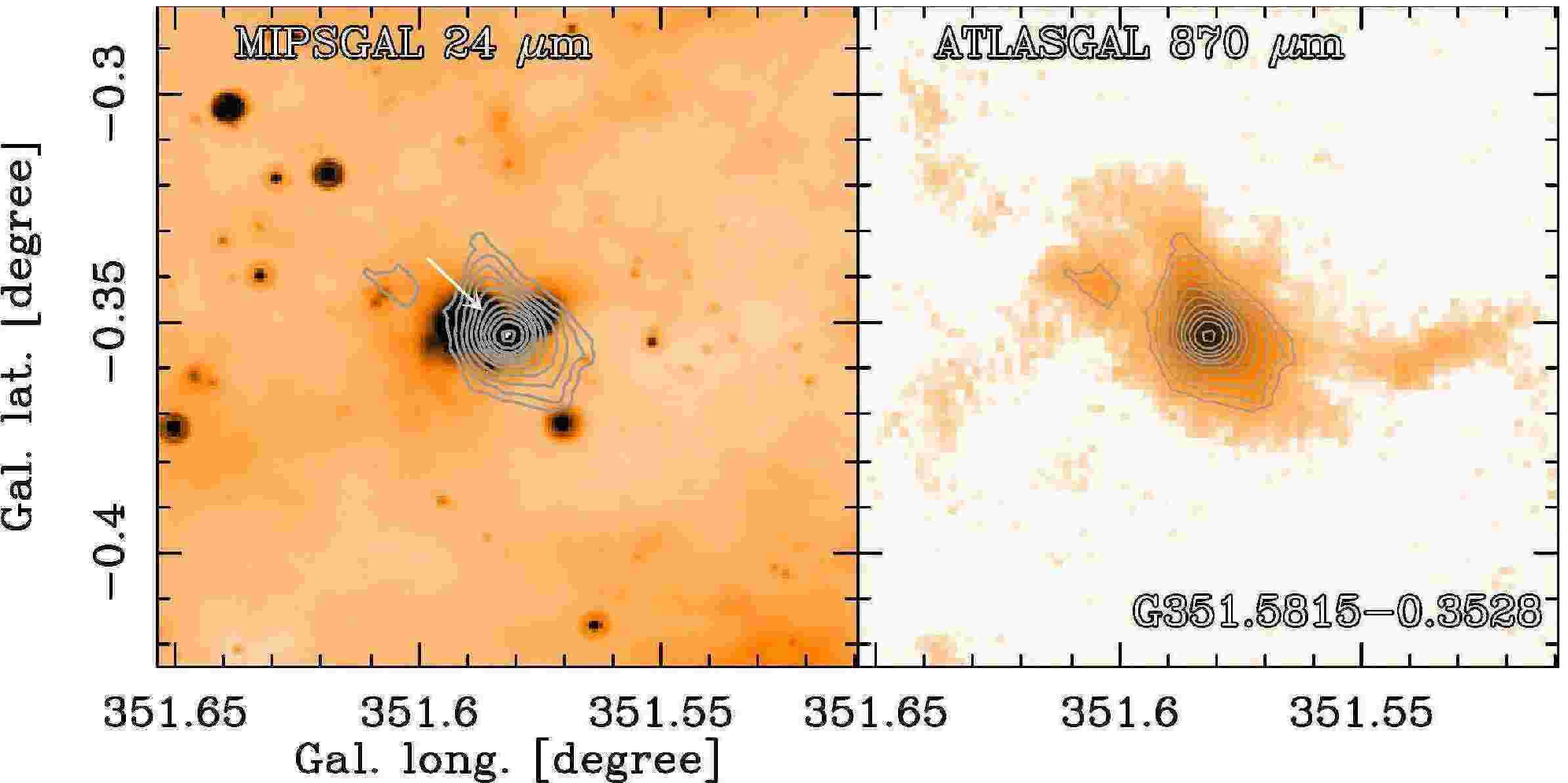} 
  \includegraphics[width=0.45\linewidth,clip]{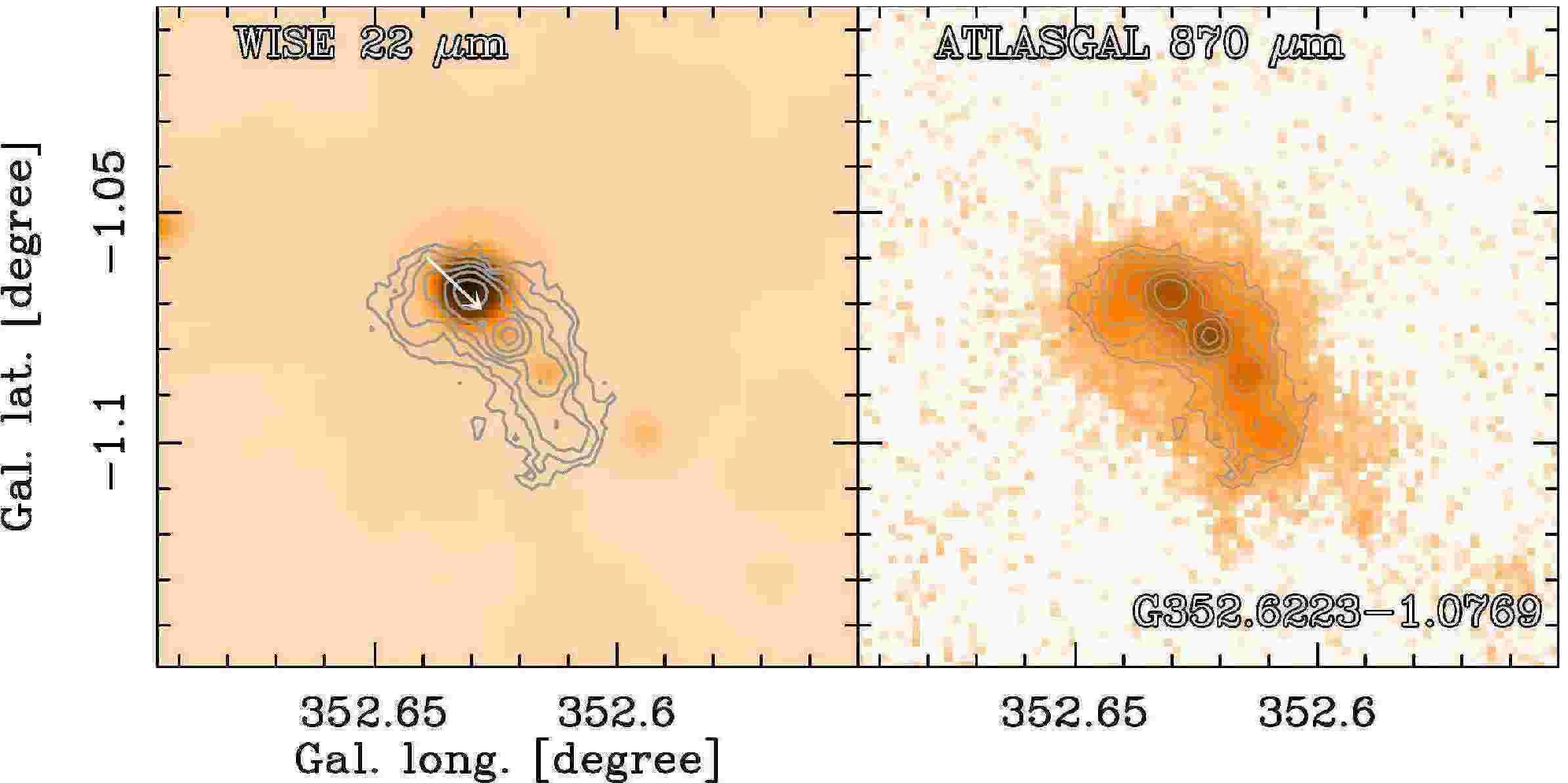} 
 \includegraphics[width=0.45\linewidth,clip]{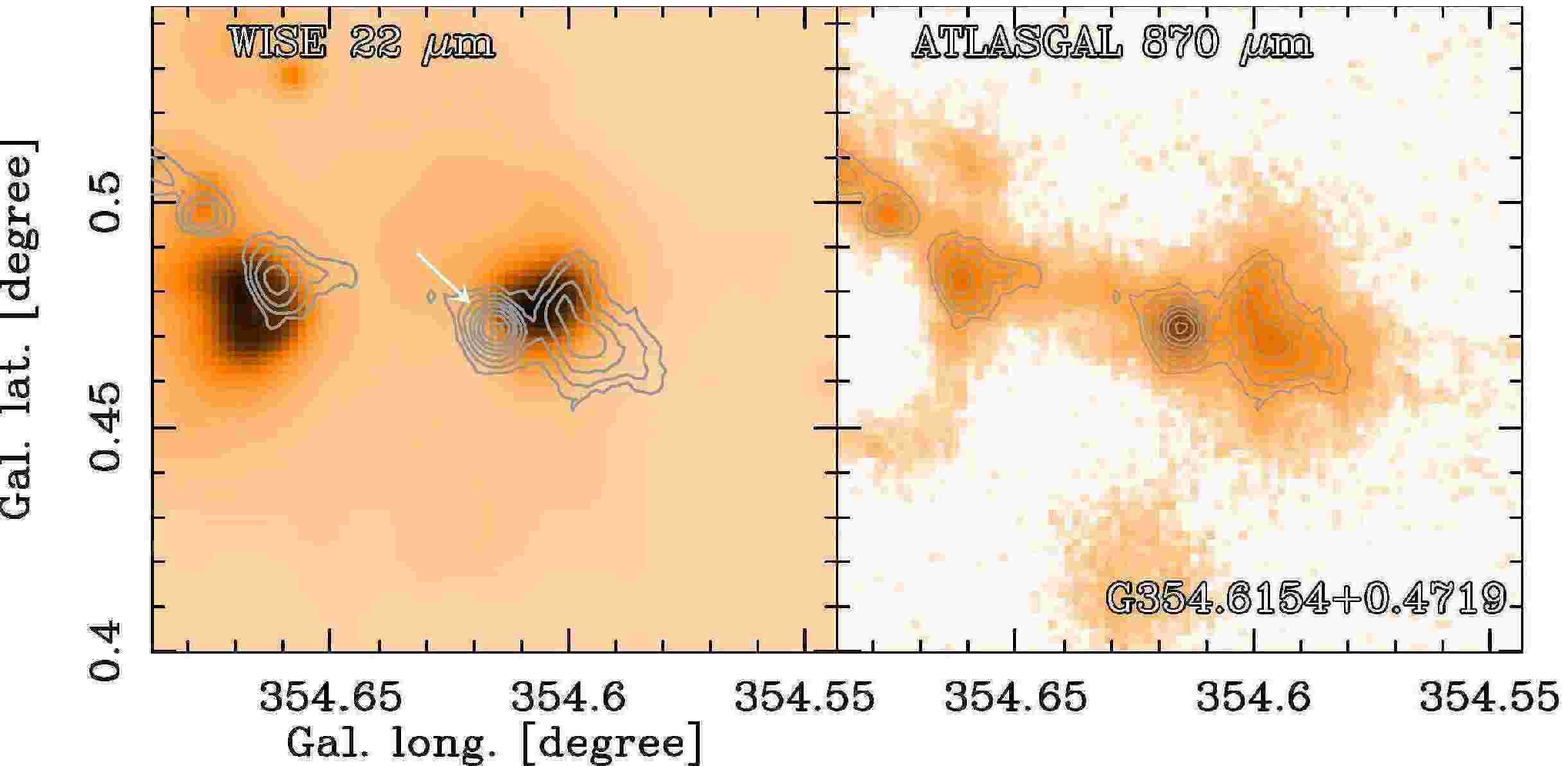} 
  \includegraphics[width=0.45\linewidth,clip]{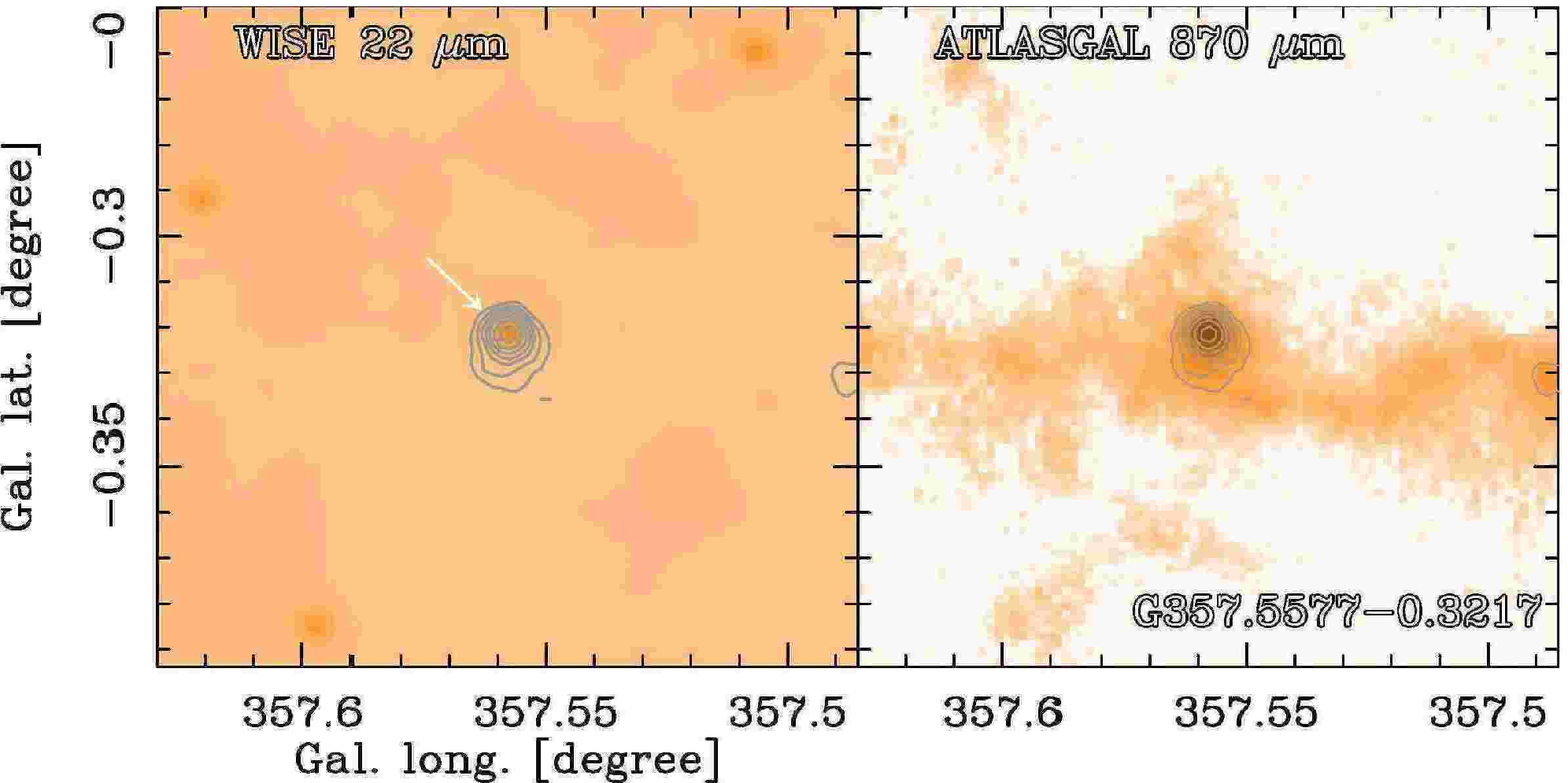} 
 \includegraphics[width=0.45\linewidth,clip]{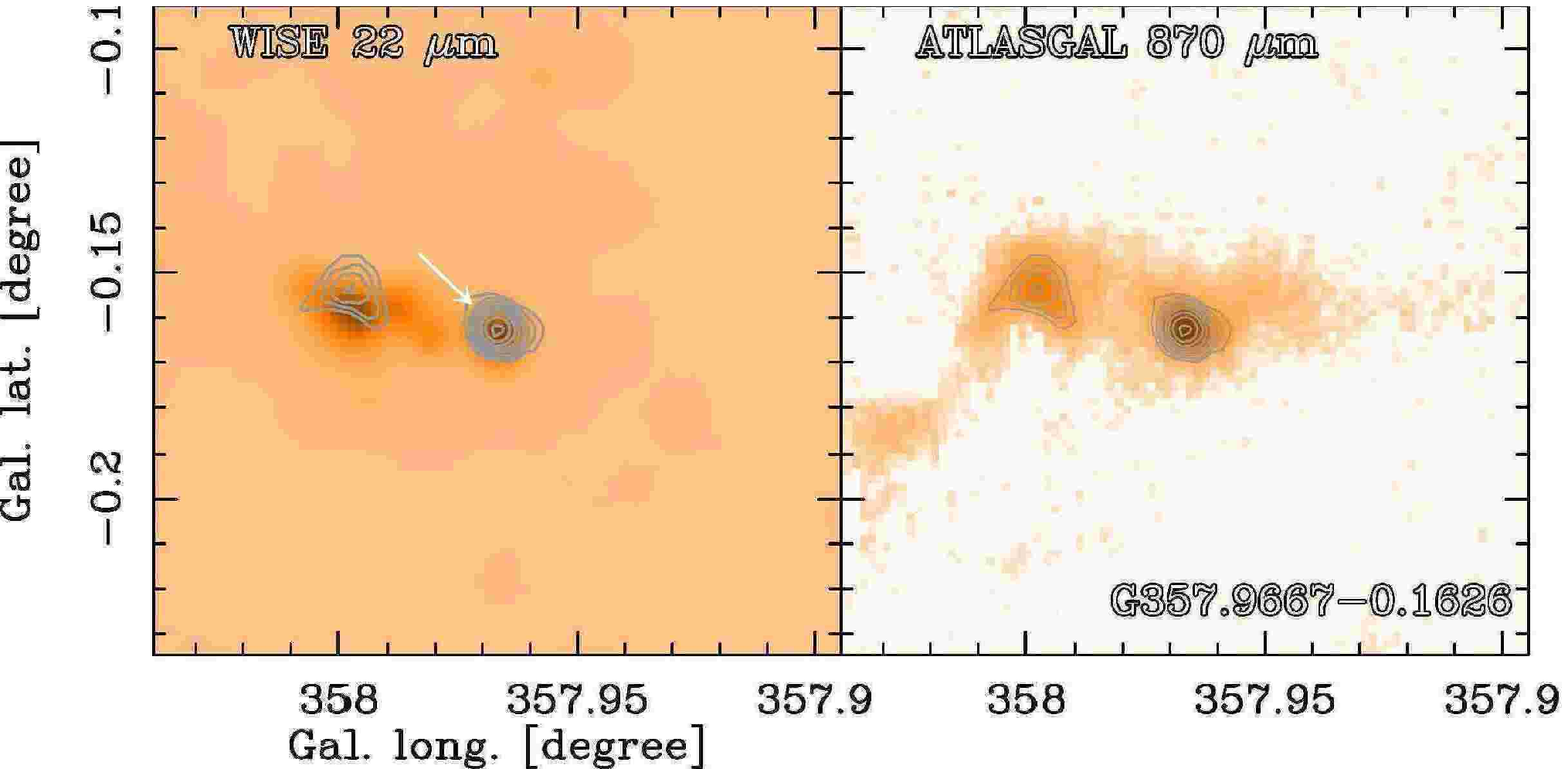} 
 \caption{Continued.}
 \end{figure*}
\begin{figure*}
\centering
\ContinuedFloat 
   \includegraphics[width=0.45\linewidth,clip]{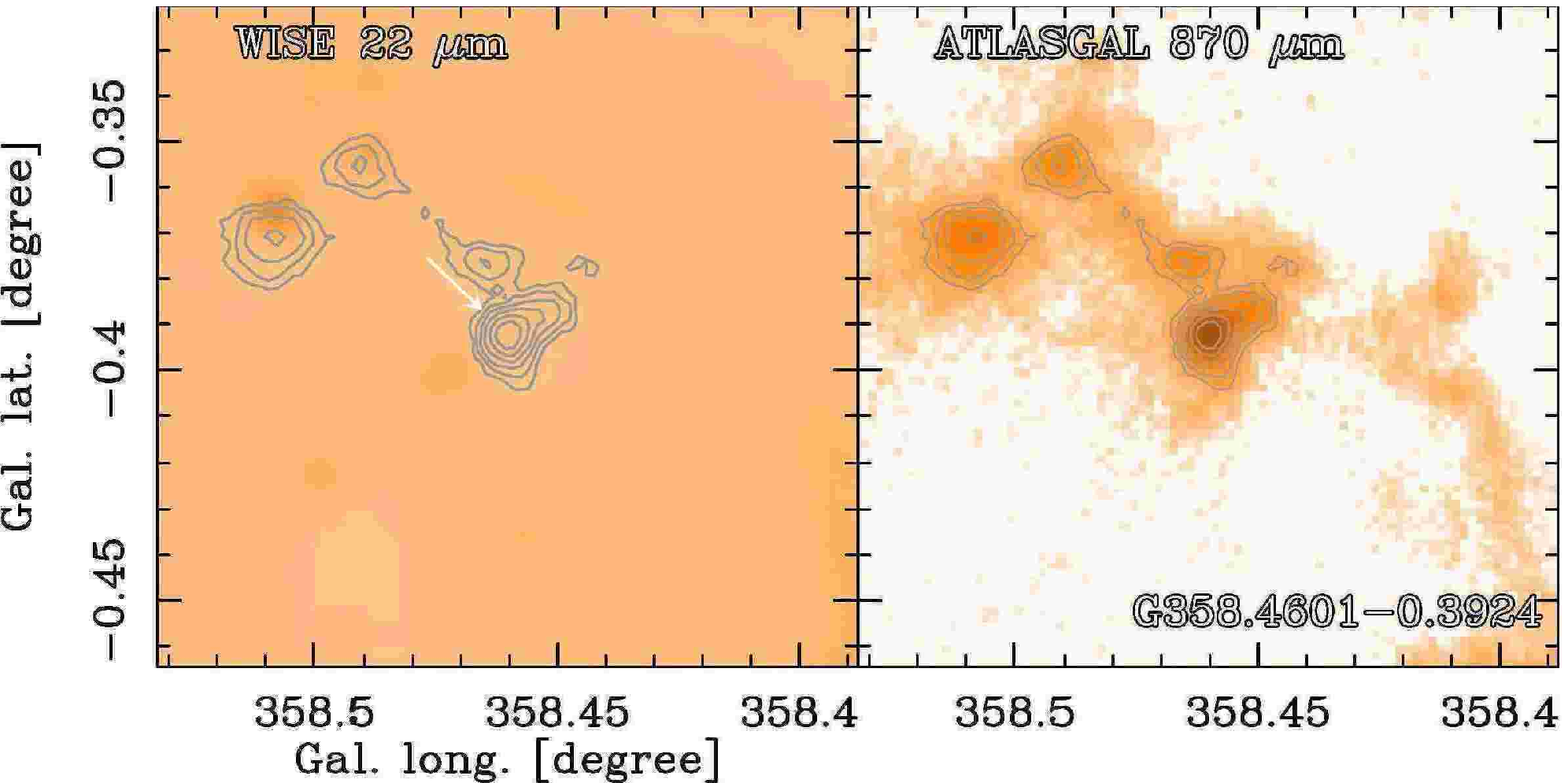} 
   \includegraphics[width=0.45\linewidth,clip]{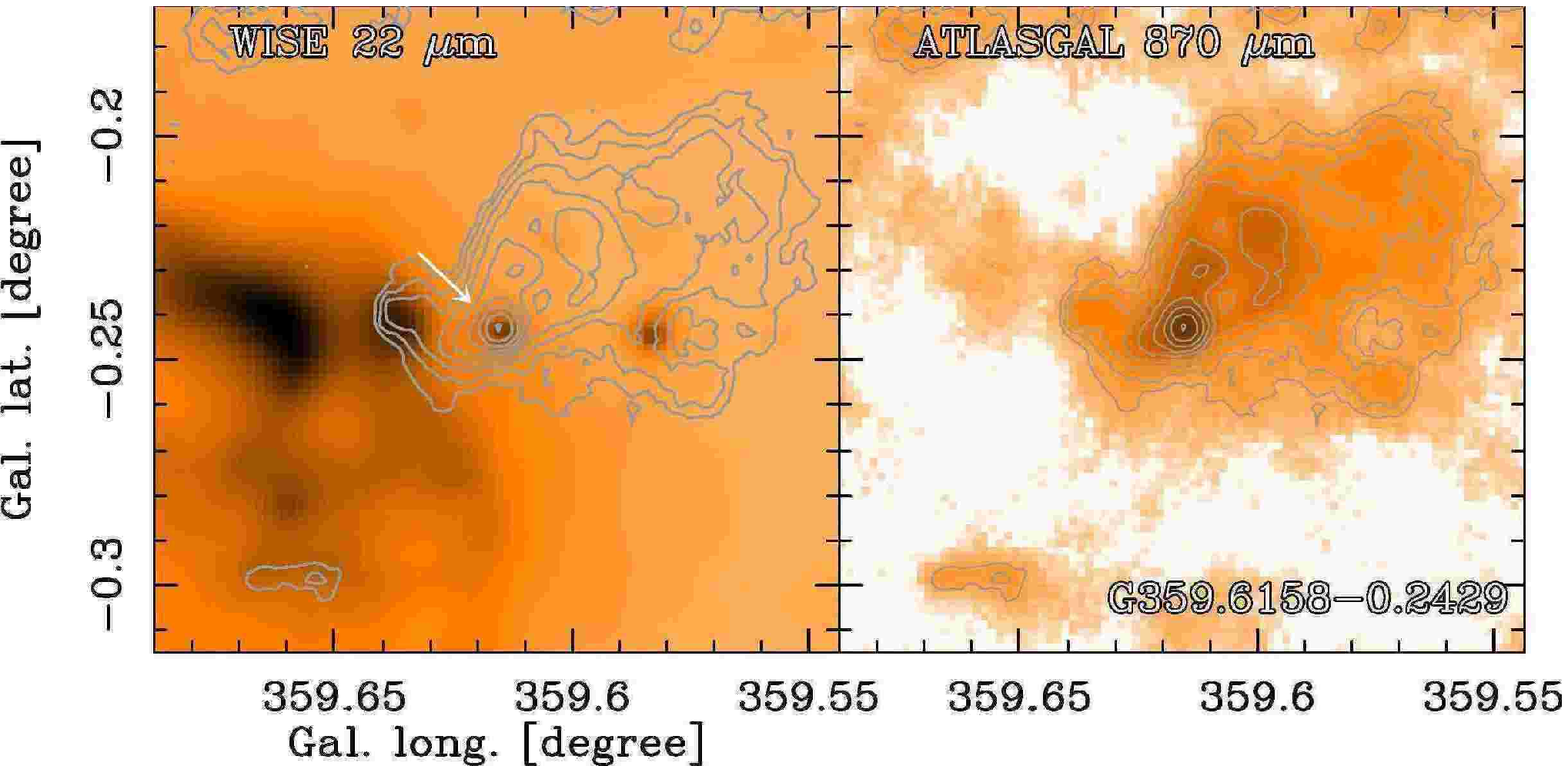} 
   \includegraphics[width=0.45\linewidth,clip]{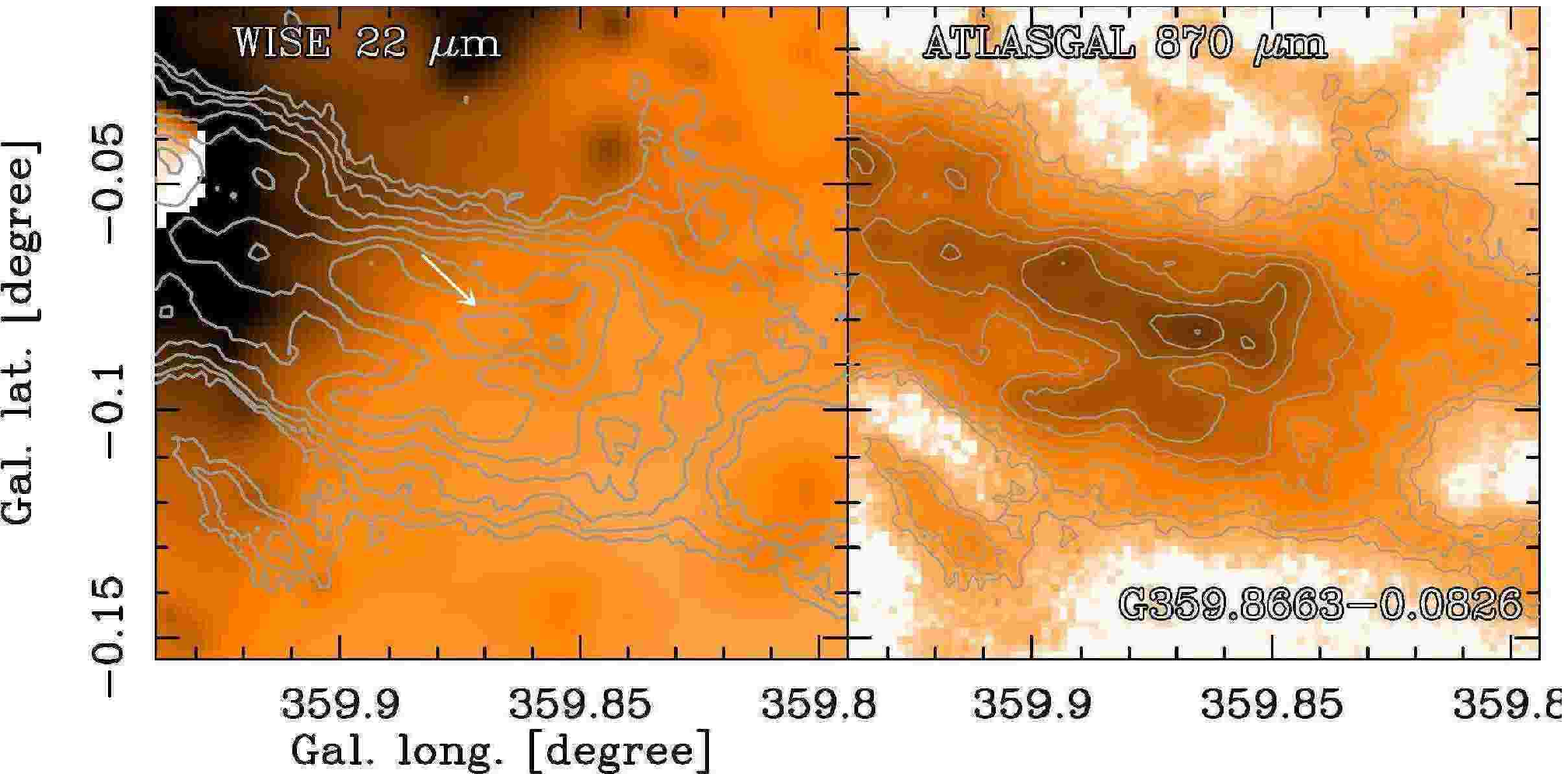} 
   \includegraphics[width=0.45\linewidth,clip]{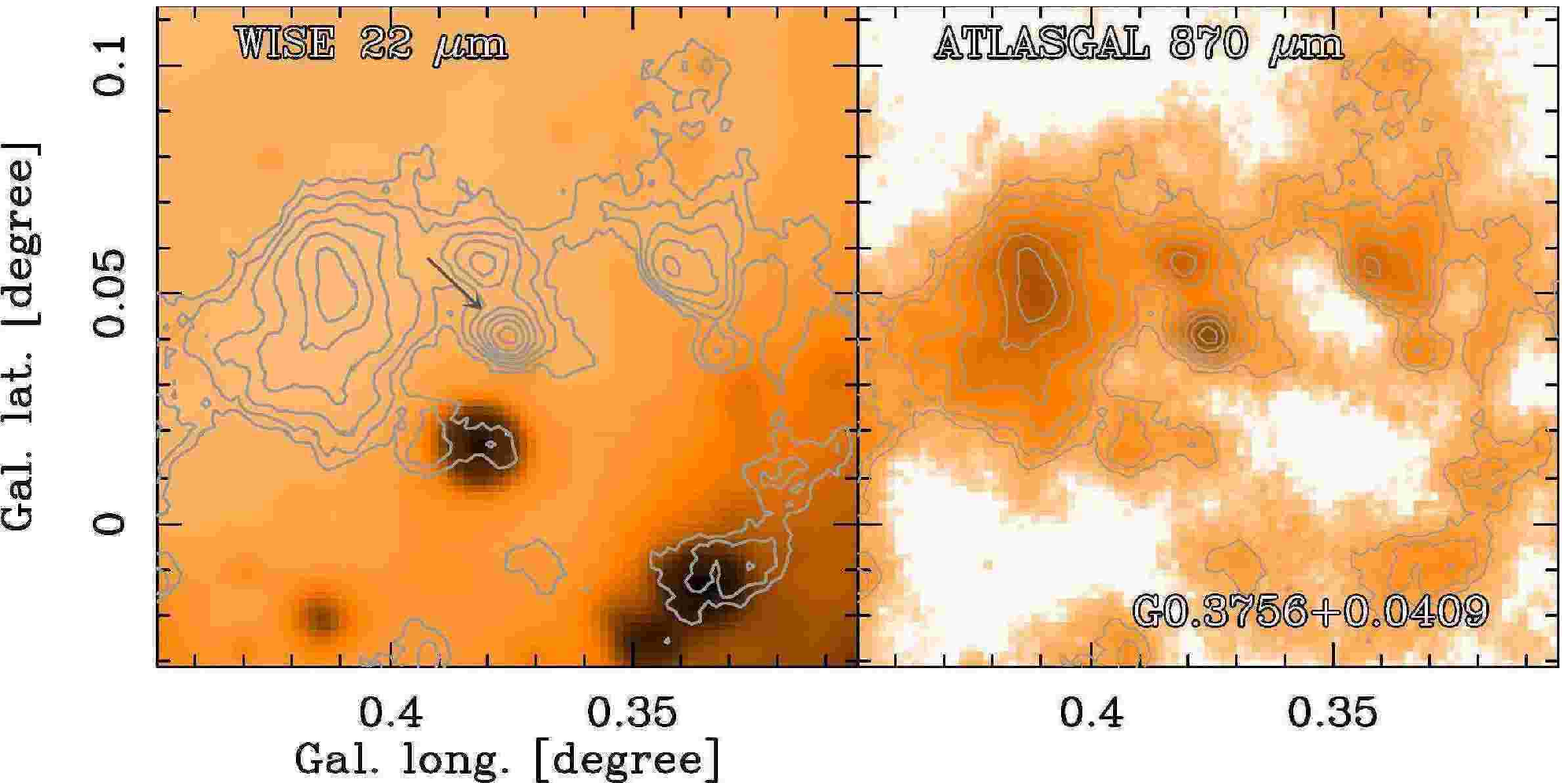} 
   \includegraphics[width=0.45\linewidth,clip]{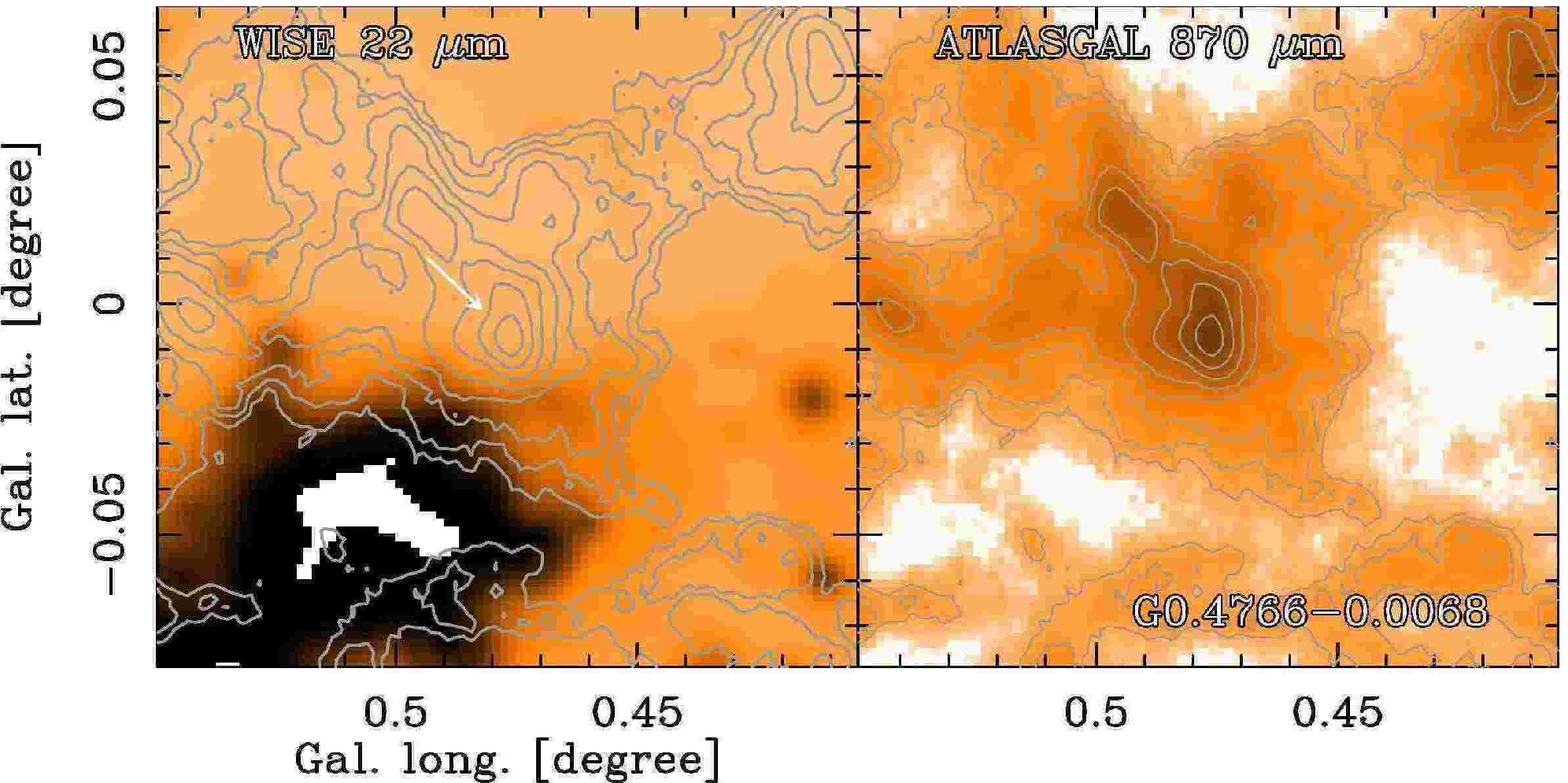} 
   \includegraphics[width=0.45\linewidth,clip]{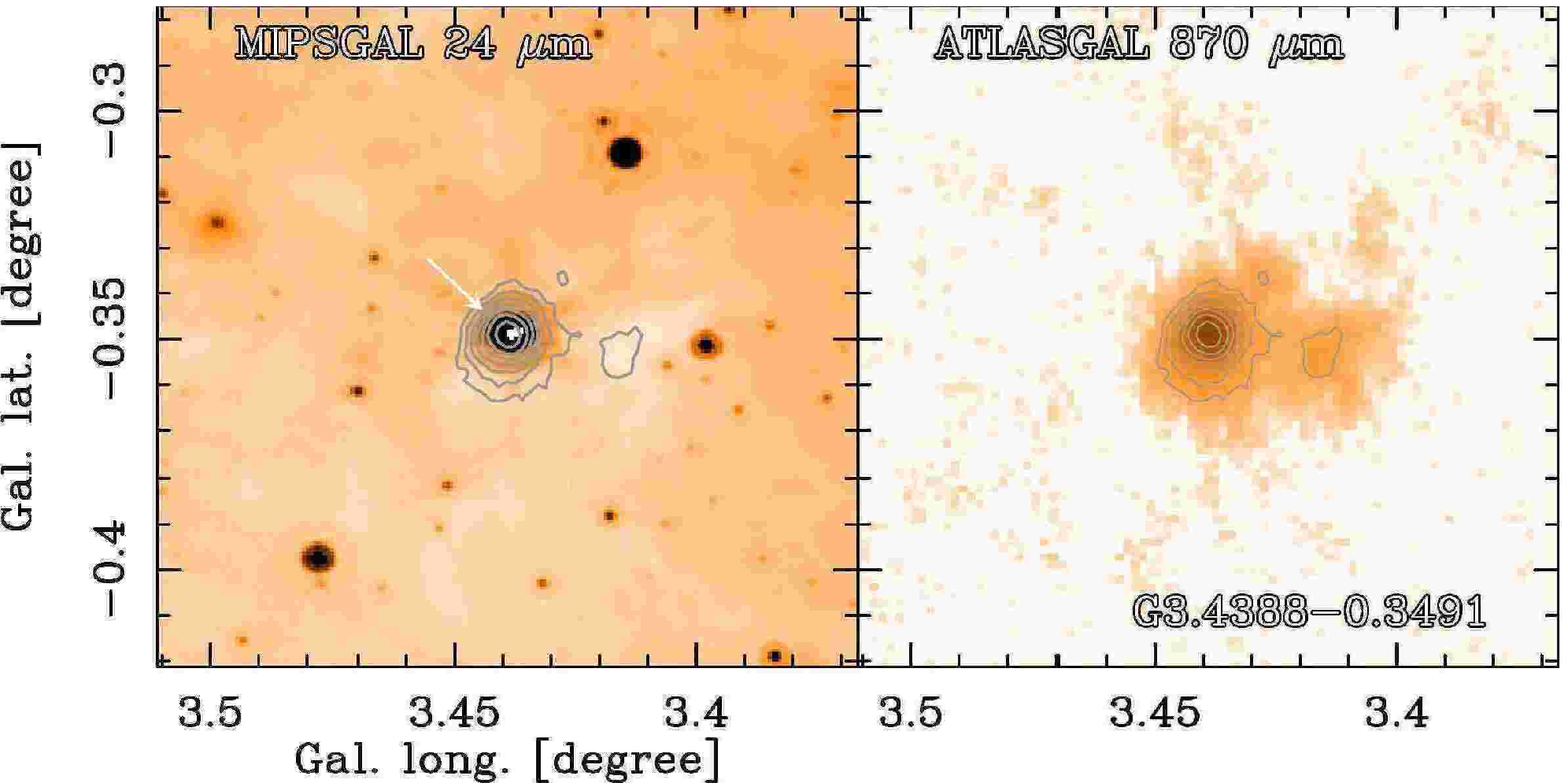} 
   \includegraphics[width=0.45\linewidth,clip]{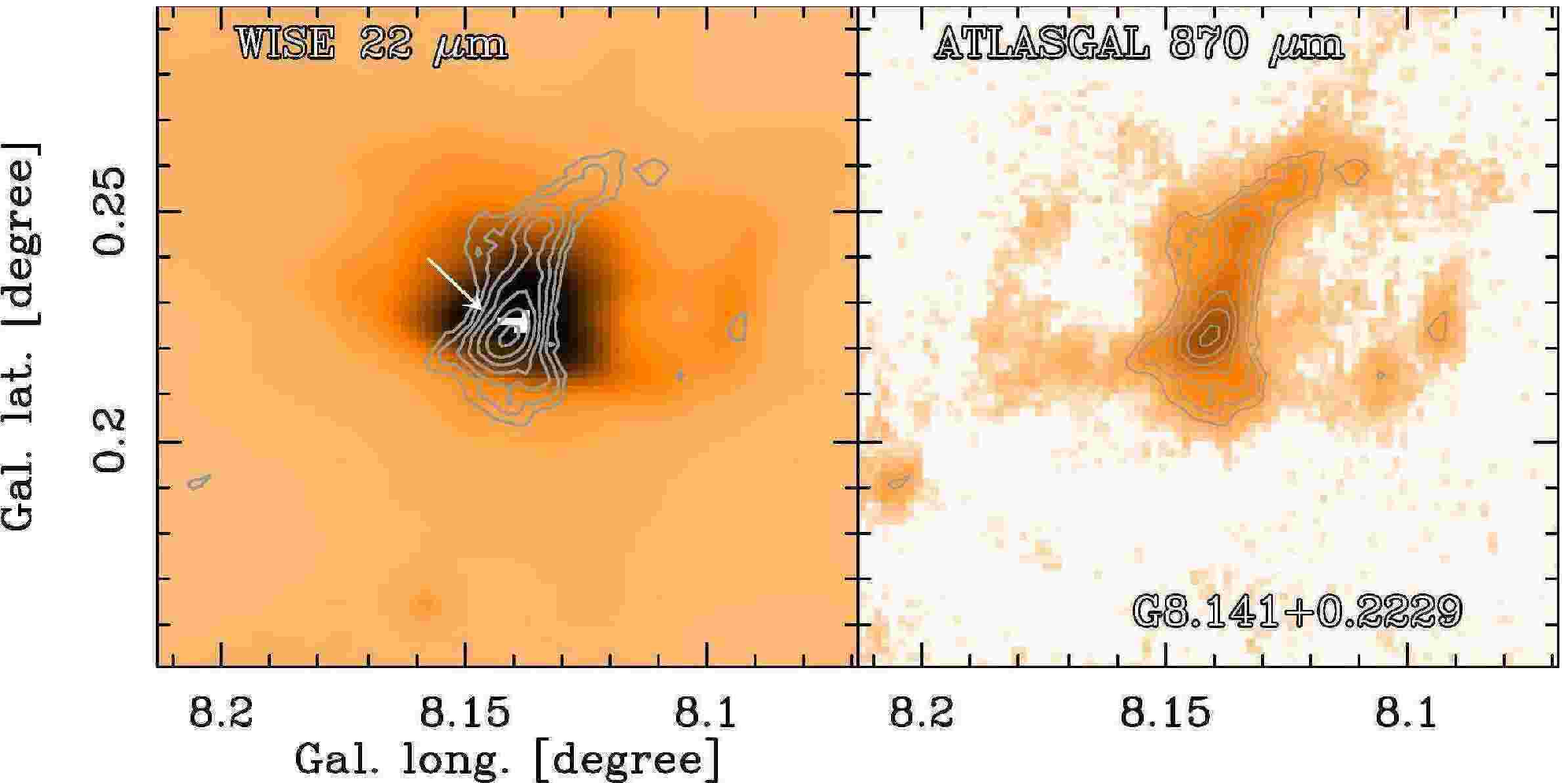} 
   \includegraphics[width=0.45\linewidth,clip]{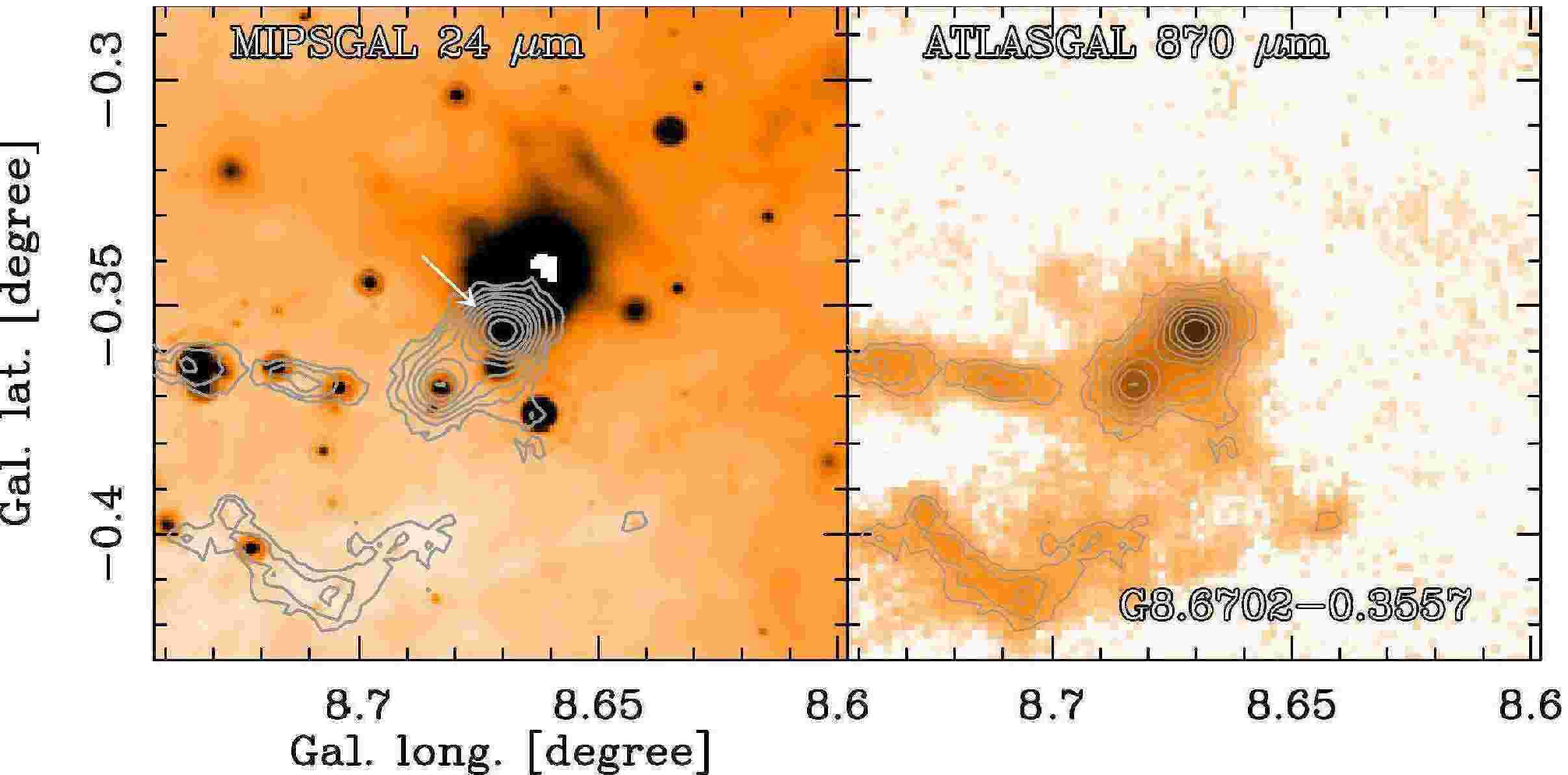} 
   \includegraphics[width=0.45\linewidth,clip]{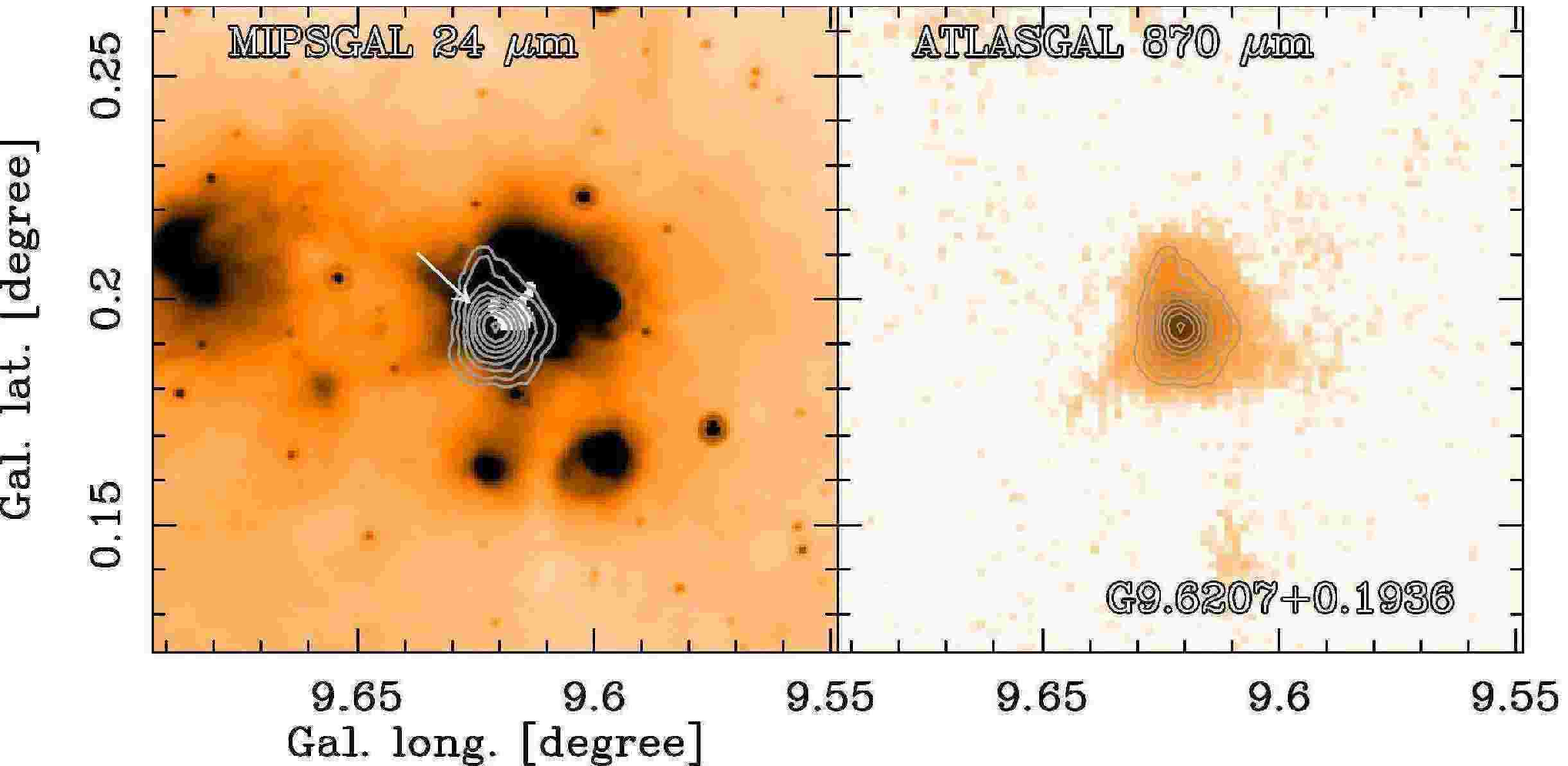} 
  \includegraphics[width=0.45\linewidth,clip]{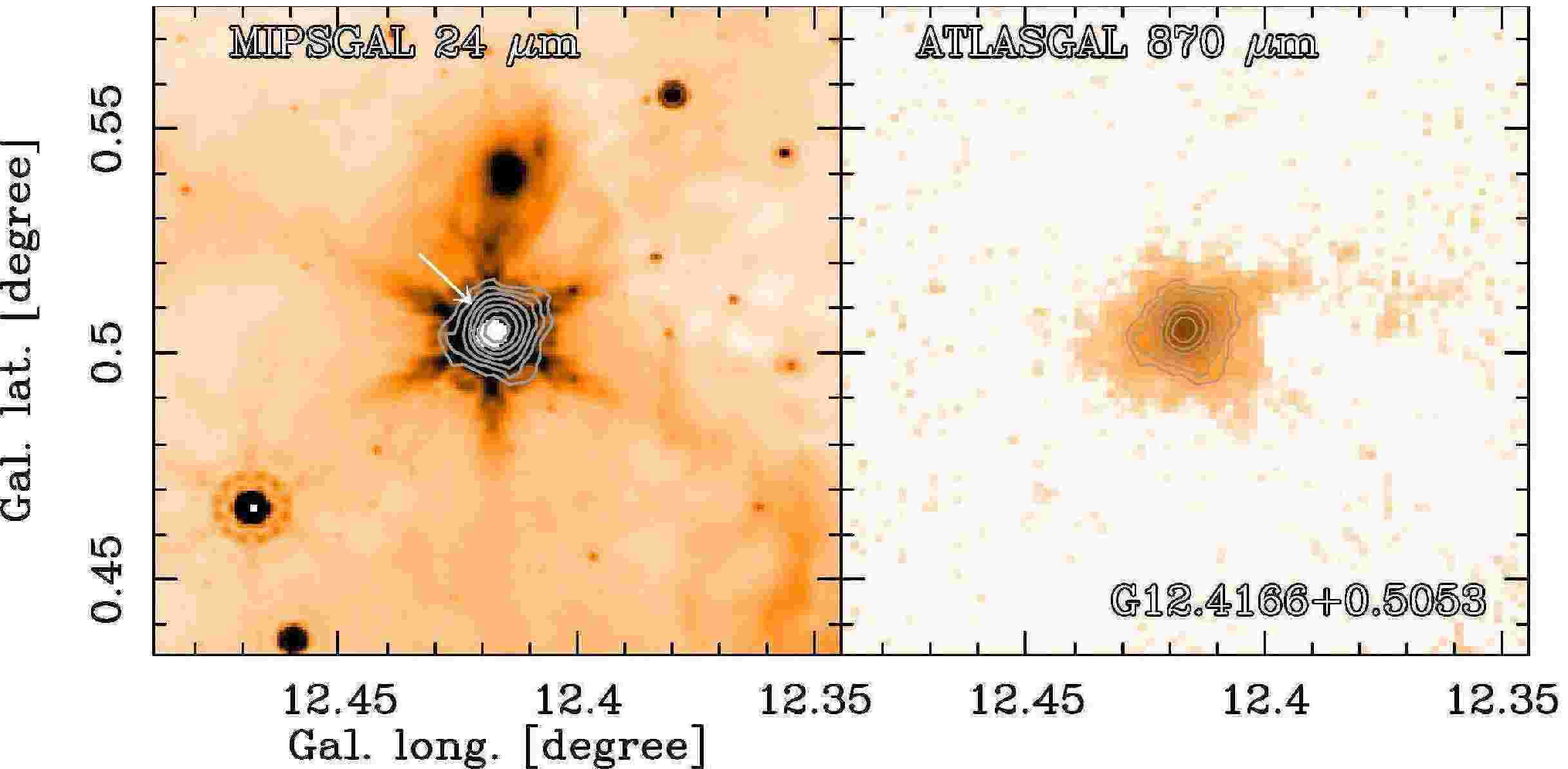} 
  \includegraphics[width=0.45\linewidth,clip]{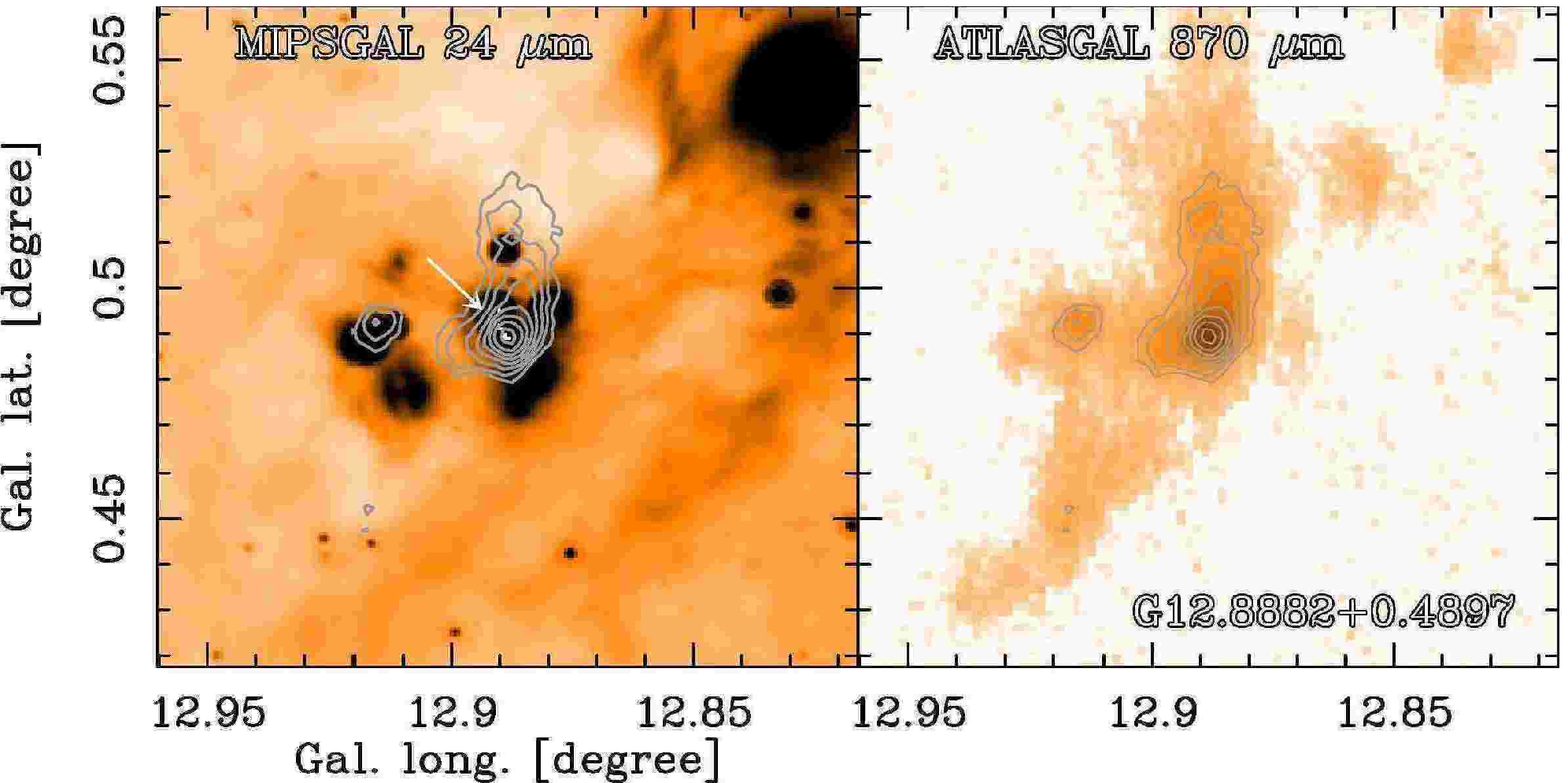} 
 \includegraphics[width=0.45\linewidth,clip]{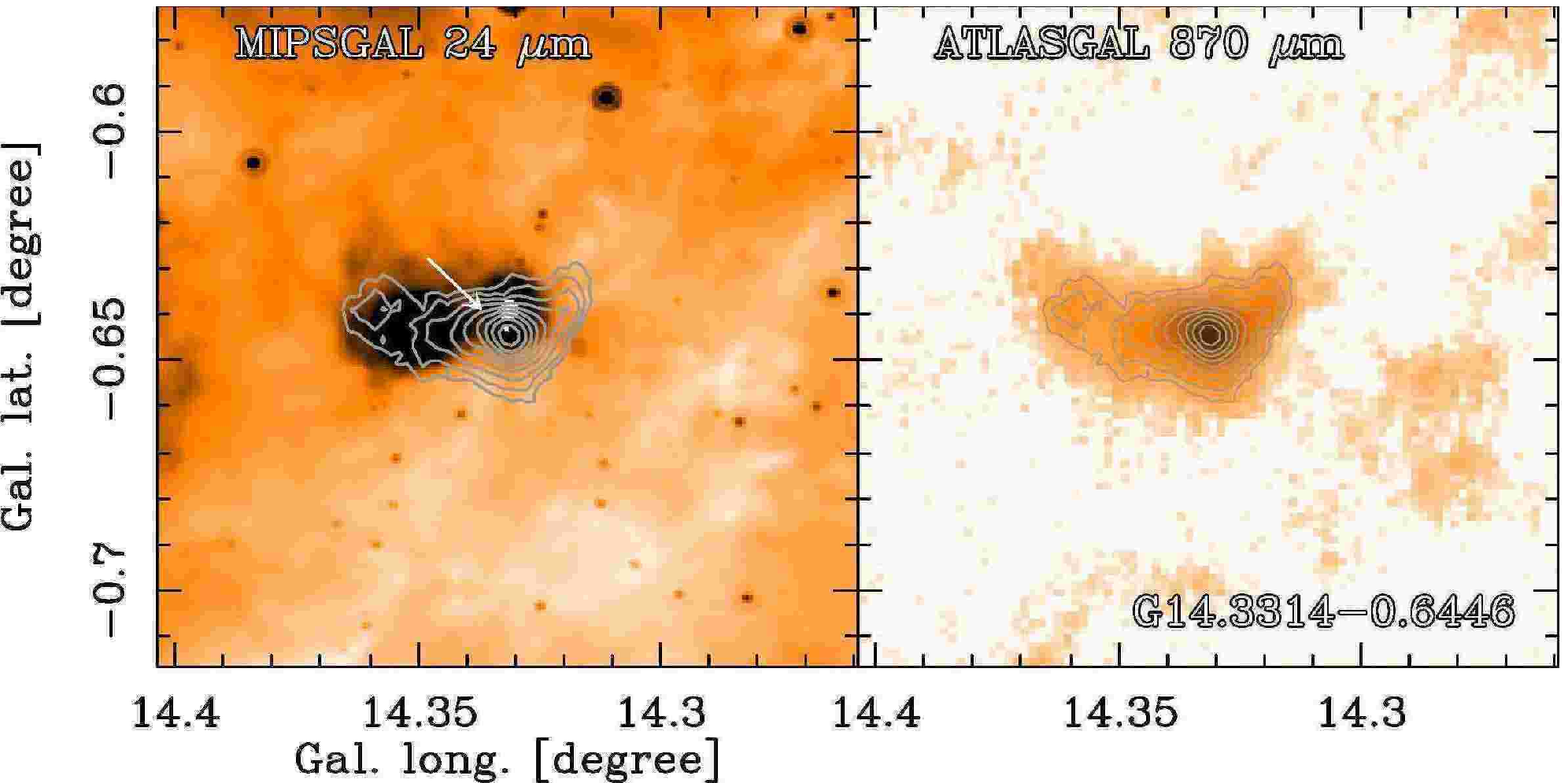} 
 \caption{Continued.}
 \end{figure*}
\begin{figure*}
\centering
\ContinuedFloat
   \includegraphics[width=0.45\linewidth,clip]{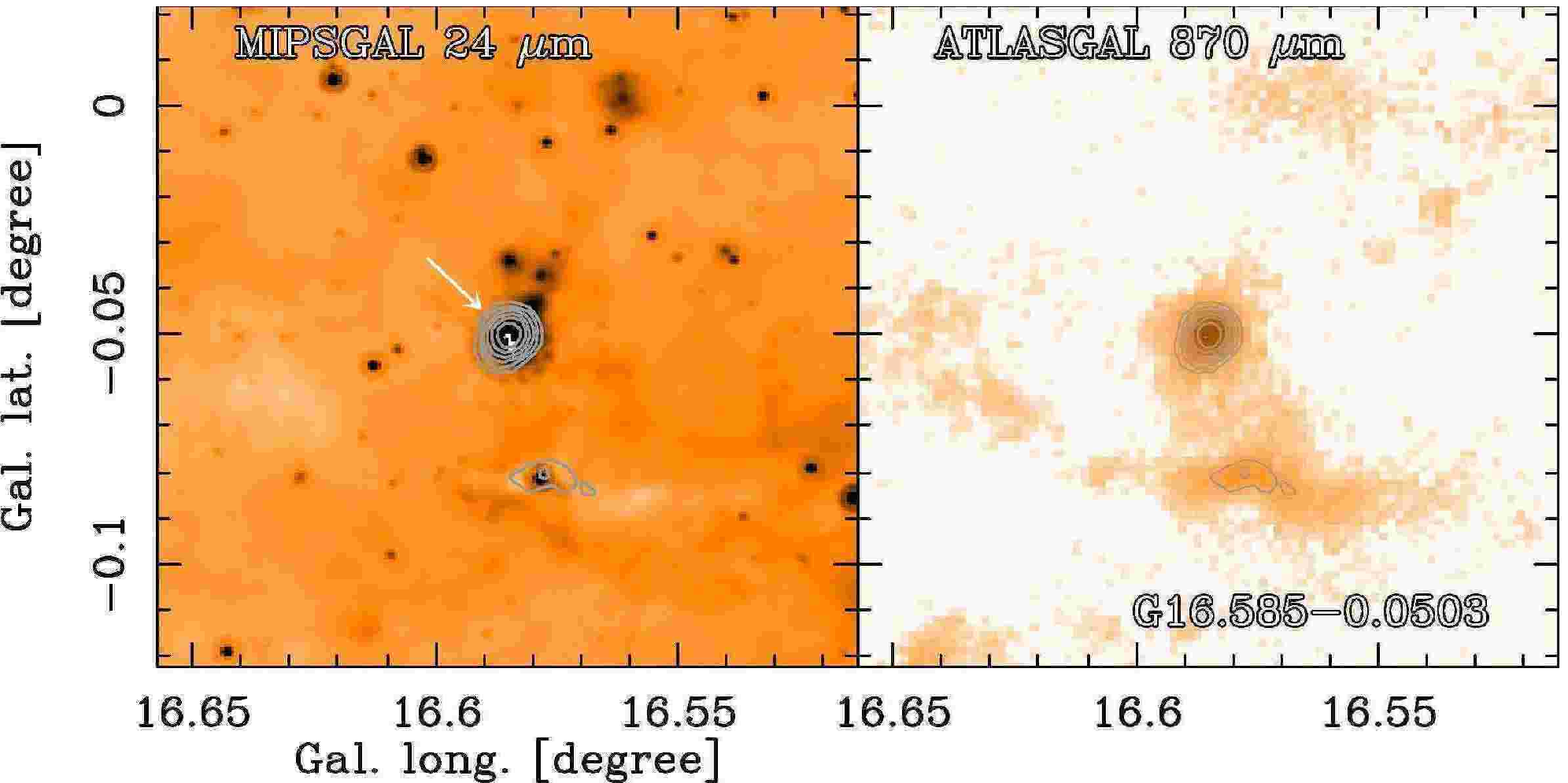} 
 \includegraphics[width=0.45\linewidth,clip]{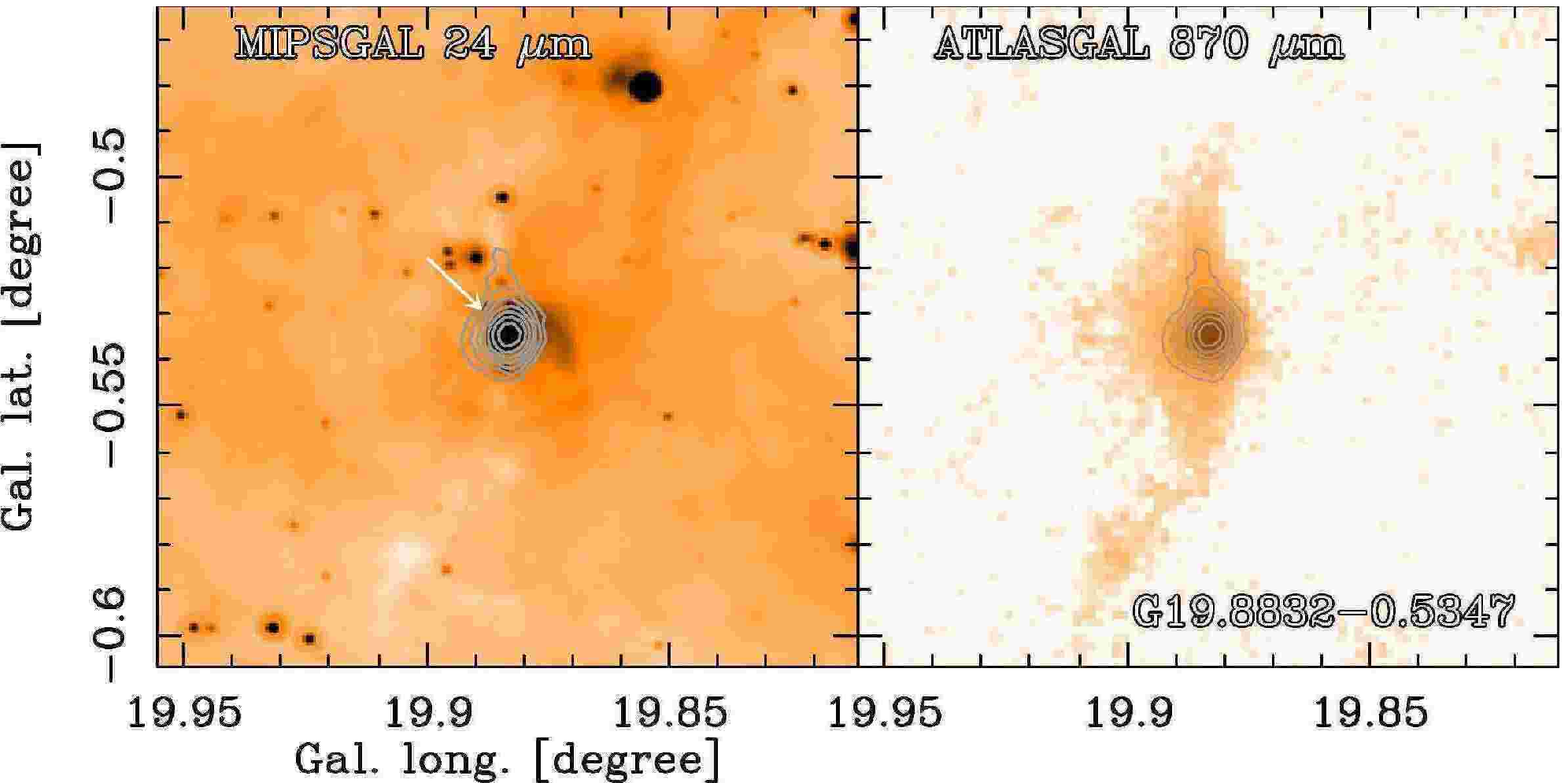} 
 \includegraphics[width=0.45\linewidth,clip]{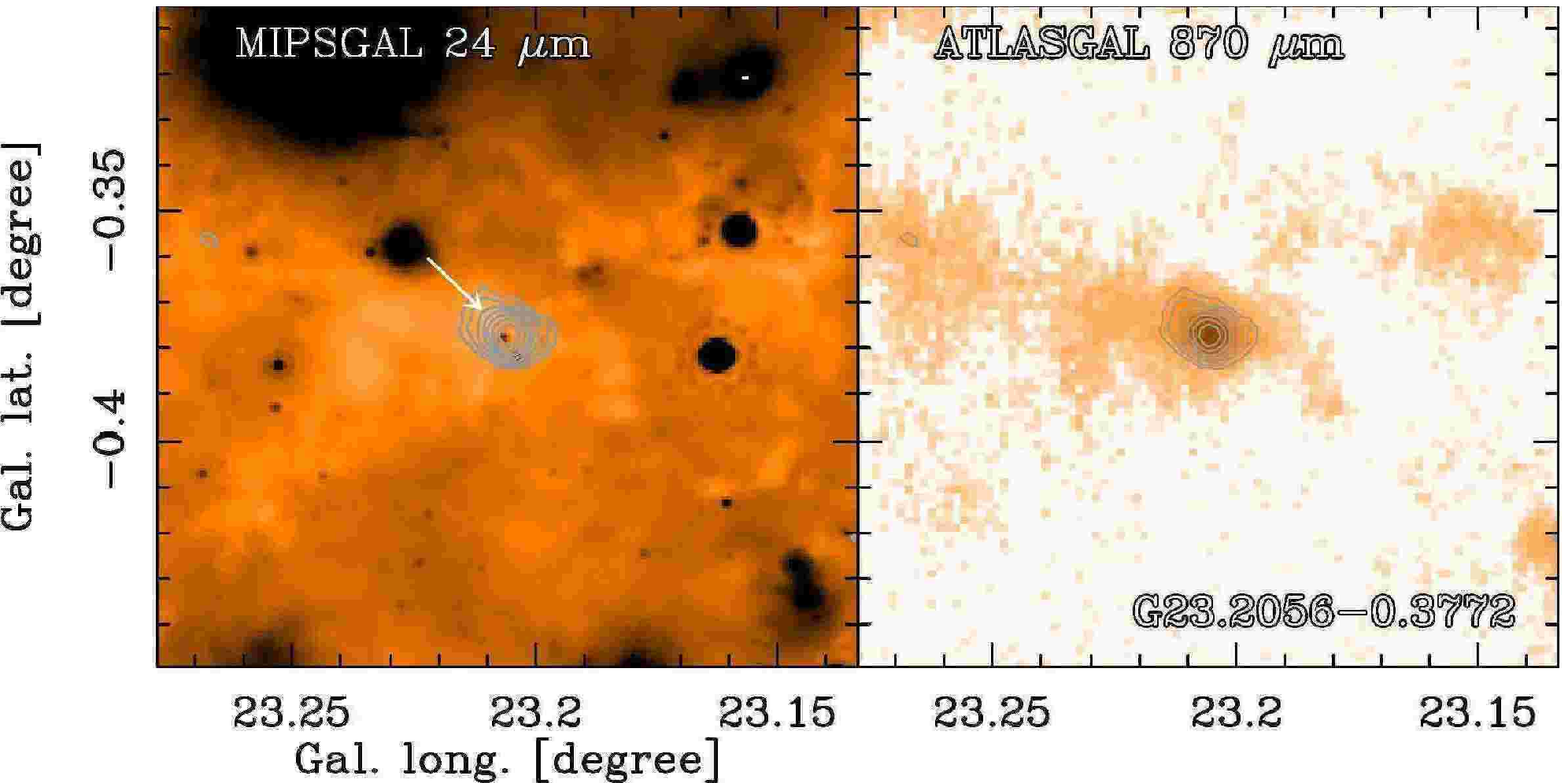} 
 \includegraphics[width=0.45\linewidth,clip]{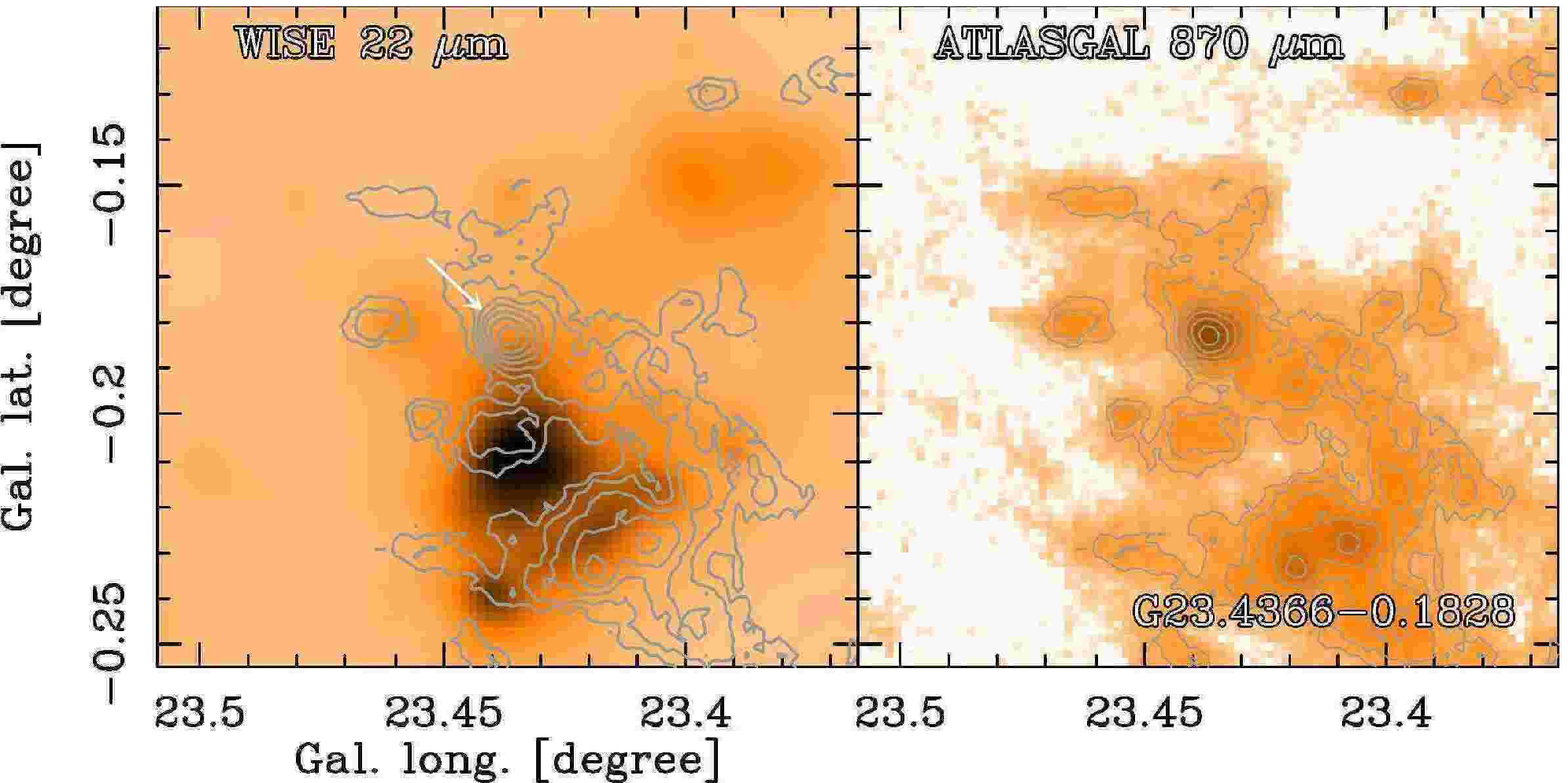} 
  \includegraphics[width=0.45\linewidth,clip]{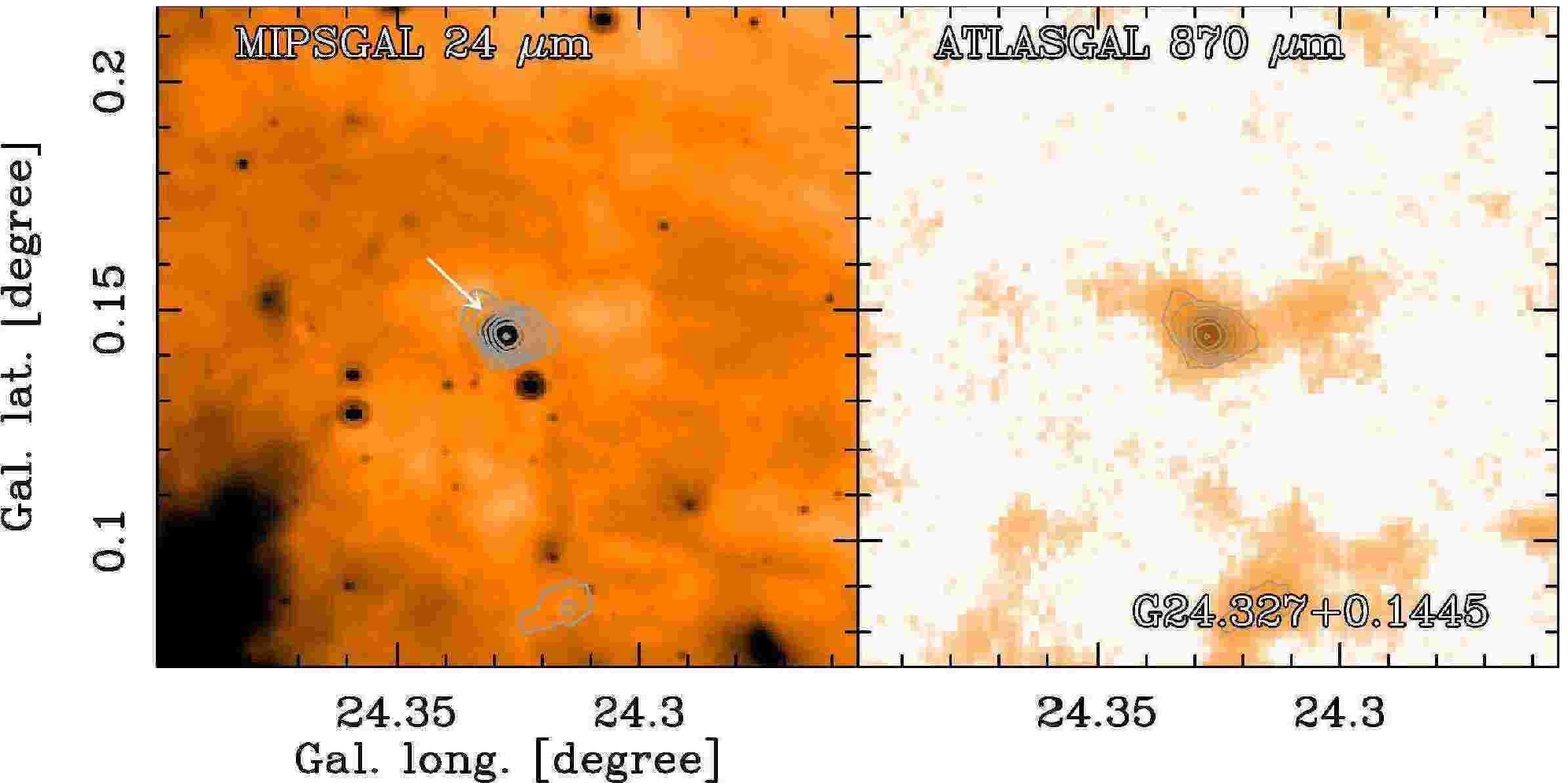} 
  \includegraphics[width=0.45\linewidth,clip]{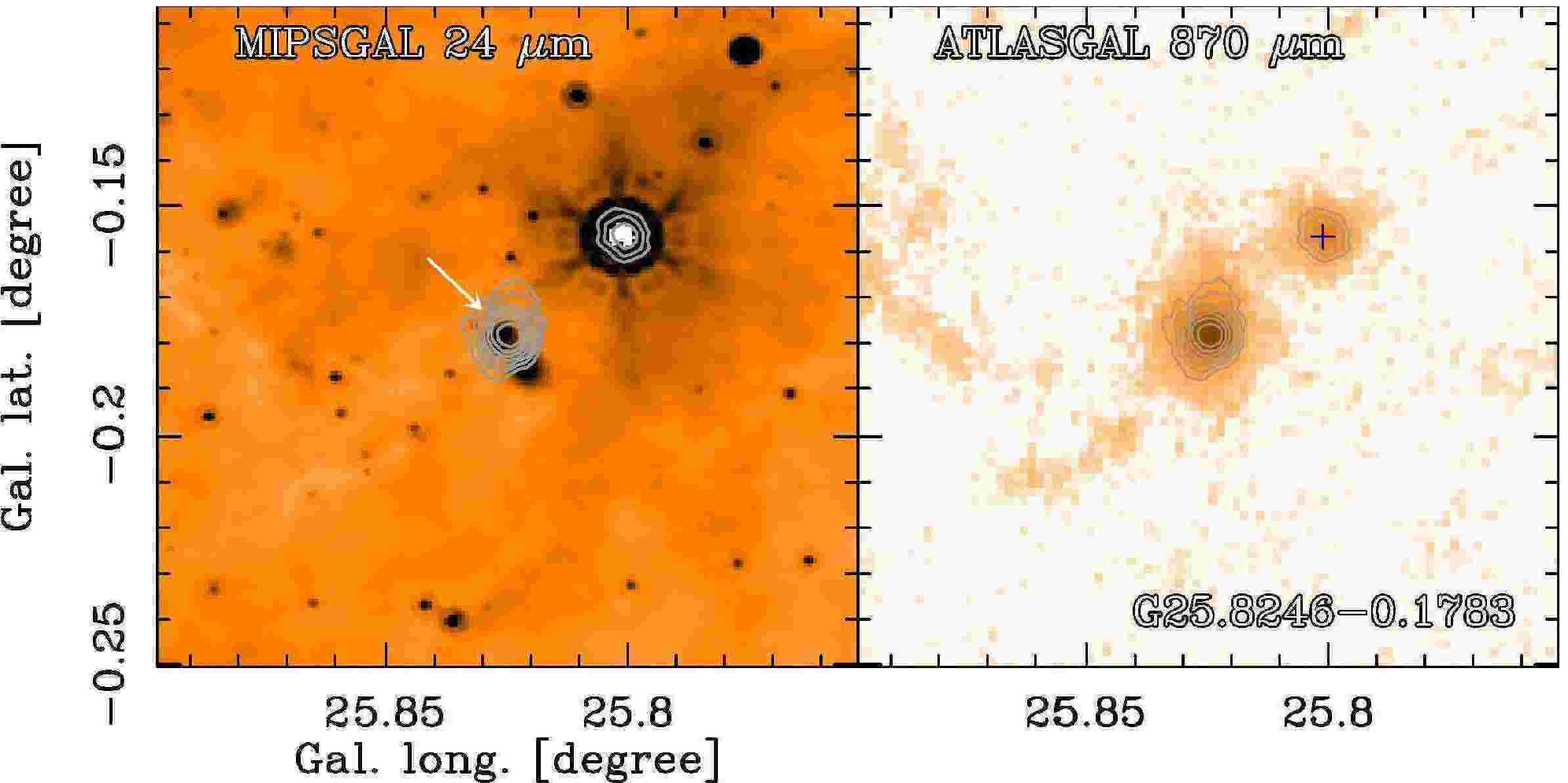} 
  \includegraphics[width=0.45\linewidth,clip]{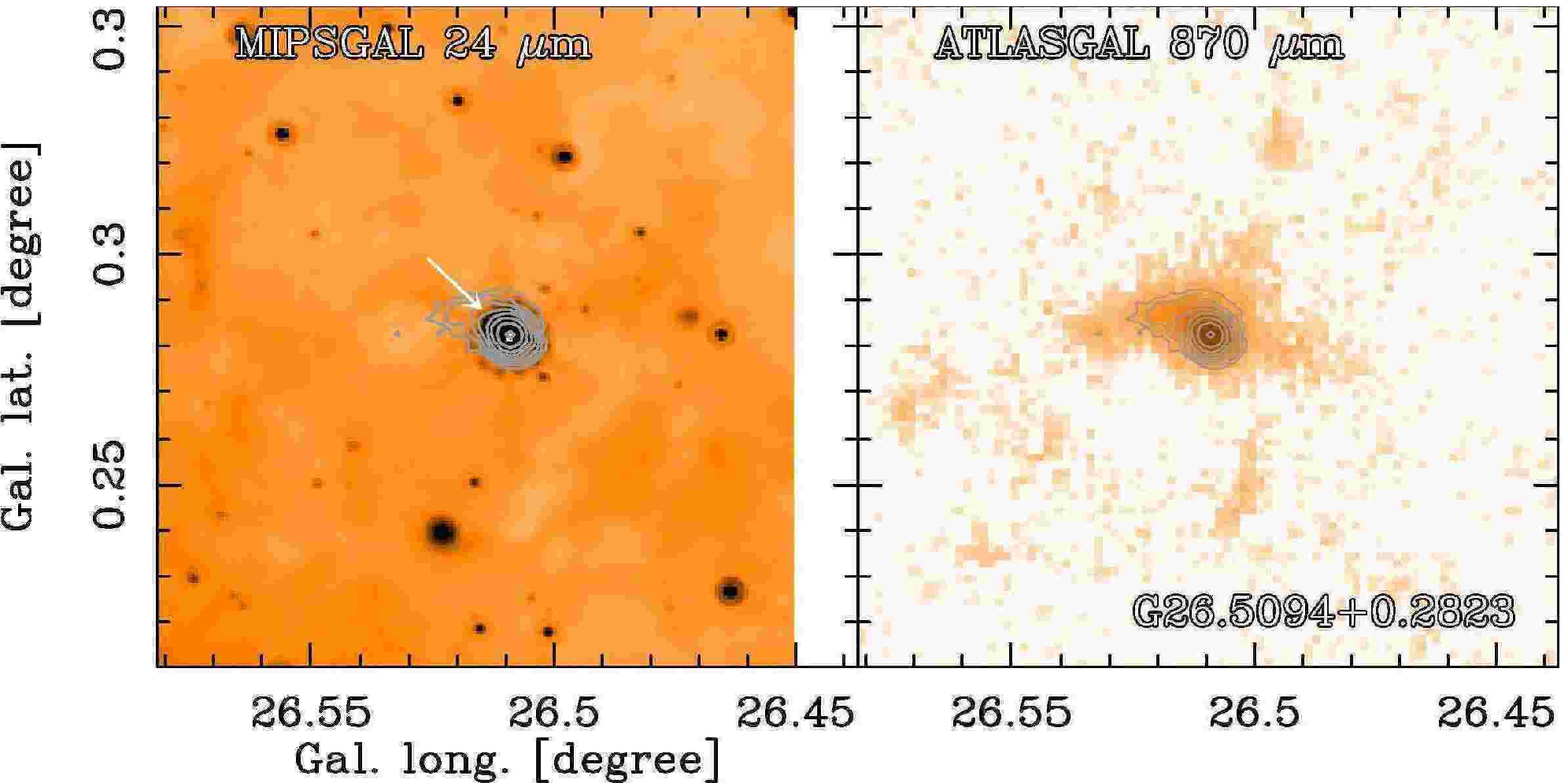} 
 \includegraphics[width=0.45\linewidth,clip]{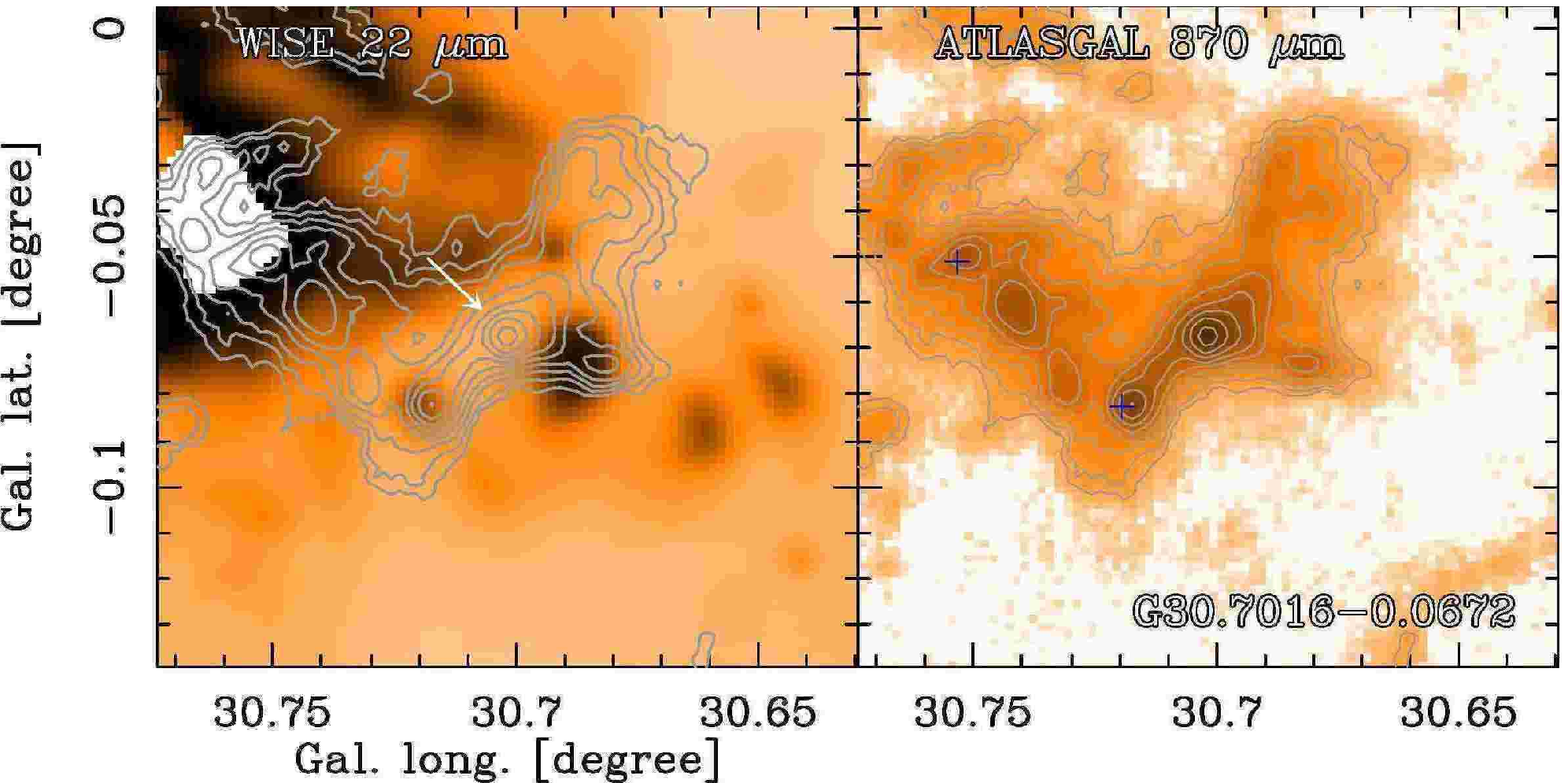} 
  \includegraphics[width=0.45\linewidth,clip]{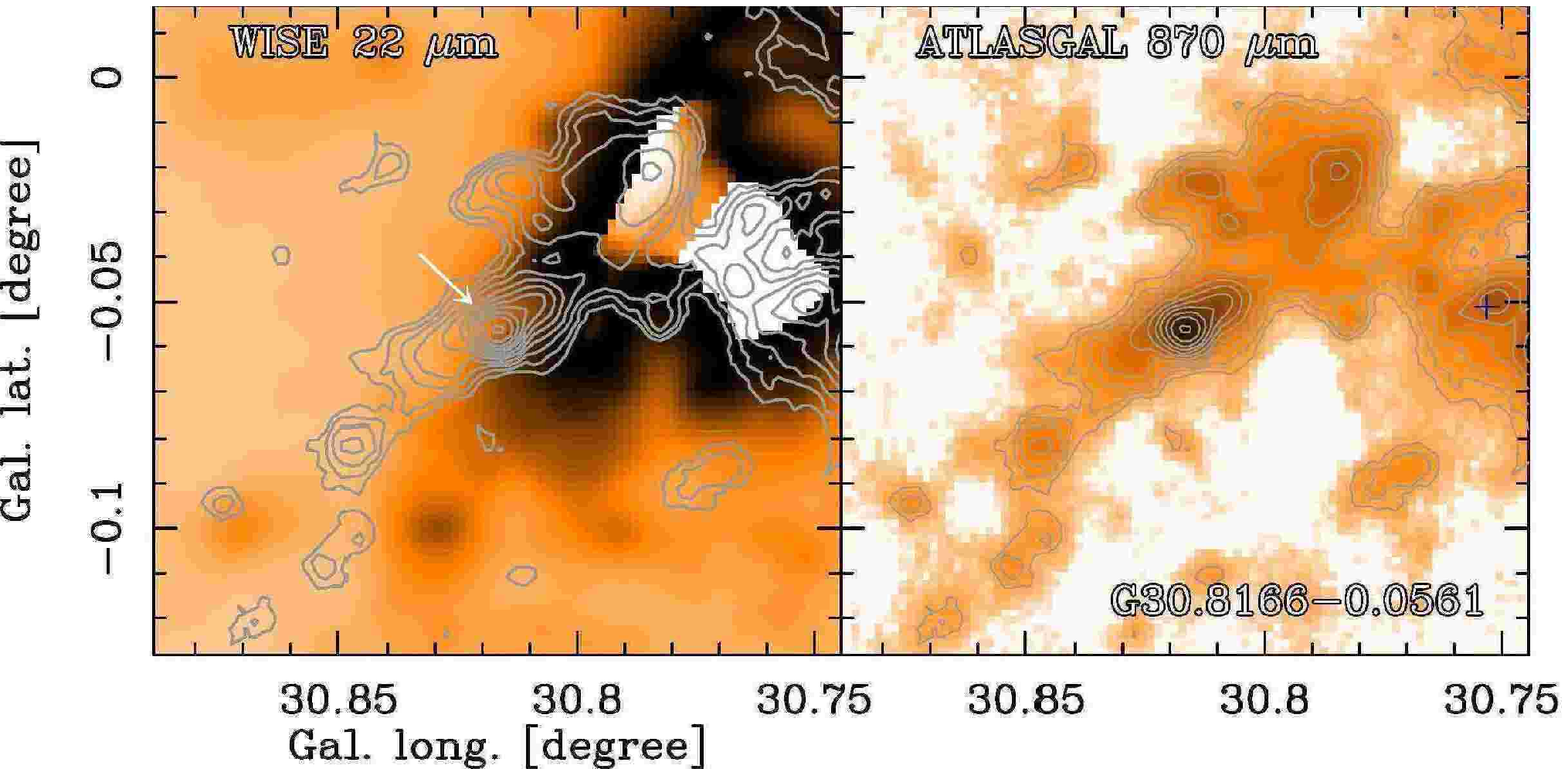} 
   \includegraphics[width=0.45\linewidth,clip]{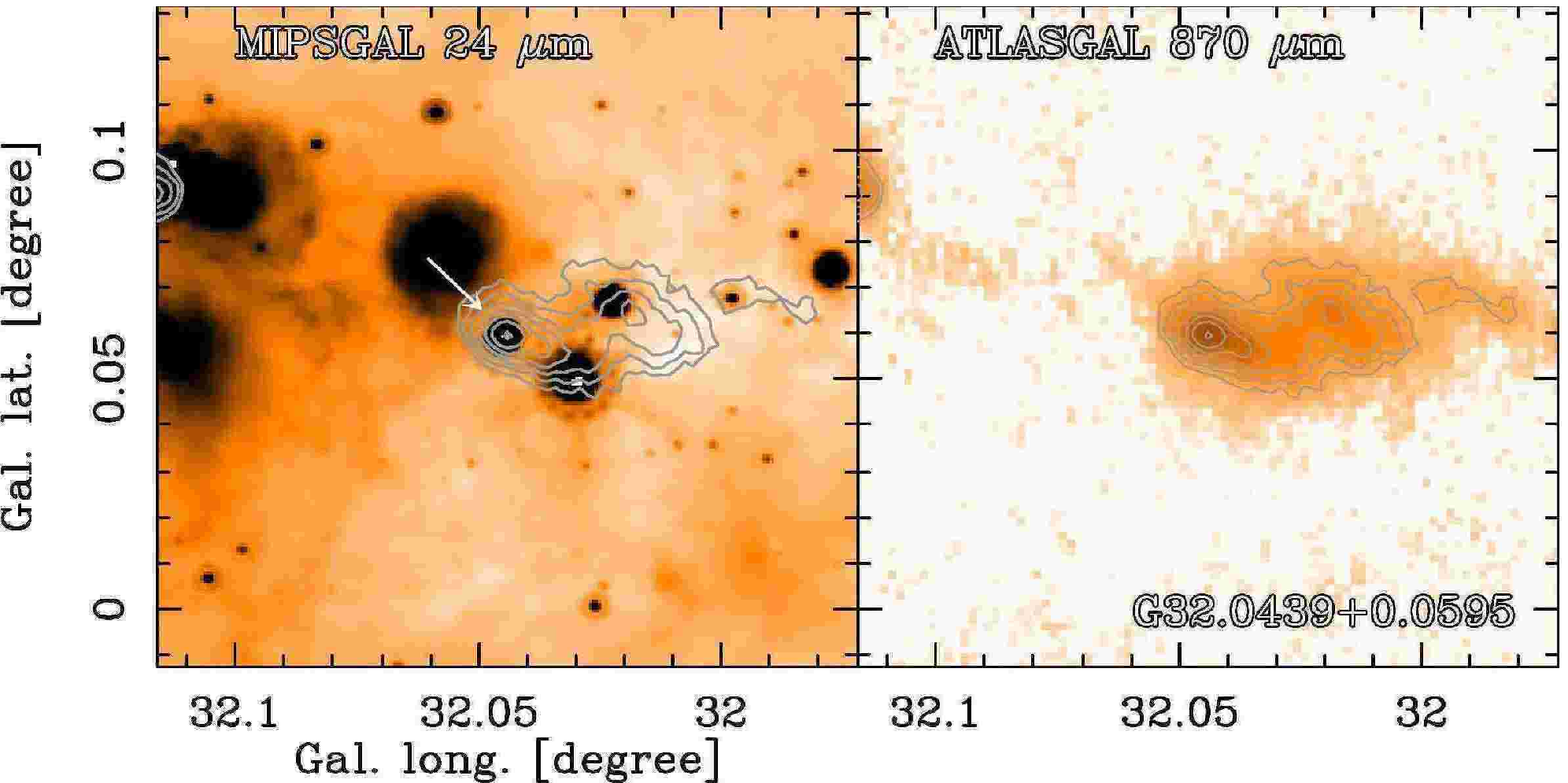} 
   \includegraphics[width=0.45\linewidth,clip]{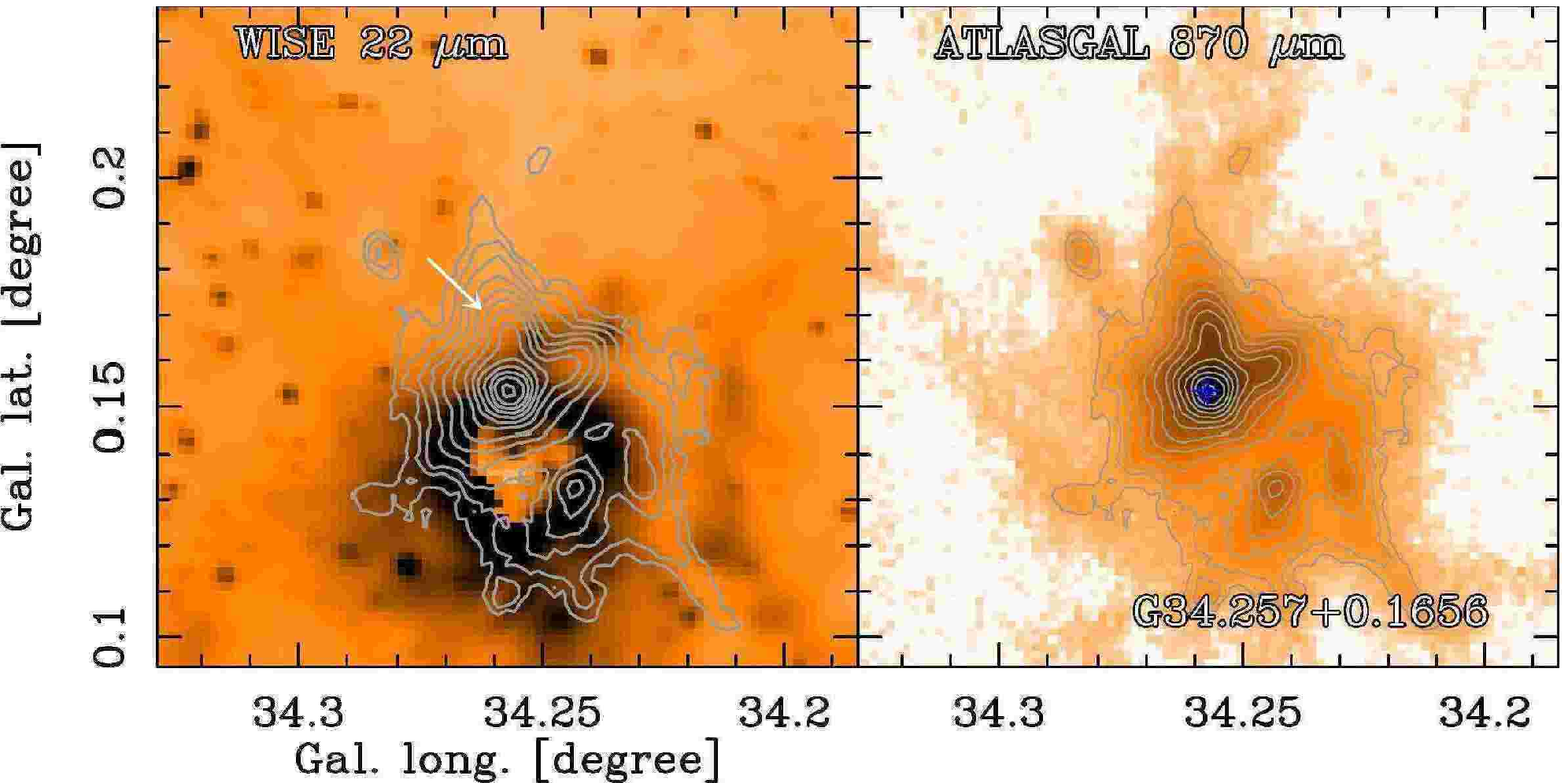} 
  \includegraphics[width=0.45\linewidth,clip]{FIG_ASTRO_PH/OUT_PAPER/165.eps} 
\caption{Continued.}
 \end{figure*}
\begin{figure*}
\centering
\ContinuedFloat
  \includegraphics[width=0.45\linewidth,clip]{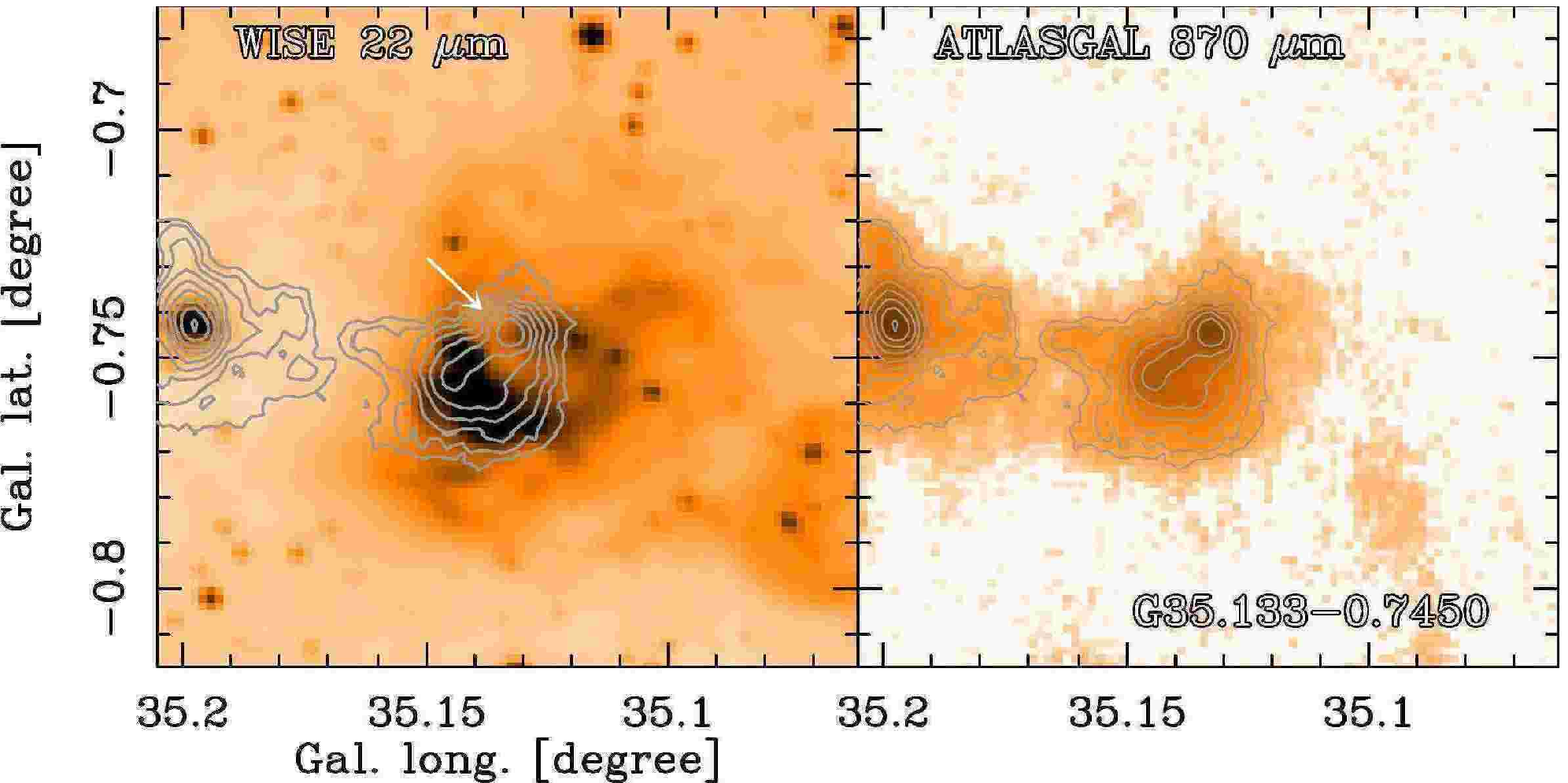} 
   \includegraphics[width=0.45\linewidth,clip]{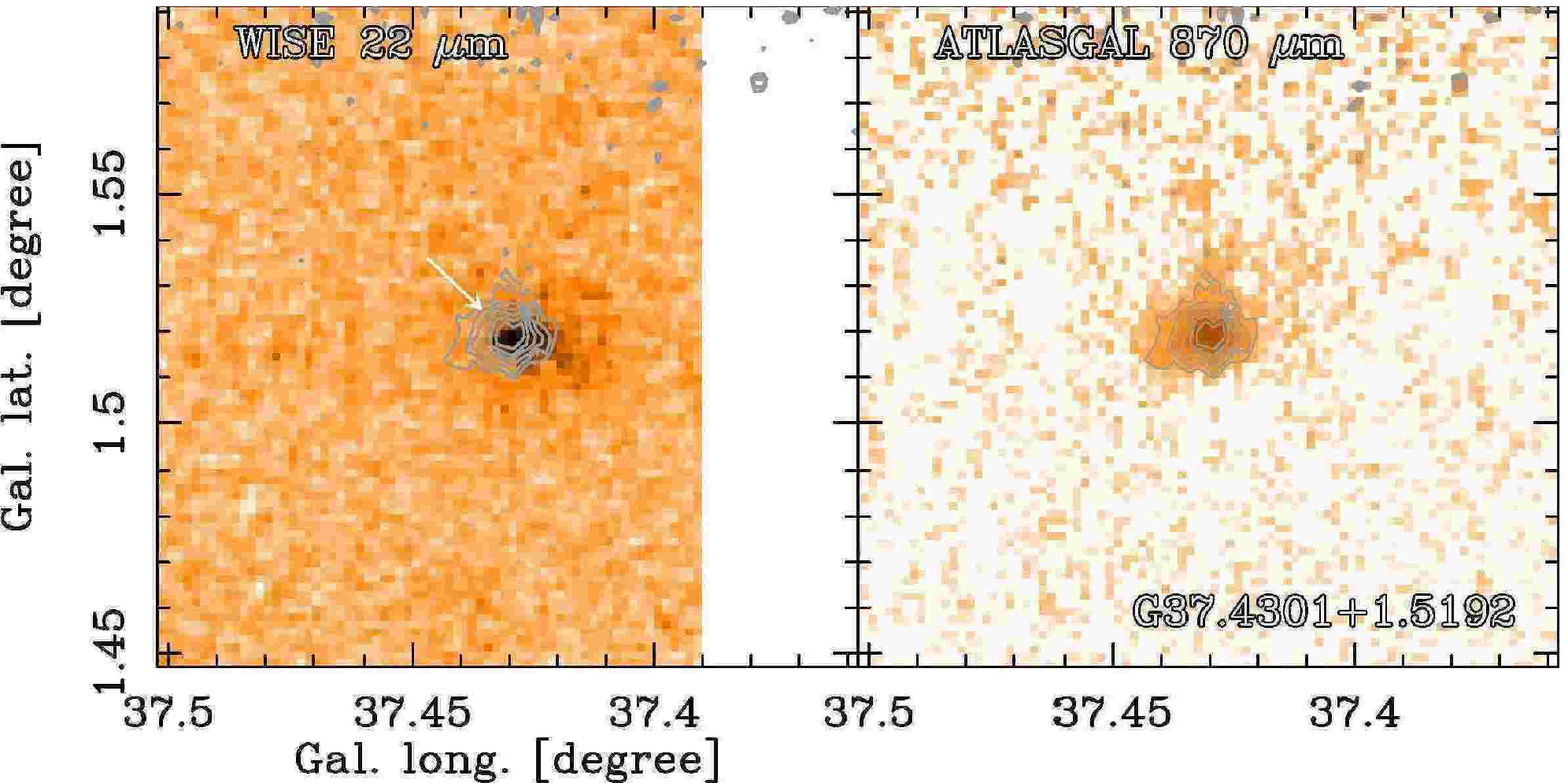} 
   \includegraphics[width=0.45\linewidth,clip]{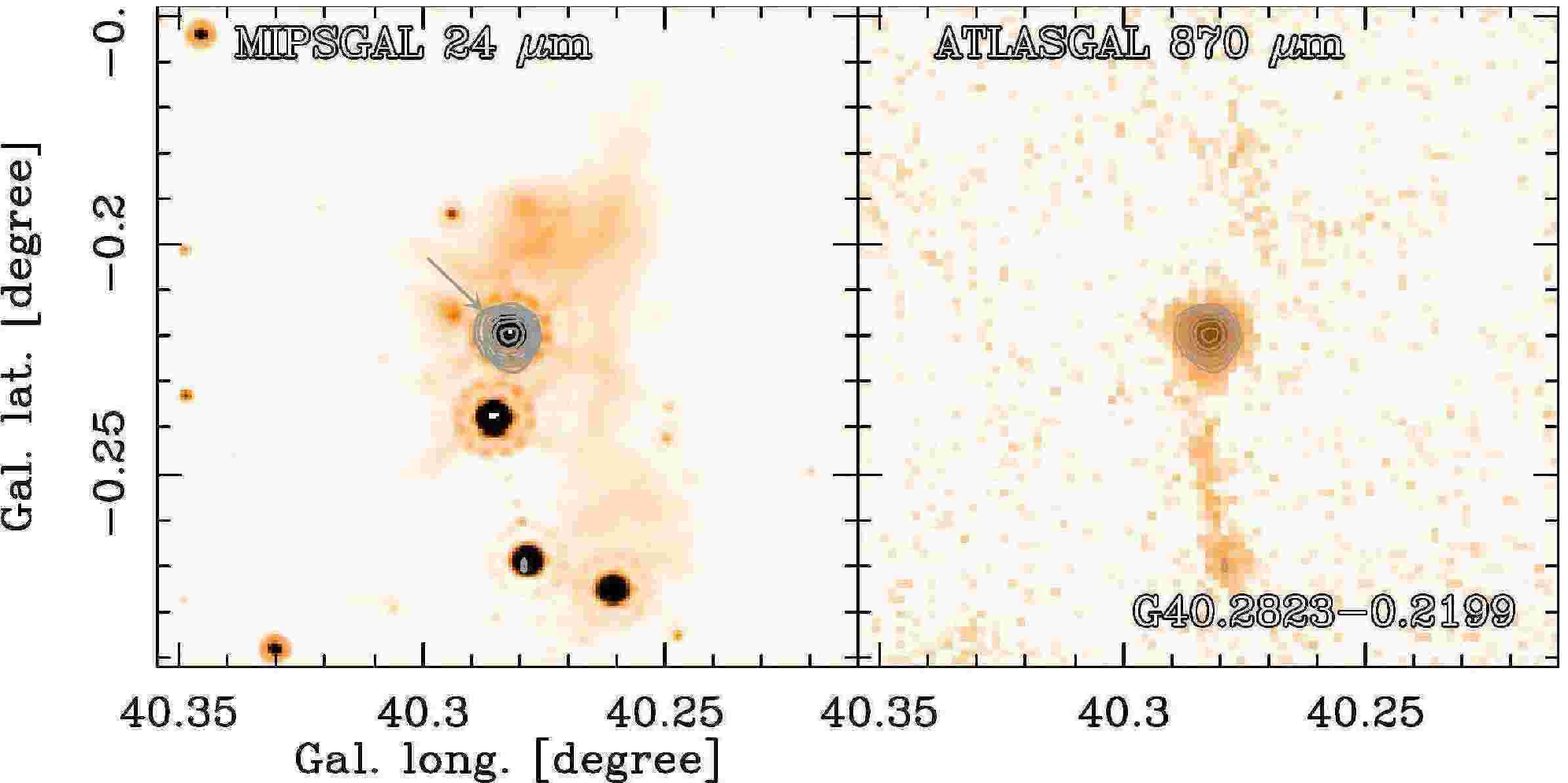} 
   \includegraphics[width=0.45\linewidth,clip]{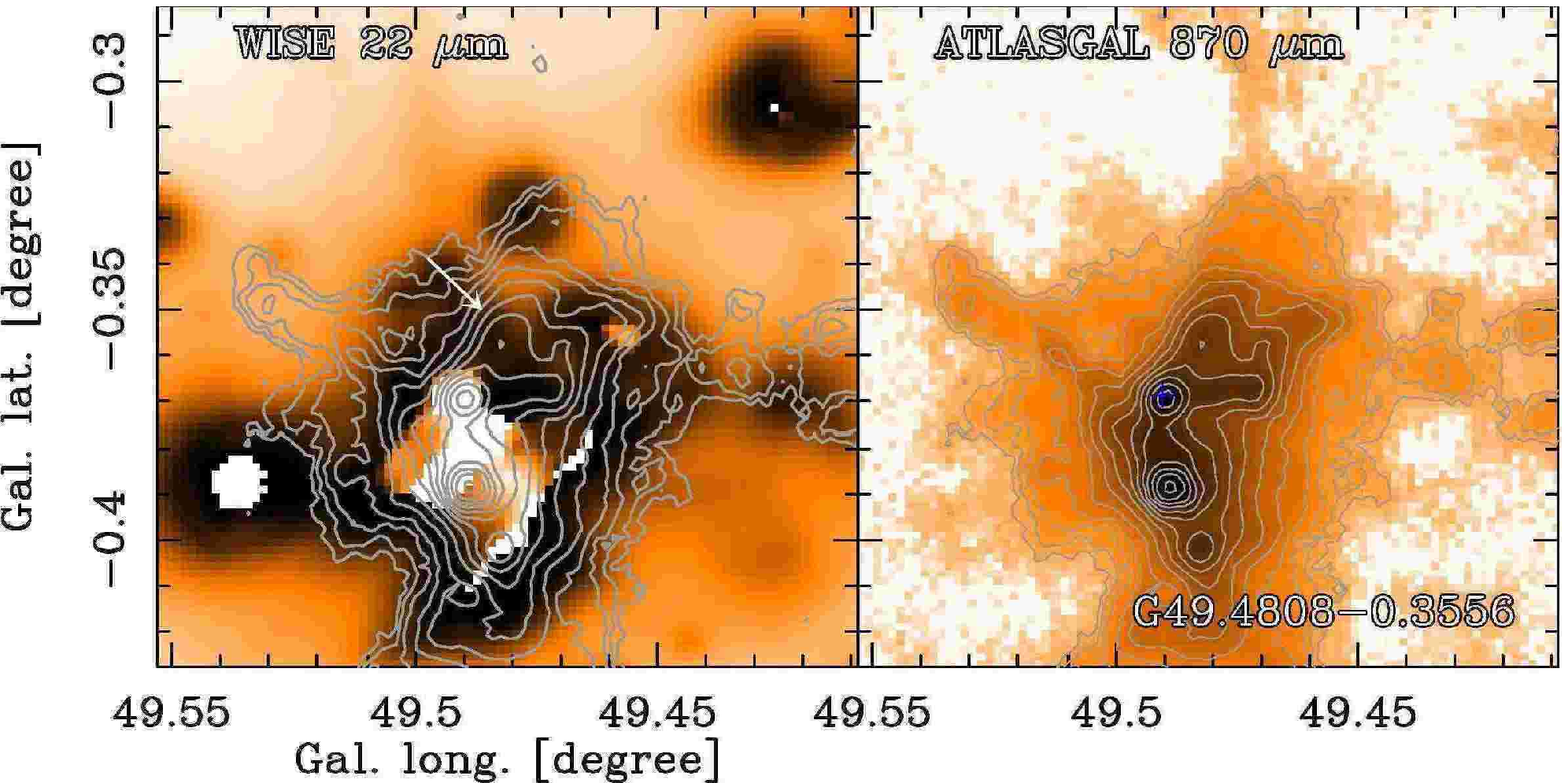} 
 \caption{Continued.}
 \end{figure*}
 
\section{SiO (8--7) lines towards infrared-quiet massive clumps with APEX}\label{app:sio}
\begin{figure*}
\centering
  \includegraphics[width=5cm,clip]{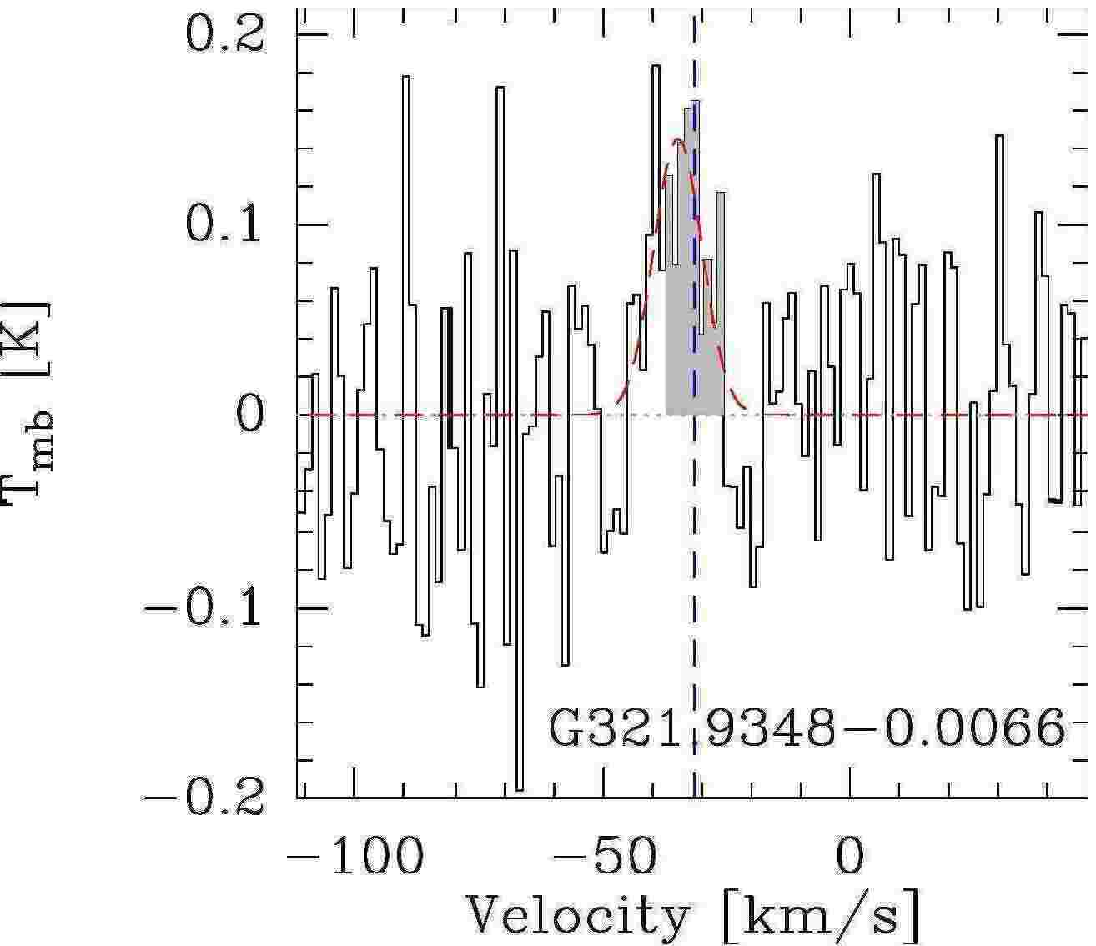}
  \includegraphics[width=5cm,clip]{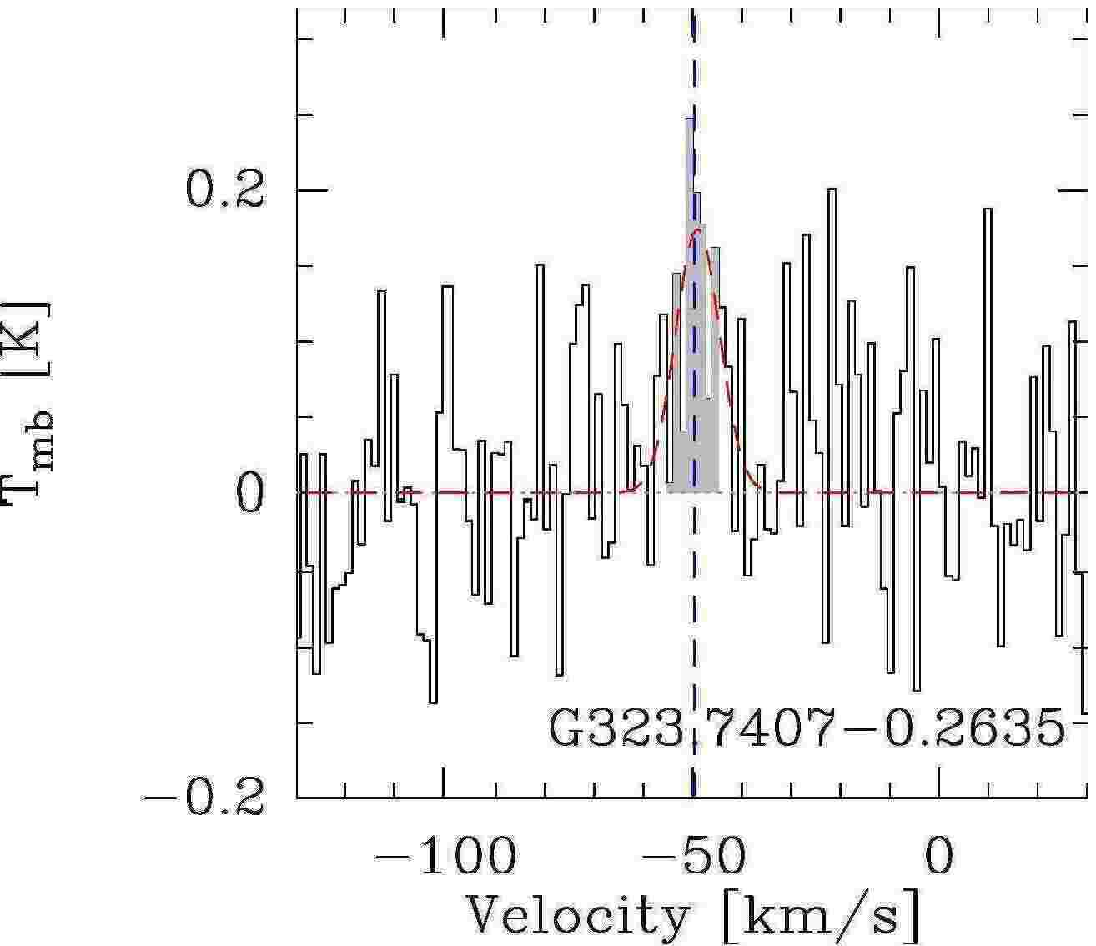}
  \includegraphics[width=5cm,clip]{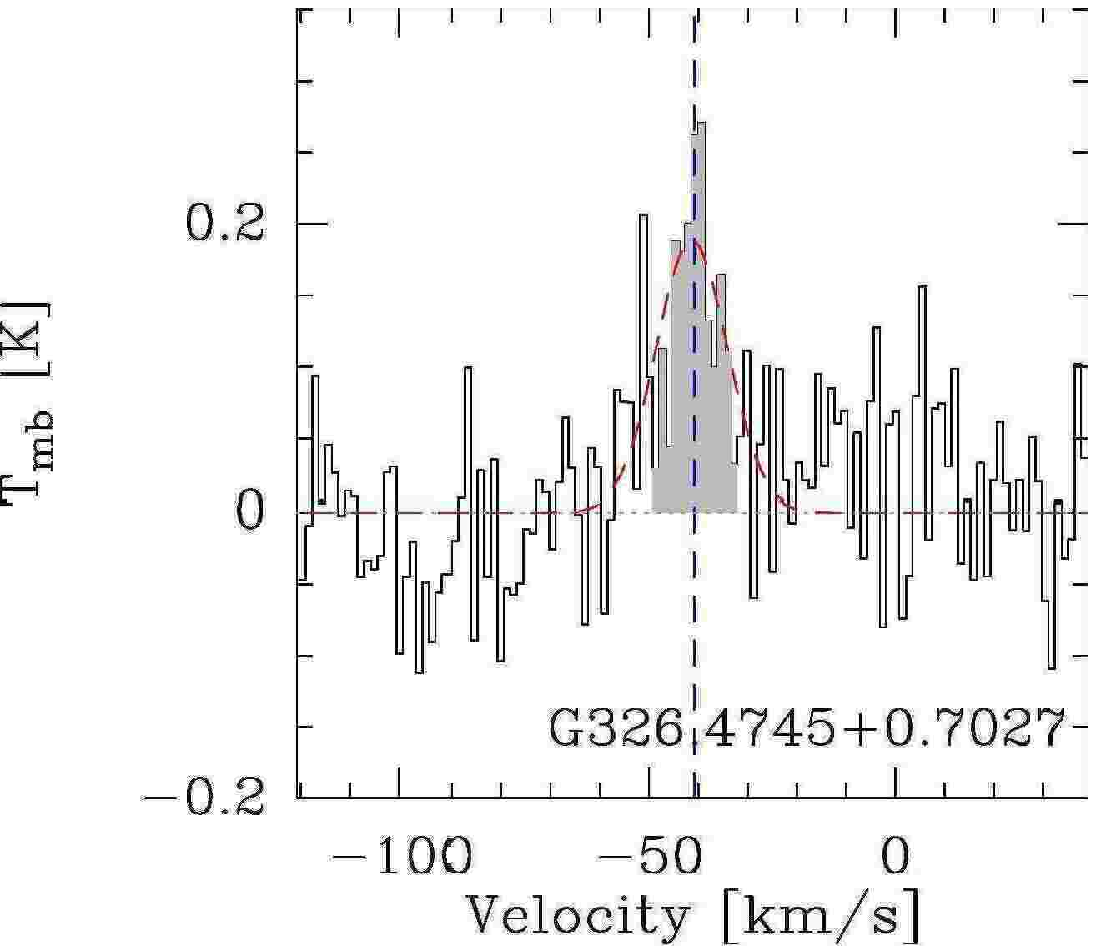}
  \includegraphics[width=5cm,clip]{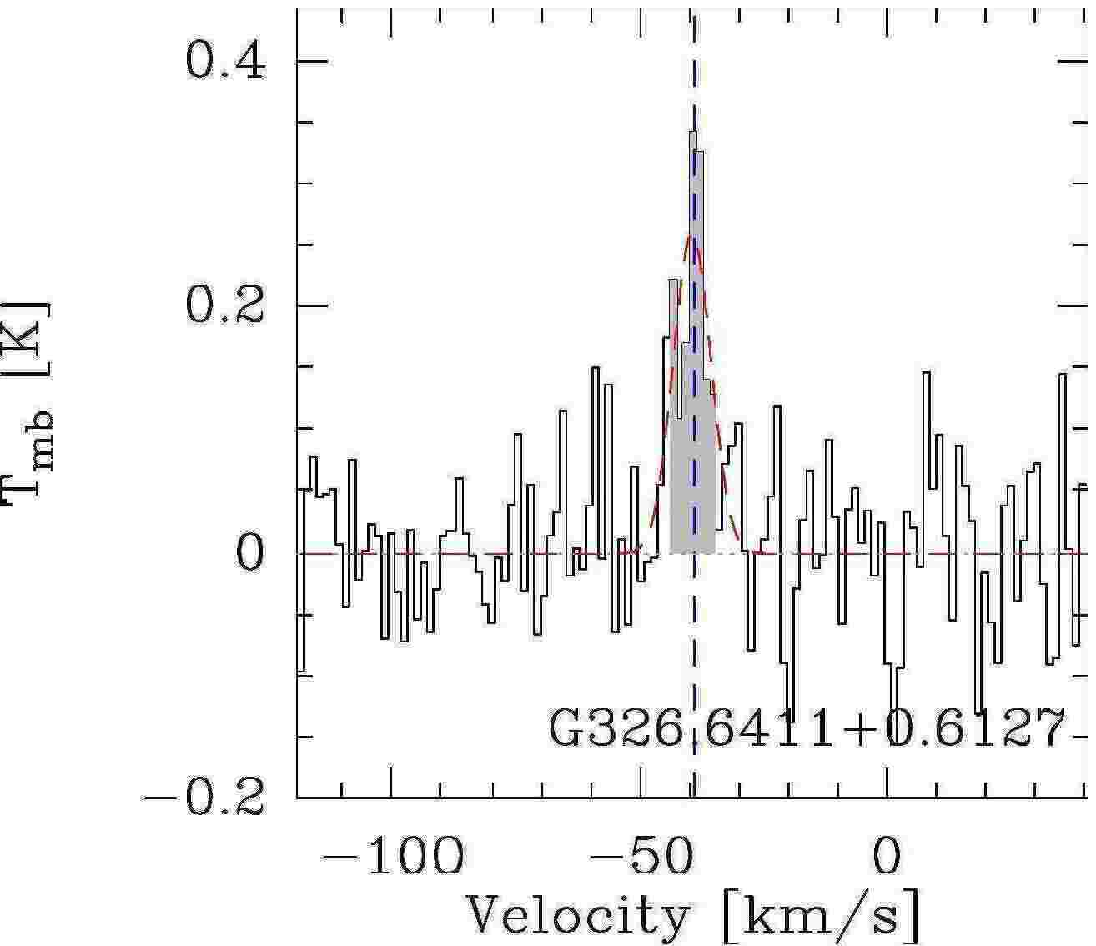}
  \includegraphics[width=5cm,clip]{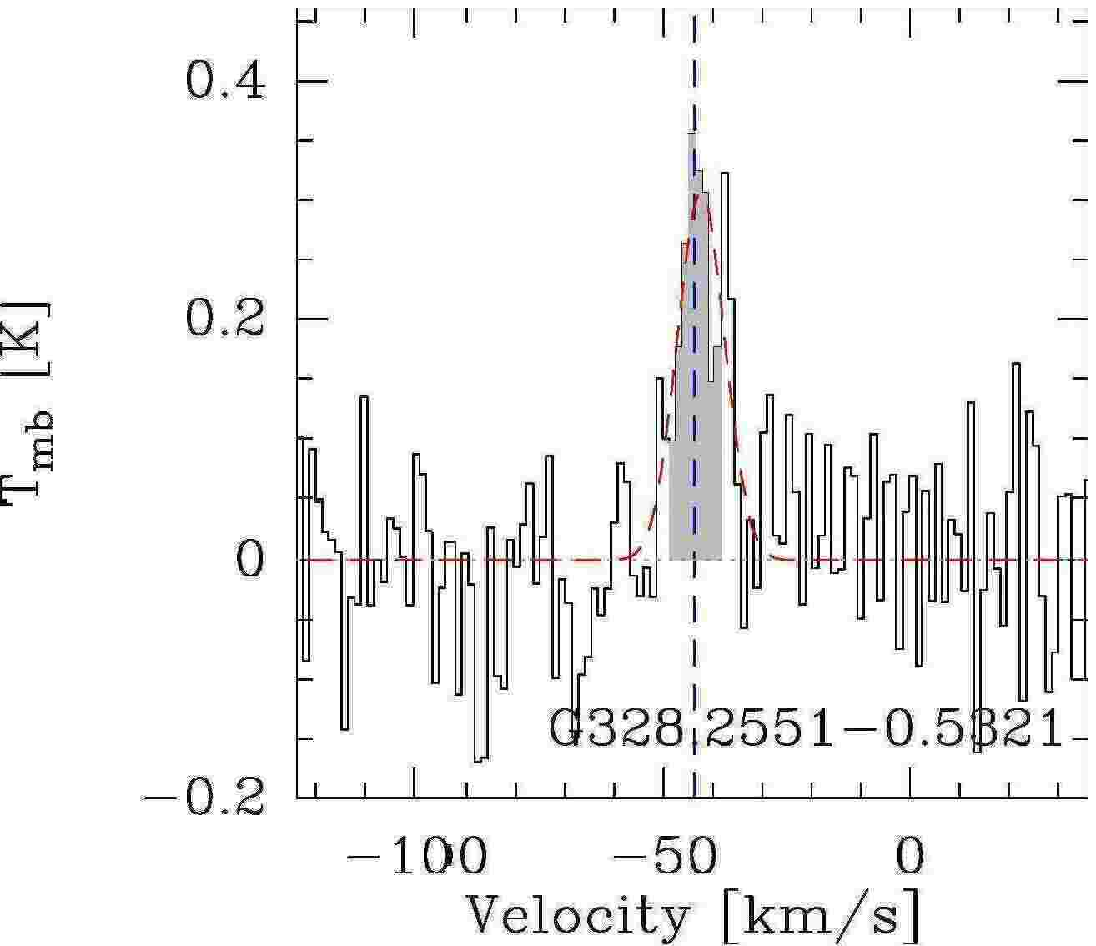}
  \includegraphics[width=5cm,clip]{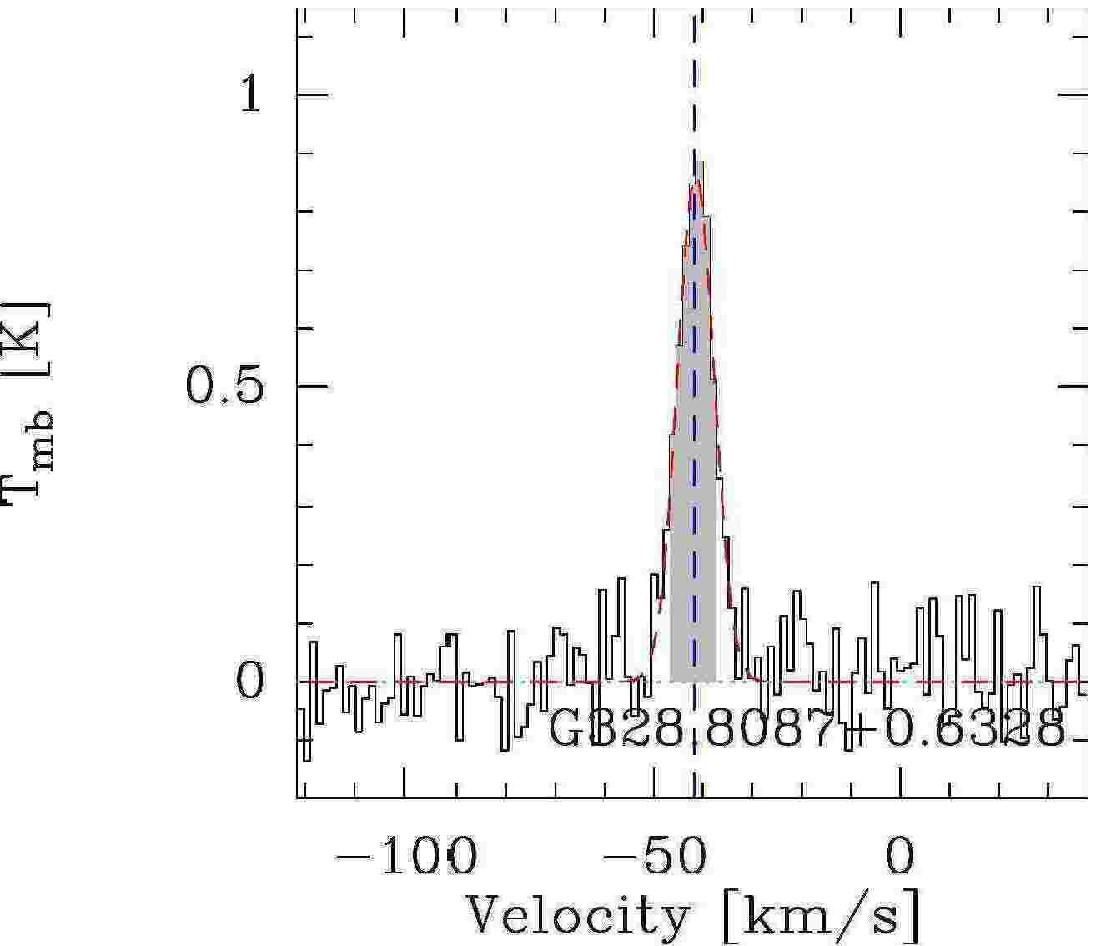}
  \includegraphics[width=5cm,clip]{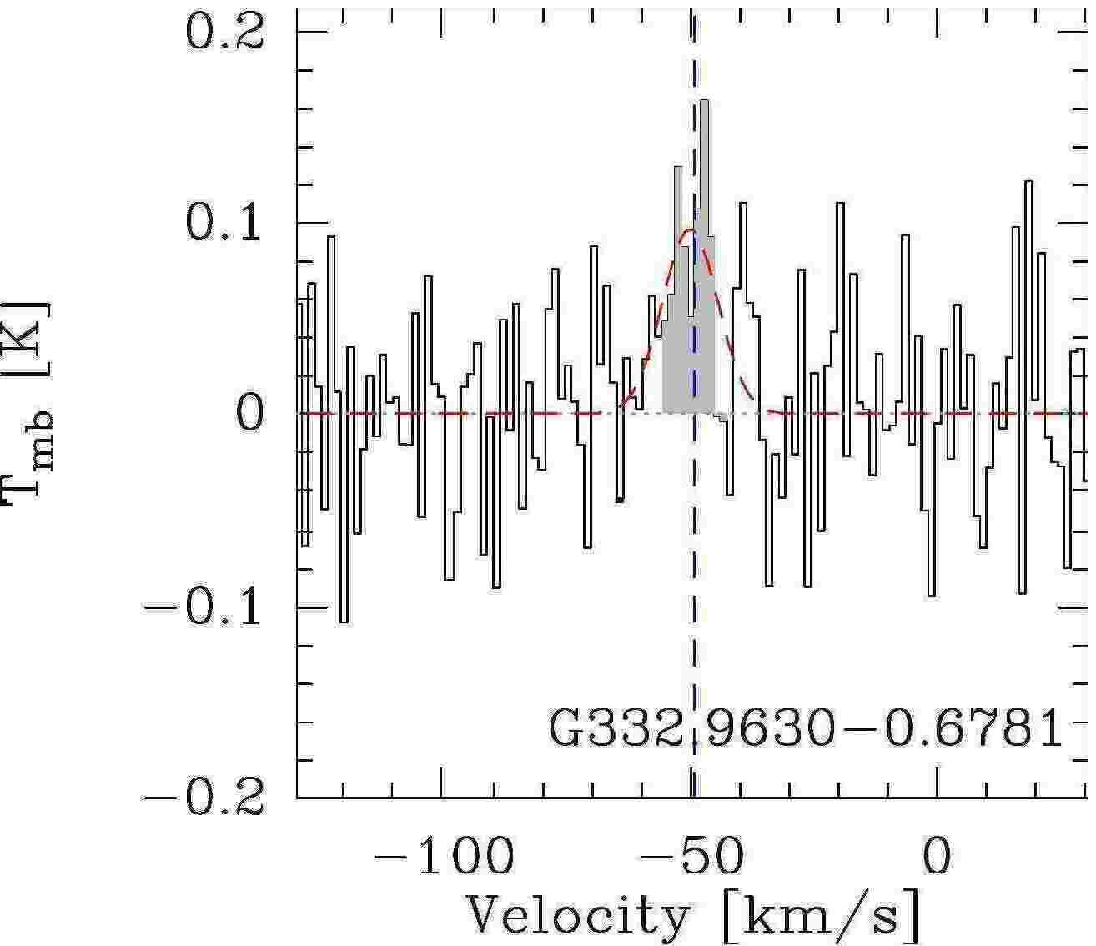}
  \includegraphics[width=5cm,clip]{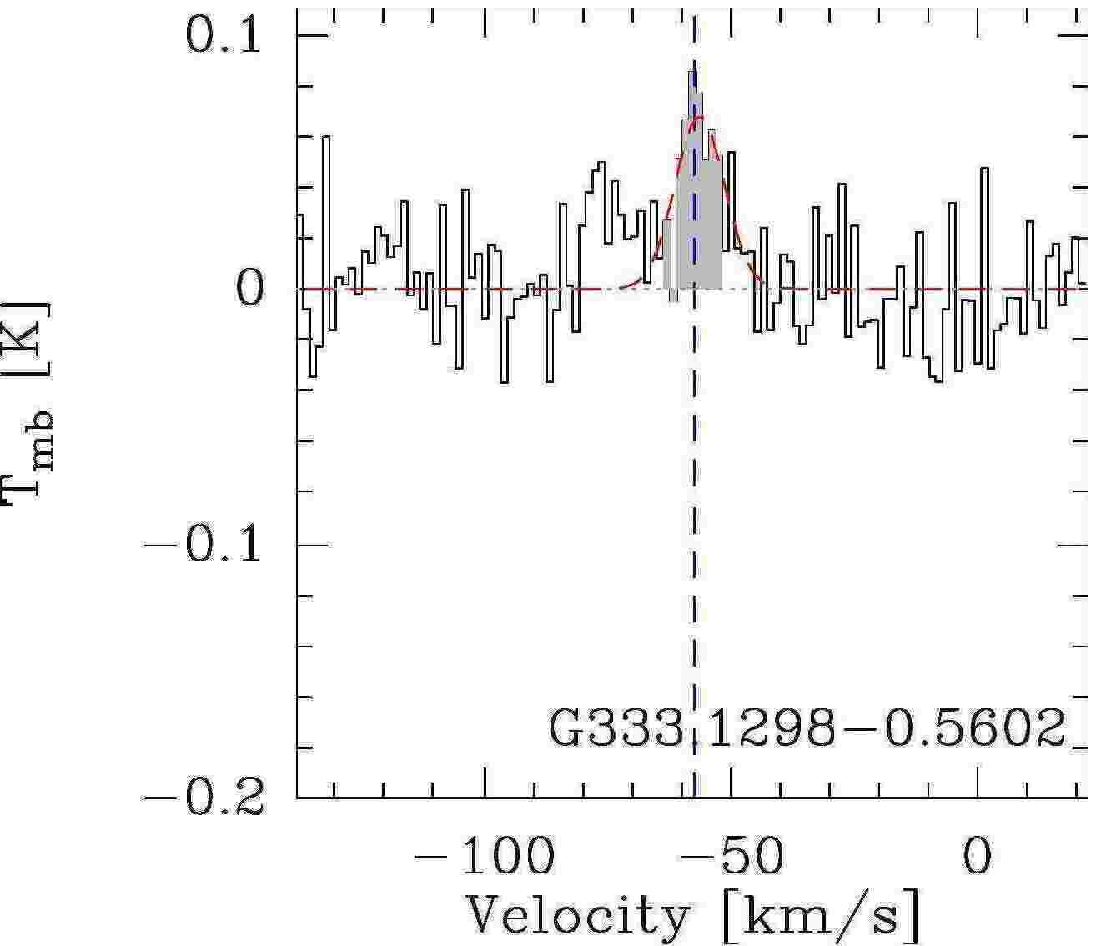}
  \includegraphics[width=5cm,clip]{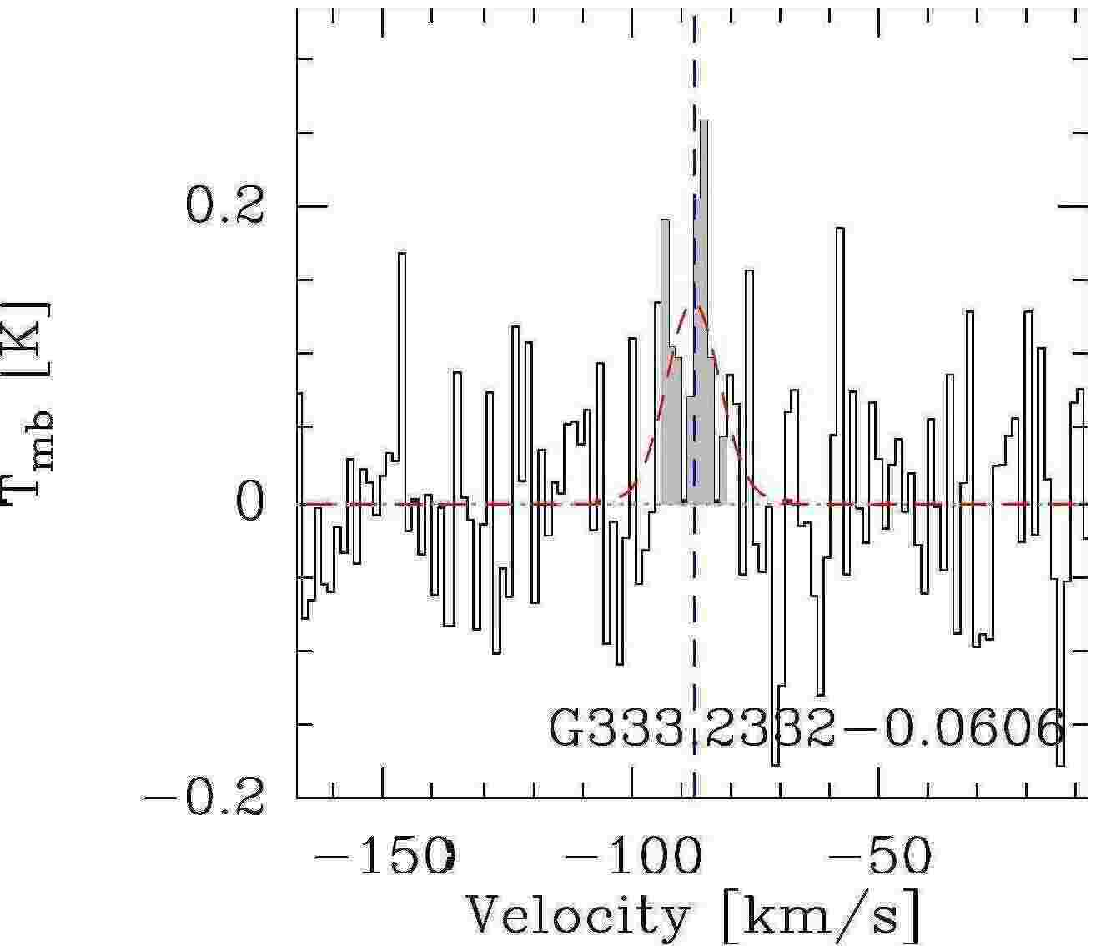}
  \includegraphics[width=5cm,clip]{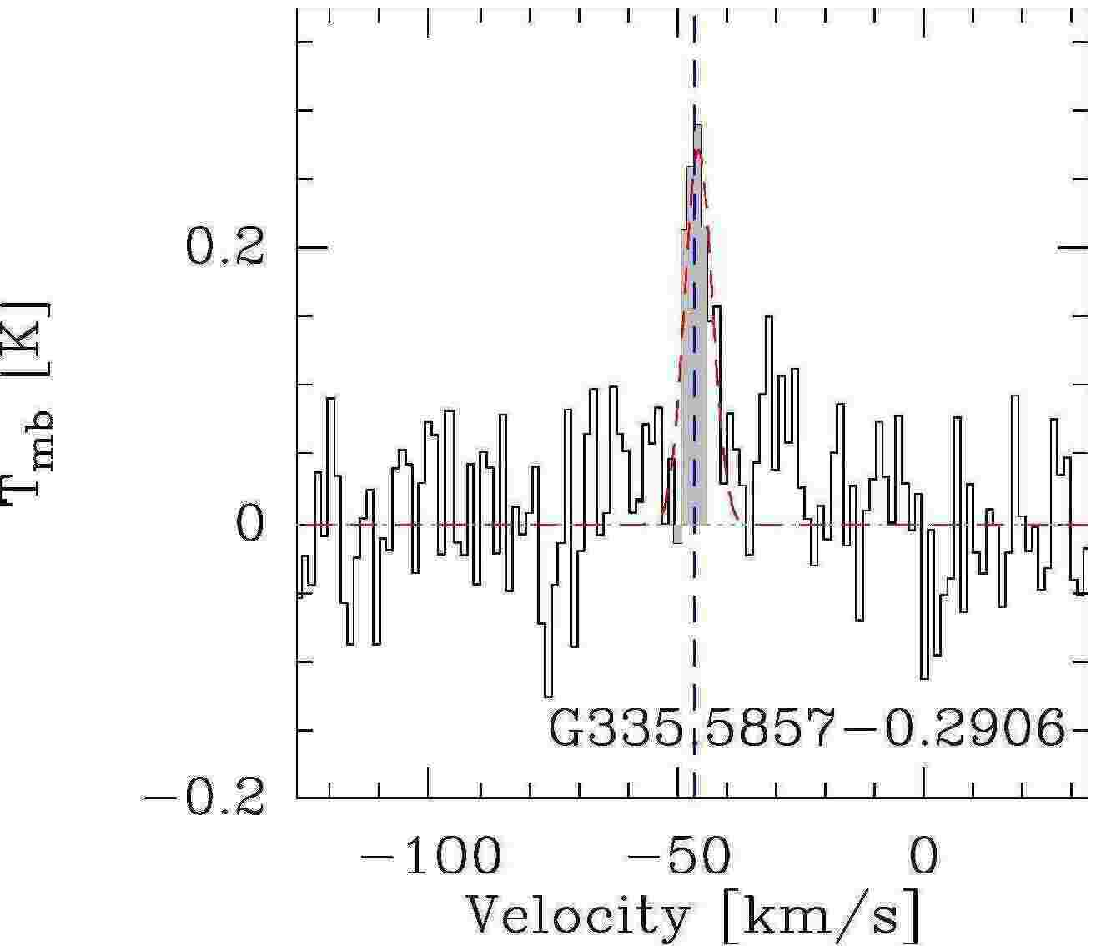}
  \includegraphics[width=5cm,clip]{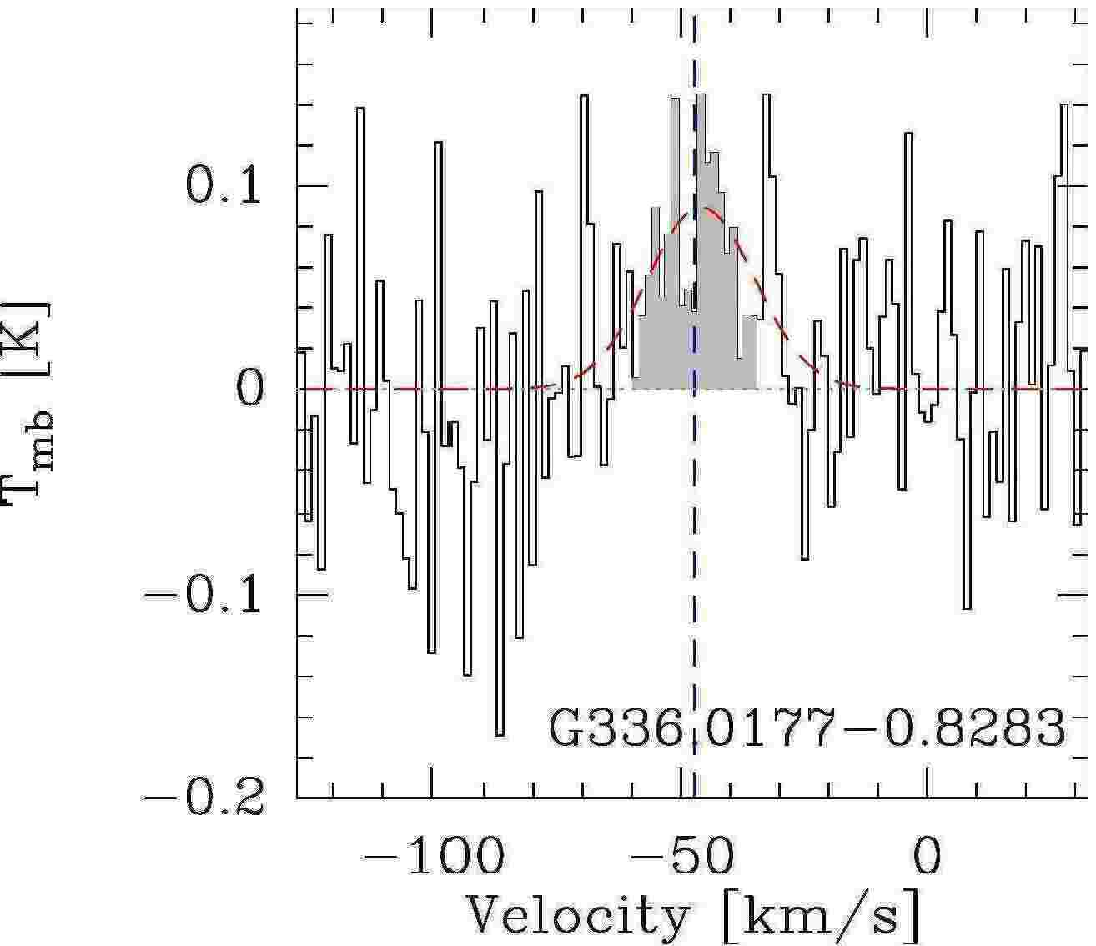}
  \includegraphics[width=5cm,clip]{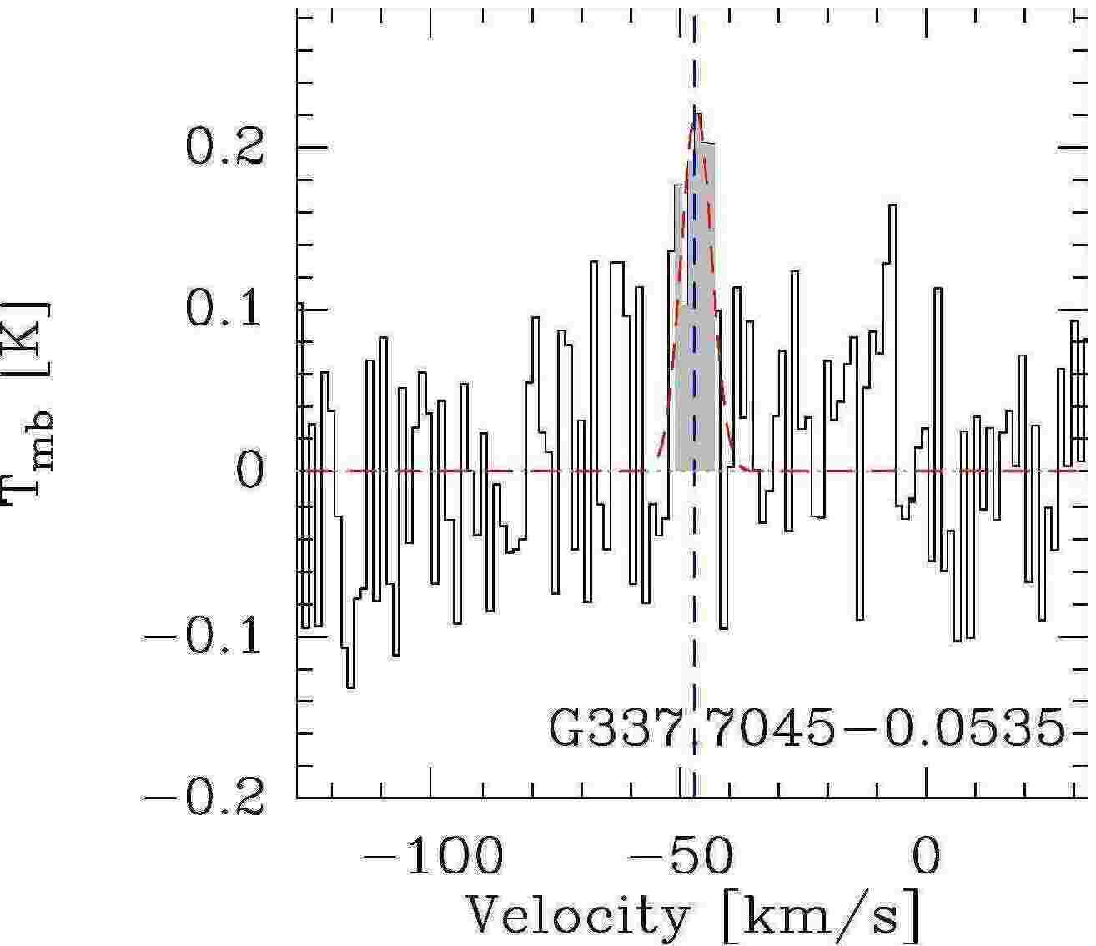}
   \caption{SiO (8--7) detections with the APEX/FLASH$^+$ receiver of a selection of infrared-quiet massive clumps. The red dashed line shows the fitted Gaussians, and the grey histogram shows the line area. The blue dashed line indicates the $v_{\rm lsr}$ of the sources.}
 \label{app:fig2}
 \end{figure*}
\begin{figure*}
\centering
\ContinuedFloat
  \includegraphics[width=5cm,clip]{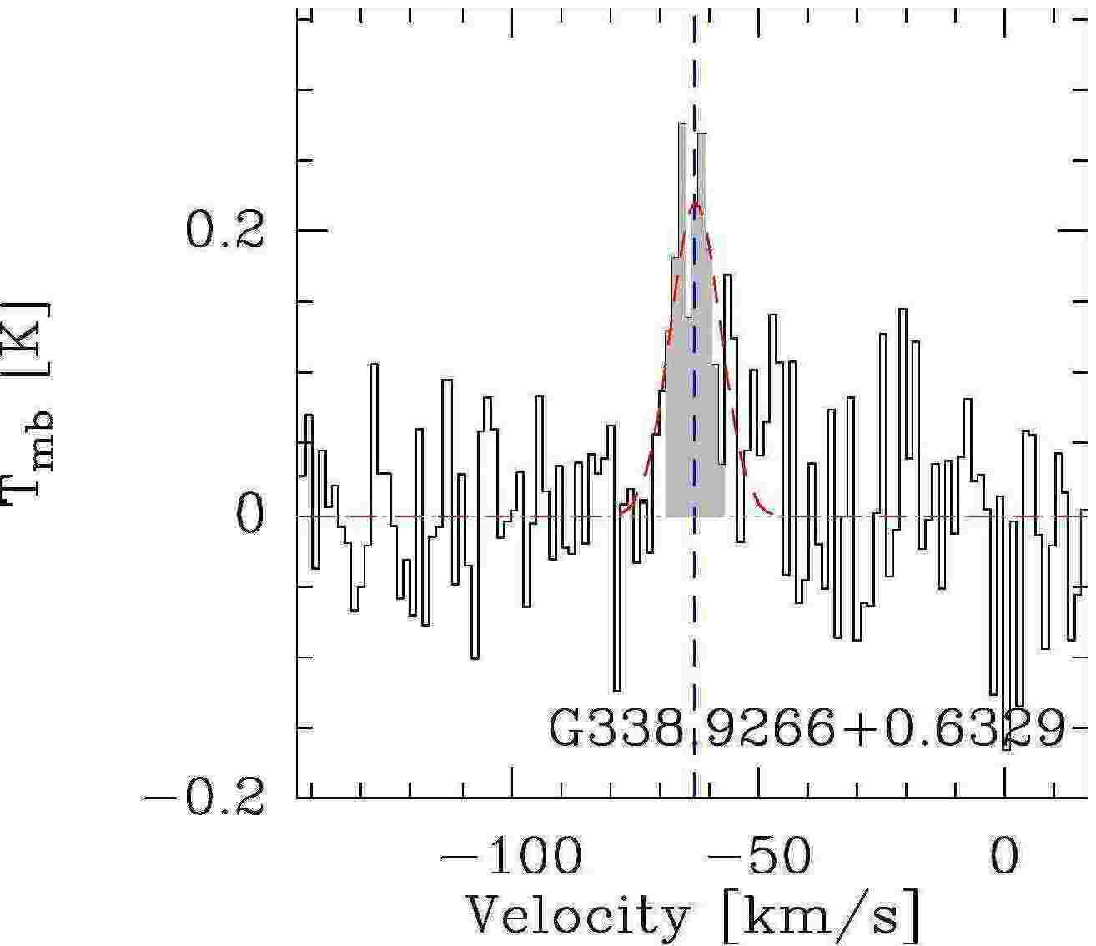}
  \includegraphics[width=5cm,clip]{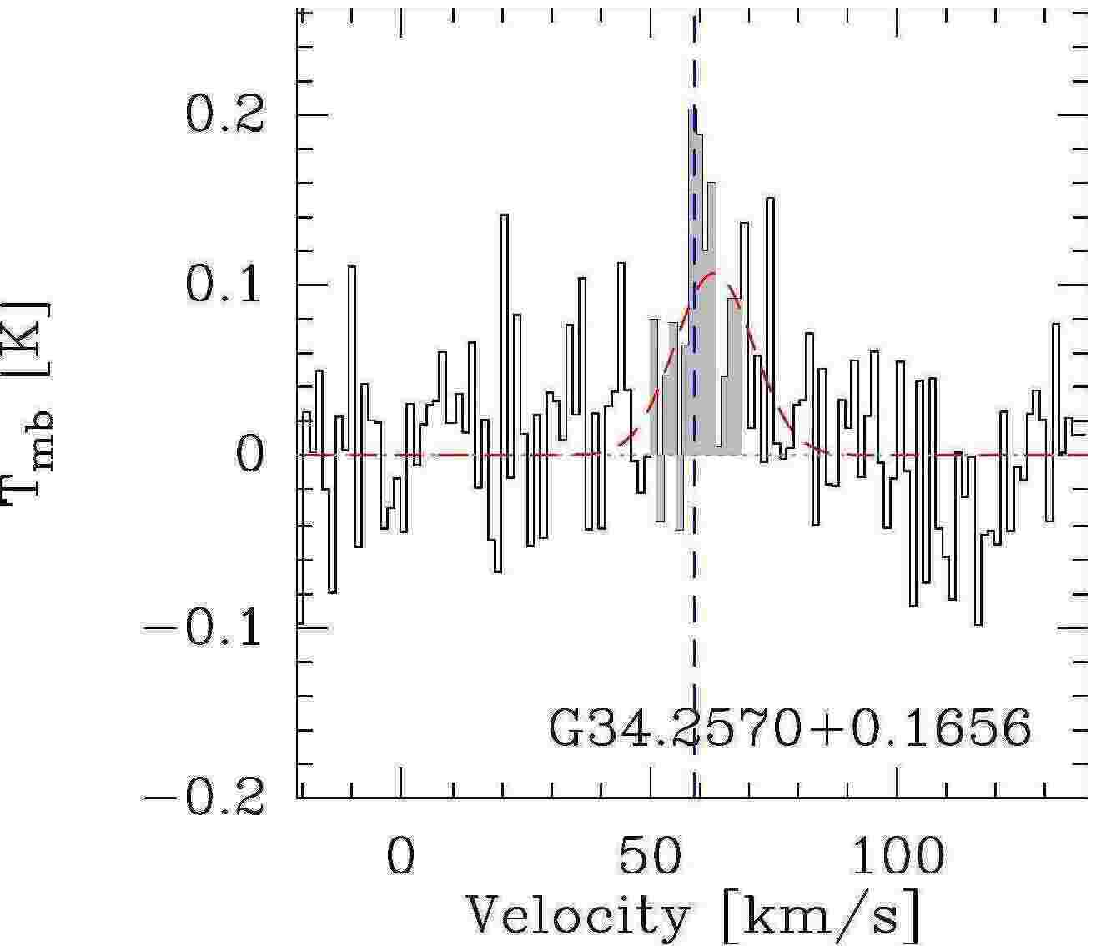}
  \includegraphics[width=5cm,clip]{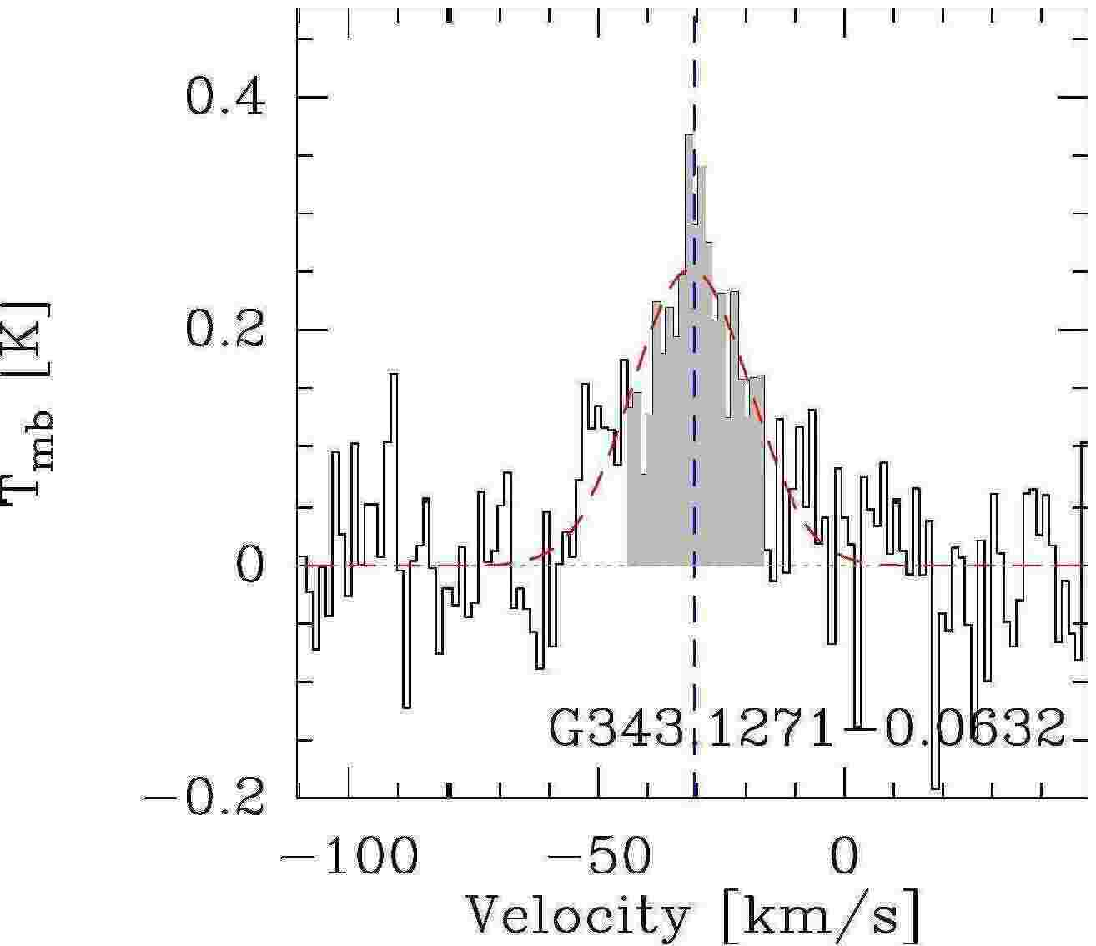}
  \includegraphics[width=5cm,clip]{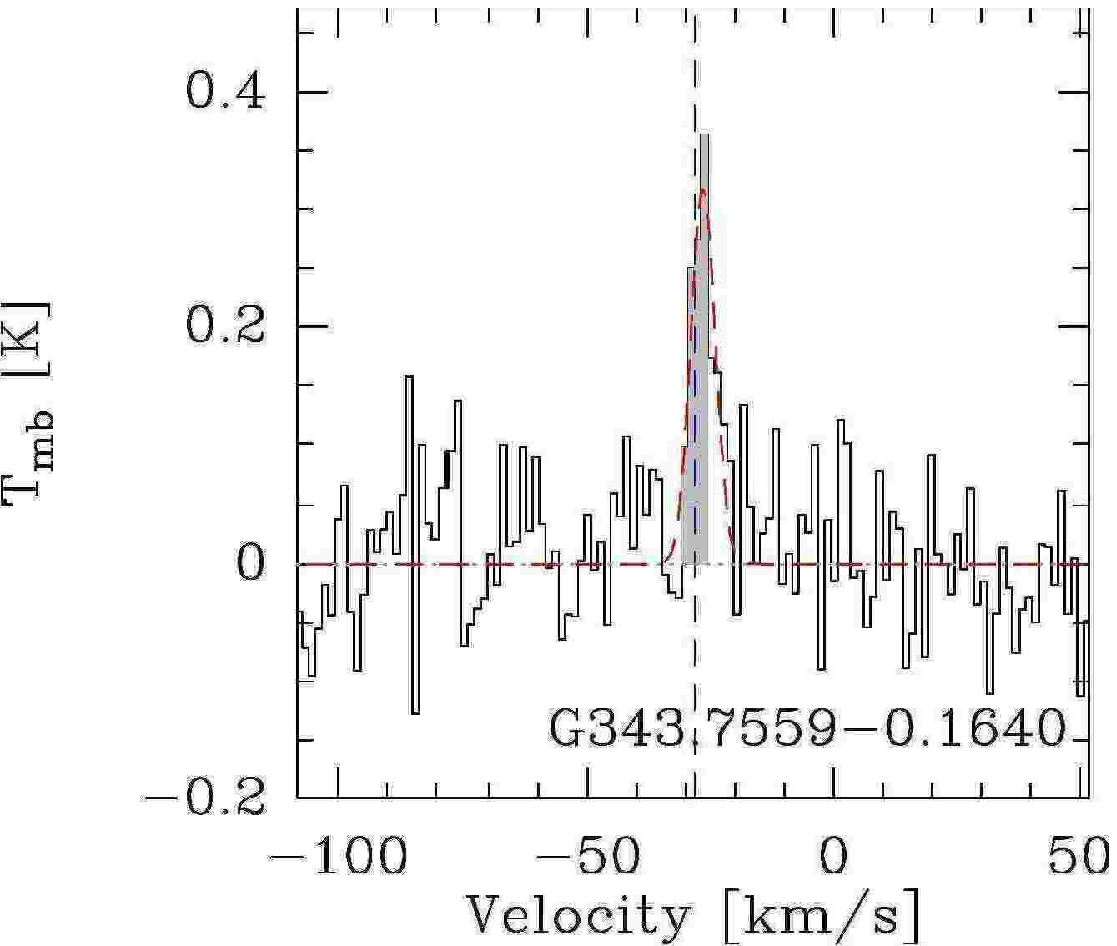}
  \includegraphics[width=5cm,clip]{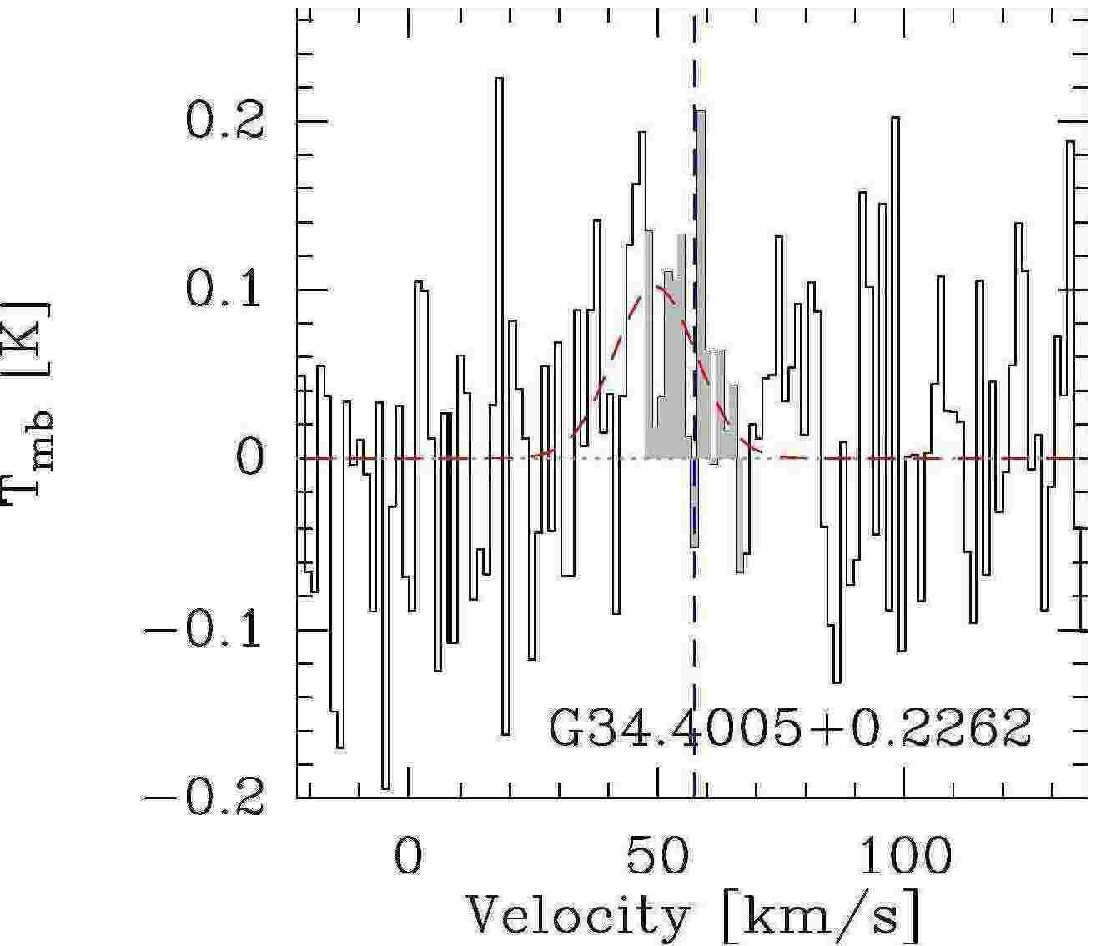}
  \includegraphics[width=5cm,clip]{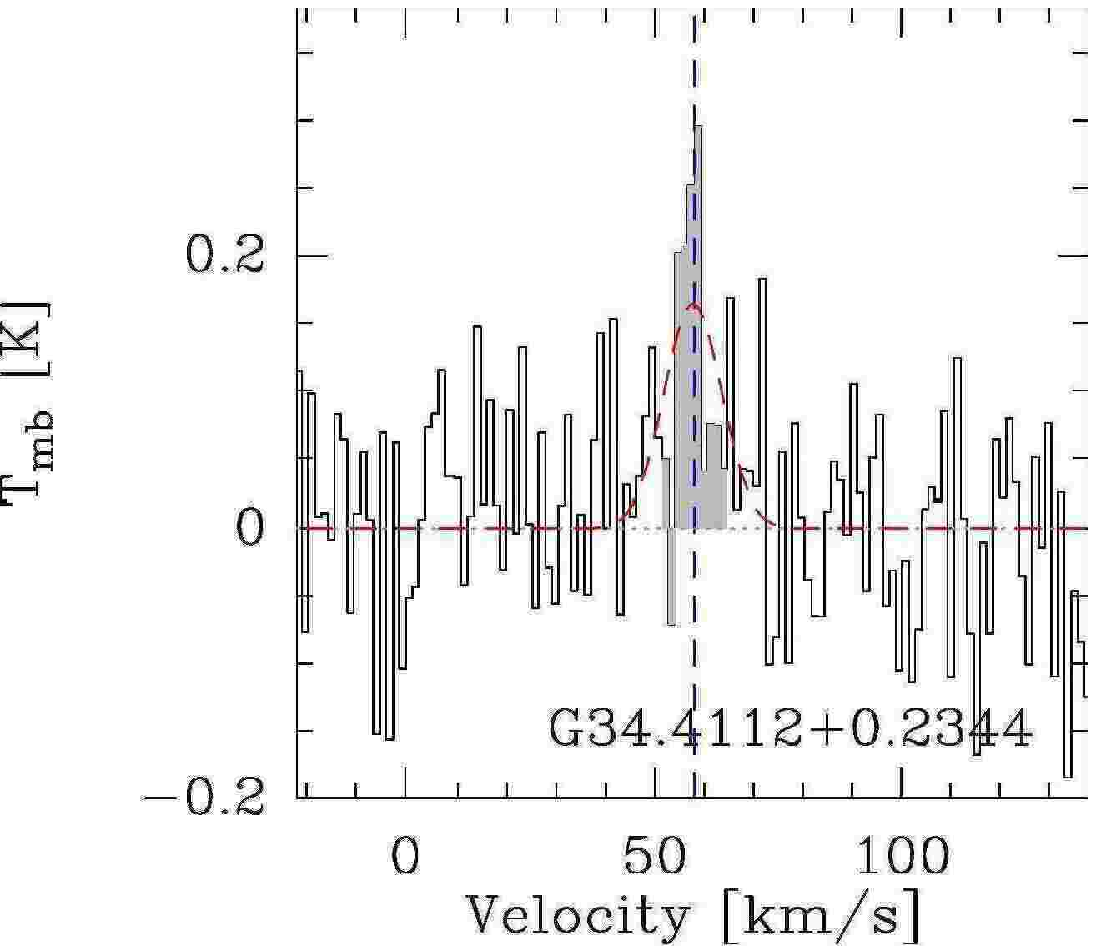}
  \includegraphics[width=5cm,clip]{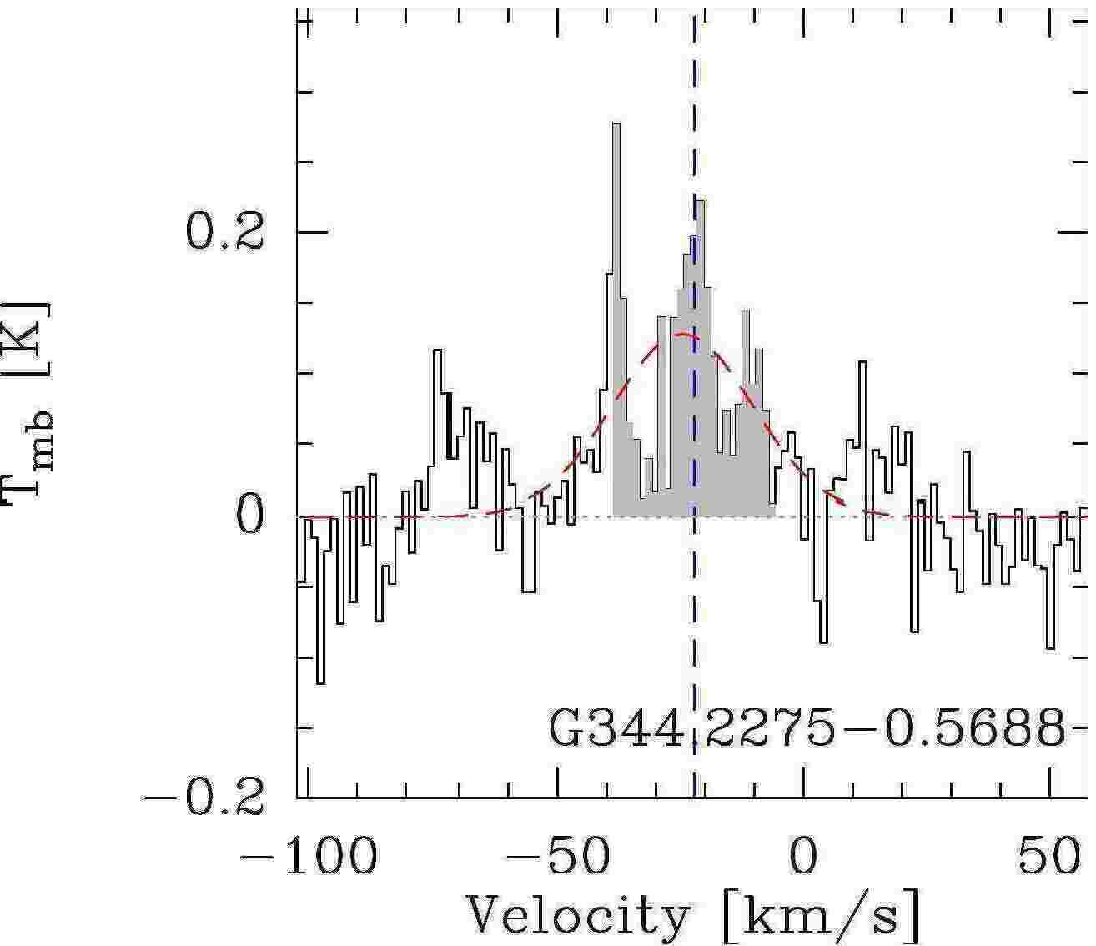}
  \includegraphics[width=5cm,clip]{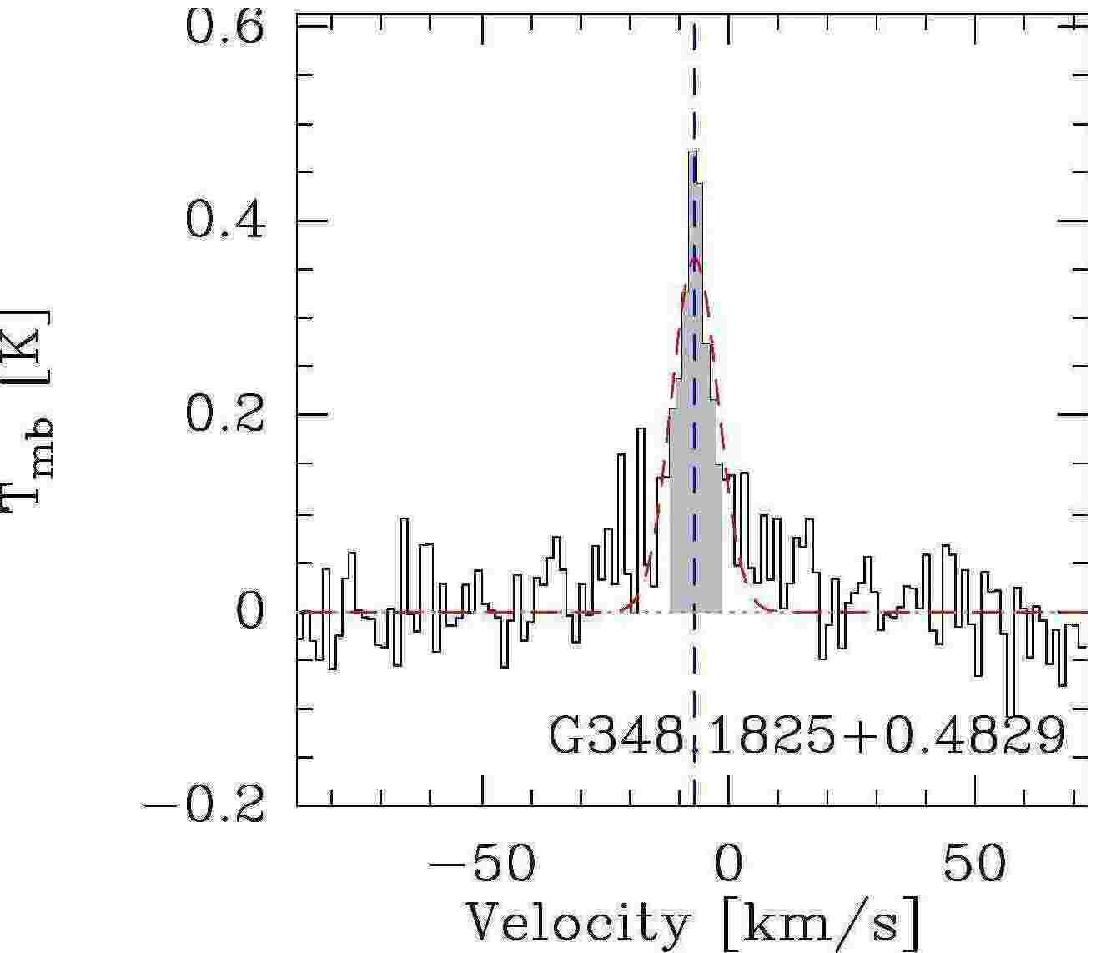}
  \includegraphics[width=5cm,clip]{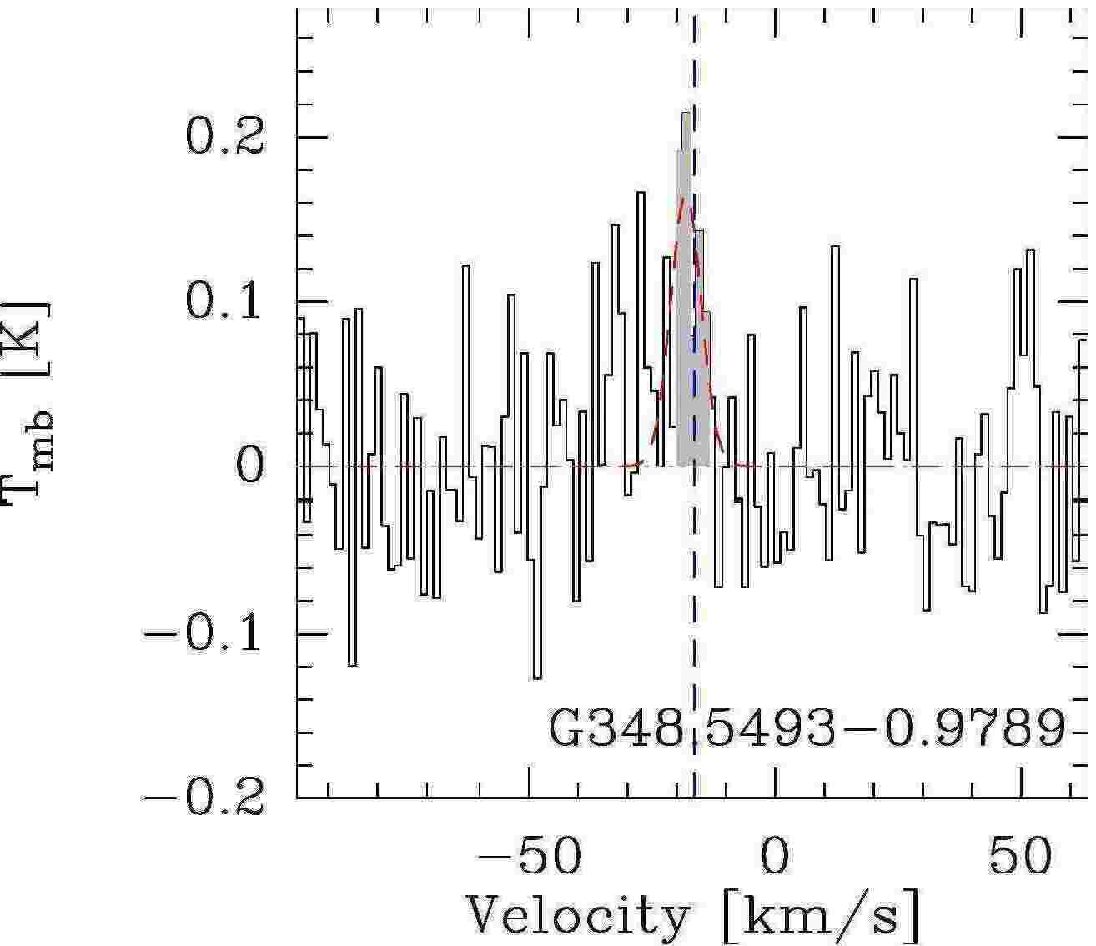}
  \includegraphics[width=5cm,clip]{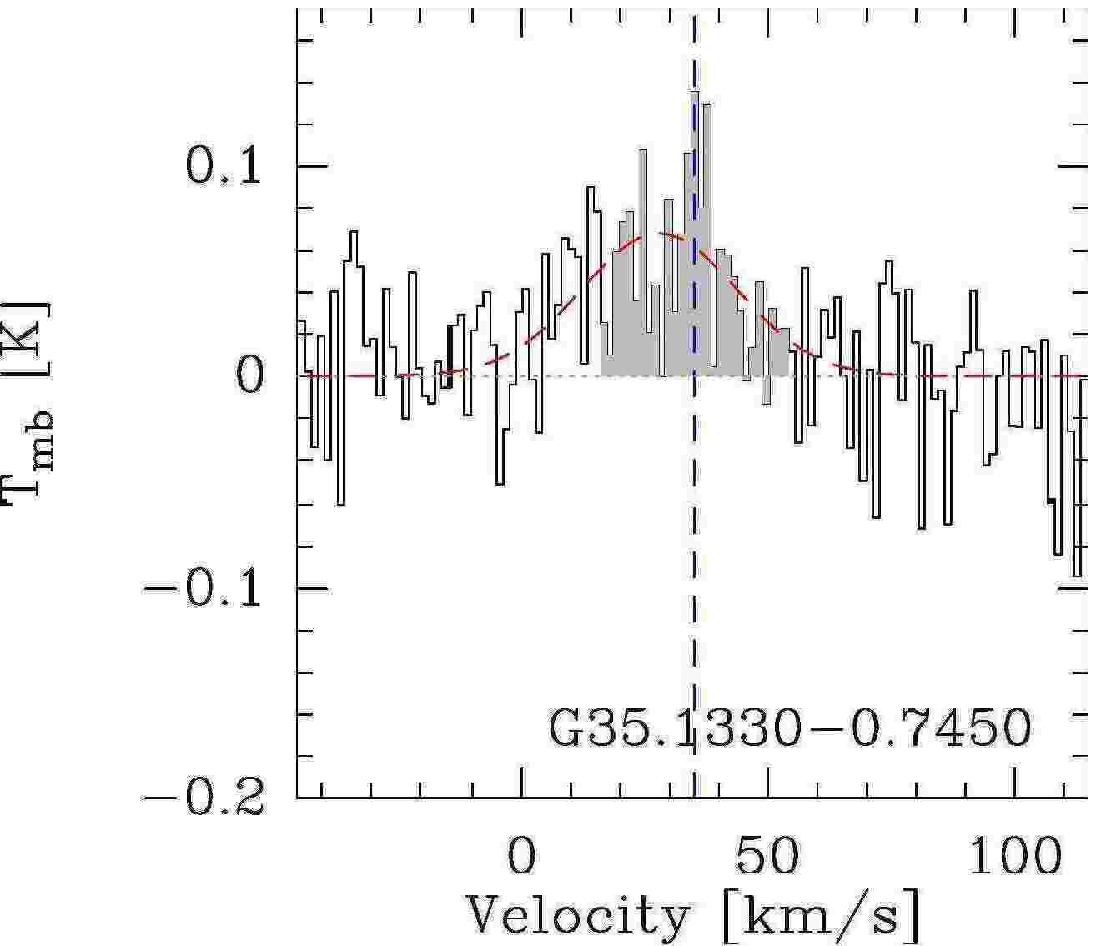}
  \includegraphics[width=5cm,clip]{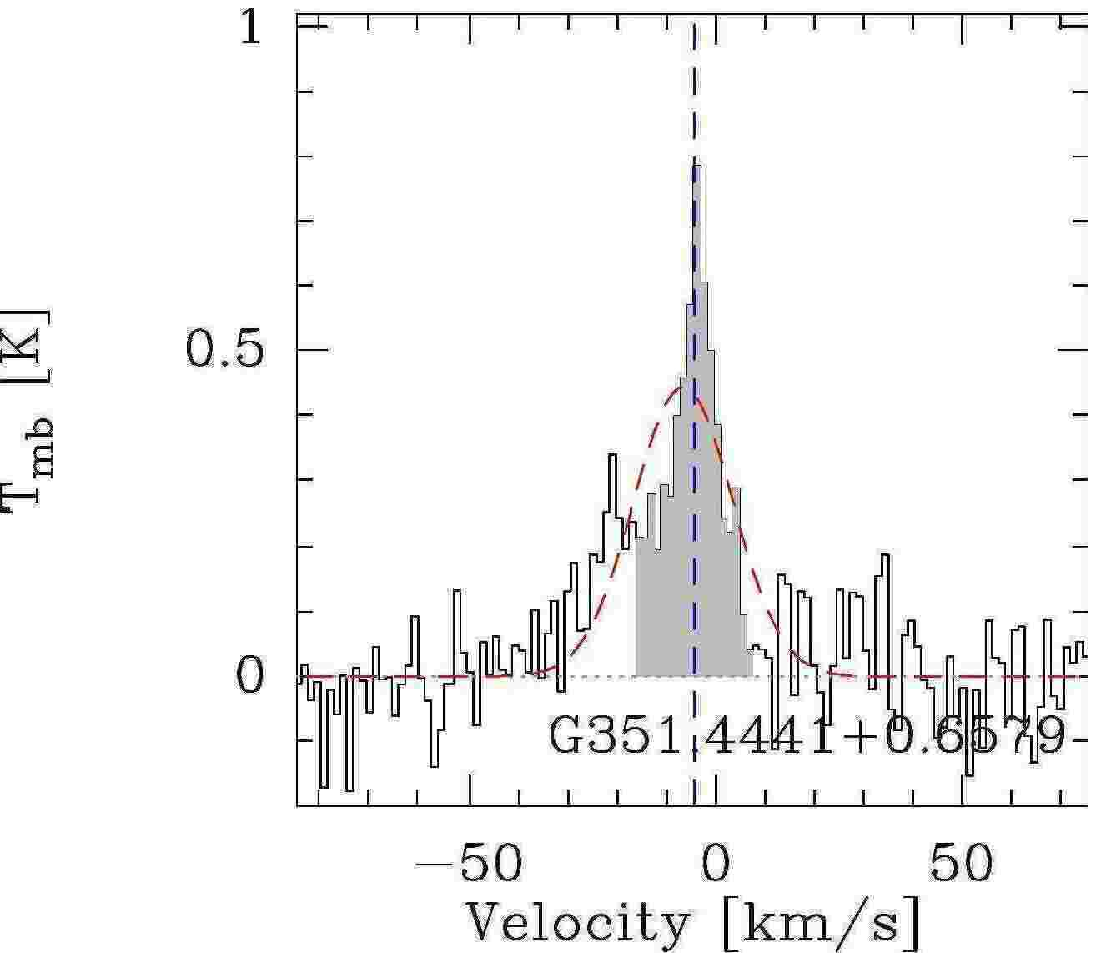}
  \includegraphics[width=5cm,clip]{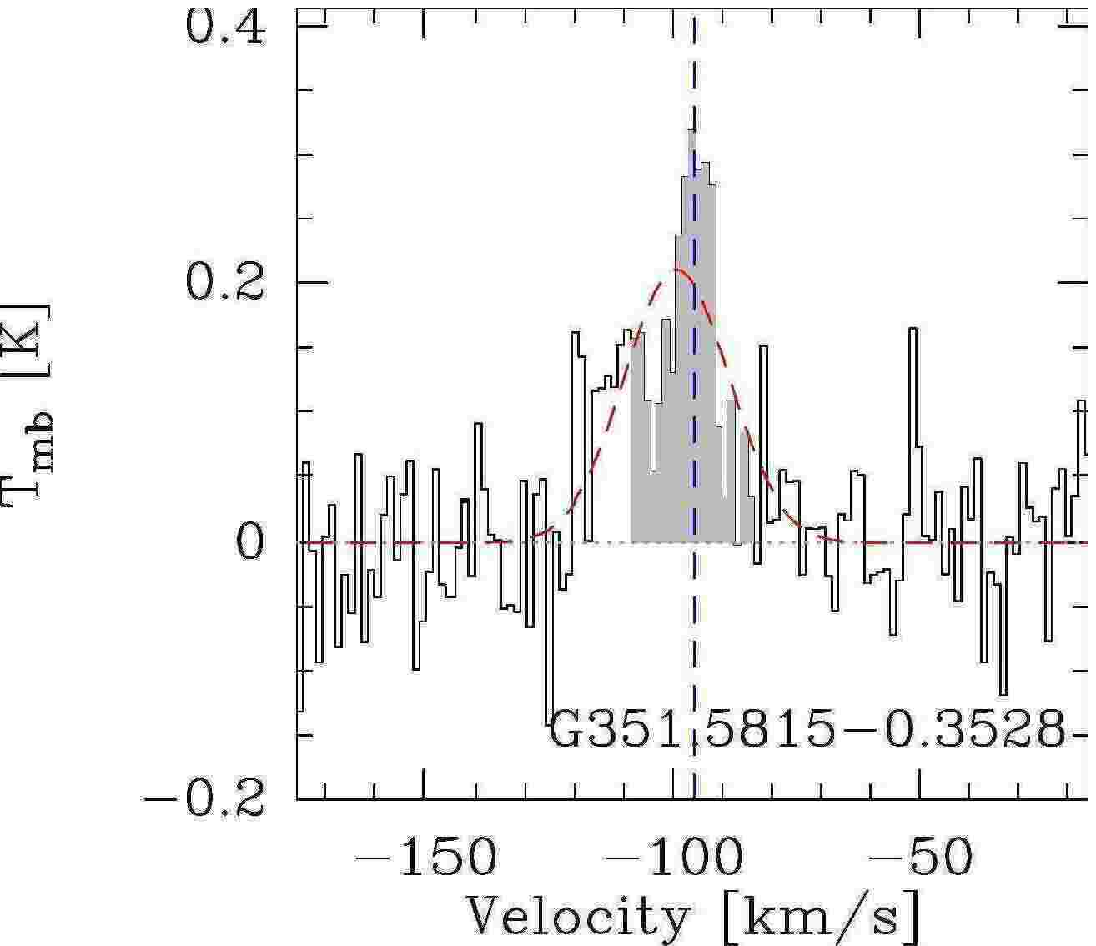}
  \includegraphics[width=5cm,clip]{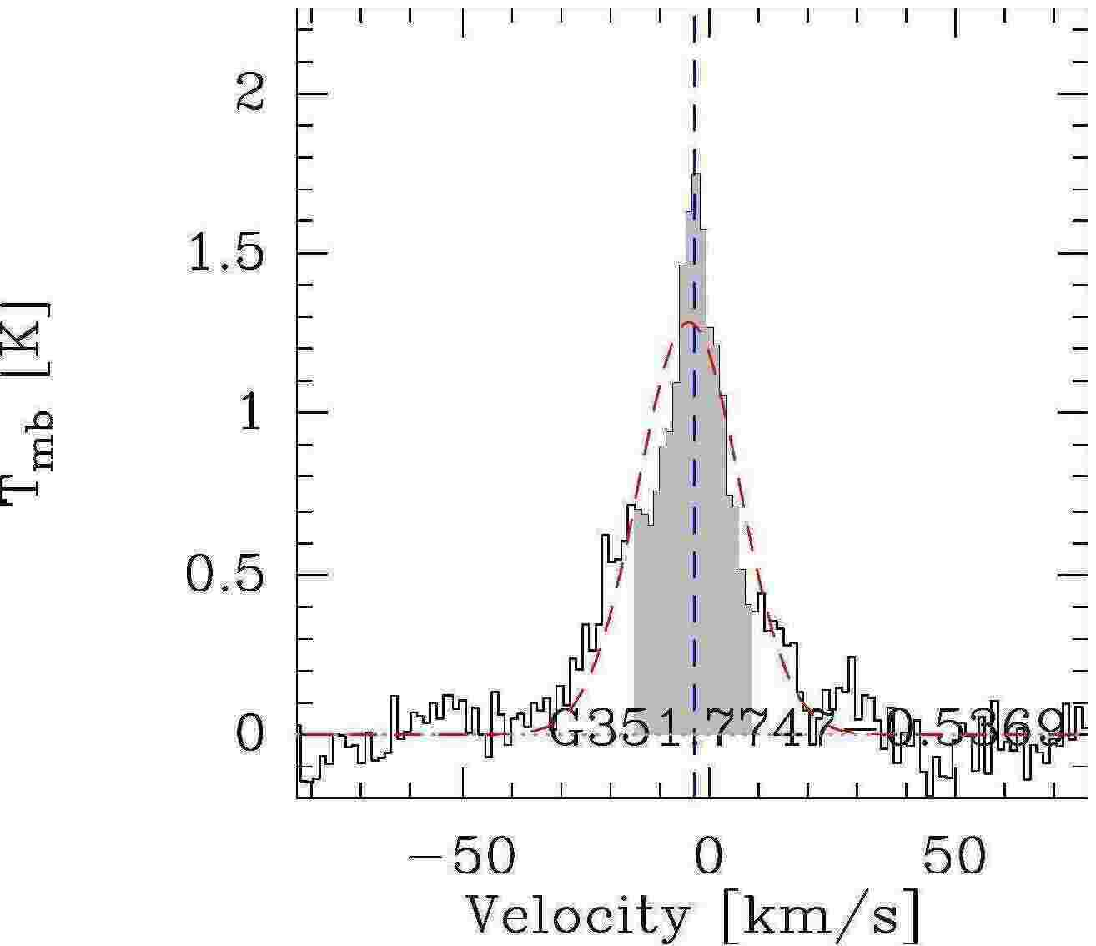}
  \includegraphics[width=5cm,clip]{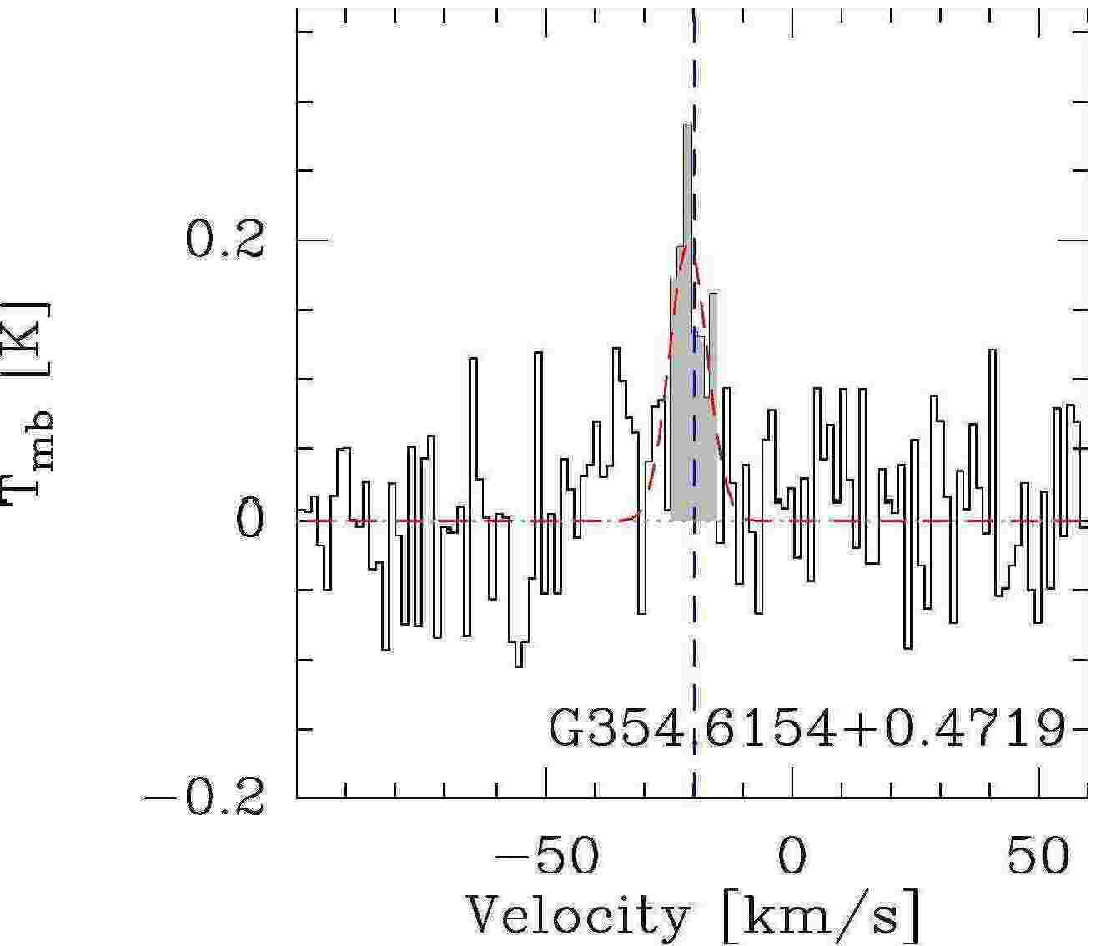}
  \includegraphics[width=5cm,clip]{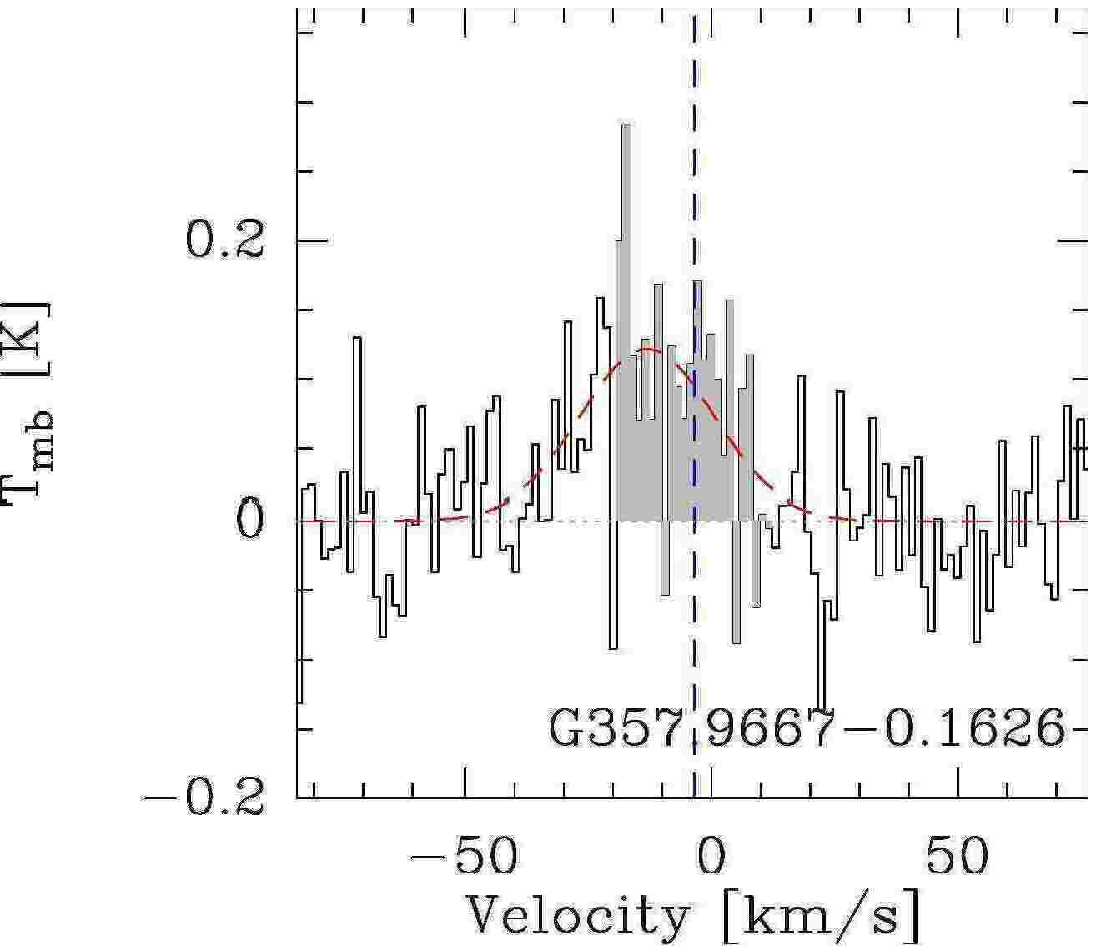}
 \caption{Continued.}
 \end{figure*}

\end{document}